\documentclass[11pt,twoside,openright]{report}

\usepackage{vmargin}
\usepackage[T1]{fontenc}
\usepackage[francais]{babel}
\usepackage{indentfirst}
\setpapersize{A4}
\usepackage{amsmath}
\usepackage{amssymb}
\usepackage{epsfig}
\usepackage{boldgreek}

\usepackage{fancyheadings}

\newcommand{\p}{\partial}
\newcommand{\bfdphi}{\bfdelta\bfphi}
\newcommand{\dphi}{\delta\phi}
\newcommand{\tvarphi}{\tilde\varphi}
\newcommand{\tmu}{\tilde\mu}
\newcommand{\tnu}{\tilde\nu}
\newcommand{\tpi}{\tilde\pi}
\newcommand{\tpsi}{\tilde\psi}
\newcommand{\tomega}{\tilde\omega}
\newcommand{\tOmega}{\tilde\Omega}
\newcommand{\tH}{\tilde H}
\newcommand{\tU}{\tilde U}
\newcommand{\tW}{\tilde W}
\newcommand{\tp}{\tilde p}
\newcommand{\tq}{\tilde q}
\newcommand{\ta}{\tilde a}
\newcommand{\tg}{\tilde g}
\newcommand{\dd}{\mbox{d}}

\newcommand{\chiT}{\chi_{_T}}
\newcommand{\phiT}{\phi_{_T}}
\newcommand{\bfphiT}{\bfphi_{_T}}
\newcommand{\rangleT}{\rangle_{_T}}

\newcommand{\eq}{$$}
\newcommand{\beq}{\begin{equation}}
\newcommand{\eeq}{\end{equation}\noindent}
\newcommand{\bear}{\begin{eqnarray*}}
\newcommand{\eear}{\end{eqnarray*}\noindent}
\newcommand{\bearn}{\begin{eqnarray}}
\newcommand{\eearn}{\end{eqnarray}\noindent}

\begin{document}
\pagestyle{plain}
\thispagestyle{empty}

ORSAY

Numéro d'ordre : 

\hspace{10.cm} LPT-ORSAY 01/27

\vspace{2.cm}

\begin{center}
{\Large UNIVERSITÉ PARIS-SUD}

\vspace{0.5cm}
{\Large LABORATOIRE DE PHYSIQUE THÉORIQUE D'ORSAY}

\vspace{2.5cm}
{\bf {\Large THÈSE DE DOCTORAT DE L'UNIVERSITÉ PARIS XI}}
\vspace{0.5cm}

présentée par

\vspace{0.5cm}
{\large \bf{Julien SERREAU}}

\vspace{1cm}
\large{Sujet :}
\vspace{0.5cm}

\Large{\bf{PHÉNOMÈNES HORS D'ÉQUILIBRE \\DANS LES COLLISIONS NUCLÉAIRES \\
           À HAUTE ÉNERGIE}}
\vspace{0.5cm}

\end{center}

\hspace{2cm}
{\large \bf{- Formation de condensats chiraux désorientés}}

\hspace{2cm}
{\large \bf{- Équilibration thermique des gluons produits}}

\vfill
\normalsize {{\noindent soutenue le $1^{\mbox{er}}$ mars 2001 }}{{ devant la commission
d'examen}\\\\
\begin{tabular}{lllc}
MM. &   \large\bf{P.}     &   \large\bf{Aurenche}    &                           \\
    &   \large\bf{R.}     &   \large\bf{Baier}       &    (Rapporteur)           \\
    &   \large\bf{J.-P.}  &   \large\bf{Blaizot}     &    (Rapporteur)           \\
    &   \large\bf{U.}     &   \large\bf{Ellwanger}   &    (Président)            \\  
    &   \large\bf{A.}     &   \large\bf{Krzywicki}   &    (Directeur de thèse)   \\
    &   \large\bf{A. H.}  &   \large\bf{Mueller}     &                           
\end{tabular}
\normalsize

\chapter*{}

\begin{flushright}

A mes parents,

\end{flushright}

\chapter*{Remerciements}
\pagenumbering{roman}
\addcontentsline{toc}{chapter}{Remerciements}

Je tiens à remercier tous les membres du Laboratoire de Physique Théorique
d'Orsay de m'avoir accueilli durant la préparation ainsi que la rédaction
de cette thèse. D'une manière ou d'une autre, volontairement ou non,
tous ont contribué à rendre ces quelques années agréables et enrichissantes. 
Un grand merci également à toute l'équipe administrative et technique du 
laboratoire, en particulier à Mireille Calvet, Nicole Cherbonnier, Mireille 
Geurts, Odile Heckenhauer, Jocelyne Puech, et bien sûr à Mme Rocher (Arlette 
de son prénom). Je suis de plus très reconnaissant envers Philippe Boucaud, 
Yves D'Aignaux et Jean-Pierre Leroy pour leur aide en ce qui concerne 
l'informatique et leur patience en ce qui concerne mes compétences dans 
ce domaine. 

C'est ici l'occasion pour moi de saluer Luc Bourhis, Jérôme Charles, 
Cyril Hugonie et Stéphane Lavignac, avec qui j'ai partagé le statut de 
thésard du LPT durant la première année, ainsi que Jean Noël Aqua, Olivier
Deloubrière, Gregorio Herdoiza, Sofiane Tafat et Nicolas Wschebor, avec qui
j'ai partagé le même sort depuis. Je tiens également à saluer Elias Khan,
Santiago Pita, Jérôme Margueron, Nicolas Sator, les (ex-)thésards de l'Institut 
de Physique Nucléaire d'Orsay ainsi que Samir Ferrag à l'école polytechnique.

Je tiens à dire un grand merci à André Krzywicki qui a guidé mes premiers pas
dans le métier de chercheur avec beaucoup d'intelligence, j'espère être à la 
hauteur de sa confiance. Un grand merci également à Dominique Schiff qui a pris
le relai avec beaucoup de patience malgré sa lourde tâche de direction du 
laboratoire. Je tiens à remercier Dominique Vautherin avec qui j'ai eu de
très enrichissantes discussions et qui a contribué dans une large mesure à 
ma formation.

Je remercie Rolf Baier et Jean-Paul Blaizot d'avoir accepté
les rôles de rapporteurs (ainsi que pour leur efficacité dans ce rôle). 
Je remercie également Patrick Aurenche, Ulrich ellwanger et Al Mueller 
d'avoir accepté de faire partie du jury.

Enfin, je tiens à remercier les organisateurs du workshop {\it Nuclear Matter
in Different Phases and Transitions} aux Houches en Avril 1998 (il s'agit de 
J.-P. Blaizot, X. Campi et M. Ploszajczak) où j'ai rencontré Cristina Volpe, 
que je remercie infiniment.....pour m'avoir appris l'italien.

\chapter*{Résumé}
\addcontentsline{toc}{chapter}{Résumé}

Le cadre général dans lequel se place cette thèse est l'étude des propriétés 
de la matière produite lors de collisions nucléaires à haute énergie,
où l'on s'attend à créer artificiellement les conditions requises pour
la formation d'un plasma de quarks et de gluons. En particulier, nous 
étudions certains aspects reliés à l'évolution intrinsèquement hors 
d'équilibre de ce type de système. 

La première partie est consacrée à l'étude de la formation possible
de condensats chiraux désorientés (DCC) lors du passage rapide de la transition 
de phase chirale. Nous calculons d'abord la toute première estimation de 
la probabilité de formation d'un champ de pion classique potentiellement observable, 
lors de l'expansion sphérique rapide d'une ``bulle'' de matière chirale chaude. Ce 
calcul nécessite la construction d'une méthode d'échantillonnage des conditions
initiales pour le champ chiral. Ensuite, en pratiquant une analyse détaillée de 
la structure d'isospin du champ classique, nous montrons que le modèle le 
plus simple utilisé jusqu'à présent ne permet pas d'expliquer le caractère 
collectif de la configuration DCC, ce qui contredit une idée très largement
admise.

Dans la seconde partie, nous étudions la thermalisation des gluons produits
dans les tout premiers instants de la collision. Nous modélisons l'effet
des collisions élastiques par une approximation de temps
de relaxation auto-cohérente, et comparons différentes conditions initiales 
proposées dans la littérature. Nous arguons que les critères utilisés dans
les travaux antécédents pour caractériser l'équilibration ne sont pas 
satisfaisant et proposons plutôt de mesurer le degré d'anisotropie de 
différentes observables. Nos conclusions contredisent celles obtenues
précédemment, nous montrons en particulier que les collisions
élastiques sont insuffisantes pour thermaliser le système aux 
énergies de RHIC.

\chapter*{Abstract}
\addcontentsline{toc}{chapter}{Abstract}

The general framework of the present thesis is the study of the properties of 
the matter produced in high energy heavy ion collisions, where one expects
to reach the conditions under which nuclear matter turns into a
quark-gluon plasma. In particular, we study some aspects related to the 
intrinsic non-equilibrium evolution of such systems.

The first part is dedicated to the study of disoriented chiral 
condensate (DCC) formation during the out of equilibrium chiral phase transition.
We compute the first estimation of the probability of forming a classical
pion field after the rapid expansion of a spherical droplet of chiral matter.
This requires the construction of a method for sampling the initial configurations
of the chiral field. Then, by performing a detailed analysis of the isospin 
structure of the classical field, we show that the simplest model, used up to now,
cannot explain the collective nature of the DCC. This contradicts a widely admitted
idea.

In the second part, we study the thermalization of initially produced gluons. 
We consider only elastic scatterings and use a self-consistent relaxation time 
approximation. We compare different initial conditions proposed in the literature.
We argue that the criteria used in previous works to characterize equilibration
are not reliable and propose instead to test the isotropy of various observables.
Our conclusions are in contradiction with those previously obtained, we show in
particular that at RHIC energies, elastic collisions are not effective enough for 
the system to thermalize.

\renewcommand{\contentsname}{Sommaire}
\tableofcontents
\addcontentsline{toc}{chapter}{Sommaire}
\pagestyle{fancy}
\chapter*{Introduction générale}
\addcontentsline{toc}{chapter}{Introduction générale}
\pagenumbering{arabic}

La matière nucléaire, telle que nous la connaissons à l'échelle
sub-nucléaire, est remarquablement bien décrite par la chromodynamique 
quantique (QCD), la théorie moderne des interactions fortes. Le bien-fondé
de la description du monde hadronique en termes de quarks et de gluons, 
les degrés de liberté fondamentaux du lagrangien de QCD, s'appuie sur
des bases expérimentales et des fondements théoriques solides. 
Une des propriétés les plus importantes des interactions fortes est la
liberté asymptotique : l'intensité effective de l'interaction entre les 
quarks et les gluons décroît avec l'échelle de distance à laquelle on la 
mesure. Ceci a entre autres pour conséquence le fait qu'à très haute
température et/ou très haute densité, les quarks et les gluons forment 
un plasma de matière déconfinée faiblement couplée. L'étude des
propriétés de la matière nucléaire dans ces conditions ``extrêmes'',
est un sujet de recherche qui progresse très rapidement. Le diagramme 
de phase des interactions fortes commence à être bien connu théoriquement, 
en particulier dans la région de potentiel chimique baryonique nulle.
L'existence de nouvelles phases de la matière est une prédiction tout
à fait remarquable de QCD.
\par
Il existe certains systèmes physiques où ces conditions extrêmes
sont naturellement réalisées, comme par exemple dans le c\oe ur des 
étoiles à neutrons, ou encore durant les premiers instant de 
l'univers primordial. De plus, on s'attend à pouvoir créer momentanément 
ces conditions ``en laboratoire'' lors de collisions entre noyaux lourds 
à très haute énergie. De nombreuses expériences ont eu lieu à l'Alternating 
Gradient Synchroton (AGS, BNL) et au Super Proton Synchrotron (SPS, CERN) 
durant ces dix dernières années, d'autres se déroulent actuellement au 
Relativistic Heavy Ion Collider (RHIC, BNL), et un programme est prévu 
à partir de 2005 au Large Hadron Collider (LHC, CERN), avec des énergies 
de plus en plus élevées\footnote{Collisions Au-Au avec $\sqrt s = 200$~GeV 
  par nucléon à RHIC et Pb-Pb avec $\sqrt s = 5.5$~TeV par nucléon à LHC !!}.
De nombreuses signatures de la matière déconfinée ont été proposées et
sont activement étudiées tant expérimentalement que théoriquement (voir
par exemple~\cite{QM99}). La dynamique d'une collision d'ions lourds est
cependant extrêmement compliquée et il est très difficile d'interpréter les
données de manière non-ambigüe. En particulier les données obtenues
au SPS, si elles fournissent de fortes indications en faveur de
la formation de matière déconfinée, ne permettent pas d'affirmer quoi que
ce soit de définitif.
\par
C'est dans ce cadre général que se place le sujet de cette thèse. Nous 
étudions certains aspects de la physique des  collisions d'ions lourds, 
reliés en particulier au fait que la matière produite n'est {\it a priori} 
pas à l'équilibre thermodynamique. Notre intérêt pour cette problématique 
est double. D'une part, certains phénomènes qui n'ont lieu que dans des 
situations hors d'équilibre pourraient donner lieu à des signatures 
uniques, fournissant ainsi une source d'informations supplémentaire 
(ou complémentaire). C'est le cas de la formation de condensats chiraux 
désorientés (DCC) lors du passage rapide de la transition de phase chirale. 
La détection de ce phénomène hypothétique pourrait par exemple être utile 
pour localiser cette transition dans le diagramme des phases.
Dans la première partie de cette thèse, nous étudions la possibilité
qu'un tel phénomène se produise. Un DCC est une région de l'espace où 
le paramètre d'ordre de la symétrie chirale oscille dans une direction 
différente de celle qu'il a dans le vide. C'est une configuration collective
classique du champ de pion. Nous calculons la toute première estimation de 
la probabilité pour qu'un champ classique potentiellement observable soit 
créé lors de l'expansion sphérique rapide du système. Pour ce faire, nous 
proposons une méthode originale d'échantillonnage des conditions initiales,
habituellement choisies de manière arbitraire dans la littérature. Nous 
obtenons une limite supérieure de cette probabilité relativement faible, 
typiquement de l'ordre de $10^{-3}$ (Chap.~\ref{PROBA}).
\par
Nous nous intéressons ensuite à l'aspect collectif du DCC : tous les modes 
du champ (classique) oscillent dans la même direction de l'espace d'isospin. 
Cette propriété est à l'origine de la signature la plus remarquable du phénomène, 
qui est aussi la plus utilisée dans les stratégies expérimentales de détection : 
la distribution anormalement large de la fraction neutre du nombre total de 
pions émis. Nous montrons que le champ produit dans le modèle microscopique 
le plus simple, utilisé jusqu'à présent, n'exhibe pas ce comportement collectif : 
les différents modes sont comme autant de DCC dont les orientations dans 
l'espace d'isospin sont indépendantes les unes des autres. Ce résultat 
contredit une idée largement admise et remet en question la possibilité de 
former un DCC dans une collision d'ions lourds (Chap.~\ref{QUENCH}).
\par 
La deuxième raison pour laquelle il est important d'étudier les aspects
hors d'équilibre dans ces collisions est le fait que les calculs actuels
concernant les signatures expérimentales reposent sur l'hypothèse selon 
laquelle le système est en équilibre thermique local. Il est important 
de savoir si cette hypothèse est justifiée dans une collision réelle. 
Si ce n'est pas le cas, il est alors nécessaire d'avoir une idée de la 
façon dont le système évolue pour pouvoir interpréter les données. Dans 
la deuxième partie de la thèse, nous nous intéressons à l'équilibration 
thermique du système de gluons initialement produits (Chap.~\ref{QGP}). 
Nous décrivons la dynamique des gluons à l'aide d'une équation de Boltzmann 
que nous modélisons par une simple approximation de temps de relaxation. 
Nous comparons différents scénarios proposés dans la littérature pour 
décrire l'état initial : les scénarios des minijets et de saturation. 
En calculant le temps de relaxation de façon auto-cohérente nous reproduisons 
de manière semi-quantitative certains résultats exacts, récemment 
obtenus numériquement dans le scénario de saturation. Pour caractériser 
l'écart à l'équilibre nous mesurons le degré d'anisotropie du système.
Nos conclusions contredisent celles de travaux précédents où des critères 
différents sont utilisés pour mesurer le degré d'équilibration. En particulier 
nous montrons que les collisions élastiques ne sont pas suffisantes pour 
thermaliser le système aux énergies de RHIC.
\par
Au cours de notre présentation, nous essayons de mettre en avant les 
idées physiques sous-jacentes aux problèmes étudiés, releguant les
détails techniques ou les dérivations non essentielles mais instructives
dans les annexes. Les différents chapitres sont relativements indépendants.
Pour des raisons pratiques et esthétiques, nous avons choisi des notations légèrement
différentes d'un chapitre à l'autre, notamment en ce qui concerne les indices
chiraux. Nous travaillons partout avec le système d'unités $\hbar = c = k_B =1$
(où $k_B$ est la constante de Boltzmann).

\part{LES CONDENSATS CHIRAUX DÉSORIENTÉS}

\chapter{DCC : introduction}
\label{INTRO}

Dans ce chapitre nous passons en revue certaines idées
pertinentes à la physique des condensats chiraux désorientés.
Après un bref rappel historique des motivations initiales, 
nous présentons les principaux développements qu'a connu le 
sujet depuis sa naissance, au début des années 1990. 
Nous présentons divers concepts et images physiques qui
constituent la toile de fond des travaux présentés dans
la première partie de cette thèse.

\section{La symétrie chirale}

La théorie des interactions fortes, ou chromodynamique quantique
(QCD) est approximativement invariante sous les opérations du 
groupe~$SU_L(2) \times SU_R(2)$ qui consistent en des 
transformations unitaires indépendantes des composantes
chirales droite et gauche du doublet de quarks légers $q=(u,d)$,
c'est la symétrie chirale. Elle est réalisée de façon spontanément 
brisée, phénomène décrit par la valeur non nulle du condensat de 
quarks\footnote{Il existe d'autres paramètres d'ordre associés à 
   la brisure spontanée de la symétrie chirale. Nous ne les considérerons 
   pas ici.}~$\langle 0 | q_L \, \bar q_R | 0 \rangle$ et 
se manifestant dans le spectre d'excitations des
interactions fortes par la faible masse
des pions~($\approx 140$~MeV) comparée à 
l'échelle de masse typique des hadrons~($\sim 1$~GeV).
Les propriétés de transformation du produit bilinéaire~$q_L \, \bar q_R$ 
sont analogues à celles d'un vecteur~$\bfphi \equiv (\sigma,\bfpi)$
sous les rotations quadri-dimensionelles du groupe~$O(4)$. Dans ce 
language, les composantes, dites d'isospin,~$\pi_1,\pi_2,\pi_3$, 
représentent les modes collectifs ou bosons de Goldstone accompagnant 
le phénomène de brisure spontanée : les excitations élémentaires
(quanta) du champ~$\bfpi$ sont les pions. La valeur moyenne dans le vide
du champ chiral $\bfphi$ pointe dans la direction~$\hat \sigma$ : on
a~$\langle 0 | \bfphi | 0 \rangle = (f_\pi,{\bf 0})$, où $f_\pi=92.5$~MeV 
est la constante de désintégration du pion. La symétrie résiduelle~$O(3)$ 
des rotations autour de cet axe est la symétrie d'isospin des interactions 
fortes. De façon générale, la valeur moyenne du champ $\bfphi$, qui
mesure l'intensité de la brisure spontanée, est appelée paramètre d'ordre
de la symétrie chirale.

\section{Un LASER à pions}

Un condensat chiral désorienté~(DCC) est un état dans lequel le
paramètre d'ordre a une orientation différente de celle qu'il a 
dans le vide. Plus précisément, c'est une configuration classique
où le champ de pion oscille de manière
cohérente dans une direction donnée de l'espace d'isospin.
Pour préciser la nature du phénomène et en comprendre
l'intérêt, il est utile de faire une petite retrospective historique.
\par
Dans le contexte des collisions hadroniques ou nucléaires
à très haute énergie, la production multiple de pions dont
les énergies transverses ne dépassent pas quelques 
centaines de~MeV est chose courante. 
Certains auteurs ont proposé d'interpréter ce phénomène 
comme résultant de la désintégration d'un état 
cohérent~\cite{Glauber,HS,BSS,Andreev}, l'idée (déjà présente
dans d'anciens articles de W.~Heisenberg~\cite{Heis})
étant de voir le processus de production multiple comme
le rayonnement d'un champ classique. On parle aussi
d'état multipions ou encore de LASER à pions.
Cette idée est longtemps restée marginale jusqu'au début des 
années 1990 où, sur des bases théoriques plus solides, elle
a été largement étudiée et développée.
\par
En effet, le développement de théories effectives de basse
énergie dans le milieu des années 80 a permis d'étudier la
dynamique des excitations de grande longueur d'onde des
interactions fortes dans le cadre de modèles basés sur
des fondements théoriques solides\footnote{Les propriétés
   de symétrie chirale étant à l'origine de l'existence
   des excitations de grande longueur d'onde (les modes de
   Goldstone), il est clair qu'elles jouent un rôle important
   quant à leur dynamique. En exploitant les propriétés de 
   symétrie chirale du lagrangien fondamental de QCD, on peut 
   écrire un lagrangien effectif de basse énergie qui se présente
   sous la forme d'un développement en gradients (qui correspond
   à un développement en puissances de l'échelle d'énergie typique 
   à laquelle on s'intéresse) et qu'il est nécessaire de tronquer en
   pratique. Le modèle~$\sigma$ non-linéaire, dont il est question
   dans la suite, est obtenu en ne retenant que le premier terme, 
   qui contient deux dérivées.}. 
L'exemple le plus simple est le modèle~$\sigma$ non-linéaire, qui 
décrit bien les propriétés des pions de très basse énergie.
En termes du champ~$\bfphi$ introduit plus haut, l'action de ce modèle
s'écrit
\beq
 \mathcal S = \frac{1}{2} \,
 \int d^4x \, \p_{\mu} \bfphi \cdot \p^{\mu} \bfphi
 \label{nonlin}
\end{equation}\noindent
avec la contrainte
\beq
 \bfphi^2 = \sigma^2 + \bfpi^2 = f_\pi^2
 \label{contrainte}
\end{equation}\noindent
La pertinence de ces modèles pour le problème de l'existence
de champs de pions classiques est double : l'approximation
classique est justifiée par le fait qu'on s'intéresse à des 
excitations de grande longueur d'onde, et l'étude des solutions
classiques permet de savoir quelles configurations sont autorisées
par la dynamique sous-jacente. C'est essentiellement ce deuxième
point qui manquait aux études précédentes. C'est dans ce cadre conceptuel, 
que différents auteurs étudient les solutions classiques du
modèle~$\sigma$ non-linéaire au début des années 1990~\cite{Ans,AR,BK1}. 
Ces études montrent en particulier qu'il est naturel de considérer 
des configurations classiques du champ de pion de grande longueur
d'onde ayant des géométries non-triviales dans l'espace d'isospin.
Illustrons ce point par un exemple. En utilisant des conditions
aux limites idéalisées modèlisant la région centrale d'une collision très
énergétique~\cite{Heis,Bjor0}, J.P.~Blaizot et A.~Krzywicki
se ramènent à un problème à une dimension : le système est invariant 
sous les boost longitudinaux et le champ ne dépend que du temps 
propre~$\tau = \sqrt{t^2 - z^2}$, où~$z$ est l'axe de la collision.
Les équations de conservation des courants vectoriel 
${\bf V}_{\mu} =  \bfpi \times \p_{\mu} \bfpi$ et axial
${\bf A}_{\mu} = \bfpi \p_{\mu} \sigma - \sigma \p_{\mu} \bfpi$
\eq
 \p_{\mu} {\bf V}^{\mu} = {\bf 0} 
 \, \, \, \, ; \, \, \, \,
 \p_{\mu} {\bf A}^{\mu} = {\bf 0}
\eq
sont facilement intégrées :
\eq
 {\bf V}_{0} =  \bfpi \times \dot \bfpi = \frac{{\bf a}}{\tau} 
 \, \, \, \,  ; \, \, \, \,
 {\bf A}_{0} = \bfpi \dot \sigma - \sigma \dot \bfpi = \frac{{\bf b}}{\tau}
\eq
où les vecteurs~${\bf a}$ et~${\bf b}$ (${\bf a} \cdot {\bf b} = 0$) 
sont des constantes d'intégration. Le champ de pion décrit donc
une trajectoire elliptique dans l'espace 
d'isospin~(${\bf c} = {\bf a} \times {\bf b}$,~$\kappa^2 = a^2 + b^2$) :
\bear
 \pi_a & = & 0 \\
 \pi_b & = & - \sin \left( \kappa \ln \frac{\tau}{\tau_0} \right) \\
 \pi_c & = & \frac{a}{\kappa} \cos \left( \kappa \ln \frac{\tau}{\tau_0} \right) 
\eear
où~$\tau_0$ est une constante délimitant l'hypersurface sur laquelle
sont localisées les sources du champ classique de pion~\cite{BK1}.
Cette solution appartient à la classe plus générale des solutions 
de type onde planes proposée par A.A.~Anselm (voir aussi~\cite{EKV}). 
\par
Parallèlement, J.D.~Bjorken suggère la possibilité de former une 
configuration particulière du champ de pions qu'il dénomme 
{\em condensat chiral désorienté}~\cite{Bjor1,Bjor2}. Il donne une
image intuitive de la collision et de la formation du DCC qu'il est bon 
d'avoir en tête car elle est très parlante, c'est le scénario 
``Baked-Alaska'' : les produits 
primaires de la collision s'éloignent de la zone d'impact avec une vitesse 
proche de celle de la lumière, formant une ``boule de feu'' en expansion 
rapide et dont l'intérieur est isolé du vide environnant. Si la densité 
d'énergie y est suffisament faible, l'intérieur ressemble de très près 
au vide. Cependant, ce vide n'étant pas en contact avec le vide physique,
l'orientation du paramètre d'ordre n'a pas de raisons d'y être la 
même, en particulier les composantes d'isospin de ce dernier peuvent 
être non nulles ; c'est un vide désorienté. Après un certain temps (de 
l'ordre de quelques fermi), la surface de la boule de feu se désagrège
(hadronisation), l'intérieur et l'extérieur 
ne sont plus séparés et le vide désorienté relaxe vers le vrai vide en 
rayonnant ses modes collectifs, les pions. Ce scénario 
phénoménologique correspond à l'idéalisation adoptée par Blaizot
et Krzywicki dans le cas d'une expansion longitudinale. Un cas particulier 
de leur solution correspond à un champ oscillant dans une direction quelconque 
de l'espace d'isospin (c'est le cas où l'intensité initiale du courant axial 
est grande comparée à celle du courant vectoriel : $b \gg a$. Le champ de 
pions oscille alors dans la direction~${\bf b}$.).
C'est le DCC. La prédiction la plus frappante de ce phénomène hypothétique 
est la distribution évènement par évènement de la proportion de pions
neutres émis par rayonnement. En effet, le nombre total de pions avec 
une composante d'isospin donnée est proportionnel (dans le cas du DCC 
uniquement, voir le Chap.~\ref{QUENCH}) à l'intégrale du carré du champ 
sur tout le volume de la bulle
\eq
 N_i \propto \int_V d^3x \, \phi_i^2 (\vec x)
\eq
Les interactions fortes étant invariantes sous les rotations 
d'isospin, toutes les directions sont équiprobables et la
proportion de pions neutres
\eq
 f = \frac{N_{\pi_0}}{N_{\pi_+} + N_{\pi_0} + N_{\pi_-}}
\eq
est alors distribuée selon la loi\footnote{Cette distribution
   est donnée explicitement dans~\cite{BK1} où les auteurs
   n'ont pourtant pas isolé la configuration DCC, ce qui n'est
   pas correct. Ceci est corrigé dans~\cite{BK2}. Notons que ce
   résultat est déja présent dans~\cite{Andreev} où l'auteur considère
   un état cohérent d'isospin total nul. Le DCC est un tel état, en effet
   le champ oscille dans une direction donnée et est, par conséquent
   parallèle à sa vitesse, ce qui signifie que l'isospin total de cette 
   configuration est nul : ${\bf V} = \bfphi \times \dot \bfphi = {\bf 0}$, et 
   ${\bf I}=\int_V {\bf V}$.}
\beq
\label{dccdist}
 \frac{dP(f)}{df} = \frac{1}{2 \sqrt f} \, ,
\end{equation}\noindent
à comparer avec la distribution binomiale très piquée autour de la valeur 
moyenne~$\bar f = 1/3$ prévue dans le cas de la production incohérente 
de pions. 
Ce phénomène pourrait expliquer les évènements Centauro et anti-Centauro
enregistrés dans le domaine des rayons cosmiques de très haute énergie,
dans lesquels des amas de pions tous chargés (Centauro) ou tous neutres
(anti-Centauro) ont été observés~\cite{centauro}.
\par
Si on comprend mieux les configurations classiques possibles,
compatibles avec la dynamique des interactions fortes, on ne sait cependant
pas comment un tel champ peut être créé lors de la collision.
En effet, dans les considérations précédentes, on a supposé
l'existence d'un champ cohérent, classique créé par des sources localisées 
sur le cône de lumière (la surface de la boule de feu). La question de la 
formation du champ classique est cachée dans l'hypothèse de l'existence 
de ces sources.

\section{Un scénario microscopique}

L'idée du DCC reçoit une impulsion considérable quand, en 1993, K.~Rajagopal
et F.~Wilczek proposent un scénario prévoyant l'émergence d'un champ fort
de pions après une collision de noyaux lourds ultra-relativistes~\cite{RW}. 
Dans ce type de collisions, la densité d'énergie par unité de volume déposée
dans la région centrale peut atteindre des valeurs très élevées (plusieurs
GeV/fm$^3$) et on s'attend à ce que la matière soit dans un état où la
symétrie chirale est restaurée\footnote{Supposons pour un instant que le
  système est décrit par un gaz de pions de masse nulle à l'équilibre
  thermodynamique. La température de ce gaz de pions (relativistes) est 
  reliée à la densité d'énergie par la loi : 
  $\epsilon = g (\pi^2/30) \, T^4$, où $g=3$ est la dégénérescence du 
  triplet de pions. A une densité d'énergie $\epsilon \approx 1$ GeV/fm$^3$,
  correspond une température $T \approx 400$ MeV. La température critique de la
  transition de phase chirale est $T_c \sim 150$ MeV.} du fait des fortes
fluctuations du champ chiral. L'expansion rapide
du système engendre une chute brutale de la densité d'énergie et par conséquent
une suppression soudaine de ces fluctuations. Le paramètre d'ordre ``se fige''
dans une direction aléatoire de l'espace d'isospin. Ce phénomène est analogue
à la formation de domaines d'aimantation lors du trempage d'un matériau
feromagnétique. C'est l'idée sous-jacente au scénario de Rajagopal et 
Wilczek, aussi appellé scénario du trempage, et que nous allons maintenant 
décrire un peu plus en détail car il est à la base des travaux décrits 
dans les chapitres suivant. En fait, ce modèle sera étudié dans le
chapitre~\ref{QUENCH} où nous aurons l'occasion d'en donner tous les 
détails. Pour le moment, il est suffisant d'en présenter les grandes 
lignes et les résultats importants.
\par
Considérons la région centrale d'une collision d'ions lourds 
ultrarelativistes, c'est à dire la région autour de $z=0$ où $z$ est 
l'axe du faisceau dans le référentiel du centre de masse de la collison.
A très haute énergie, les nucléons des noyaux incidents 
ne sont pratiquement pas ralentis et ont complètement évacué la région 
centrale après un temps très bref. Cette dernière est alors formée
de matière non-baryonique~\cite{CYWONG} (la densité de baryons est 
égale à celle des anti-baryons). Nous nous intéressons à la
dynamique des modes de grande longueur d'onde et nous considèrerons
le champ de pions dans cette région. Nous négligeons en particulier
les effets possibles dus à la présence d'autres mésons ou baryons ainsi
qu'à celle de matière déconfinée. Nous supposerons de plus que le système
a atteint un état d'équilibre thermique local après un certain temps.
Si la densité d'énergie dans la région centrale est suffisante, la symétrie
chirale est restaurée. Pour décrire l'évolution ultérieure du champ de pion,
nous devons utiliser un modèle capable de prendre en compte la possibilité
d'un état symétrique~$O(4)$. Il est clair, du fait de la 
contrainte ~(\ref{contrainte}) que ce n'est pas le cas du modèle $\sigma$ 
non-linéaire. L'extension la plus simple consiste à relâcher 
la contrainte : c'est le modèle $\sigma$-linéaire\footnote{Le 
   modèle~$\sigma$-linéaire a été proposé par M.~Gell-Mann et M.~Lévy 
   en 1960~\cite{Levy} (en fait avant, par Schwinger, voir dans~\cite{Levy}) 
   pour décrire les interactions fortes entre nucléons. Ces derniers 
   étaient représentés par le doublet d'isospin~$(p,n)$ et l'interaction
   pion-nucléon par un couplage de Yukawa.}
\beq
 \mathcal S = \int d^4x \, 
 \left\{ \frac{1}{2} \, \p_{\mu} \bfphi \cdot \p^{\mu} \bfphi -
	\frac{\lambda}{4} \, \left(\bfphi \cdot \bfphi - v^2 \right)^2 + 
	H \sigma \right\} \, .
 \label{sigma}
\end{equation}\noindent
Le terme~$H \sigma$ brise explicitement la 
symétrie et rend compte de la masse non-nulle des  pions. 
Au niveau classique, on a
\bear
 H & = & f_\pi m_\pi^2 \\
 m_\pi^2 & = & \lambda \left( f_\pi^2 - v^2 \right) \\
 m_\sigma^2 & = & \lambda \left( 3 f_\pi^2 - v^2 \right)
\eear
On retrouve le modèle non-linéaire en prenant la limite 
chirale~($m_\pi \rightarrow 0$) et en intégrant le degré
de liberté lourd~$\sigma$. 
\par
Rajagopal et Wilczek ont étudié numériquement la dynamique du paramètre 
d'ordre dans le cadre du modèle $\sigma$ linéaire classique en supposant
un état initial symétrique : les composantes $\phi_j$ 
du champ sont des variables aléatoire gaussiennes de même variance
et distribuées indépendemment sur les différents n\oe uds d'un réseau cubique. 
Bien qu'ayant en tête un système en expansion, Rajagopal et Wilczek travaillent
dans une configuration statique. Le trempage des fluctuations initiales est
modélisé par la faible valeur de la variance dans l'état initial.
Durant l'évolution ultérieure du champ de pion, les modes de Fourier
de grande longueur d'onde sont fortement amplifiés en comparaison des 
modes de plus petite longueur d'onde, et oscillent dans le 
temps de manière cohérente avec une fréquence $2 \pi / \sqrt{m_\pi^2 + k^2}
\approx 2 \pi / m_\pi$. L'énergie est ensuite répartie de manière équivalente 
entre les modes (équipartition) par la dynamique non linéaire. Cependant 
dans une situation en expansion, la dilution du système entraine le
découplage des modes (freeze-out). S'il a lieu suffisament tôt, celui-ci peut
empêcher le système d'atteindre l'équipartition. Dans ce cas, la configuration
finale du champ consiste en une superposition de modes de grande longueur d'onde
et {\em de grande amplitude}. C'est le champ fort dont nous avions besoin.
Le résultat de la Réf.~\cite{RW} est reproduit sur la Fig.~\ref{fig_RW1} où 
l'on voit l'évolution temporelle du module au carré des composantes de 
Fourier de $\phi_3$, moyenné sur des bins de largeur $\delta k$ centrés 
en $||\vec k||=k$ pour diffférentes valeurs de $k$.

\begin{figure}[htbp]
\epsfxsize=4.in \centerline{ \epsfbox{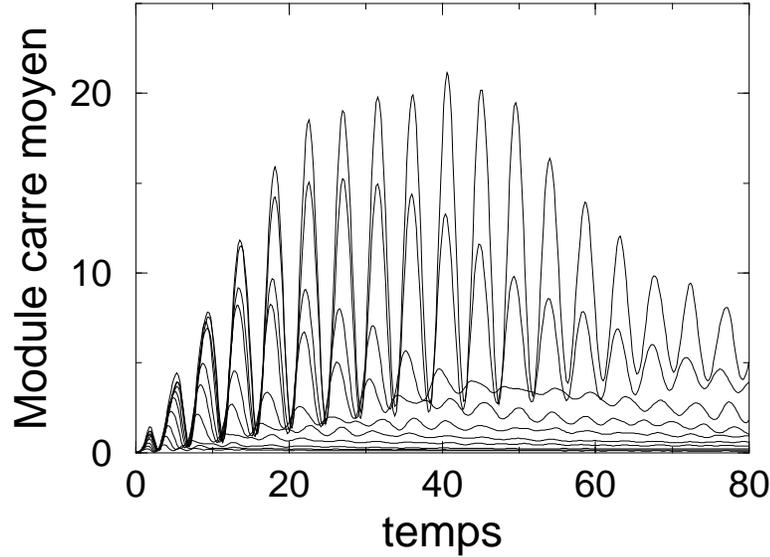}}
\caption{\small Le module au carré moyen des composantes de Fourier de
   $\phi_3$ en fonction du temps (en unités du pas du réseau $a$) dans 
   le scénario du trempage de la Réf.~\cite{RW}. On a calculé la moyenne 
   du module au carré de tous les modes tels que $k=||\vec k||$ est dans 
   le bin de largeur $\delta k =0.057a^{-1}$ centré autour de (de haut en 
   bas sur la figure) $ka=0.20$, $0,26$, $0,31$, $0,37$, $0,48$, $0,60$, 
   $0,71$, $0,94$, $1,16$ et $1,39$.} 
\label{fig_RW1}
\end{figure}
Le phénomène décrit plus haut peut être compris qualitativement à l'aide 
d'une simple approximation~\cite{RW,Rajrev,BKrev}. Les équations du mouvement s'écrivent
\beq
 \left( \p^2 + \lambda (\phi^2 - v^2) \right) \bfphi = H {\bf n}_\sigma \, ,
 \label{eom}
\end{equation}\noindent
où l'on a utilisé la notation $\phi^2 = \bfphi \cdot \bfphi$, et
où ${\bf n}_\sigma$ est un vecteur unitaire dans la direction $\sigma$ 
de l'espace chiral. En remplaçant le terme non linéaire $\phi^2$ par sa 
valeur moyenne sur le volume (dans ce qui suit, nous dénotons la valeur 
moyenne spatiale par des crochets : 
$\langle \mathcal O \rangle = (1/V) \int_V d^3x \, \mathcal O(\vec x)$) et 
en prenant la transformée de Fourier de l'équation obtenue, on 
obtient\footnote{Cette approximation est 
   équivalente à la limite $N \rightarrow \infty$ où $N$ est le nombre
   de composante du champ ($N=4$), c'est une approximation de type champ 
   moyen. Nous aurons l'occasion d'en parler d'avantage au chapitre~\ref{PROBA}.}
\beq
\label{MFeom}
 \left( \frac{d^2}{dt^2} + k^2 + m_{eff}^2(t) \right) \bfpi(\vec k,t) = {\bf 0} \, ,
\end{equation}\noindent
où 
\beq 
\label{effmass} 
 m_{eff}^2(t) = \lambda \left( \langle \phi^2 \rangle (t) - v^2 \right)
\end{equation}\noindent

est la masse effective (au carré) des excitations du champ à 
l'instant $t$. Elle représente la courbure instantanée du potentiel 
effectif vu par le champ.
\begin{figure}[htbp]
\epsfxsize=4.in \centerline{ \epsfbox{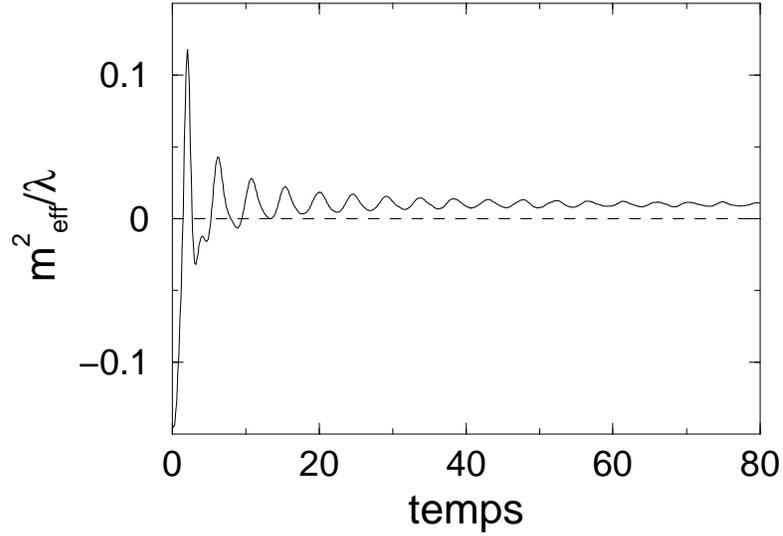}}
\caption{\small La masse effective au carré $m_{eff}^2(t)$ en fonction 
   du temps (en unité du pas du réseau $a$) dans le scénario du trempage 
   de la Réf.~\cite{RW}.} 
\label{fig_RW2}
\end{figure}

L'évolution temporelle de $m_{eff}^2(t)$ dans le scénario de la Réf.~\cite{RW}
est représenté sur la Fig.~\ref{fig_RW2}. Quand $\langle \phi^2 \rangle (t) < v^2$
cette masse effective est imaginaire pure, ce qui se traduit par une croissance
exponentielle des amplitudes des modes de Fourier tels que $k^2 < -m_{eff}^2$
(cf.Eq.~(\ref{MFeom}). Ce mécanisme est bien connu dans 
divers domaines de la physique, comme la physique des solides ou encore 
la fragmentation nucléaire, sous le nom d'instabilité spinodale. Mentionnons 
le fait que le mécanisme de l'instabilité spinodale n'explique pas toute
l'amplification observée en résolvant numériquement les équations~(\ref{eom}). 
En effet, on peut voir sur la Fig.~\ref{fig_RW2} que le carré de la masse 
effective~(\ref{effmass}) n'est négatif que pour des temps $t \lesssim 10$ 
(en unité de pas du réseau) alors que l'amplification des modes de basse 
fréquence se poursuit jusqu'à $t \sim 50$. On peut toutefois 
comprendre  qualitativement l'amplification ultérieure à l'aide du mécanisme 
de résonance parametrique dû aux oscillations régulières de la valeur 
moyenne $\langle \sigma \rangle$ autour de sa valeur asymptotique
$f_\pi$~\cite{param1,param2,param3}.

\section{Expansion et conditions initiales}

Le scénario du trempage fournit donc un scénario microscopique simple
pour la formation possible d'un champ fort de pion lors d'une collision entre 
noyaux très énergétiques. De ce fait, il permet d'obtenir des informations 
au moins qualitatives sur les propriétés du champ classique ainsi 
produit. Le résultat de Rajagopal et Wilczek a par la suite été 
confirmé~\cite{Bialas,GGP} et le modèle du trempage a reçu un intérêt 
croissant. Diverses améliorations y ont été apportées, les plus importantes 
étant l'inclusion de certains effets quantiques et l'abandon de l'hypothèse
concernant les conditions initiales (trempage ``à la main'' des fluctuations
initiales) en faveur d'un scénario prenant explicitement en compte l'expansion. 
\par
La modélisation de l'expansion consiste simplement à se placer dans le 
référentiel en co-mouvement, qui se déplace avec un élément de volume du 
système. Par exemple dans le cas d'une expansion longitudinale à la vitesse 
de la lumière, le temps mesuré dans le repère en co-mouvement est le temps 
propre $\tau=\sqrt{t^2 - z^2}$ : au temps $t$, le système a une extension
spatiale $-t \le z \le t$. Autrement dit, on spécifie les conditions aux bords,
nécessaires à la résolution des équations du mouvement, non pas sur un
hyperplan $t=$cte, mais sur une hypersurface $\tau=$cte (voir le modèle
de Blaizot et Krzywicki décrit plus haut). En définissant la variable de 
rapidité $\eta=(1/2) \, \ln(t-z)/(t+z)$, on peut réécrire 
dans~(\ref{eom}) $\p_t^2 - \p_z^2 = \p_\tau^2
+ (1/\tau) \, \p_\tau - (1/\tau^2) \, \p_\eta^2$. En plus du terme habituel
d'accélération $\p_\tau^2$, on voit apparaître un terme de friction 
$\propto \p_\tau$ qui traduit le fait que la densité d'énergie dans un 
co-volume décroît du fait de l'expansion. Différents auteurs~\cite{HuWang,Asakawa,Bialas} ont étudié numériquement
les solutions de~(\ref{eom}) dans une géometrie en expansion longitudinale,
avec un état initial stable (où la masse effective~(\ref{effmass})
est initialement positive). Il ressort de ces études que, pour des 
conditions initiales appropriées, l'expansion entraine le système dans la région 
d'instabilité et le champ fort est créé. Cependant, le phénomène 
dépend fortement des conditions initiales. 
\par
Dans la Réf.~\cite{RAN0}, J.~Randrup propose une méthode systématique
d'échantillonnage des configurations initiales du champ chiral $\bfphi(\vec x)$
à partir d'un ensemble thermique. La fonction de partition du système en 
équilibre thermique à la température $T$ dans le volume $V$ est calculée
en traitant les fluctuations thermiques du champ autour de sa valeur
moyenne spatiale $\langle \bfphi \rangle$ dans une approximation
de champ moyen (l'approximation de Hartree). On peut alors 
calculer la distribution de probabilité des valeurs possibles 
de $\langle \bfphi \rangle$ ainsi que celle des fluctuations 
$\bfdelta \bfphi(\vec x) = \bfphi (\vec x) - \langle \bfphi \rangle$.
Dans un article ultérieur~\cite{RAN1}, Randrup étudie les trajectoires 
classiques générées par les équations~(\ref{eom}) avec des conditions initiales
aléatoires, échantillonnées selon cette méthode. La masse effective
(Eq.~(\ref{effmass})) s'écrit ($\langle \bfdelta \bfphi \rangle = {\bf 0}$)
\beq 
 m_{eff}^2 = \lambda \left( \langle \phi \rangle^2 + 
 \langle \delta \phi^2 \rangle - v^2 \right) \, ,
 \label{effmass2} 
\end{equation}\noindent
où $\langle \phi \rangle^2 = \langle \bfphi \rangle \cdot \langle \bfphi \rangle$.
La région d'instabilité $m_{eff}^2 < 0$ est représentée sur la 
Fig.~\ref{fig_traj} dans le plan
$(\langle \phi \rangle,\langle \delta \phi^2 \rangle^{1/2})$.
Randrup représente ses trajectoires dans ce même plan et étudie l'incursion
du système dans la région d'instabilité en fonction de la température initiale
d'une part et du taux d'expansion d'autre part. En effet, supposons une
expansion $D$-dimensionnelle, le temps mesuré dans le repère en co-mouvement
est alors : $\tau = \sqrt{t^2 - x^i x_i}$, où $i=1...D$, et le terme de friction
dans les équations du mouvement devient $(D/\tau) \p_\tau$ : l'expansion 
est d'autant plus efficace que $D$ est grand ($D \le 3$).

\begin{figure}[htbp]
\epsfxsize=4.6in \centerline{ \epsfbox{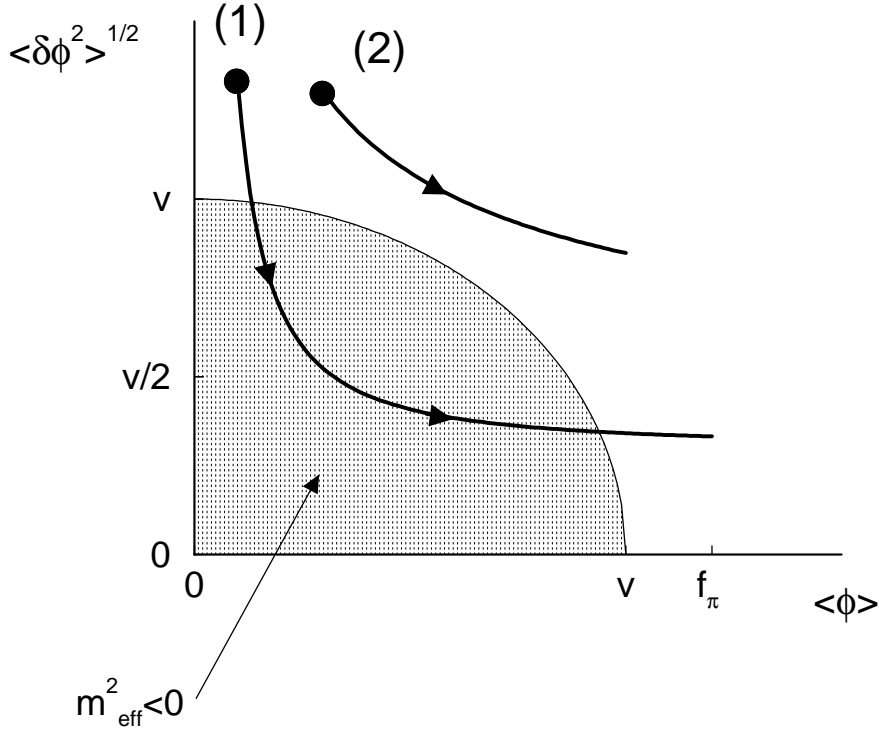}}
\caption{\small Représentation shématique de deux trajectoires typiques
   dans le plan $(\langle \phi \rangle,\sqrt{\langle \dphi^2 \rangle})$. 
   L'évènement (1) traverse la région spinodale ($m_{eff}^2<0$) et subit 
   une forte amplification. Dans la région d'instabilité, après le 
   refroidissement brutal des fluctuations (trempage), la valeur des 
   fluctuations du champ est $\sim v/2$, comme dans le modèle {\it ad hoc}
   de~\cite{RW}.} 
\label{fig_traj}
\end{figure}

Randrup observe que lorsque le système
est préparé à une température $T=400$~MeV, il n'entre dans la région d'instabilité
que pour~$D>1$. Pour $D=3$, l'instabilité apparait pour des températures
allant de $200$~MeV à $500$~MeV (au moins). Au début de l'évolution, l'expansion
rapide supprime fortement les fluctuations thermiques tandis que 
$\langle \phi \rangle$, initialement petit pour des températures supérieures
à la température critique $T_c \sim 150$~MeV, ne change que très peu. Le système
entre dans la région d'instabilité. Les modes de grande longueur d'onde,
et en particulier $\langle \phi \rangle$, le mode zéro, sont alors fortement
amplifiés et le système est ``poussé'' hors de la région d'instabilité.
Si la température initiale est trop importante, les fluctuations n'ont pas 
le temps de décroître suffisamment avant que $\langle \phi \rangle$ n'augmente
(vers sa valeur dans la phase brisée) et le système n'est jamais instable. 
A l'inverse, si la température est trop 
faible, on est trop proche de la température critique et $\langle \phi \rangle$ 
est si grand que l'on passe a coté de la région intéressante. 
Il est intéressant de remarquer que, dans le cas où le système entre dans
la région d'instabilité, la valeur typique des fluctuations après la brève 
période de refroidissement (trempage) est $\langle \delta \phi^2 \rangle^{1/2} 
\sim v/2$, ce qui correspond à la configuration initiale {\it ad hoc} de 
Rajagopal et Wilczek~\cite{Raj,RAN1}.  
\par
Nous comprenons donc comment, à partir d'une configuration chaotique du champ,
où les différents modes sont initialement indépendants, la dynamique 
hors d'équilibre du système génère momentanément une configuration formée par 
la superposition de modes de grande longueur d'onde et de grande 
amplitude. 

\section{Effets quantiques}

Le phénomène d'amplification des modes de grande longueur d'onde dans 
notre système en expansion est analogue\footnote{Dans le cas du DCC il s'agit
   cependant d'un problème en couplage fort} à celui de la création 
massive de particules durant la phase d'inflation de l'univers primordial.
Cette analogie a conduit différents auteurs à appliquer au premier problème
des méthodes développées pour le second, notamment en ce qui concerne la
prise en compte de certaines corrections quantiques à la dynamique. 
Dans la Réf.~\cite{Boyanovsky}, les auteurs travaillent
avec un système statique en modélisant le trempage ``à la main'' de 
la même façon que dans~\cite{RW}. Les fluctuations thermiques et quantiques 
sont décrites par une matrice densité supposée gaussienne, ce qui conduit 
à des équations du mouvement dont la structure est similaire à celle des
Eqs.~(\ref{MFeom}) et~(\ref{effmass2}), où $\langle ... \rangle$ doit être 
maintenant compris comme la moyenne sur ces fluctuations. 
Les auteurs des Réfs.~\cite{Cooper,Lampert,Lamperthesis} prennent en compte
les effets quantiques dans l'approximation de champ moyen (équivalente
à l'ansatz gaussien de~\cite{Boyanovsky}) dans un modèle avec expansion.
Les résultats de ces études sont similiaires à ceux obtenus dans le cas 
classique\footnote{Dans le modèle avec expansion, l'inclusion des effets
   quantiques donne des résultats quantitativement similaires à ceux du
   cas classique~\cite{ABL}} : on observe une augmentation importante du 
nombre de quanta de basse énergie durant la période d'instabilité. 
Le champ est alors essentiellement une superposition de modes classiques 
de grande longueur d'onde. De même que dans le cas classique, l'ingrédient 
crucial est la dynamique fortement hors d'équilibre due à l'expansion et 
le phénomène d'amplification dépend fortement des valeurs initiales des 
moyennes du champ et de sa dérivée temporelle. Ce dernier point fera 
l'objet de l'étude présentée au Chap.~\ref{PROBA}, où nous construirons 
une méthode originale d'échantillonnage des conditions initiales, que
nous utiliserons pour calculer la probabilité de former un champ
classique potentiellement observable.
\par
Dans le language de la mécanique quantique, le champ classique ainsi
formé n'est autre que la valeur moyenne du champ quantique dans l'état 
final du système, au moment du découplage. Autrement dit, l'état final 
est un état quasi-classique, c'est à dire un état cohérent~\cite{Glauber}. 
De façon générale, indépendante de toute dynamique sous-jacente, on peut 
construire différents types d'états cohérents multi-pions prenant 
en compte les contraintes imposées par les lois de conservation, par 
exemple de la charge électrique ou encore de l'isospin~\cite{HS,BSS,Andreev}. 
La structure de l'état formé dans un scénario réaliste dépend de
la dynamique microscopique sous-jacente. Par exemple, il est facile
de voir que l'approximation de champ moyen correspond à une description
du système en termes des états dits compressés\footnote{Ceci est directement
   relié aux transformations de Bogoliubov connectant les opérateurs de
   création et d'anihilation de quasi-particule à différents instants
   (voir Chap.~\ref{PROBA}).} 
(``squeezed state'') qui sont une généralisation des états 
cohérents~\cite{squeeze}. Le DCC est un état particulier où tous les 
modes du champ sont orientés dans la même direction de l'espace d'isospin 
(voir par exemple~\cite{ABL}) : c'est un état d'isospin total nul, ce qui 
est à l'origine de la loi~(\ref{dccdist})~\cite{Andreev}. Pour décrire
la formation d'un tel état collectif dans un scénario microscopique donné, 
il faut prendre en compte les non-linéarités de la dynamique. Au niveau 
quantique, cela recquiert de considérer des corrections au delà de 
l'approximation de champ moyen. En fait, nous verrons au Chap.~\ref{QUENCH} 
que le problème se pose déjà au niveau classique. En effet, nous 
montrerons que dans le modèle de Rajagopal et Wilczek, où l'état initial est 
supposé incohérent, l'état final n'est pas un état collectif, les 
modes amplifiés sont indépendants les uns des autres. Contrairement à 
une idée largement admise, le caractère collectif supposé du DCC n'a 
pas encore trouvé d'explication microscopique.
\par
Jusqu'ici nous avons considéré le modèle le plus simple possible en faisant
des approximations parfois très grossière, ce qui nous a permis de dégager
l'essentiel de la physique du phénomène. Nombre de corrections ont été
considérées dans la littérature (très abondante) sur le 
sujet\footnote{Voir la {\it `DCC home page'} http://wa98.web.cern.ch/WA98/DCC}.
Mentionnons par exemple les effets dus aux interactions du champ de pion 
avec son environement dans une collision d'ions lourds~\cite{KRZ,RAN3,quarks},
la modélisation d'effets quantiques au delà du champ moyen 
par une source stochastique dans les équations du mouvement du paramètre
d'ordre~\cite{Stock}, ou encore l'étude de modèles alternatifs au modèle
$\sigma$-linéaire, comme les modèles de Nambu-Jona-Lasinio~\cite{NJL} ou de
Gross-Neveu~\cite{GrossNeveu}. Ces études n'altèrent pas l'image de la 
formation d'un champ classique que nous avons décrite ici, qui s'avère 
donc être relativement générique.
\par
La détection éventuelle d'un DCC dans des expériences de collisions
hadroniques ou nucléaires à haute énergie pourrait permettre d'obtenir
des informations précieuses sur la structure du vide des interactions
fortes, ainsi que sur les propriétés de la transition de phase chirale.
Jusqu'à maintenant deux groupes ont tenté de détecter ce phénomène :
T864 (minimax), une expérience de collisions proton-antiproton à 
Fermilab~\cite{minimax} et WA98 au SPS du CERN (collisions d'ions 
lourds)~\cite{WA98}, toutes deux reportent des résultats négatifs.
En fait à ce jour les seuls ``signaux positifs'' sont ceux reportés dans le 
domaine des rayons cosmiques au début des années 1980.
Une partie du programme expérimental du détecteur 
STAR\footnote{Voir la {\it `STAR home page'} http://www.star.bnl.gov/STAR} 
à RHIC (BNL) est dédiée à la recherche de grandes fluctuations dans la 
distribution évènement-par-évènement de la fraction de pions neutres.
Diverses autres signatures ont été proposées dans la littérature, 
mentionnons par exemple l'augmentation du  nombre de photons de basse
énergie~\cite{RAN4,gamma} par désintégration des pions neutres provenant
du DCC, ou encore, plus récemment, l'anomalie dans les abondances de baryons
multi-étrange $\Omega$ et $\bar\Omega$~\cite{omega}.
\par
Pour conclure ce chapitre d'introduction, reprenons les premières
questions-réponses de la `DCC trouble list'' de J.~D.~Bjorken, 
qui résument bien l'idée sous-jacente aux études présentées
dans les deux chapitres suivant,

\vspace{0.2cm}
\begin{center}
 Existence of DCC :
\end{center}
\vspace{0.2cm}
\begin{center}
\begin{tabular}{ll}
Must it exist?      &     NO                          \\
Should it exist?    &     MAYBE                       \\
Might it exist?     &     YES                         \\
Does it exist?      &     IT'S WORTH HAVE A LOOK      \\
\end{tabular}
\end{center}

\chapter{La probabilité de formation \\
         d'un champ de pion classique}
\label{PROBA}

Un scénario plausible pour la formation d'un champ de pion classique 
lors d'une collision nucléaire à haute énergie
a été identifié~\cite{RW} : l'expansion rapide du système entraine la 
suppression brutale des fluctuations thermiques (trempage), générant une 
amplification importante des modes de grande longueur d'onde. 
Ce mécanisme microscopique permet de faire des prédictions 
qualitatives utiles pour l'élaboration de stratégies expérimentales,
comme par exemple le fait que le signal soit à rechercher dans les pions
de basse énergie. Un paramètre crucial pour la phénoménologie
est la probabilité pour qu'un tel champ fort soit produit. En effet
on choisira une méthode expérimentale plutôt qu'une autre selon
que le nombre typique de collisions donnant lieu au phénomène
cherché est de une sur dix ou de une sur un million.
\par
Dans ce chapitre, nous nous proposons d'estimer l'ordre de 
grandeur de ce nombre. Il est clair que les modèles actuels ne permettent 
pas de dire quoi que ce soit de fiable au niveau quantitatif, néanmoins
il est légitime de chercher à avoir une estimation, même grossière,
de cette probabilité. En fait, nous en obtiendrons une limite supérieure.
Le travail exposé dans ce chapitre a été réalisé en collaboration avec 
André Krzywicki et a fait l'objet d'une publication dans la revue 
Physics Letters B~\cite{KRZJS}.

\section*{Cadre théorique et image physique}

Le scénario du trempage 
est relativement robuste en ce qui concerne les détails de la dynamique. 
Ici, nous utiliserons le modèle $\sigma$-linéaire en tant qu'approximation 
de la dynamique des excitations de basse énergie de QCD que sont les pions. 
Le point essentiel est la forte dépendance du phénomène d'amplification 
vis-à-vis des conditions initiales. Notre but étant d'estimer l'ordre de 
grandeur de la probabilité d'avoir une amplification donnée, il nous faut 
donner un poids statistique à chacune des configurations initiales possibles. 
\par
Pour ce faire, nous aurons recours à des approximations, parfois
très simplificatrices, mais qui ``capturent'' l'essentiel
de la physique du phénomène. Ainsi nous traiterons les effets
quantiques à l'ordre dominant dans un développement
en $1/N$, où $N$ est le nombre de composantes du champ chiral $\bfphi$ ($N=4$).
C'est une approximation de type champ moyen qui, nous l'avons vu dans
le chapitre précédent (Eqs.~(\ref{MFeom}) et (\ref{effmass})), décrit
la physique de l'instabilité spinodale, responsable du phénomène 
d'amplification. De plus, nous considèrerons le cas d'une expansion sphérique,
la plus efficace pour le trempage des fluctuations initiales. De manière 
générale, nous nous placerons 
toujours dans le cas le plus favorable pour la formation d'un champ fort 
de pion, de façon à avoir une limite supérieure à la probabilité cherchée.
En effet on s'attend d'ores et déjà à ce que celle-ci soit faible~\cite{BK2} 
et, notre objectif étant d'avoir une estimation qualitative, une limite 
supérieure est un paramètre pertinent pour la phénoménologie.
\par
Nous focalisons notre attention sur une petite bulle de 
matière chirale chaude, formée lors de la collision, qui subit
une expansion très rapide et qui, selon son état initial, subit 
éventuellement une période d'instabilité durant laquelle les modes
de grande longueur d'onde sont fortement amplifiés. Nous supposons 
que les fluctuations statistiques du champ de pion à l'intérieur
de la bulle sont décrites par un ensemble thermique local. Exploitant
cette hypothèse, nous proposons une méthode originale d'échantillonnage
des configurations initiales du champ, habituellement choisies de façon 
arbitraire dans la littérature. En combinant cette méthode avec le 
formalisme de la Réf.~\cite{Lampert} (voir aussi~\cite{Lamperthesis}), 
nous calculons la toute première estimation de la probabilité pour 
qu'un champ fort (classique), potentiellement observable, soit formé 
dans une collision d'ions lourds.
\par
En toute rigueur, la probabilité obtenue doit être interprétée comme 
une probabilité conditionnelle, à multiplier par la probabilité pour 
que notre bulle initiale soit formée dans une collision réelle. Le 
calcul de cette dernière quantité recquiert cependant un modèle pour 
l'ensemble de la collision, ce qui dépasse de loin le cadre de l'étude 
présentée ici. Néanmoins, l'hypothèse d'équilibre thermique local
peut être vue, non pas comme une approximation de l'état réel du 
système, mais comme une paramétrisation
de la distribution des configurations initiales possibles. 
Dans ce sens, le modèle décrit ci-dessus est suffisamment 
générique pour pouvoir, malgré sa  simplicité, donner une 
estimation raisonnable de la probabilité cherchée.

\newpage

Le chapitre est organisé comme suit : tout d'abord nous écrivons
les équations dynamiques à l'ordre dominant en $1/N$ dans une géométrie
en  expansion (nous reprenons pour l'essentiel le formalisme des
Refs.~\cite{Lampert,Lamperthesis}). Nous introduisons ensuite
la notion de champ interpolant qui nous permettra de relier les nombres de
particules dans les états initial et final. La méthode d'échantillonage
des conditions initiales est exposée en détails. Enfin nous présentons 
les résultats de notre calcul numérique, suivis d'une discussion et de la 
conclusion.


\section{Le formalisme}

La densité de Lagrangien du modèle $\sigma$-linéaire dans une 
géométrie quelconque, caractérisée par sa métrique $g_{\mu \nu}$, s'écrit
(cf.~(\ref{sigma}))
\eq
 \mathcal L = \sqrt{-g} \left( \frac{1}{2} g_{\mu \nu} \p^\mu \bfphi \cdot
 \p^\nu \bfphi - \frac{\lambda}{4} \left( \bfphi \cdot \bfphi - v^2 \right)^2
 + H \sigma \right)
\eq
où $g$ est le déterminant de la métrique. L'action
$\mathcal S = \int d^4x \mathcal L$ est, comme il se doit, un scalaire
sous les transformations du système de coordonées. 
\par
Nous cherchons à étudier l'évolution hors d'équilibre d'un système quantique
avec un nombre infini de degré de liberté. Le formalisme adapté est
la théorie quantique des champs hors d'équilibre. La théorie quantique des
champs a été initialement développée pour décrire des problèmes de diffusion :
on prépare le système à $t=-\infty$ et on fait une mesure à $t=+\infty$, où
$\pm\infty$ signifient ``très longtemps avant/après l'interaction''.
Il s'agit donc de calculer des amplitudes de transition entre états 
asymptotiques. La problématique qui nous occupe ici est très différente :
ayant préparé le système dans un certain état initial, décrit 
par la matrice densité $\rho (t_0)$, on cherche à suivre son évolution 
au cours du temps.
On veut calculer des objets du type $\langle \mathcal O \rangle (t) =
\mbox{Tr}(\rho (t) \mathcal O )$, l'évolution temporelle étant donnée
par les équations du mouvement $\dot \rho = \left[ H , \rho \right]$, où $H$ 
est le hamiltonien du système. Il existe un formalisme très élégant 
qui permet d'utiliser les méthodes développées pour les problèmes 
de diffusion dans le cas des problèmes hors d'équilibre, c'est le formalisme
dit de ``chemin fermé'' (Closed Time Path)~\cite{CTP0,CTP1,CTP2}. 
En particulier, cette formulation assure la causalité des équations obtenues.
Les particularités propres à ce formalisme, comme par exemple la 
forme matricielle des propagateurs, ne se manifestant pas dans 
l'approximation de champ moyen, nous n'en parlerons plus dans la suite.

\subsection{Approximation de champ moyen}

L'approximation dite ``de champ moyen'' consiste à remplacer 
les interactions entre quanta (les modes de Fourier du champ, 
en nombre infini) par une interaction moyenne effective. Les modes sont 
{\em effectivement} découplés les uns des autres en ce sens que chacun d'eux
ne ``voit'' les autres que par l'intermédiaire du champ moyen qu'ils 
créent : chaque mode évolue dans un potentiel effectif créé par tous les 
autres modes. Nous avons déjà vu un exemple de ce genre d'approximation
dans le chapitre précédent (Eqs.~(\ref{MFeom}) et~(\ref{effmass})). 
Rappelons-le, il nous sera utile dans la suite. Les équations du mouvement
pour le champ quantique $\hat\bfphi$ s'écrivent (nous travaillerons dans 
le système de coordonnées de Minkowski tout au long de cette section :
$g_{\mu\nu}=\mbox{diag} (1,-1,-1,-1)$)
\beq
\label{eom1}
 \left( \p^2 + \lambda (\hat\phi^2 - v^2) \right) \hat\bfphi = H {\bf n}_\sigma \, .
\end{equation}\noindent
En remplaçant le terme $\phi^2$ de la partie non-linéaire par sa valeur
moyenne sur l'état considéré, et en décomposant le champ en la somme de
sa valeur moyenne $\bfphi=\langle \hat\bfphi \rangle$ et des fluctuations
autour de celle-ci : $\hat\bfphi = \bfphi + \bfdphi$, on obtient 
($\langle \bfdphi \rangle = {\bf 0}$)
\eq
 \left( \p^2 + \lambda ( \phi^2 + \langle \dphi^2 \rangle - v^2 ) \right) 
 (\bfphi + \bfdphi) = H {\bf n}_\sigma \, .
\eq
La valeur moyenne de cette équation nous donne une équation pour la valeur
moyenne du champ. En soustrayant cette dernière à l'équation ci-dessus,
on obtient l'équation pour les fluctuations :
\bearn
\label{condensat}
 ( \p^2 + \chi (x) ) \bfphi (x) & = &  H {\bf n}_\sigma \, ,  \\
\label{fluctuation}
 ( \p^2 + \chi (x) ) \bfdphi (x) & = & {\bf 0} \, ,
\eearn
où
\beq
\label{meanfield}
 \chi (x) = 
 \lambda \left( \phi^2 (x) + \langle \dphi^2 (x) \rangle - v^2 \right) \, .
\end{equation}\noindent
On voit bien ce qui se passe sur ces équations : le condensat (ou paramètre
d'ordre) $\bfphi$ évolue dans un 
potentiel effectif qui n'est autre que le potentiel classique, corrigé
par l'effet des fluctuations quantiques (le terme $\langle \dphi^2 \rangle$),
la dynamique desquelles est gouvernée par une équation de type Klein-Gordon
avec une masse qui dépend du temps. Les fluctuations du champ sont donc 
effectivement découplées et évoluent dans un potentiel effectif quadratique
dont la courbure est déterminée d'une part par la valeur du condensat, et 
d'autre part
par l'effet {\em moyen} des fluctuations elles-même. La masse effective $\chi(t)$
représente la courbure locale du potentiel effectif à l'instant $t$.
\par
Il existe différent types d'approximations de champ moyen. Dans ce chapitre,
nous travaillerons avec l'approximation dite ``grand $N$'' qui consiste à 
ne retenir que l'ordre dominant dans un développement en $1/N$ où $N$ est
le nombre de composantes du champ ($N=4$). Les équations ainsi obtenues
sont précisément les Eqs.~(\ref{condensat})-(\ref{meanfield}).
Bien entendu la dérivation présentée ci-dessus ne constitue pas une preuve 
de cette affirmation et est seulement un moyen rapide d'écrire les équations. 
Il existe plusieurs manières d'obtenir l'ordre dominant en 
$1/N$~(voir par exemple~\cite{CJP,CHKMPA,Jackiw,DJ,CJT,Boyanovsky}). 
Ces différentes approches, bien qu'équivalentes, 
sont toutes aussi instructives les unes que les autres car elles éclairent 
le problème sous des angles différents et mettent ainsi en lumière divers 
aspects de l'approximation grand $N$ en particulier, mais aussi, et plus 
généralement, de l'approximation de champ moyen. Dans la dérivation ci-dessus, 
nous avons sacrifié le problème de la lumière à celui du plus court chemin, 
et cela pour plus de clarté. Nous procèderons suivant cette ligne tout au long
de ce chapitre, de façon à ne pas alourdir notre discours. Cependant, certains
compléments utiles à la compréhension sont détaillés dans les Annexes, 
auxquelles le lecteur est renvoyé le cas échéant. 
\par
Ici, nous partons des Eqs.~(\ref{condensat})-(\ref{meanfield})
et ne faisons que mentionner les aspects de l'approximation de
champ moyen pertinents pour notre étude. Tout d'abord, et il est facile 
de le voir à partir des équations ci-dessus, il s'agit d'une approximation
non-perturbative, dans ce sens qu'elle correspond à la resommation d'une 
classe infinie de termes de la série 
perturbative\footnote{Dans le cas présent, il s'agit des diagrammes dits 
   ``tadpoles'' ou encore ``daisy'' et ``super-daisy''. Un autre exemple 
   est donné par la resommation des boucles thermiques dures (HTL) dans les 
   théories de jauge à haute température, où le champ moyen décrit les
   effets collectifs de grande longueur d'onde 
   (voir par exemple~\cite{LeBellac}).}. 
C'est de plus une approximation semi-classique, ce dont on peut se convaincre
si on réfléchit à ce qu'est le champ créé par l'ensemble des particules
du milieu. De façon plus formelle, plaçons-nous un instant dans l'image
de Schrödinger : c'est maintenant la matrice densité du système qui dépend
du temps. L'approximation de champ moyen correspond à imposer que celle-ci
soit une gaussienne à chaque instant (voir par 
exemple~\cite{Boyanovsky})\footnote{Ceci est 
   intimement lié au fait que les opérateurs de création et d'anihilation 
   des excitations (ou quasi-particules) du système à un instant $t$ sont 
   reliés aux opérateurs à l'instant initial par une transformation de 
   Bogoliubov (voir plus loin, voir aussi l'Annexe~\ref{EXPANSION}).}.
Par exemple, si le système est initialement dans un état cohérent
(c'est à dire semi-classique), il reste dans un état cohérent à chaque 
instant : dans l'approximation de champ moyen, la dynamique est 
essentiellement classique. Ces deux aspects des Eqs.~(\ref{condensat})-(\ref{meanfield}) sont tout à fait pertinents
pour le problème qui nous intéresse : la formation d'un champ classique
dans une théorie de couplage fort. Remarquons enfin que dans 
l'Eq.~(\ref{fluctuation}), les fluctuations dans les différentes directions 
d'isospin sont traitées sur un pied d'égalité. Cet artefact de 
l'approximation nous simplifie la tâche. En effet, dans ce 
chapitre, nous nous intéressons à l'intensité {\em totale} du champ de pion, 
c'est à dire sommée sur les directions d'isospin\footnote{L'étude
   de l'intensité dans les directions d'isospin individuelles fera l'objet
   du Chap.~\ref{QUENCH}.}. 
Le champ $\bfdphi$ doit être vu comme décrivant la fluctuation moyenne 
dans l'espace d'isospin.

\subsection{Expansion sphérique}

Pour tenir compte de l'expansion du système, nous nous plaçons dans
un petit volume en co-mouvement, c'est à dire qui se déplace avec le ``fluide''
en expansion. Dans la suite, nous considérons le cas d'une expansion
tri-dimensionnelle à symétrie sphérique à la vitesse de la lumière.
Le temps mesuré par un observateur dans un co-volume est
le temps propre $\tau = \sqrt{t^2 - r^2}$, où $r$ est la distance à l'origine. 
La symétrie du problème est telle que la valeur du condensat ne dépend que de 
$\tau$. Vue dans le référentiel où notre bulle en expansion est au repos,
l'hypersurface $\tau=$~cte est une sphère dont le rayon croît très vite : 
à l'instant $t$, celui-ci vaut $r = \sqrt{t^2 - \tau^2}$. En passant d'une
hypersurface de temps propre constant à une autre, on décrit les différentes 
couches sphériques à l'intérieur de notre bulle, elles-même en expansion. Imaginez
un oignon dont les différentes couches s'étendent les unes derrière les autres,
comme dans le shéma de la Fig.~\ref{fig_oignon}.

\begin{figure}[htbp]
\epsfxsize=7.in \centerline{ \epsfbox{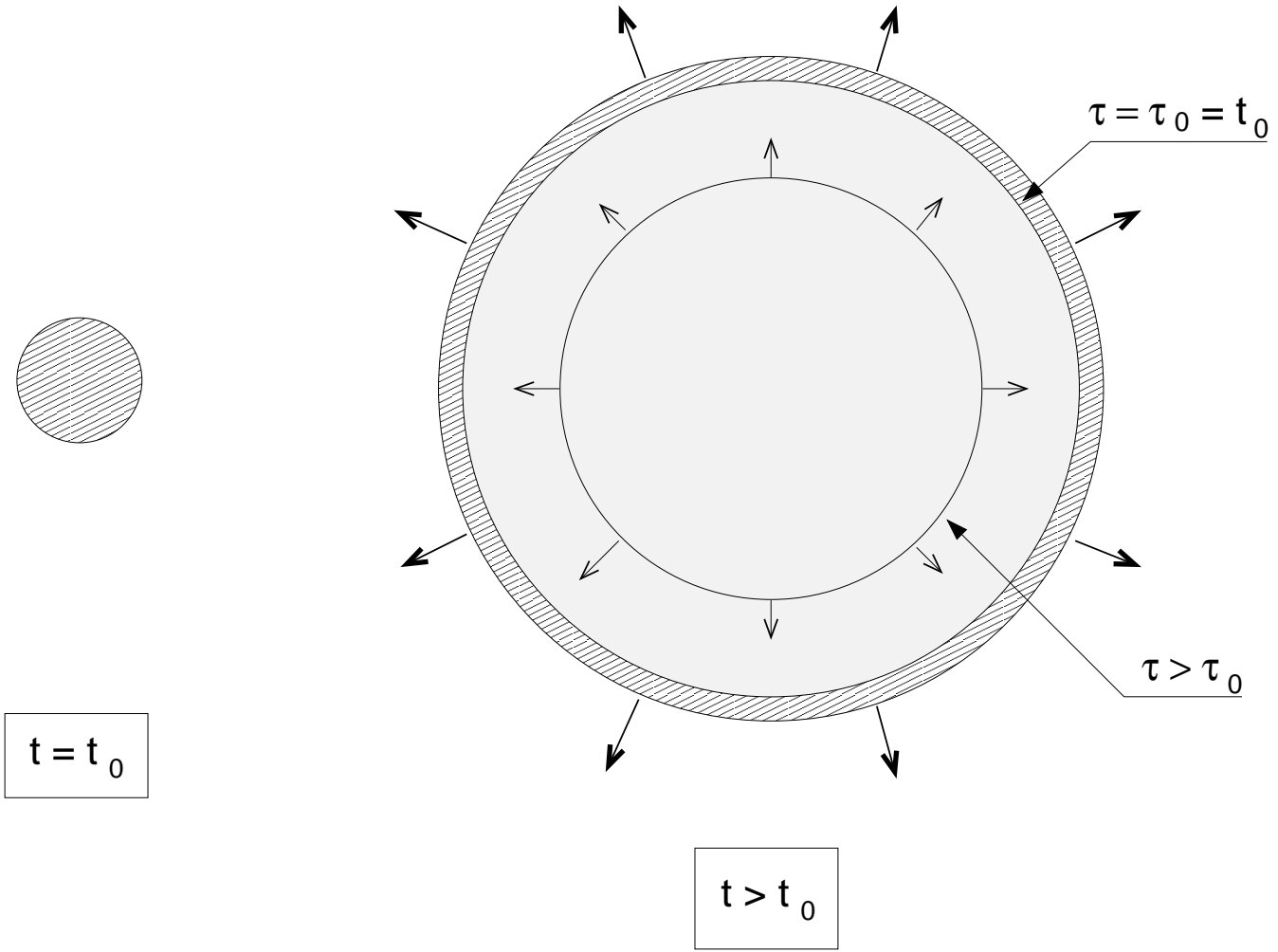}}
\caption{\small Schéma de la bulle de matière chirale en expansion
   vue dans le référentiel où elle est globalement au repos. A l'instant
   initial $t=t_0$, on a une bulle de rayon $R_0$ à l'équilibre thermique
   local. Celle-ci ``explose'' : aux instants ultérieurs $t>t_0$, la 
   surface externe de la bulle s'étend à la vitesse de la lumière,
   sa surface interne décrit l'hyperboloïde $\tau=\sqrt{t^2-r^2}=\tau_0=t_0$.
   Les hyperboloïdes successives $\tau>\tau_0$ décrivent l'intérieur de
   la bulle : ce sont des couches successives, en expansion les unes derrière
   les autres.} 
\label{fig_oignon}
\end{figure}
\par
Commençons par introduire quelques notations et définitions.

\subsubsection*{Coordonnées}

L'élément de longueur infinitésimal s'écrit
\eq
 \dd s^2 = \dd t^2 - \dd r^2 + 
 r^2 ( \dd \theta^2 - \sin^2 \theta \, \dd \varphi^2) \, .
\eq
Les coordonnées adaptées à la description de notre co-volume sont
\eq
 \tau = \sqrt{t^2 - r^2} \, \, , \, \, 
 \eta = \frac{1}{2} \ln \left( \frac{t+r}{t-r} \right) \, \, , \, \, 
 \theta \, \, , \, \, \varphi \, \, ,
\eq
en termes desquelles l'élément de longueur infinitésimal 
s'écrit\footnote{Il s'agit d'une métrique de
   Robertson-Walker~\cite{Birrell,Weinbook}.}
\beq
\label{metrique}
 \dd s^2 = g_{\mu \nu} \, \dd x^\mu \dd x^\nu = 
 \dd \tau^2 - \tau^2 \left( h_{ij} \, \dd x^i \dd x^j \right) \, , 
\end{equation}\noindent
où $h_{ij}$ est la métrique tri-dimensionnelle décrivant l'hypersurface
$\tau=1$ (encore appellée pseudo-sphère de rayon unité)
\eq
 h_{ij} = \mbox{diag} \, (1 \, , \, \sinh^2 \eta \, , \, 
 \sinh^2 \eta \, \sin^2 \theta) \, ,
\eq
et ($g$ et $h$ sont les déterminants de $g_{\mu \nu}$ et $h_{ij}$ respectivement)
\beq
\label{determinant}
 \sqrt{-g} = \tau^3 \, \sqrt{h} \, \, \, , \, \, 
 \sqrt{h} = \sinh^2 \eta \, \sin \theta \, .
\end{equation}\noindent

\subsubsection*{Laplaciens}

Le Laplacien quadri-dimensionnel est défini par
\bear
 \Delta^{(4)} \equiv \nabla_\mu \nabla^\mu & = & \frac{1}{\sqrt{-g}} \,
 \p_\mu \, \sqrt{-g} \, g^{\mu\nu} \, \p_\nu \\ 
 & = & \frac{1}{\tau^3} \p_\tau \, \tau^3 \, \p_\tau - 
 \frac{1}{\tau^2} \Delta^{(3)}\, ,
\eear
où $\nabla_\mu$ désigne la dérivée covariante et $\Delta^{(3)}$ est le 
Laplacien tri-dimensionnel sur la pseudo-sphère unité
\eq
 \Delta^{(3)} = \frac{1}{\sqrt{h}} \, \p_i \, \sqrt{h} \, h^{ij} \, \p_j \, .
\eq
Afin d'exploiter la symétrie sphérique du problème, il est
utile de décomposer le champ sur l'ensemble complet des fonctions
propres de $\Delta^{(3)}$, définies par\footnote{On peut montrer que 
   $\mathcal Y_{\vec s} (\vec x) =
   u_{sl} (\eta) \, Y_l^m (\theta,\varphi)$, où $Y_l^m (\theta,\varphi)$ sont
   les harmoniques sphériques usuelles, et $u_{sl} (\eta)$ forment un ensemble
   complet de fonctions réelles, réguliéres à l'origine.}~\cite{Fulling}
\beq
\label{eigenfunc}
 \Delta^{(3)} \mathcal Y_{\vec s} (\vec x) = 
 - (s^2 + 1) \mathcal Y_{\vec s} (\vec x) \, ,
\end{equation}
où $\vec x \equiv (\eta , \theta , \varphi)$, $\vec s \equiv (s,l,m)$ ; s est 
une variable continue, sans dimension $0 < s < + \infty$, $l$ et $m$ sont 
des nombres entiers $0 \leq l < + \infty$ et $-l \leq m \leq l$. 
Les ``harmoniques hyperboliques'' $\mathcal Y_{\vec s}$ sont bien connues~\cite{BANDER}, elles forment 
une base orthogonale de l'espace des fonctions sur la pseudo-sphère unité. 
Nous choisirons leur normalisation comme suit
\beq
\label{ortho1}
 \int d^3x \, \sqrt{h} \, \mathcal Y_{\vec s \, '}^* (\vec x) \,
 \mathcal Y_{\vec s} (\vec x) = \delta^{(3)} (\vec s - \vec s \, ') \equiv
 \delta (s - s') \, \delta_{ll'} \, \delta_{mm'} \, .
\end{equation}\noindent
Dans la suite nous utiliserons les propriétés 
suivantes~\cite{Lamperthesis,BANDER}
\beq
\label{ortho2}
 \int d^3s \, \mathcal Y_{\vec s}^* (\vec x \, ') \, \mathcal Y_{\vec s} (\vec x) =
 \frac{\delta^{(3)} (\vec x - \vec x \, ')}{\sqrt{h}}  \equiv 
 \frac{\delta (\eta - \eta') \, \delta (\theta - \theta') \, 
 \delta (\varphi - \varphi')}{\sqrt{h}} \, ,
\end{equation}\noindent
\beq
\label{propriete}
 \mathcal Y_{\vec s}^* = (-1)^m \, \mathcal Y_{-\vec s} \, \, \, \, , \, \, \, \,
 \sum_{lm} \, | \mathcal Y_{\vec s} (\vec x) |^2 = \frac{s^2}{2 \pi^2} \, ,
\end{equation}\noindent
avec les notations
\eq
 \int d^3s \equiv \int_0^{+\infty} ds \, \, \sum_{lm} \, \, \, \, , \, \, 
 -\vec s = (s,l,-m) \, .
\eq
\par
Notre observateur en co-mouvement mesure des observables physiques, comme
par exemple le nombre de particules, à des ``instants de temps propre'' $\tau$.
Pour décrire son système, il lui faut donc quantifier la théorie sur les
hypersurfaces $\tau=\mbox{cte}$. Le formalisme adapté à ce problème est celui
de la théorie quantique des champs en espace courbe~\cite{Birrell}. Nous allons
suivre ici les Refs.~\cite{Lampert,Fulling} 
(voir aussi~\cite{Lamperthesis}). 

\subsubsection*{Quantification}

Les équations dynamiques à l'ordre dominant en $1/N$ dans la
géométrie en expansion avec $\bfphi (x) = \bfphi (\tau)$ s'écrivent
\bearn
\label{mean}
 \ddot\bfphi (\tau) + \frac{3}{\tau} \dot\bfphi (\tau) + 
 \chi (\tau) \, \bfphi (\tau) & = & H {\bf n}_\sigma \, ,  \\
\label{fluc} 
 \left( \Delta^{(4)} + \chi (\tau) \right) \, \bfdphi (\tau,\vec x) & = & 
 {\bf 0} \, ,
\eearn
où le point dénote la dérivée temporelle, et où
\beq 
\label{chi}
 \chi (\tau) = \lambda \left( \phi^2 (\tau) + 
 \langle \dphi^2 \rangle (\tau) - v^2 \right) \, . 
\end{equation}\noindent
Les différentes composantes chirales $\dphi_a$ de la fluctuation étant
traitées sur un pied d'égalité dans cette approximation, nous concentrons
notre attention sur une d'entre elles et omettons les indices
chiraux pour simplifier les notations. Nous mentionnerons au fur et à mesure
les modifications (évidentes) de nos formules quand ces indices sont réintroduits.
\par
Tout d'abord, il est judicieux d'introduire les variables adimensionnées
\beq
\label{reduite}
 \varphi (\tau,\vec x) = \tau \, \dphi (\tau,\vec x) \, \, \, , \, \, \,
 \pi (\tau,\vec x) = \tau \, \dot\varphi (\tau,\vec x) \, .
\end{equation}\noindent
En effet, dans la suite nous parlerons des représentations
de Schrödinger et de Heisenberg, et nous aurons besoin de définir
l'opérateur unitaire qui connecte ces deux représentations. Or, on
peut montrer (voir Annexe~\ref{EXPANSION}) qu'en termes des variables
(\ref{reduite}) le problème s'exprime comme un ensemble (infini)
d'oscillateurs harmoniques dont les fréquences dépendent du temps 
(voir Eq.~(\ref{disp}) plus bas). En définissant la représentation 
de Schr\"odinger à l'instant de référence $\tau_0$, et en dénotant 
par $U (\tau,\tau_0)$ l'opérateur d'évolution, on a
\bearn
\label{HSphi}
 \varphi (\tau) & = & U (\tau,\tau_0) \, \varphi (\tau_0) \, 
 U^{-1} (\tau,\tau_0) \, , \\
\label{HSpi}
 \pi (\tau) & = & U (\tau,\tau_0) \, \pi (\tau_0) \, 
 U^{-1} (\tau,\tau_0) \, , 
\eearn
où l'on a omis la dépendance spatiale. 
De là on tire les relations 
\bear
 \dphi (\tau) & = & \frac{\tau_0}{\tau} \, U (\tau,\tau_0) \, \dphi (\tau_0) \, 
 U^{-1} (\tau,\tau_0) \, , \\
 \dot{\dphi} (\tau) & = & \frac{\tau_0^2}{\tau^2} \, U (\tau,\tau_0) \, 
 \dot{\dphi} (\tau_0) \, U^{-1} (\tau,\tau_0) \, . \\
\eear
On peut construire explicitement\footnote{Il s'agit d'une construction
   formelle faisant intervenir les solutions de l'équation du mouvement
   (cf. Eq.~(\ref{OH}). Il est toutefois intéressant de noter que l'on
   peut choisir arbitrairement les conditions initiales, pourvu que
   celles-ci satisfassent la contrainte de Wronskien (Eq.~(\ref{wronsk}).} 
l'opérateur d'évolution $U$ pour un potentiel quadratique général~\cite{Combescure}.
Cependant, pour notre propos, nous n'avons pas besoin de cette construction,
seules les lois de transformation ci-dessus nous seront utiles.
\par
La décomposition du champ sur les modes $\vec s$ s'écrit
\eq
 \varphi (\tau,\vec x) = 
 \int d^3s \, \left( a_{\vec s} \, \psi_s (\tau) \, \mathcal Y_{\vec s} (\vec x) +
 a_{\vec s}^{\dagger} \, \psi_s^* (\tau) \, \mathcal Y_{\vec s}^* (\vec x) \right)
 \, .
\eq
Définissons les projections $\varphi_{\vec s}$ et $\pi_{\vec s}$ 
du champ et de son moment conjugué\footnote{Les fonctions modes, ainsi que 
   les opérateurs de création et d'anihilation portent un indice 
   chiral : $\psi_s^a$, et $a_{a,\vec s}$.
   Les relations de commutation sont : 
   \eq
    \left[ a_{a,\vec s} \, ; \, a_{b,\vec s'}^{\dagger} \right] = 
    \delta_{ab} \, \delta^{(3)} (\vec s - \vec s') \, .
   \eq }
\bearn
\label{modephi}
 \varphi_{\vec s} (\tau) & = & \int d^3 x \, \sqrt{h} \, 
 \mathcal Y_{\vec s}^* (\vec x) \, \varphi (\tau,\vec x) = 
 \psi_s (\tau) \, a_{\vec s} + 
 \psi_s^* (\tau) \, (-1)^m \, a_{-\vec s}^{\dagger} \, \, ,\\
\label{modepi}
 \pi_{\vec s} (\tau) & = & \int d^3 x \, \sqrt{h} \, 
 \mathcal Y_{\vec s}^* (\vec x) \, \varphi' (\tau,\vec x) = 
 \psi_s' (\tau) \, a_{\vec s} + 
 \psi_s^{*'} (\tau) \, (-1)^m \, a_{-\vec s}^{\dagger} \, \, ,
\eearn
où le prime désigne la dérivation par rapport à $u = \ln (m_\pi \tau)$ : 
$\psi' = \tau \dot\psi$. Les opérateurs de création et d'anihilation,
$a_{\vec s}^{\dagger}$ et $a_{\vec s}$ satisfont aux relations de commutation
\beq
\label{commut}
 \left[ a_{\vec s} \, ; \, a_{\vec s \, '}^{\dagger} \right] = 
 \delta^{(3)} (\vec s - \vec s \, ') \, ,
\end{equation}\noindent
lea autres commutateurs étant nuls.
Les ``fonctions mode'' $\psi_s$ satisfont aux équations 
différentielles
\beq
\label{mode}
 \ddot\psi_s (\tau) + \frac{1}{\tau} \, \dot\psi_s (\tau) +
 \omega_s^2 (\tau) \, \psi_s (\tau) = 0 \, ,
\end{equation}\noindent
que l'on peut réécrire
\beq
\label{OH}
 \psi_s'' (\tau) + \tomega_s^2 (\tau) \, \psi_s (\tau) = 0 \, ,
\end{equation}\noindent
où $\tomega_s = \tau \omega_s$ et
\beq
\label{disp}
 \omega_s (\tau) = \sqrt{ \frac{s^2}{\tau^2} + \chi (\tau)} \, .
\end{equation}\noindent
On voit que le système se réduit à un ensemble d'oscillateurs Eq.(\ref{OH}),
couplés entre eux d'une part, et avec le condensat d'autre part, à travers 
le terme de masse $\sqrt{\chi (\tau)}$. L'effet de ce couplage se traduit
uniquement par la dépendance en temps de cette masse. Les fréquences physiques 
sont données par (\ref{disp}) (voir Annexe~\ref{EXPANSION}).
Le champ de fluctuation $\dphi$ et son moment conjugué $\Pi = \sqrt{-g} \, 
\dot{\dphi}$ (voir Annexe~\ref{EXPANSION}) satisfont aux relations de 
commutation canoniques à temps propres égaux $\left[ \dphi (\tau,\vec x) \, ; \, 
\Pi (\tau,\vec x') \right] = i \, \delta^{(3)} (\vec x - \vec x')$. Celles-ci
sont équivalentes à (\ref{commut}) si les fonctions mode satisfont la condition
de Wronskien
\beq
\label{wronsk}
 W_s = \psi_s (\tau) \, \psi_s^{*'} (\tau) - 
 \psi_s^* (\tau) \, \psi_s' (\tau) = i \, .
\end{equation}\noindent
\par
Pour compléter la procédure de quantification, il faut spécifier les
conditions initiales pour les fonctions modes, ainsi que pour leur dérivées.
En effet, différents choix pour les $\psi_s$ correspondent à différentes 
définitions des quasi-particules, ou, de façon équivalente, à différentes
définitions du vide.
Nous suivons ici les auteurs de la Réf.~\cite{Lampert}, en
choisissant le vide adiabatique d'ordre zéro~\cite{Fulling,Birrell} (voir
Annexe~\ref{EXPANSION}), c'est à dire
\eq
 \psi_s (\tau_0) = g_s (\tau_0) \, \, \, , \, \, \,
 \psi_s' (\tau_0) = g_s' (\tau_0) \, ,
\eq
avec
\beq
\label{adiab}
 g_s (\tau) = \frac{1}{\sqrt{2 \tomega_s (\tau)}} \, 
 \exp \left( \int_{\tau_0}^\tau d\tau' \, \omega_s (\tau') \right) \, ,
\end{equation}\noindent
Il est commode pour la suite d'introduire la notation
\bearn
\label{adiab0}
 \psi_s^{(0)} (\tau) & = & \frac{1}{\sqrt{2 \tomega_s (\tau)}} \, , \\
\label{adiab1}
 \psi_s^{(1)} (\tau) & = & 
 - \left[ \frac{\tomega_s \, ' (\tau)}{2 \tomega_s (\tau)} +
 i \, \tomega_s (\tau) \right] \, \psi_s^{(0)} (\tau) \, .
\eearn
La condition initiale pour les fonctions mode s'écrit alors
\beq
\label{init}
 \psi_s (\tau_0) = \psi_s^{(0)} (\tau_0) \, \, \, , \, \, \,
 \psi_s' (\tau_0) = \psi_s^{(1)} (\tau_0) \, .
\end{equation}\noindent
L'approximation adiabatique peut être vue comme une définition du vide physique.
Le problème de la définition d'un ``bon'' vide vient de la dépendance en temps 
de la métrique, Eq.~(\ref{metrique}). Cette dernière, couplée au champ quantique 
par l'intermédiaire du terme $\sqrt{-g}$ dans l'action, agit comme un champ 
classique qui rayonne des particules. Un état initalement vide, se remplit
de quanta au fur et à mesure que le temps passe. Ce problème est brièvement
discuté dans l'Annexe~\ref{EXPANSION}. Pour résumer, l'approximation 
adiabatique consiste à minimiser le taux de production de particules dû
à l'expansion. Le choix des conditions initiales (\ref{init}) n'est cependant
pas essentiel pour notre problème, ce que nous avons vérifié en essayant
différents choix. Mentionnons toutefois que la condition adiabatique permet
de renormaliser la théorie de manière non-ambig\"ue, indépendante du temps.

\subsubsection*{Renormalisation}

L'équation de gap (\ref{chi}) fait intervenir la valeur moyenne du 
produit de deux opérateurs au même point $\langle \dphi^2 (\tau,\vec x) \rangle$.
Cette quantité diverge et il est nécessaire de régulariser la théorie,
par exemple en introduisant une coupure ultraviolette $\Lambda$.
Il est bien connu que ce type de modèle devient trivial (la constante
de couplage tend vers zéro) quand $\Lambda \rightarrow \infty$,
on doit donc garder $\Lambda$ fini. Comme nous l'avons discuté plus haut, 
le modèle $\sigma$-linéaire est une approximation de la dynamique des 
excitations bosoniques de grande longueur d'onde de QCD. C'est un modèle 
effectif de basse énergie qui n'a de sens physique que s'il est défini avec
une coupure finie $\Lambda \lesssim 1$~GeV, qui doit être vue comme un paramètre.
Cependant, il est utile d'isoler les divergences et de les réabsorber
dans la définition de quantités physiques de façon à minimiser
la dépendance des résultats avec $\Lambda$. C'est dans ce sens qu'il
faut comprendre la ``renormalisation'' décrite ci-dessous.
\par
Afin d'isoler et de traiter les divergences de l'équation de gap,
il est nécessaire de spécifier l'état du système sur lequel est prise
la valeur moyenne. Anticipant sur la suite, nous écrivons
\eq
 \langle a_{a,\vec s}^{\dagger} \, a_{b,\vec s \, '} \rangle =
 n_s \, \delta_{ab} \, \delta^{(3)} (\vec s - \vec s \, ') \, \, \, , \, \, \,
 \langle a_{a,\vec s} \, a_{b,\vec s \, '} \rangle = 0 \, ,
\eq
où l'on a réintroduit les indices chiraux. Nous supposerons un état 
d'équilibre thermique local à l'instant initial, de sorte que 
$n_s$ décroit exponentiellement vite pour les grandes valeurs de $s$.
En utilisant les Eqs.~(\ref{reduite}) et (\ref{modephi}), ainsi que
les propriétés (\ref{propriete}) et les relations de commutation
(\ref{commut}), on obtient
\eq
 \langle \dphi^2 (\tau,\vec x) \rangle = 
 \sum_{a=1}^N \langle \dphi_a^2 (\tau,\vec x) \rangle = 
 \frac{N}{\tau^2} \int_0^{+\infty} \frac{s^2 ds}{2 \pi^2} \, 
 ( 2 n_s + 1 ) \, |\psi_s (\tau)|^2 \, .
\eq
La partie divergente de cette intégrale vient de la contribution du vide
$\sim \int s^2 ds \, |\psi_s|^2$. Pour $s^2 \gg 1$, on écrit, 
en ligne avec l'approximation adiabatique\footnote{L'approximation
   adiabatique est exacte pour $s \rightarrow +\infty$, en effet, dans ce cas 
   on a $\tomega_s \simeq s$, et la solution de l'Eq.~(\ref{OH}) ayant pour 
   condition initiale (\ref{init}) s'écrit 
   \eq
    \psi_s \simeq \frac{1}{\sqrt{2s}} \, \exp [ -i \, s \ln (\tau/\tau_0) ] = 
    g_s (\tau) \, .
   \eq
   Remarquons que si le système, au cours de son évolution temporelle,
   traverse une période d'instabilité ($\chi < 0$), certaines fréquences
   ont des valeurs négatives pour cette période, et 
   $\int_{\tau_0}^{\tau} \omega_s$ n'est pas un nombre réel. Pour ces modes,
   l'ansatz adiabatique ne peut être utilisé en tant qu'approximation de 
   la fonction mode car alors, la relation (\ref{wronsk}) n'est pas satisfaite. 
   Cette objection ne s'applique cependant pas pour les modes de haute énergie
   qui nous intéressent ici, et dont les fréquences sont toujours réelles.},
\eq
 \psi_s (\tau) = g_s (\tau) + \mbox{reste,}
\eq
où $g_s$ est donnée par l'Eq.~(\ref{adiab}), et où le ``reste'' ne donne 
pas de contribution singulière. Ceci nous permet d'isoler
la partie divergente\footnote{La coupure $\Lambda$ s'interprete comme
   l'impulsion maximale permise. Ici l'impulsion est $s/\tau$ 
   (cf. Eq.~(\ref{disp})), l'intégrale sur $s$ est donc coupée à 
   $s_{max} = \Lambda \tau$.}
\bear
 \frac{1}{\tau^2} \,\int^{\Lambda \tau} s^2 ds \, | g_s (\tau) |^2 & = & 
 \frac{1}{\tau^2} \,\int^{\Lambda \tau} 
 \frac{s^2 ds}{2 \sqrt{s^2 + \tau^2 \chi (\tau)}} \\
 & = & \int^{\Lambda} \frac{k^2 dk}{2 \sqrt{k^2 + \chi (\tau)}} \, .
\eear
Cette dernière intégrale est calculée dans l'Annexe~\ref{INTEGRALE}, on a
\eq
 \langle \dphi^2 \rangle (\tau) = \frac{N}{8 \pi^2} \, 
 \left( \Lambda^2 - \chi \, \ln \frac{\Lambda}{\sqrt{|\chi|}} \right) + ... \, ,
\eq
où $...$ représente des termes dépendant faiblement de $\Lambda$. Ceci est
valable quel que soit le signe de $\chi$.
La dépendance quadratique dans la coupure est indépendante du temps.
Elle est facilement éliminée en soustrayant à la masse effective au 
carré $\chi$ sa valeur dans le vide, qui n'est autre que le carré de la masse 
physique du pion. En effet, dans le vide on a
$\bfphi (\tau) \equiv (f_\pi,{\bf 0})$ et l'équation de gap s'écrit
\eq
 m_\pi^2 = 
 \lambda \left( \, f_\pi^2 + \langle \, 0 \, | \, \dphi^2 \, | \, 0 \, \rangle - 
 v^2 \, \right) \, ,
\eq
avec
\bear
 \langle \, 0 \, | \, \dphi^2 \, | \, 0 \, \rangle & = & 
 N \, \int_0^{\Lambda} \frac{k^2 dk}{2 \pi^2} \, \frac{1}{2 \sqrt{k^2 + m_\pi^2}} \\
 & = & \frac{N}{8 \pi^2} \, 
 \left( \Lambda^2 - m_\pi^2 \, \ln \frac{\Lambda}{m_\pi} \right) + ... \, .
\eear
En exprimant $\chi$ en fonction de $m_\pi^2$, ce qui revient à absorber
la divergence dans le paramètre $v \equiv v (\Lambda)$, on s'affranchit 
du terme quadratique en $\Lambda$ :
\eq
 \frac{1}{\lambda} (\chi (\tau) - m_\pi^2) = \phi^2 (\tau) - f_\pi^2 \, + \,
 \langle \, \dphi^2 \, \rangle (\tau) - 
 \langle \, 0 \, | \, \dphi^2 \, | \, 0 \, \rangle \, .
\eq
La dépendance logarithmique restante est éliminée en définissant la
constante de couplage renormalisée $\lambda_R$, telle que 
(voir Annexe~\ref{INTEGRALE})~\cite{CJP,Lamperthesis}
\bear
 \frac{1}{\lambda_R} & = & \frac{1}{\lambda} + \frac{N}{8 \pi^2} \,
 \int_0^{\Lambda} \frac{k^2 dk}{( k^2 + m_\pi^2 )^{3/2}} \\
 & = & \frac{1}{\lambda} + \frac{N}{8 \pi^2} \, \ln \frac{\Lambda}{m_\pi} + ...
\eear
Comme promis, la procédure de renormalisation est indépendante du temps.
L'équation de gap est entièrement exprimée en terme de quantités physiques
mesurables, et donc indépendantes de la coupure $\Lambda$, qui doivent
être tirées de l'expérience. Nous utiliserons les valeurs proposées dans
les Réfs.~\cite{Lampert,Lamperthesis}\footnote{Dans ces Réfs. 
   (voir aussi~\cite{Cooper}), la constante de couplage renormalisée 
   $\lambda_R$ est extraite de l'amplitude de diffusion 
   $\pi \pi \rightarrow \pi \pi$ dans l'onde $s$ (moment angulaire $l=0$), 
   dans la voie isoscalaire (isospin total $I=0$).}
de façon à pouvoir vérifier nos calculs en les comparant aux leurs
\bear
 m_\pi & = & 139.5 \, \mbox{MeV} \, , \\
 f_\pi & = & 92.5 \, \mbox{MeV} \, , \\
 \lambda_R & = & 7.3 \, .
\eear 
\par
Nous concluons cette partie par une remarque concernant les valeurs possibles
de la coupure. La constante de couplage renormalisée dépend de l'échelle
d'énergie $\mu$ à laquelle on la mesure, ici nous avons choisi $\mu=m_\pi$.
Réécrivons la relation entre la constante de couplage nue 
$\lambda (\Lambda)$ et la constante de couplage renormalisée $\lambda_R (\mu)$
sous la forme
\eq
 \lambda (\Lambda) = \frac{\lambda_R (\mu)}{1 - \lambda_R (\mu)\frac{N}{8 \pi^2}
 I_3(\Lambda,\mu)} \, ,
\eq
où $I_3(\Lambda,\mu) \equiv I_3(\Lambda/\mu)$ est l'intégrale définie plus 
haut (voir aussi Annexe~\ref{INTEGRALE}). On voit que pour 
$\lambda_R > 0$\footnote{Si $\lambda_R < 0$, $\lambda < 0$ et la théorie
   est instable},
il existe une valeur critique finie $\Lambda_c$ de la coupure pour laquelle la 
constante de couplage nue diverge. En supposant que $\Lambda_c/m_\pi \gg 1$,
on obtient $I_3 \simeq \ln (2\Lambda_c/m_\pi)-1$, c'est à dire ($N=4$,
$\lambda_R(m_\pi)=7.3$)
\eq 
 \frac{\Lambda_c}{m_\pi} \simeq \frac{1}{2} \, \mbox{e}^{1 + 8 \pi^2/N \lambda_R}
 \approx 20 \, .
\eq
C'est le pôle de Landau de la théorie $\Phi^4$ : pour $\Lambda > \Lambda_c$, 
la constante de couplage nue $\lambda (\Lambda)$ est négative, la théorie est 
instable. On doit donc prendre une valeur de la coupure inférieure au 
pôle de Landau. La valeur numérique de $\Lambda_c$ est de l'ordre
de $3$~GeV, ce qui est bien au delà de l'échelle typique d'énergie que
nous cherchons à décrire ici ($\lesssim 1$~GeV). Toujours en suivant les Réf.~\cite{Lampert,Lamperthesis}, 
nous prendrons
\eq
 \Lambda = 800 \, \mbox{MeV} \, .
\eq

\subsection{Particules physiques, champ interpolant}

Notre but est de calculer le nombre de pions produits dans l'état
final. Pour ce faire, il nous faut définir précisément ce qu'est
cette quantité. C'est le sujet de cette partie.
\par
Nous avons choisi, comme instant de référence, où les représentations de
Schr\"odinger et de Heisenberg coïncident, l'instant initial $\tau=\tau_0$.
Dans la représentation de Schr\"odinger, on a la décomposition suivante
\bearn
\label{modephischro}
 \varphi_{\vec s} & \equiv & \varphi_{\vec s} (\tau_0) = 
 \psi_s (\tau_0) \, a_{\vec s} + 
 \psi_s^* (\tau_0) \, (-1)^m \, a_{-\vec s}^{\dagger} \, \, ,\\
\label{modepischro}
 \pi_{\vec s} & \equiv & \pi_{\vec s} (\tau_0) = 
 \psi_s' (\tau_0) \, a_{\vec s} + 
 \psi_s^{*'} (\tau_0) \, (-1)^m \, a_{-\vec s}^{\dagger} \, \, .
\eearn
Avec la condition adiabatique (\ref{init}),  $a_{\vec s}^{\dagger}$
($a_{\vec s}$) s'interprète comme la représentation de Schr\"odinger de 
l'opérateur qui crée (anihile) un quanta de fréquence $\omega_s (\tau_0)$.
Ce sont les quantas appropriés à la description de l'état initial où le
système est un plasma de pions interagissant fortement les uns avec les 
autres. Dans l'approximation de champ moyen ce système en interaction
est remplacé par un ensemble d'excitations effectives libres de masse 
$\sqrt{\chi_0}$, en d'autres termes on ``diagonalise'' le problème en 
identifiant les excitations ou modes physiques du plasma. L'information
concernant l'interaction entre les pions est entièrement contenue dans 
la masse effective de ces quasi-particules. Il est bien clair que, du fait 
de l'expansion la définition des quasi-particules change avec le temps : 
les excitations physiques du système à l'instant $\tau_0$ ne sont plus
les excitations {\em physiques} (c'est à dire qu'elles ne diagonalisent 
plus le hamiltonien) à l'instant $\tau_0 + \delta\tau$. En fait 
les opérateurs de création et d'anihilation correspondant
aux quasi-particules à l'instant $\tau$ sont reliés à ceux correspondant 
aux quasi-particules à l'instant $\tau_0$ par le biais d'une transformation
unitaire dite de Bogoliubov (cf. Eq.~(\ref{App1_bogo}) dans
l'Annexe~\ref{EXPANSION}).
\par
Il existe une infinité de manière de décomposer le champ sous la forme
(\ref{modephischro})-(\ref{modepischro}), chacune correspondant à un
choix particulier d'opérateurs de création et d'anihiliation ou, de façon
équivalente, à un choix de la fonction mode et de sa dérivée (la seule
contrainte sur ces fonctions est qu'elle doivent satisfaire la relation
de Wronskien (\ref{wronsk})). Les quasi-particules physiques du système à 
l'instant final $\tau_f$ sont des quanta de fréquence $\omega_s (\tau_f)$. 
En dénotant par $b_{\vec s}^{\dagger}$ et $b_{\vec s}$ les opérateur de 
création et d'anihilation correspondant, dans la représentation de 
Schr\"odinger, on a
\bear
 \varphi_{\vec s} & = & \xi_s (\tau_0) \, b_{\vec s} + 
 \xi_s^* (\tau_0) \, (-1)^m \, b_{-\vec s}^{\dagger} \, \, ,\\
 \pi_{\vec s} & = & \xi_s' (\tau_0) \, b_{\vec s} + 
 \xi_s^{*'} (\tau_0) \, (-1)^m \, b_{-\vec s}^{\dagger} \, \, .
\eear
Les fonctions modes pour des quanta de fréquence $\omega_s (\tau_f)$
sont données par la condition adiabatique (\ref{adiab0})-(\ref{adiab1})
\bear
 \xi_s (\tau_0) & = & \psi_s^{(0)} (\tau_f) \, , \\
 \xi_s' (\tau_0) & = & \psi_s^{(1)} (\tau_f) \, .
\eear
Dans la représentation de Heisenberg, les opérateurs de champ (\ref{modephi}) 
et (\ref{modepi}) peuvent se réécrire (cf. (\ref{HSphi})-(\ref{HSpi}))
\bearn
\label{modephib}
 \varphi_{\vec s} (\tau) & = & \xi_s (\tau_0) \, b_{\vec s} (\tau) + 
 \xi_s^* (\tau_0) \, (-1)^m \, b_{-\vec s}^{\dagger} (\tau) \, \, ,\\
\label{modepib}
 \pi_{\vec s} (\tau) & = & \xi_s' (\tau_0) \, b_{\vec s} (\tau) + 
 \xi_s^{*'} (\tau_0) \, (-1)^m \, b_{-\vec s}^{\dagger} (\tau) \, \, ,
\eearn
où $b_{\vec s} (\tau) = U(\tau,\tau_0) \, b_{\vec s} \, U^{-1}(\tau,\tau_0)$
est la représentation de Heisenberg de $b_{\vec s}$. En utilisant les 
Eqs.~(\ref{modephi})-(\ref{modepi}), (\ref{modephib})-(\ref{modepib}), et 
(\ref{wronsk}), on trouve aisément la transformation de Bogoliubov connectant
les opérateurs $a_{\vec s}$, $a_{\vec s}^\dagger$ et $b_{\vec s} (\tau)$, 
$b_{\vec s}^\dagger (\tau)$ :
\eq
 b_{\vec s} (\tau) = \alpha_s (\tau) \, a_{\vec s} + 
 \beta_s (\tau) \, (-1)^m \, a_{-\vec s}^\dagger \, ,
\eq
avec
\bear
 i \, \alpha_s^* (\tau) & = & \xi_s (\tau_0) \, \psi_s^{*'} (\tau) -
 \xi_s' (\tau_0) \, \psi_s^* (\tau) \, , \\
 i \, \beta_s^* (\tau) & = & \xi_s (\tau_0) \, \psi_s' (\tau) -
 \xi_s' (\tau_0) \, \psi_s (\tau) \, .
\eear
On vérifie, à l'aide de (\ref{wronsk}), que cette transformation est
unitaire à chaque instant\footnote{La définition des quasi-particules que 
   nous avons adoptée ici diffère de celle de la Réf.~\cite{Lampert}, où la
   base dite ``adiabatique'' est utilisée de manière abusive. En effet,
   cette base n'a de sens que si les fréquences de tous les modes sont
   toujours réelles~\cite{Parker}, ce qui n'est pas le cas ici. 
   Quand $\chi<0$, la transformation reliant les opérateurs de la base 
   adiabatique à différents instants n'est pas unitaire pour les modes 
   $s^2<\tau^2\chi$, puisque le facteur exponentiel entrant dans la définition
   de la base adiabatique n'est pas une phase pure. Nous comprenons que,
   dans leur calcul, le facteur d'amplification a été obtenu à partir de la
   formule dérivée pour des fréquences réelles (le facteur de phase n'apparait
   pas dans la formule du facteur d'amplification). Avec la dérivation 
   présentée ici, nous obtenons la même formule pour le facteur 
   d'amplification dans le cas général (cf. Eq.~(\ref{amplification}).} :
\eq
 | \alpha_s (\tau) |^2 -| \beta_s (\tau) |^2 = 1 \, .
\eq
Bien sûr, bien que les formules ci-dessus soient valables pour tout $\tau$,
elles n'ont de signification physique que pour $\tau=\tau_f$. En effet,
les quanta créés par l'opérateur $b_{\vec s}^\dagger$ ne correspondent
aux excitations physiques du système qu'à cet instant (ce sont des excitations
de fréquence $\omega_s (\tau_f)$). Avec les notations
introduites ici, on peut écrire une formule valable pour toute valeur
de $\tau_f$ :
\beq
\label{bogolubov}
 b_{\vec s} (\tau_f) = \alpha_s (\tau_f) \, a_{\vec s} + 
 \beta_s (\tau_f) \, (-1)^m \, a_{-\vec s}^\dagger \, ,
\end{equation}\noindent
avec
\bearn
\label{bogoalpha}
 i \, \alpha_s^* (\tau_f) & = & \psi_s^{(0)} (\tau_f) \, \psi_s^{*'} (\tau_f) -
 \psi_s^{(1)} (\tau_f) \, \psi_s^* (\tau_f) \, , \\
\label{bogobeta}
 i \, \beta_s^* (\tau_f) & = & \psi_s^{(0)} (\tau_f) \, \psi_s' (\tau_f) -
 \psi_s^{(1)} (\tau_f) \, \psi_s (\tau_f) \, .
\eearn
\par
Le nombre de quasi-particules physiques à l'instant $\tau_f$ est facilement
obtenu. Nous verrons, dans la prochaine partie, que dans l'état initial
\eq
 \langle a_{a,\vec s}^{\dagger} \, a_{b,\vec s \, '} \rangle =
 n_s^a \, \delta_{ab} \, \delta^{(3)} (\vec s - \vec s \, ') \, \, \, , \, \, \,
 \langle a_{a,\vec s} \, a_{b,\vec s \, '} \rangle = 0 \, ,
\eq
où l'on a rétabli les indices chiraux. On obtient alors 
\beq
 \langle b_{a,\vec s}^{\dagger} (\tau_f) \, b_{b,\vec s \, '} (\tau_f) \rangle =
 n_s^a (\tau_f) \, \delta_{ab} \, \delta^{(3)} (\vec s - \vec s \, ')
 \, \, \, , \, \, \,
 \langle b_{a,\vec s} (\tau_f) \, b_{b,\vec s \, '} (\tau_f) \rangle = 0 \, ,
\end{equation}\noindent
et le nombre moyen de quasi-particules de composante chirale $a$ dans
l'état final est donné par
\beq
\label{number_final}
 n_s^a (\tau_f) + \frac{1}{2} = 
 \mathcal A_s (\tau_f) \, \left( n_s^a + \frac{1}{2} \right) \, ,
\end{equation}\noindent
où l'on a défini le facteur d'amplification
\beq
\label{amplification}
 \mathcal A_s (\tau_f) = 2 \, |\beta_s (\tau_f)|^2 + 1 \, 
\end{equation}\noindent
Nous verrons dans la suite que pour $\tau_f \gtrsim 10$ fm, le système
a atteint un régime stationnaire où $\chi (\tau_f) \simeq m_\pi^2$ : les
excitations physiques sont des pions libres, dont nous calculerons 
la multiplicité à l'aide de la formule (\ref{number_final}) ci-dessus.

\section{Échantillonnage des conditions initiales}

Cette partie constitue le c\oe ur de notre travail. Nous considérons une
petite bulle sphérique de matière chirale chaude, de rayon initial $R_0$, 
que nous supposons avoir été formée à un instant $t_0$ dans la région centrale 
d'une collision nucléaire à haute énergie, et qui subit une expansion sphérique
rapide. Nous supposons de plus que dans ce volume initial $V_0$, le champ
de pion fluctue dans l'ensemble thermique local. Les effets combinés des 
fluctuations initiales et de l'expansion rapide du système peuvent générer 
un ``accident'' : une période d'instabilité résultant dans l'amplification 
importante des modes de basse fréquence. Notre but est d'estimer la fréquence 
avec laquelle cet accident arrive, et plus précisément la probabilité d'avoir 
une amplification donnée. Le fait que cet accident ait lieu ou non dépend de 
la configuration initiale du champ de pion à l'intérieur de la bulle. Il nous 
faut donc échantillonner les configurations initiales possibles.
\par
Le modèle présenté dans les parties précédentes décrit la dynamique du 
champ de pion à l'intérieur de la bulle, c'est à dire dans le quadri-volume 
$\tau \ge \tau_0 = t_0$. Notre modélisation de l'expansion sphérique suppose
qu'à chaque instant $t$, le paramètre d'ordre est constant sur toute
la surface de la sphère. Ceci est assuré par le fait que la valeur du
paramètre d'ordre sur cette surface est la même que ce qu'elle était dans 
le petit volume initial, quand les différents points de la surface étaient 
en contact causal. Les conditions ``initiales'' sur l'hypersurface $\tau=\tau_0$,
dont nous avons besoin pour résoudre les équations du mouvement, sont donc
données par la configuration du champ de pion dans le volume $V_0$ au temps
$t_0$.
\par
Remarquons que l'idéalisation adoptée ici revient à supposer que la
surface de la bulle est indéfiniment couplée à un bain thermique
à la température $T$. Nous discuterons plus loin les implications
du fait que cette source d'énergie extérieure ne dure qu'un temps 
fini dans une situation plus réaliste. 

\subsection{Les conditions initiales}

Comment est caractérisée une configuration initiale du champ de pion ? 
Il est facile de se convaincre (cf. Eq.~(\ref{fluc}), voir aussi la 
Réf.~\cite{Boyanovsky}), que l'approximation de champ moyen correspond 
à une matrice densité $\rho (\tau_0) \equiv \rho_0$ gaussienne 
et qui plus est, diagonale dans l'espace des impulsions. L'état du système 
est donc complètement spécifié par la donnée des valeurs moyennes 
et ``largeurs'' de cette matrice densité :
\eq
 \phi_a (\tau_0) \, , \, \dot \phi_a (\tau_0) \, ,
\eq
et
\eq 
 \langle a_{a,\vec s}^{\dagger} \, a_{b,\vec s \, '} \rangle =
 N_s^a \, \delta_{ab} \, \delta^{(3)} (\vec s - \vec s \, ') \, \, \, , \, \, \,
 (-1)^m \, \langle a_{a,-\vec s} \, a_{b,\vec s \, '} \rangle = 
 P_s^a \, \delta_{ab} \, \delta^{(3)} (\vec s - \vec s \, ')\, ,
\eq
où $a,b=1,...,N$ ($N=4$). En principe les valeurs de $N_s^a$ et $P_s^a$ 
fluctuent d'un évènement à l'autre et doivent être tirées aléatoirement
dans l'ensemble thermique local. Cependant, en ce qui concerne 
l'évolution temporelle du système, ces quantités ne jouent un rôle
que dans la formule de la masse effective (\ref{chi}), où elles
apparaissent sous une intégrale sur tous les modes. Les fluctuations
statistiques sont alors moyennées à zéro et on peut remplacer ces
quantités par leurs valeurs moyennes
\eq
 \bar{N_s^a} = n_s \, \, \, , \, \, \, \bar{P_s^a} = 0 \, ,
\eq
où $n_s$ est la distribution de Bose-Einstein pour les quanta de 
fréquence $\omega_s (\tau_0)$ et est indépendant de la direction 
d'isospin dans l'approximation grand $N$ :
\eq
 n_s = \frac{1}{\mbox{e}^{\omega_s (\tau_0)/T} - 1} \, .
\eq
Les quantités $N_s^a$ et $P_s^a$ interviennent aussi dans le calcul). 
Ici, nous sommes intéressés à la multiplicité moyenne et nous remplacerons 
partout $N_s^a$ et $P_s^a$ par leurs valeurs moyennes.
\par
Les quanta de l'état initial sont définis par la donnée des fonctions 
modes $\psi_s (\tau_0)$ et de leur dérivées $\dot\psi_s (\tau_0)$, 
qui sont à leur tour complètement déterminées par les valeurs de 
la masse effective initiale $\chi (\tau_0) \equiv \chi_0$ et de 
sa dérivée $\dot\chi (\tau_0) \equiv \dot\chi_0$. On obtient ces
deux quantités à partir de l'équation de gap, en utilisant 
l'Eq.~(\ref{adiab}) (ci-dessous, on fait le changement de variable 
$k=s/\tau$) : $\chi_0$ est la solution de
\beq
\label{chi0}
 \chi_0 = \lambda \, \left[ \phi_0^2 - v^2  + N \, A_1 (\chi_0) \right] \, ,
\end{equation}\noindent
et $\dot\chi_0$ satisfait l'équation algébrique
\beq
\label{chi0dot}
 \dot\chi_0 \left[ \frac{1}{\lambda} + \frac{N}{4} A_3 (\chi_0) \right] = 
 2 \, \bfphi_0 \cdot \dot\bfphi_0 - 
 \frac{N}{2 \tau_0} \left[ 2 A_1 (\chi_0) + \chi_0 A_3 (\chi_0) -
 \frac{\Lambda^3}{2 \pi^2} 
 \frac{\coth [ E_\Lambda (\chi_0) / 2T ]}{E_\Lambda (\chi_0)} \right] \, ,
\end{equation}\noindent
avec $\bfphi_0 \equiv \bfphi (\tau_0)$ et 
$\dot\bfphi_0 \equiv \dot\bfphi (\tau_0)$, et où l'on a introduit 
les notations suivantes :
\beq
\label{notation1}
 E_k (\mu^2) = \sqrt{k^2 + \mu^2} \, 
\end{equation}\noindent
\beq
\label{notation2}
 A_p(\mu^2) = \int_0^\Lambda \frac{k^2 dk}{2 \pi^2} \, 
 \frac{1}{[E_k (\mu^2)]^p} \, \coth \left( \frac{E_k (\mu^2)}{2T} \right) \, .
\end{equation}\noindent
Il est facile de vérifier que l'Eq.~(\ref{chi0dot}), exprimée en termes
de la constante de couplage renormalisée $\lambda_R$, ne contient aucune
dépendance forte dans la coupure $\Lambda$, comme il se doit.
\par
En  conclusion, on voit que la configuration initiale du système est 
complètement spécifiée par la donnée des huit nombres réels
$\{ \bfphi_0 \, , \, \dot\bfphi_0 \}$. Un exemple en est donné dans la 
Fig.~\ref{fig_chi}, qui représente la variation temporelle de la masse
effective au carré $\chi (\tau)$ pour un des choix des valeurs initiales du 
paramètre d'ordre et de sa dérivée faits dans~\cite{Lamperthesis}. 
Dans cet exemple $\chi$ prend des valeurs négatives et on assiste à une 
forte amplification des modes de basse fréquence. La Fig.~\ref{fig_amplif} 
montre le facteur d'amplification $\mathcal A_s (\tau_f)$ correspondant
(cf. Eq.~(\ref{amplification})) en fonction de $s$ pour $\tau_f=5$, $10$ et 
$15$~fm.

\begin{figure}[htbp]
\epsfxsize=4.in \centerline{ \epsfbox{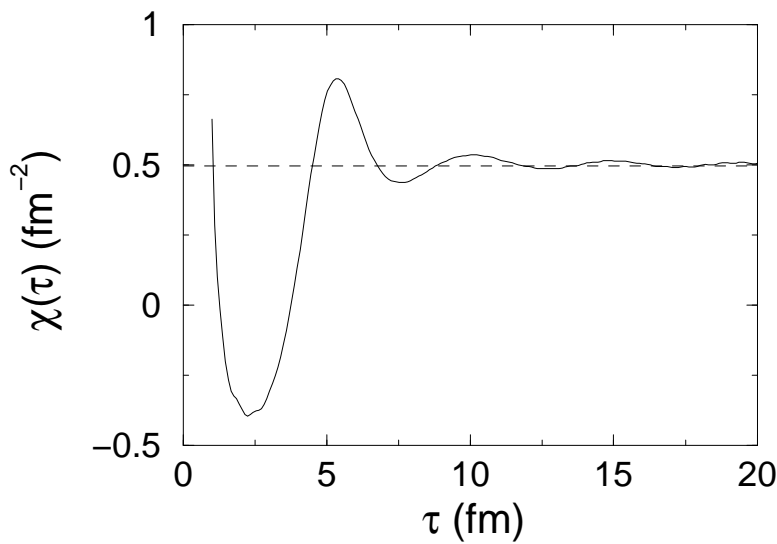}}
\caption{\small Evolution temporelle de $\chi (\tau)$ (en fm$^{-2}$)
pour la consition initiale (cf.~\cite{Lamperthesis,Lampert}) 
$\bfphi (\tau_0) = (0.3,{\bf 0})$ (en fm$^{-1}$), 
$\dot\bfphi (\tau_0) = (-1,{\bf 0})$ (en fm$^{-2}$), à $\tau_0=1$~fm. 
Les paramètres sont $\Lambda=800$ MeV, $m_\pi=139.5$ MeV, $f_\pi=92.5$ MeV, $\lambda_R=7.3$. La ligne en tirets montre la valeur de $m_\pi^2=0.5$~fm$^{-2}$.} 
\label{fig_chi}
\end{figure}
\begin{figure}[htbp]
\epsfxsize=4.in \centerline{ \epsfbox{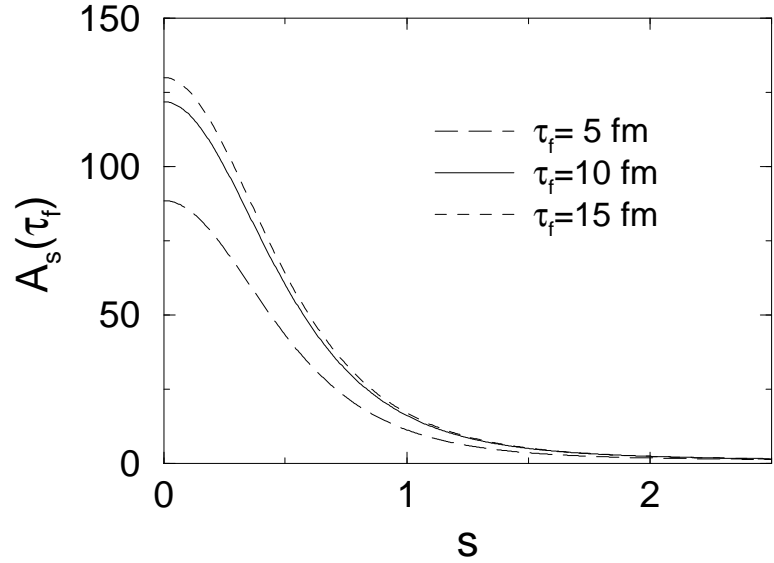}}
\caption{\small Le facteur d'amplification $\mathcal A_s (\tau_f)$
correspondant à l'évènement de la Fig.~\ref{fig_chi}, pour différentes valeurs
de $\tau_f$.} 
\label{fig_amplif}
\end{figure}

Dans la littérature, les valeurs des conditions initiales sont 
généralement choisies de manière arbitraire. Dans la partie suivante, 
nous allons montrer que l'hypothèse, largement utilisée, d'un état 
initial en équilibre thermique local est suffisante pour déterminer
la probabilité pour que ces nombres prennent des valeurs données.

\subsection{Équilibre thermique local}

Comme nous l'avons vu plus haut, les conditions sur l'hypersurface 
$\tau=\tau_0$ sont données par la configuration du champ à l'intérieur 
de la bulle de rayon $R_0$, en équilibre thermique local à l'instant $t_0$. 
Dans cette partie, nous focalisons notre attention sur cette dernière. 
Nous utiliserons des notations conventionnelles, à ne pas confondre avec 
les notations utilisées dans le reste du chapitre : $\vec x$ désigne la 
position dans le système de coordonnées cartésiennes, $\vec x = (x,y,z)$, 
$d^3x = dx dy dz$. Nous omettrons d'écrire la variable temporelle, partout
égale à $t_0$.
\par
Pour exposer notre argument de façon claire, nous négligeons dans un 
premier temps les complications dues à la mécanique quantique et 
considérons l'exemple d'un champ scalaire classique, sans structure 
d'isospin, en équilibre thermique local à la température $T$ dans une
petite boule de volume $V_0$. Cela signifie que tout se passe 
comme si cette boule était une sous-partie d'un grand système de volume 
$V \gg V_0$ en équilibre thermodynamique à la même température $T$. 
Dans ce grand système fictif, en l'absence de corrélations à longue 
distance (ce qui est le cas pour des interactions à courte portée) 
la variance des fluctuations statistiques de la moyenne spatiale du 
champ sur le volume total $V$ est très petite, d'ordre $1/V$ (elle 
est nulle dans la limite thermodynamique $V \rightarrow \infty$, cf.
Eq.~(\ref{varphiapprox})). La variance correspondante pour la moyenne 
spatiale du champ {\em sur le volume $V_0$} est plus grande d'un 
facteur $V/V_0$ (pourvu que le rayon de la boule soit au moins de 
l'ordre de la longueur de corrélation thermique\footnote{Considérons 
   le système de volume $V$, et divisons le en $k=V/V_0$ sous-parties 
   identiques de volume $V_0$.
   La moyenne spatiale du champ dans chacune de ces sous-parties est
   \eq
    \bar\phi_i = \frac{1}{V_0} \int_{V_i} d^3x \, \phi(\vec x) \, \, , \, \,
    i=1,2,...,k \, .
   \eq
   Choisissons le rayon $R_0$ des cellules au moins égal à la 
   longueur de corrélation du champ à l'équilibre thermodynamique à la 
   température $T$, $R_0 \gtrsim \lambda_T$. Ainsi, par construction,
   les $\bar\phi_i$ sont des variables aléatoires indépendantes, 
   ayant toutes la même distribution de probabilité. La somme de
   ces variables aléatoires, $S = \sum_{i=1}^k \, \bar\phi_i$, est
   une variable aléatoire dont la moyenne statistique est égale à
   la somme des moyennes statistiques des $\bar\phi_i$, et dont
   la variance est égale à la somme des variances des $\bar\phi_i$ : 
   $\mbox{E}[S] = k \mbox{E}[\bar\phi]$, et 
   $\mbox{Var}[S] = k \mbox{Var}[\bar\phi]$. La somme $S$ n'étant 
   autre que la moyenne spatiale du champ sur le volume total $V$,
   à un facteur $k$ près : $\bar\phi_V = S/k$, on en déduit
   \eq
    \mbox{E}[\bar\phi] = \mbox{E}[\bar\phi_V] \, \, \, \, , \, \, \, \,
    \mbox{Var}[\bar\phi] = \frac{V}{V_0} \mbox{Var}[\bar\phi_V] \, .
   \eq}) $\lambda_T = 1/\sqrt{\chiT}$. Le même raisonnement s'applique 
à la dérivée temporelle du champ moyennée sur le volume $V_0$. Le point 
essentiel est que, dans le système en  équilibre local, les valeurs du 
champ et de sa dérivée, moyennés sur le volume $V_0$ de la bulle, fluctuent 
dans l'ensemble canonique et donc de manière parfaitement prédictible, 
pourvu que le rayon initial $R_0$ ne soit pas trop petit.
\par
Appliquons ces idées à notre problème. Dans la théorie quantique, on définit
les opérateurs ($a$ désigne les composantes chirales)
\bear
 \bar\Phi_a  & = & \frac{1}{V_0} \int_{V_0} d^3x \, \Phi_a (\vec x) \, ,\\
 \bar{\dot\Phi}_a & = & \frac{1}{V_0} \int_{V_0} d^3x \, \dot\Phi_a (\vec x) \, .
\eear
Ceux-ci ne reçoivent pas de contribution de la part des modes de petite
longueur d'onde, moyennés à zéro. Un observateur vivant dans la boule
de volume $V_0$, et mesurant ces observables, identifiera donc le résultat 
de sa mesure avec les valeurs du paramètre d'ordre et de sa dérivée 
respectivement. La distribution des valeurs possibles $\bar\phi_a$ et 
$\bar{\dot\phi}_a$ des résultats de cette mesure est donnée par la
distribution de Wigner\footnote{La distribution de probabilité de la valeur
   mesurée d'une observable $\mathcal  O$ est déterminée par la fonction
   caractéristique $\langle \mbox{e}^{i k \mathcal O} \rangle$.} 
(voir Annexe~\ref{THERMAL}), caractérisée par les moyennes
\bearn
\label{moyphi}
 \mbox{E} \left[ \bar\phi_a \right] & = &  
 \langle \, \bar\Phi_a \, \rangleT \, , \\
\label{moyphidot}
 \mbox{E} \left[ \bar{\dot\phi}_a \right] & = &  
 \langle \, \bar{\dot\Phi}_a \, \rangleT \, ,
\eearn
les variances\footnote{Tous les termes non-diagonnaux dans les indices 
    chiraux sont nuls dans l'approximation grand $N$}
\bearn
\label{varphi}
 \mbox{Var} \left[ \bar\phi_a \right] & = &  
 \langle \, \bar\Phi_a^2 \, \rangleT  - 
 \langle \, \bar\Phi_a \, \rangleT^2 \, , \\
\label{varphidot}
 \mbox{Var} \left[ \bar{\dot\phi}_a \right] & = &  
 \langle \, \bar{\dot\Phi}_a^2 \, \rangleT  - 
 \langle \, \bar{\dot\Phi}_a \, \rangleT^2 \, ,
\eearn
et par la covariance
\beq
\label{cov}
 \mbox{Cov} \left[ \bar\phi_a , \bar{\dot\phi}_a \right] = \frac{1}{2} \, 
 \langle \, \bar\Phi_a \bar{\dot\Phi}_a + \bar{\dot\Phi}_a \bar\Phi_a \, \rangleT -
 \langle \, \bar\Phi_a \, \rangleT \, \langle \, \bar{\dot\Phi}_a \, \rangleT
\end{equation}\noindent
où $\langle \mathcal O \rangleT = Tr ( \rho \mathcal O )$ désigne la
moyenne sur l'ensemble thermique ($\rho \propto \exp (-H/T)$, où $H$ est 
le hamiltonien du système). Le calcul des quantités ci-dessus est détaillé
dans l'Annexe~\ref{THERMAL}. L'état d'équilibre thermodynamique est homogène et 
invariant par translation dans le temps. En dénotant par $\bfphiT$ la valeur du 
paramètre d'ordre dans cet état et $\chiT$ celle de la masse effective au 
carré, on obtient, à l'aide de l'Eq.~(\ref{mean}),
\beq
\label{thermik}
 \bfphiT = \frac{H}{\chiT} \, {\bf n}_\sigma \, \, , \, \, \mbox{et} \, \, \,
 \dot\bfphiT = {\bf 0} \, ,
\end{equation}\noindent
d'où on déduit
\bearn
\label{moyphi1}
 \mbox{E} \left[ \bar\phi_a \right] & = &  \frac{H}{\chiT} \, \delta_{a0} \, , \\
\label{moyphidot1}
 \mbox{E} \left[ \bar{\dot\phi}_a \right] & = &  0 \, .
\eearn
L'équation de gap thermique, dont $\chiT$ est la solution, s'écrit 
(cf. Eq.~(\ref{notation2}))
\beq
\label{gapthermik}
 \chiT = \lambda \, \left[ \phiT^2 - v^2  + N \, A_1 (\chiT) \right] \, ,
\end{equation}\noindent
où $\phiT^2$ est donné par l'Eq.~(\ref{thermik}).
Dans l'approximation de champ moyen, le système en interaction est remplacé
par un ensemble d'excitations libres de masse effective $\sqrt{\chiT}$,
c'est à dire que le hamiltonien effectif est quadratique et la matrice 
densité de l'ensemble thermique $\rho \propto \exp (-H/T)$ est gaussienne. 
La distribution de probabilité des valeurs possibles de $\bar\bfphi$ et
$\bar{\dot\bfphi}$ est donc complètement déterminée par les paramètres
(\ref{moyphi})-(\ref{varphidot}), la covariance étant nulle. 
\par
Quand le rayon $R_0$ est grand devant la longueur de corrélation
dans l'ensemble thermique à température $T$, $\lambda_T = 1/\sqrt{\chiT}$,
les fluctuations à l'intérieur de la bulle sont indépendante de ce qui 
se passe à l'extérieur, et on peut donner une expression analytique
pour les variances :
\bearn
\label{varphiapprox}
 \mbox{Var} \left[ \bar\phi_a \right] & \simeq & \frac{1}{2 V_0 \sqrt{\chiT}} \, 
 \coth \left( \frac{\sqrt{\chiT}}{2T} \right) \, , \\
\label{varphidotapprox}
 \mbox{Var} \left[ \bar{\dot\phi}_a \right] & \simeq & \frac{\sqrt{\chiT}}{2 V_0} \, 
 \coth \left( \frac{\sqrt{\chiT}}{2T} \right) \, .
\eearn
La dispersion de la variable $\bar\phi_a$, calculée numériquement (voir
Annexe~\ref{THERMAL}), est plus petite d'un facteur $2$ ($1.5$) que ce 
que donne la formule ci-dessus pour $R_0=\lambda_T$ ($R_0=2\lambda_T$), 
pour des températures allant de $T=200$ à $400$ MeV. Pour la variable
$\bar{\dot\phi}_a$, l'écart entre le résultat analytique ci-dessus et 
le calcul exact est de $20\%$ ($8\%$). Cet écart augmente rapidement 
pour $R_0 < \lambda_T$. Les fluctuations sont approximativement 
indépendantes de ce qui se passe hors de la bulle, pourvu que 
$R_0 \gtrsim \lambda_T$.
\par
En résumé, la configuration initiale du système est complètement déterminée
par la donnée des quatres composantes chirales du paramètre d'ordre et de 
sa dérivée à l'instant $\tau_0$ : $\bfphi (\tau_0)$ et $\dot\bfphi (\tau_0)$. 
Ces huit nombres sont des variables gaussiennes indépendantes dont les moyennes
sont données par les Eqs.~(\ref{moyphi1}) et (\ref{moyphidot1}), et dont les 
variances sont approximativement données par les Eqs.~(\ref{varphiapprox}) 
et (\ref{varphidotapprox}).

\section{Résultats et discussion}

Dans cette partie nous appliquons la méthode d'échantillonnage des 
conditions initiales décrite ci-dessus au calcul de la probabilité
d'avoir une amplification donnée dans l'état final. Nous résumons tout 
d'abord la stratégie du calcul et examinons la question du choix des 
paramètres. Ceux-ci sont contraints par la cohérence physique du modèle 
et des approximations utilisées. De plus, ils doivent ètre choisis en 
correspondance avec la question posée. Enfin, nous présentons et 
discutons les résultats du calcul, puis nous définissons les valeurs 
de l'amplification correspondant à un phénomène obsevable dont nous 
calculons la probabilité d'occurence.

\subsection{La stratégie : récapitulatif}

Les paramètres de notre modèle, outre ceux déjà fixés : $m_\pi$, $f_\pi$, 
$\lambda_R$ et $\Lambda$, sont l'instant $t_0 = \tau_0$, le rayon $R_0$ 
et la température $T$ initials. Nous discuterons plus loin le choix
des valeurs de ces paramètres. Commençons par énumérer les différentes 
étapes du calcul :
 
\begin{enumerate}

\item
Nous calculons d'abord la 
longueur de corrélation $\lambda_T$ du champ dans l'ensemble thermodynamique 
à la température $T$. Celle-ci est donnée par l'inverse de la masse effective
des excitations (quasi-particules) du système, dont le carré est solution
de l'équation d'auto-cohérence (Eqs.~(\ref{thermik}) et (\ref{gapthermik})) 
à l'équilibre thermique :
\eq
 \frac{\chiT}{\lambda} = \left( \frac{H}{\chiT} \right)^2 - v^2 + 
 N \int_0^\Lambda \frac{k^2 dk}{2 \pi^2} \,
 \frac{2 N_k(\chiT) + 1}{2 E_k (\chiT)} \, .
\eq
Puis nous calculons les moyennes et dispersions ($\sigma_1$ et $\sigma_2$) 
des distributions de probabilité des valeurs possibles du paramètre d'ordre 
$\bfphi (\tau_0)$ et de sa dérivée $\dot\bfphi (\tau_0)$ à l'aide 
des formules (\ref{TH_moyphi}), (\ref{TH_moyphidot}), (\ref{TH_sigma1}) 
et (\ref{TH_sigma2}) de l'Annexe~\ref{THERMAL} : 
\bear
 V_0^2 \, \sigma_1^2 & = & 4 \, R_0^4 \, \int_0^\Lambda
 \frac{dk}{E_k} \, \coth \left( \frac{E_k}{2T} \right) \, \mathcal F (k R_0) \, , \\
 V_0^2 \, \sigma_2^2 & = & 4 \, R_0^4 \, \int_0^\Lambda
 dk \, E_k \, \coth \left( \frac{E_k}{2T} \right) \, \mathcal F (k R_0) \, ,
\eear
où $F (y) = ( \sin y - y \, \cos y )^2/y^4$ et $E_k = \sqrt{k^2 + \chiT}$.

\item
\label{firststep}
Nous tirons aléatoirement les valeurs initiales des composantes chirales
du paramètre d'ordre et de sa dérivée : 
$\left\{ \bfphi (\tau_0) ; \dot\bfphi (\tau_0) \right\}$, avec lesquelles
nous initialisons complètement le système. Nous calculons tout d'abord
la masse effective des excitations physiques du milieu à l'aide de l'équation
d'auto-cohérence à l'instant initial (équilibre thermique local) (\ref{chi0}) :
\eq
 \chi_0 = \lambda \, \left[ \phi_0^2 - v^2  + N \, A_1 (\chi_0) \right] \, ,
\eq
et dont la dérivée temporelle à l'instant initial est donnée par 
l'Eq.~(\ref{chi0dot}).
Avec ces deux nombres, nous initialisons l'ensemble des fonctions modes
$\psi_s (\tau_0)$ (Eqs.~(\ref{adiab1})-(\ref{init}))
\bear
 \psi_s (\tau_0) & = & \frac{1}{\sqrt{2 \tomega_s (\tau_0)}} \, , \\
 \psi_s (\tau_0) & = & 
 - \left[ \frac{\tomega_s' (\tau_0)}{2 \tomega_s (\tau_0)} +
 i \, \tomega_s (\tau_0) \right] \, \psi_s (\tau_0) \, ,
\eear
avec $\tomega_s (\tau_0) = \tau_0 \omega_s (\tau_0) = 
\sqrt{s^2 + \tau_0^2 \, \chi_0}$ 
et $\tomega_s' (\tau_0) = (2 \tau_0 \, \chi_0 + \tau_0^2 \, \dot\chi_0)/
2\omega_s (\tau_0)$.

\item
On peut alors résoudre les équations du mouvement (\ref{mean}), (\ref{mode}) et
(\ref{chi})
\bear
 \ddot\bfphi (\tau) + \frac{3}{\tau} \dot\bfphi (\tau) + 
 \chi (\tau) \, \bfphi (\tau) & = & H {\bf n}_\sigma \, ,  \\
 \ddot\psi_s (\tau) + \frac{1}{\tau} \, \dot\psi_s (\tau) +
 \omega_s^2 (\tau) \, \psi_s (\tau) & = & 0 \, ,
\eear
et
\eq 
 \frac{\chi (\tau)}{\lambda} = \phi^2 (\tau) - v^2 +  
 \frac{N}{\tau^2} \, \int_0^{\Lambda \tau} \frac{s^2 ds}{2 \pi^2} \, 
 (2 n_s + 1) \, | \psi_s (\tau) |^2 \, .
\eq

\item
\label{laststep}
On obtient ainsi les valeurs des fonctions modes et de leur dérivées à 
l'instant final : $\psi_s (\tau_f)$ et $\dot\psi_s (\tau_f)$, 
ainsi que celles de la masse des excitations physiques et sa dérivée : 
$\chi (\tau_f) \equiv \chi_f$ et $\dot\chi (\tau_f) \equiv \dot\chi_f$,
à l'aide desquelles on calcule la multiplicité finale (\ref{number_final})
\eq
 n_s^a (\tau_f) + \frac{1}{2} = 
 \mathcal A_s (\tau_f) \, \left( n_s^a + \frac{1}{2} \right) \, ,
\eq
où le facteur d'amplification est donné par
\eq
 \mathcal A_s (\tau_f) = 2 \, |\beta_s (\tau_f)|^2 + 1 \, ,
\eq
avec (\ref{bogobeta})
\eq
 |\beta_s^* (\tau_f)|^2 = |\psi_s^{(0)} (\tau_f) \, \psi_s' (\tau_f) -
 \psi_s^{(1)} (\tau_f) \, \psi_s (\tau_f)|^2 \, .
\eq
et (\ref{adiab0})-(\ref{adiab1})
\bear
 \psi_s^{(0)} (\tau_f) & = & \frac{1}{\sqrt{2 \tomega_s (\tau_f)}} \, , \\
 \psi_s^{(1)} (\tau_0) & = & 
 - \left[ \frac{\tomega_s' (\tau_f)}{2 \tomega_s (\tau_f)} +
 i \, \tomega_s (\tau_f) \right] \, \psi_s^{(0)} (\tau_0) \, ,
\eear
où $\tomega_s (\tau_f) = \sqrt{s^2 + \tau_f^2 \, \chi_f}$ 
et $\tomega_s' (\tau_f) = (2 \tau_f \, \chi_f + \tau_f^2 \, \dot\chi_f)/
2\omega_s (\tau_f)$.

\item
On répète les pas~\ref{firststep} à~\ref{laststep} autant de fois que possible, 
de façon à avoir une bonne statistique.

\item
Enfin, on recommence avec d'autres valeurs des paramètres.
\end{enumerate}

\subsection{Cohérence physique du modèle et choix des paramètres}

Il est clair que le modèle effectif de basse énergie que nous utilisons
ici n'a de sens que pour des températures $T \ll \Lambda$. De façon plus 
fine, le modèle $\sigma$-linéaire n'a pas de sens pour des températures 
trop élevées devant la température critique de la transition
de phase chirale $T_c \approx 160$ MeV. En effet, à ces températures,
on doit tenir compte des degrés de libertés colorés que sont les quarks
et les gluons, la température critique de la transition de déconfinement 
étant du même ordre de grandeur. Pour la physique qui nous intéresse, 
on doit cependant démarrer dans la phase symétrique ($T > T_c$), et
on néglige le phénomène du déconfinement, sans toutefois aller
trop haut en température.
\par
Voyons d'abord quelles sont les limites imposées par la cohérence physique
du modèle sur les paramètres $R_0$ et $\tau_0=t_0$. En ce qui concerne la 
taille de la bulle initiale, nos hypothèses ne sont valables que pour 
$R_0 \gtrsim \lambda_T$. En effet, l'hypothèse d'équilibre local n'a de 
sens que si les degrés de liberté du système fluctuent plus ou moins 
indépendamment de l'environnement de la bulle. Dit d'une autre façon, 
nous avons vu que l'approximation de champ moyen consiste à voir le 
système comme un ensemble d'excitations indépendantes de masse effective
$\sqrt{\chi_0} \sim \sqrt{\chiT}$. Cette image n'a de sens que si le 
volume dans lequel vivent ces excitations est au moins supérieur à 
leur extension spatiale $\sim 1/\sqrt{\chiT}$. Nous n'excluons bien 
évidemment pas la possibilité que l'instabilité que nous cherchons à 
décrire se développe dans une bulle plus petite, mais ce scénario est 
en dehors du domaine d'applicabilité de notre modèle. L'instant initial 
$t_0$ est le temps de formation de notre bulle lors de la collision. 
Celui-ci doit être supérieur au temps de formation d'une 
quasi-particule qui est au moins de l'ordre du temps caractéristique 
des interactions fortes $\sim 1$~fm. De même, il ne fait aucun sens de 
parler d'équilibre pour des temps inférieurs à cette échelle. Enfin, 
on a $t_0 \ge R_0 \gtrsim  \lambda_T \sim 1$~fm pour les températures 
considérées. 
\par
Revenons maintenant à la question de départ : nous voulons estimer l'ordre
de grandeur de la limite supérieure\footnote{Notre description repose
   implicitement sur l'hypothèse que cette probabilité est rare.
   En effet, dans le cas contraire il faudrait s'inquièter de la 
   possibilité que plusieurs bulles développant des instabilités soient
   formées lors de la collision, et prendre en compte les interactions
   entre des bulles voisines. En particulier, l'hypothèse selon laquelle 
   la valeur du paramètre d'ordre est constant sur la surface de la bulle 
   n'aurait alors aucun sens.} de la probabilité d'avoir une amplification 
notable. Nous nous plaçons donc dans les conditions les plus favorables 
pour qu'une instabilité se développe. Nous avons déjà vu que le trempage 
des fluctuations initiales est d'autant plus efficace que le terme de 
friction dans les équations du mouvement est grand. Celui-ci étant 
inversement proportionnel au temps, nous commencerons l'expansion le 
plus tôt possible, c'est à dire que nous prendrons\footnote{Remarquons 
   de plus que pour que l'hypothèse d'équilibre local à l'instant initial 
   ait un sens, nous devons choisir $R_0$ aussi grand que possible, c'est 
   à dire $R_0=t_0$.} 
$t_0 = R_0 = \lambda_T$. Par souci de comparaison, nous étudierons aussi
le scénario $t_0 = R_0 = 2 \, \lambda_T$.
\par
Il pourrait sembler qu'une température initiale élevée (qui correspond à 
une agitation thermique plus grande, et donc à une longueur de corrélation 
plus faible) soit préférable. Cependant de trop fortes fluctuations initiales 
sont plus difficiles à supprimer par trempage. On peut le voir sur l'équation 
de gap (\ref{chi}) : il est d'autant plus facile d'entrainer le carré de la 
masse effective dans la région des valeurs négatives que la contribution des 
fluctuations $\langle \, \dphi^2 \, \rangle$ est petite (voir aussi la
Fig.~\ref{fig_traj} dans le Chap.~\ref{INTRO}).
Dans la suite, nous considèrerons deux valeurs de la température initiale :
$T=200$ MeV et $T=400$ MeV.
\par
Le dernier point concerne le temps $\tau_f$ auquel on mesure la
multiplicité finale. Pour l'ensemble des évènements considérés, le
système atteint un régime stationnaire où $\chi(\tau) \approx m_\pi^2$
pour $\tau \gtrsim 10$~fm. On a alors un gaz de pions libres en
expansion. Nous prendrons $\tau_f=10$~fm.

\subsection{Résultats}

Nous résolvons numériquement les équations du mouvement en utilisant
un algorithme de Runge-Kutta d'ordre quatre~\cite{NumRec}, les intégrales
sur les modes étant calculées par la méthode de Simpson, corrigée par une 
méthode de trapèzes lorsque le nombre de points sous l'intégrale est pair 
(rappelons que la borne d'intégration dépend du temps). Les équations de 
gap à l'équilibre thermodynamique et à l'instant initial sont résolues par 
une simple méthode de bissection. Nous reproduisons les résultats
des Réfs.~\cite{Lampert,Lamperthesis}\footnote{Nous remercions M. A. Lampert,
   de nous avoir communiqué les résultats de ses calculs} (voir 
Fig.~\ref{fig_chi}). 
\par
Quand le système entre dans la région d'instabilité spinodale $\chi < 0$,
les modes dont la fréquence $\omega_s$ devient imaginaire pure sont très 
fortement amplifiés. Il est clair que l'amplification est d'autant plus 
importante que $s$ est petit : le facteur d'amplification est une 
fonction monotone décroissante de $s$. Dans la suite nous focalisons 
notre attention sur le  mode $s=0^+$, le plus amplifié. Nous
appelons $P(A)$ la probabilité pour que le facteur d'amplification 
correspondant, calculé à $\tau_f = 10$ fm, ait la valeur $\mathcal A_0 = A$. 
La Fig.~\ref{fig_histo} représente les histogrammes des probabilités $P(A)$ 
pour les quatres scénarios correspondant aux valeurs $T=200$ et $400$ MeV et $R_0/\lambda_T=1$ et $2$. 

\begin{figure}[htbp]
\epsfxsize=6.in \centerline{ \epsfbox{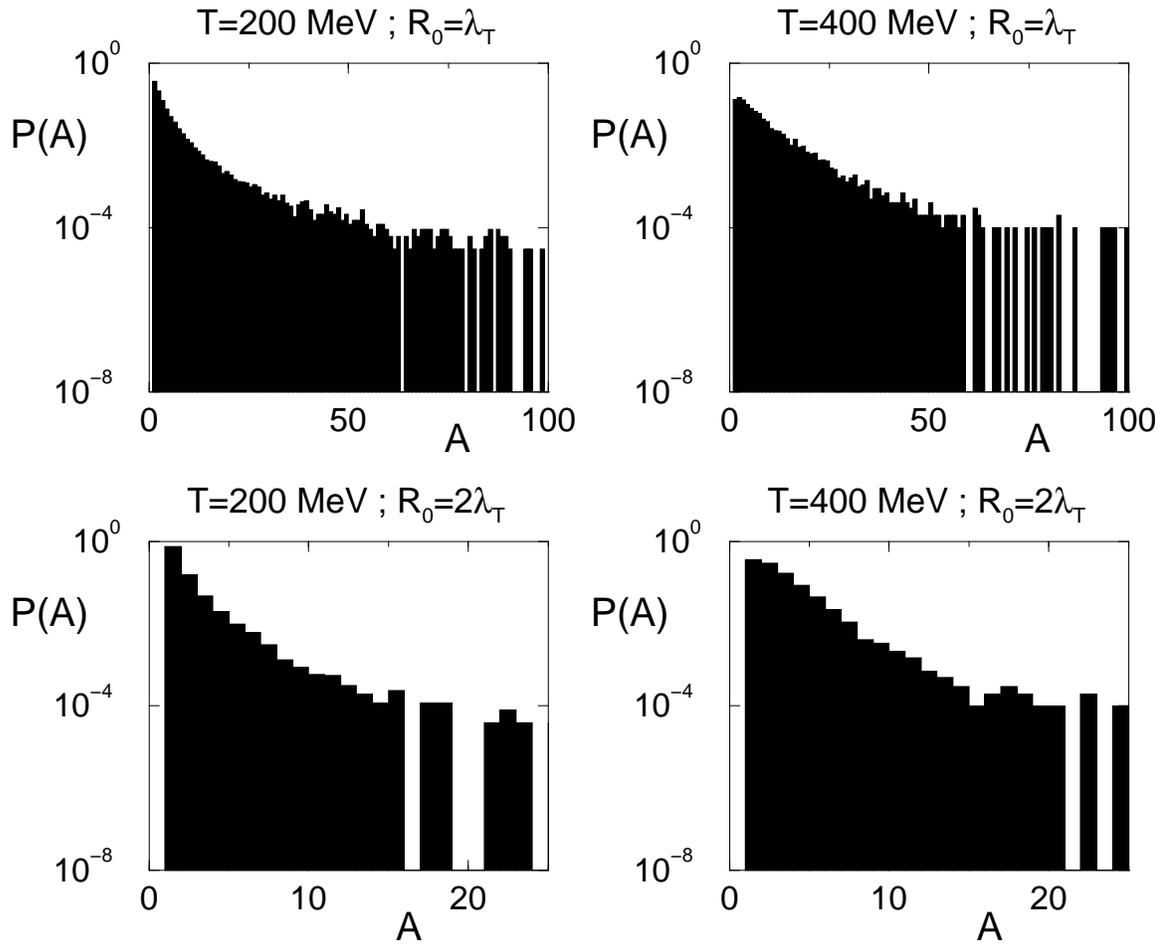}}
\caption{\small Le facteur d'amplification du ``mode zéro'' dans les quatres 
    scénarios suivant : $T=200$ MeV et $R_0=\lambda_T=1.17$ fm ($3.2 \times 10^4$
    évènements) ; $T=400$ MeV et$R_0=\lambda_T=0.68$ fm ($10^4$ évts) ; 
    $T=200$ MeV et $R_0=2\lambda_T$ ($2.5 \times 10^4$ évts) ; 
    $T=400$ MeV et $R_0=2\lambda_T$ ($10^4$ évts).} 
\label{fig_histo}
\end{figure}

Comme on s'y attend (voir la discussion plus haut), le cas $R_0=2\lambda_T$
est le plus défavorable en ce qui concerne la possibilité de fortes 
amplifications. Dans tous les cas, on voit clairement que seule une petite 
fraction des évènements correspond à une amplification importante. 
Les histogrammes correspondant au cas $R_0=\lambda_T$ sont assez bien
représentés par 
\eq
 P(A) = \frac{0.805}{(1 + 0.27 A)^{3.21}} \, ,
\eq
pour $T=200$ MeV, et
\eq
 P(A) = \frac{0.213}{(1 + 0.032 A)^{6.959}} \, ,
\eq
pour $T=400$ MeV. Il est impossible de reproduire la queue des 
distributions par une loi exponentielle du type $P(A) \propto \exp (-aA^b)$. 
\par
Il est intéressant de chercher à caractériser les états initials destinés
à être fortement amplifiés. Il semble qu'une condition soit la petitesse
relative de l'intensité du courant iso-vectoriel classique, 
c'est à dire calculé avec le paramètre d'ordre $\bfphi$, à l'instant initial. 
Les courants iso-vectoriel et iso-axial classiques sont définis par
\bear
 {\bf V_{\mu}} & =&  \bfpi \times \p_{\mu} \bfpi \, ,\\
 {\bf A_{\mu}} & =&  \bfpi \p_{\mu} \sigma - \sigma \p_{\mu} \bfpi \, ,
\eear
où l'on a écrit $\bfphi \equiv (\bfpi,\sigma)$.
Les intensités initiales ${\bf V^{\mu}} \cdot {\bf V_{\mu}}$ et 
${\bf A^{\mu}} \cdot {\bf A_{\mu}}$ de ces courants sont 
représentées en fonction de l'amplification finale $A$ sur la 
Fig.~\ref{fig_courant} (dans le cas $T=200$ MeV, $R_0=\lambda_T$).
Il est intéressant de remarquer que dans le modèle 
classique en 1+1 dimensions (expansion longitudinale) des Réfs.~\cite{BK1,BK2}, 
la petitesse du rapport des valeurs initiales des intensités des courants 
vectoriel et axial est la condition pour qu'aux temps longs le paramètre 
d'ordre oscille dans une direction bien déterminée de l'espace d'isospin
(voir aussi le Chap.~\ref{INTRO}). Autrement dit, c'est la condition
pour avoir une configuration DCC. Dans le cas présent cependant, le
champ classique formé dans l'état final est une superposition du paramètre
d'ordre (le mode zéro) et des modes de grande longueur d'onde. Il n'est pas 
clair que tous oscillent collectivement dans la même direction de l'espace 
d'isospin. Cette question sera examinée dans le Chap.~\ref{QUENCH}. 

\begin{figure}[htbp]
\epsfxsize=4.5in \centerline{ \epsfbox{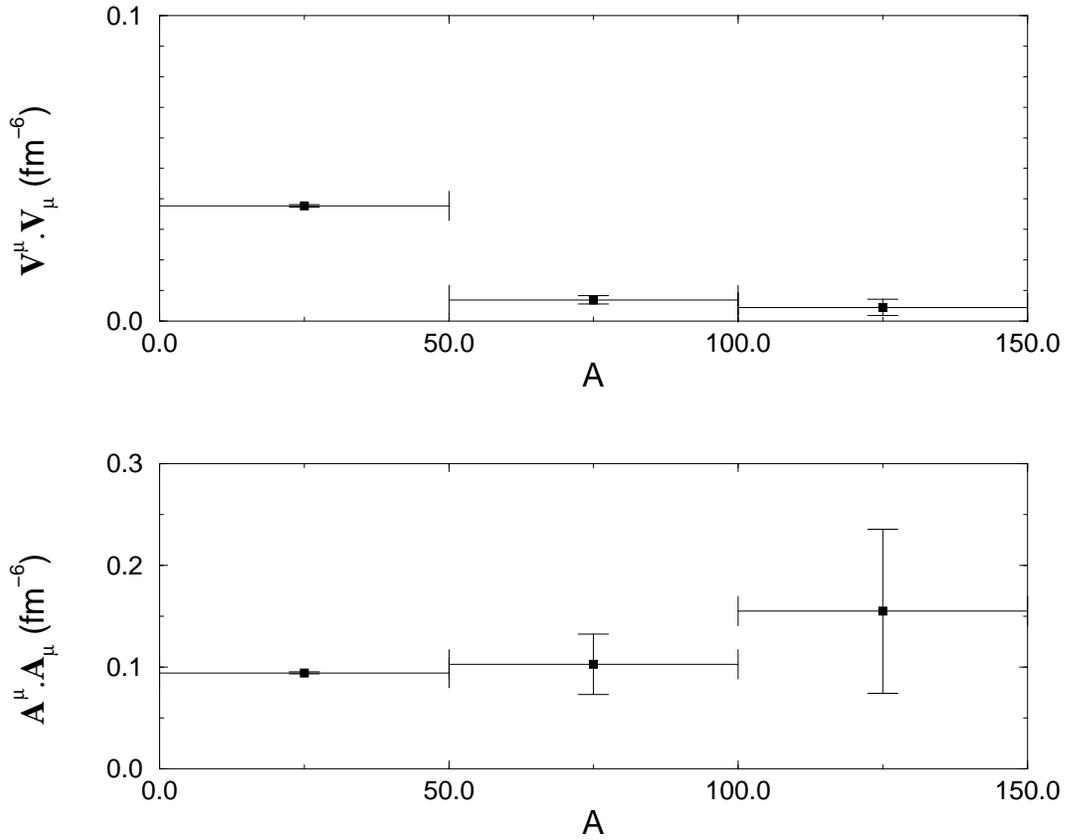}}
\caption{\small Les valeurs initiales des intensités des courants iso-vectoriel 
et iso-axial classiques correspondant aux amplifications observées dans l'état
final. On a calculé la valeur moyenne des intensités sur l'ensemble des
évènements dont l'amplification est comprise entre $0$ et $50$, $50$ et $100$,
et $100$ et $150$. Les barres d'erreurs représentent les incertitudes 
statistiques. En moyenne, les évènements ayant subit une forte amplification 
se distinguent des autres par la petitesse relative de l'intensité initiale 
du courant vectoriel.} 
\label{fig_courant}
\end{figure}

Nous voulons maintenant faire le lien avec la phénoménologie et
déterminer quelles sont les valeurs de l'amplification qui donnent
lieu à phénomène (potentiellement) observable. Estimons la densité
$E dn/d^3p$ de pions produits par unité invariante d'espace des phases. 
Ces pions sont produits par les différents éléments de volume en co-mouvement,
c'est à dire qu'ils sont émis sur l'hypersurface $\tau=\tau_f$. 
Les pions émis par l'élément de volume $\tau_f^3 \, \sinh^2 \eta \, \sin \theta
\, d\eta \, d\theta \, d\varphi = \tau_f^3 \, \sqrt h \, d^3x$, centré sur
le point de coordonnées hyperbolique $(\eta,\theta,\varphi)$ ont, dans
le référentiel où celui-ci est au repos, une énergie $\omega_s (\tau_f)$.
Dans un référentiel quelconque, cet élément de volume est en mouvement
avec la quadri-vitesse $u^\mu = d x^\mu / d\tau$, et la quadri-impulsion
$p^\mu$ des pions émis est telle que $p^\mu u_\mu = \omega_s (\tau_f)$.
Cette relation nous dit quelles sont les valeurs de l'impulsion $k=s/\tau_f$ 
qui contribuent à la quadri-impulsion $p^\mu = (E,\vec p)$ ($p^2 = \chi_f$), 
compte tenu du mouvement de l'élément de fluide. Par exemple, dans le 
référentiel où la bulle est globalement au repos, on a 
$p^\mu u_\mu = E \cosh \eta - p \cos \theta \sinh \eta = \omega_s (\tau_f)$.
La contribution de cet élément de volume à la densité invariante $E dn/d^3p$ 
est proportionnelle à $n_s (\tau_f) \, p^\mu u_\mu \, d^4x \, \sqrt{-g } \, 
\delta (\tau-\tau_f)$, où l'on a écrit l'élément de volume sous une forme
manifestement invariante. Au total, on a donc~\cite{CooperFrye}
\beq
\label{density}
 E\frac{d n}{d^3p} (p) = \int d^4x \, \sqrt{-g} \, \delta (\tau-\tau_f) \,
 f(x,p) \, p^\mu u_\mu \, ,
\end{equation}\noindent
où\footnote{La formule de Planck pour le rayonnement du corp noir est obtenue,
   à partir de l'Eq.~(\ref{density}), en remplaçant la contrainte $\tau = \tau_f$ 
   par $t=$~cte, de sorte que $p^\mu u_\mu = E$, et en prenant 
   $f(x,p)= 2(2 \pi)^{-3}[\exp (E/T) -1]^{-1}$, le facteur $2$ étant le nombre
   d'états de polarisation possibles du photon. Cet exemple permet de fixer
   les normalisations.}
\eq
 f(x,p) = \frac{N}{(2 \pi)^3} \, n_s (\tau) \, .
\eq
L'intégrand dans (\ref{density}) ne dépendant que de la quadri-impulsion 
$p^\mu$, $E d n / d^3 p$ est une fonction de l'invariant
$p^2=\chi=$~cte et est, par conséquent, constante : on a un spectre 
plat\footnote{Le spectre décroissant obtenu dans les
   Réfs.~\cite{Lampert,Lamperthesis} est incorrect. Ceci a été confirmé
   en privé par F.~Cooper.}.
Ceci est dû à l'idéalisation que nous avons adopté, qui consiste à
considérer que le système est couplé à un bain thermique pendant une
durée infinie ou, en d'autres termes, que l'expansion dure un temps infini.
Par exemple, dans le référentiel où la bulle est globalement au repos,
aussi grand que soit le module $p$ de l'impulsion considérée, il existe
toujours un ensemble d'éléments de volume qui se déplacent suffisament
vite (avec une grande rapidité radiale) et qui contribuent à la 
densité invariante. On peut modéliser grossièrement l'effet de la durée
finie du régime d'expansion en limitant l'intégrale sur la variable de 
rapidité dans l'Eq.~(\ref{density}) aux valeurs $\eta < \eta_{sup}$, où
$\eta_{sup}$ reflète l'influence de l'environnement de la bulle. Dans
ce cas, chaque élément de volume en co-mouvement émettant essentiellement
des pions de basse énergie (dans son référentiel propre), le spectre
(\ref{density}) est coupé pour des valeur de l'impulsion $p_{sup} \sim \chi_f \,
\sinh \eta_{sup}$. Une discussion de ces aspects nécessite la modélisation
de la collision dans son ensemble, ce qui dépasse le cadre de notre étude.
La multiplicité totale (irréalistement infinie dans notre cas)
dépend fortement de cette coupure, et ne peut être estimée de façon 
fiable dans ce modèle. Cependant, la formule (\ref{density}) (voir aussi
(\ref{estimate}) plus loin) est probablement une estimation raisonnable 
de la densité invariante de pions de basse énergie. On écrit
\eq
 E\frac{d n}{d^3p} \, (\vec p) = E\frac{d n}{d^3p} \, (\vec p = \vec 0) = 
 \frac{N}{\chi_f} \int_0^{\Lambda \tau_f} \frac{s^2 ds}{2 \pi^2} \, n_s (\tau_f) \, .
\eq
Pour les évènements ayant subi une forte amplification, la multiplicité 
est approximativement proportionnelle au facteur d'amplification du mode
le plus amplifié. Sur la Fig.~\ref{fig_specamplif} on a représenté la 
quantité $c$ définie par :
\beq
\label{estimate}
 E\frac{\dd n}{\dd^3p} = c \, A \, .
\end{equation}\noindent
en fonction de $A$, pour un ensemble d'évènements tels que $A > 20$, dans le
cas $T=200$~MeV et $R_0=\lambda_T$. Le nombre $c$ fluctue relativement peu 
d'un évènement à l'autre pour $A \gtrsim 30$. On observe la même chose pour
$T=400$~MeV. On a $c \simeq 5$~GeV$^{-2}$ dans les deux cas. 

\begin{figure}[htbp]
\epsfxsize=5.2in \centerline{ \epsfbox{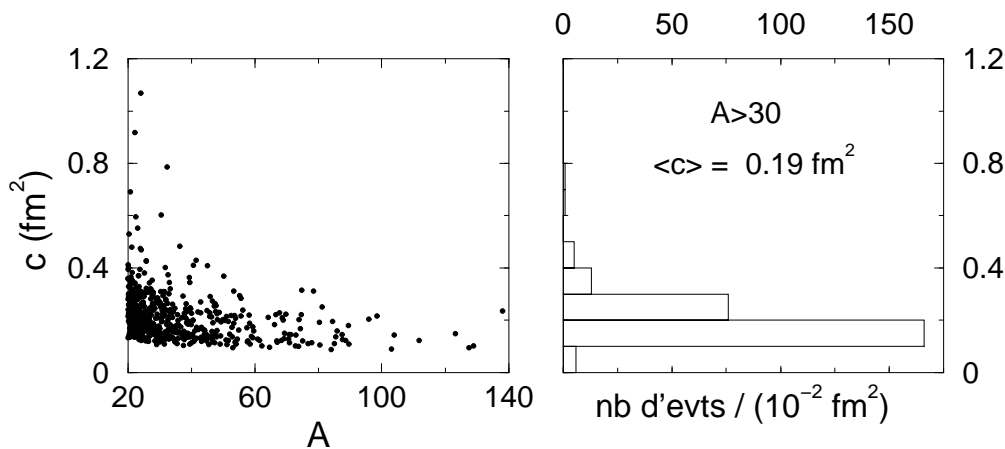}}
\caption{\small La constante $c$ (Eq.~(\ref{estimate})) en fm$^2$ pour 
   quelques ($268$) évènements dans le scénario $T=200$~MeV, $R_0=\lambda_T$. 
   A gauche, on a représenté $c$ en fonction de l'amplification $A$ pour $A>20$,
   à droite, l'histogramme correspondant, restreint aux évènements tels que
   $A>30$. Pour ces derniers on observe $< c > = 0.19$~fm$^2$, avec
   $\Delta c = (<c^2> - <c>^2)^{1/2} = 8 \times 10^{-2}$~fm$^2$. On obtient 
   un résultat similaire dans le cas $T=400$~MeV.} 
\label{fig_specamplif}
\end{figure}

Cette estimation est à comparer avec la densité de pions mous produits par 
unité d'espace des phases lors d'une collision d'ions lourds. Il est 
instructif de considérer l'exemple simple suivant : dans une collision 
typique, le spectre de pions émis est relativement plat dans la région 
centrale de rapidité $y\simeq0$, où $y=(1/2) \ln [(E+p_z)/(E-p_z)]$ est
la rapidité longitudinale ($z$ est l'axe de la collision). Dénotons par 
$h$ la hauteur de ce plateau. Dans la région centrale, on peut 
parametriser le spectre inclusif par 
\eq
 \left. E\frac{d n}{d^3p} \right|_{incl} = 
 \left. \frac{d n}{dy \, d^2p_t} \right|_{incl} =
 \frac{h}{\pi \langle p_t^2 \rangle} \, 
 \exp \left( - \frac{p_t^2}{\langle p_t^2 \rangle} \right)\, .
\eq
Dans les collisions centrales Pb-Pb au SPS du CERN, on observe~\cite{PbPbSPS} 
un nombre de l'ordre de $200$ $\pi^-$ par unité de rapidité, c'est à dire, 
pour tous les pions, $h = dn/dy \approx 600$. Donc, la densité
de pions de très faible impulsion transverse $p_t \simeq 0$, vaut
approximativement $1900$~GeV$^{-2}$, où on a pris 
$\langle p_t^2 \rangle = 0.1$~GeV$^2$. Les fluctuations
de cette densité sont $\sim \sqrt{h}/ \pi \langle p_t^2 \rangle 
\approx 75$~GeV$^{-2}$. Pour que notre signal soit observable, il doit
se détacher de ces fluctuations statistiques. D'après l'estimation 
(\ref{estimate}), on voit que le signal est supérieur à trois fois
la fluctuation si $A \gtrsim 45$. Dans le cas $R_0=\lambda_T$,
la probabilité correspondante est approximativement
\beq
\label{probabilité}
 \mbox{Proba} (A>45) \approx 4 \times 10^{-3} \, ,
\end{equation}\noindent
aussi bien pour $T=200$~MeV que pour $T=400$~MeV. Cette estimation
nous donne l'ordre de grandeur de la limite supérieure de la probabilité
d'avoir une amplification observable. Il s'agit, bien sûr, d'une probabilité 
conditionnelle, puisque notre calcul repose sur l'hypothèse selon laquelle
notre bulle est initialement dans un état d'équilibre thermique local 
(ou, ce qui a un sens plus faible, que les fluctuations initiales sont 
correctement décrites par un ensemble thermique local). 

\section{Conclusion et commentaire}

Résumons-nous : dans le cadre le plus courament utilisé dans la littérature
sur les DCC (modèle $\sigma$-linéaire, approximation de champ moyen, 
équilibre local à l'instant initial), nous avons construit une méthode
originale d'échantillonnage des valeurs initiales du paramètre d'ordre
et de sa dérivée temporelle, qui spécifient complètement l'état initial
du système.
Nous plaçant dans le cadre le plus favorable possible pour générer une
forte amplification des modes de grande longueur d'onde (expansion sphérique,
taille initiale de la bulle aussi petite que possible), et combinant
le formalisme des Réfs.~\cite{Lampert,Lamperthesis,Fulling} avec notre
méthode, nous avons calculé la distribution des valeurs possibles du
facteur d'amplification (Eq.~(\ref{amplification})) du mode le plus amplifié.
Une analyse phénoménologique simple nous permet alors d'estimer la probabilité
d'avoir une amplification donnant lieu à un signal détectable 
(Eq.~(\ref{probabilité})). Celle-ci s'avère assez faible, grossièrement 
de l'ordre de $10^{-3}$ pour une collision centrale Pb-Pb au SPS du CERN. 
Ce résultat doit encore être multiplié par la probabilité pour que 
nos hypothèses soient vérifiées dans une collision réelle. Le calcul de 
cette dernière nécessite une description de l'ensemble de la collision, 
ce qui va bien au-delà de l'étude présentée ici. 
\par
Il est raisonnable de penser que notre estimation, même si elle est très
approximative, donne l'ordre de grandeur auquel il faut s'attendre et doit
donc être prise en compte aussi bien par les expérimentateurs chasseurs
de DCC que par les théoriciens. En particulier, il ressort de notre étude 
que la probabilité pour que plusieurs bulles subissant une forte 
amplification soient formées lors d'une collision est 
faible\footnote{Si plusieurs DCC sont formés durant une collision, leurs
   orientations respectives étant indépendantes, la distribution en charge
   du nombre total de pions émis (cf.~Eq.~(\ref{dccdist})) devient de plus en 
   plus piquée autour de la valeur moyenne $1/3$ à 
   mesure que le nombre d'émetteurs augmente~\cite{Amado}.}.
De plus, du point de vue expérimental, les analyses globales
sont à proscrire, au bénéfice d'analyses ``évènement-par-évènement''.
\par
Nous terminerons ce chapitre par un commentaire concernant l'attitude
générale que nous avons adopté quant à sa présentation : nous avons
calculé la probabilité d'avoir une amplification donnée, ayant en tête 
le fait que ce phénomène correspond à la formation 
d'un DCC~\cite{RW}. Cependant, une étude réalisée ultérieurement 
(cf. Chap.~\ref{QUENCH}) indique que la situation n'est pas aussi claire.
En effet, nous verrons que si le mécanisme du trempage est responsable de
la formation d'un champ classique de pion - ce qui est une caractéristique
essentielle de la configuration DCC - il ne suffit pas à générer la structure 
d'isospin du DCC. Le calcul présenté dans ce chapitre, initialement 
conçu comme un calcul de la probabilité de formation d'un DCC~\cite{KRZJS},
correspond donc plutôt au calcul de la probabilité de formation d'un champ  
de pion classique. Ceci n'affecte cependant en rien la pertinence de notre 
résultat.

\chapter{Description microscopique de la formation d'un DCC}
\label{QUENCH}

Nous avons vu dans les chapitres précédents que le scénario du trempage
prévoit la formation d'un champ de pion classique lors du passage rapide
de la transition de phase chirale dans une collision d'ions lourds 
ultra-relativistes. Ce scénario, initialement proposé par Rajagopal et 
Wilczek en 1993~\cite{RW}, a été largement admis comme une 
description microscopique de la formation d'un condensat chiral 
désorienté, un état hypothétique du champ de pion proposé au début
des années 1990~\cite{AR,BK1,Bjor2}. La correspondance entre la 
configuration du champ issue du trempage d'un état thermique et 
celle proposée au début des années 1990 n'est cependant pas claire. 
Par exemple, les fortes fluctuations de la fraction de pions neutres 
prévues dans l'hypothèse du DCC n'ont jamais été observées 
dans le scénario du trempage~\cite{GGP,Raj,RAN2}.
\par
La configuration DCC est caractérisée par trois points 
essentiels~\cite{BK2,ABL,Bjor3} :
\begin{enumerate}
\item c'est une excitation cohérente du champ, une
configuration essentiellement classique,
\item chaque composante de Fourier du champ oscille dans une direction
bien déterminée de l'espace d'isospin,
\item les orientations des différents modes sont alignées dans une seule
et même direction.
\end{enumerate}
L'objet de ce chapitre est l'analyse statistique détaillée de la 
configuration du champ de pion produit par trempage, le but
étant de déterminer si la structure décrite ci-dessus est 
réalisée de façon générique dans ce scénario.
\par
Pour ce faire, nous reprenons le modèle original de~\cite{RW} dans 
lequel le système, initialement dans un état symétrique évolue selon
les équations du mouvement du modèle $\sigma$-linéaire classique.
Les motivations de ce choix sont multiples. D'abord, les effets quantiques
ne sont pas de première importance, le système étant essentiellement
classique~\cite{ABL}. De plus, nous ne savons traiter ceux-ci de façon
simple que dans l'approximation de champ moyen où les différents modes
sont indépendants les uns des autres. Ce traitement n'est donc pas
adapté à l'étude des corrélations entre les modes. Nous avons vu dans le 
chapitre précédent que dans un scénario plus réaliste, où le trempage est 
dû à l'expansion rapide, le système n'entre que très rarement dans la 
région d'instabilité, ce qui nous oblige à avoir une statistique 
importante pour l'étude des configurations intéressantes. Dans le 
modèle statique de~\cite{RW} où le trempage est réalisé de manière 
artificielle, le système démarre dans la région spinodale et tous les 
évènements sont amplifiés. Enfin, l'étude que nous nous proposons 
de réaliser remet en question une idée largement admise. Il est donc 
préférable de raisonner dans le cadre le plus simple de façon à ne pas 
obscusrcir notre argumentation.
\par
Nous commençons par rappeler en détails le modèle de Rajagopal
et Wilczek, puis nous construisons les outils nous permettant
d'étudier la structure d'isospin du champ. Nous présentons 
ensuite les résultats de l'analyse statistique de l'état final,
que nous comparons à des travaux antécédents. Enfin nous
discutons les implications de notre étude tant du point de
vue phénoménologique qu'en ce qui concerne la compréhension
microscopique de la formation d'un DCC. Ce travail
à été publié dans la revue Phys. Rev. D~\cite{JS}.

\newpage
\section{Le formalisme}
\subsection{Le modèle de Rajagopal et Wilczek}

Nous avons déjà introduit ce modèle dans le Chap.~\ref{INTRO}.
Nous le décrivons ici en détail. Le paramètre d'ordre de la symétrie
chirale est représenté par un champ vectoriel à quatre composantes :
$\bfphi = (\bfpi,\sigma)$, et son évolution temporelle est
gouvernée par les équations du mouvement du modèle $\sigma$-linéaire
classique (cf. Eq.~(\ref{sigma}))
\beq
\label{motion}
 \left( \partial^2  - \lambda v^2 + \lambda\phi^2 (\vec x,t) \right) 
 \bfphi (\vec x,t) = H \mbox{\bf n}_\sigma \, \, ,
\end{equation}\noindent
où les paramètres $v$, $\lambda$ et $H$ sont déterminés à partir
des quantités physiques\footnote{Nous suivons ici le choix de~\cite{RW}
   pour les valeurs des quantités physiques, de façon à tester nos résultats
   en les comparant aux leurs.}
$m_\pi=135$~MeV, $f_\pi=92.5$~MeV et $m_\sigma=600$~MeV à travers les relations
\bear
 H & = & f_\pi m_\pi^2 \, , \\
 m_\pi^2 & = & \lambda \left( f_\pi^2 - v^2 \right) \, , \\
 m_\sigma^2 & = & \lambda \left( 3 f_\pi^2 - v^2 \right) \, .
\eear
Nous résolvons numériquement ces équations sur un réseau cubique de maille
$a$ et de volume $V=L^3$, où $L=Na$. Pour ce faire, nous devons spécifier 
les conditions aux bords, ainsi que l'ensemble des valeurs du champ $\bfphi$ 
et de sa dérivée temporelle $\p_t \bfphi \equiv \dot\bfphi$ en chacun 
des n\oe uds du réseau à l'instant initial. 
\par
Suivant~\cite{RW}, la configuration initiale est échantillonnée à partir
d'une distribution symétrique reflétant un état thermique à une température
$T>T_c$ : les valeurs de $\bfphi$ et $\dot\bfphi$ aux différents noeuds
du réseau sont des variables aléatoires gaussiennes indépendantes de moyennes
$\bfphi=\dot\bfphi={\bf 0}$ et de variances respectives\footnote{Dans~\cite{RW}, 
   les variances sont données pour les longueurs des vecteurs $\bfphi$ et
   $\dot\bfphi$, par exemple $\langle \phi^2 \rangle = \sum_{j=1}^4 \langle 
   \phi_j^2 \rangle = v^2/4$.}
$\langle \phi_j^2 \rangle = v^2/16$ et $\langle \dot\phi_j^2 \rangle = v^2/4$
($j=1,2,3,4$), en unités du pas du réseau. L'indépendance statistique des valeurs
du champ et de sa dérivée en différents n\oe uds du réseau signifie que le pas 
$a$ a un sens physique : c'est la longueur de corrélation du champ (dégénérée
en $\bfpi$ et $\sigma$) à la température $T$. Nous prendrons, avec Rajagopal
et Wilczek, $a=(200 \, \mbox{MeV})^{-1}=1$~fm. Nous avons déjà mentionné, dans
le Chap.~\ref{INTRO}, le fait que bien que les valeurs des variances du 
champ et de sa dérivée soient choisies arbitrairement dans~\cite{RW}, ce choix 
a reçu une justification avec les travaux ultérieurs de Randrup~\cite{RAN1} : 
la valeur donnée ci-dessus est approximativement égale à la valeur typique 
du carré des fluctuations, après que l'expansion ait emmené le système au c\oe ur
de la région d'instabilité spinodale (cf. Fig.~\ref{fig_traj}). Ce trempage
artificiel permet de nous limiter aux configurations qui subissent une forte
amplification, les seules qui nous intéressent ici.
\par
Nous dénotons par $\bfvarphi (\vec k,t)$ et $\dot\bfvarphi (\vec k,t)$ les
composantes de Fourier du champ et de sa dérivée temporelle à l'instant $t$
et choisissons des conditions aux bords de Neumann, de sorte que ces 
composantes soient réelles. Ce n'est pas là un point essentiel, mais ce choix
s'avère plus judicieux\footnote{Avec des conditions aux bords périodiques, 
   les composantes de Fourier sont des nombres complexes dont les parties
   réelle et imaginaire jouent le rôle de deux degrés de liberté distincts.
   C'est un artefact des conditions périodiques que ces deux degrés 
   de liberté soient réunis pour décrire un seul et même mode de Fourier
   (à ce propos, voir l'Annexe~\ref{neutral}). Pour discuter les questions 
   des trajectoires des ondes $\vec k$ dans l'espace d'isospin, ainsi que 
   des corrélations entre celles-ci, il est préférable que les composantes 
   de Fourier soient des nombres réels.}\label{ftnote_bc}
que des conditions périodiques pour ce qui est de discuter le 
problème de la polarisation\footnote{Dans la suite, nous
   appelerons (iso-)polarisation de l'onde $\vec k$, la trajectoire dans 
   l'espace d'isospin décrite au cours du temps par le vecteur 
   $\bfvarphi (\vec k,t)$.} 
des ondes $\bfvarphi (\vec k,t)$ dans l'espace d'isospin. 
On a (cf. Annexe~\ref{neumann})
\bearn
\label{ampl}
 \bfvarphi (\vec k,t) & = & \left(\frac{2}{L}\right)^{3/2} \, 
 \int_V d^3x \, \cos (k_x x) \, \cos (k_y y) \, \cos (k_z z) \, 
 \bfphi(\vec x,t) \, , \\
\label{ampldot}
 \dot\bfvarphi (\vec k,t) & = & \left(\frac{2}{L}\right)^{3/2} \,
 \int_V d^3x \, \cos (k_x x) \, \cos (k_y y) \, \cos (k_z z) \,  
 \dot\bfphi(\vec x,t) \, .
\eearn
L'espace de Fourier est un réseau discret de pas $\Delta k = \pi/Na$ et, 
les fonctions modes étant périodiques et paires dans chacune des directions, 
on se limite au cube $0 \le k_i \le \pi/a$.
\par
L'évolution hors d'équilibre du système génère une amplification
importante de l'amplitude des oscillations des ondes
$\bfvarphi (\vec k,t)$ correspondant au modes de grande longueur d'onde
(cf. Fig.~\ref{fig_ellipse}), et ce dans chacune des directions 
d'isospin. Ce phénomène explique la formation possible d'un champ de pion fort
(classique), c'est la première condition pour avoir une configuration DCC.
Pour que les deux autres conditions énumérées dans l'introduction soient
satisfaites, les amplifications de différentes composantes d'isospin d'un 
même mode doivent être telles que les ondes $\bfvarphi (\vec k,t)$
soient linéairement polarisées dans l'espace d'isospin (c'est à dire que
le vecteur $\bfvarphi$ oscille le long d'un axe dans cet espace), et 
les amplifications de différents modes doivent être corrélées
entre elles, de façon à ce que les modes oscillent tous le long du même
axe. Nous reviendrons sur ces aspects de manière plus précise dans la suite
de notre exposé, le point important ici est le fait que, dans une configuration
DCC, ou plus exactement, dans un ensemble statistique dont une configuration
générique est de type DCC, les différents modes sont fortement corrélés.
\par
Le point de départ de notre étude est la remarque suivante : les valeurs 
de $\bfphi$ et $\dot\bfphi$ en différents n\oe uds du réseau sont supposées 
être des nombres aléatoires gaussiens indépendants. Il est facile de voir 
(cf. Annexe~\ref{neumann}) que cela implique que les valeurs 
des composantes de Fourier $\bfvarphi$ et $\dot\bfvarphi$ sur les n\oe uds 
du réseau réciproque sont aussi des nombres gaussiens {\em indépendants}. 
En d'autres termes, il n'existe pas de corrélations entre les modes 
dans l'état initial. Pour créer une configuration DCC, le mécanisme du 
trempage doit, non seulement être efficace en ce qui concerne l'amplification 
des modes (et il l'est), mais il doit aussi être capable de {\em générer 
des corrélations} entre les modes amplifiés, ce qui est loin d'être une 
évidence (cf. Eq.~(\ref{motion})).

\subsection{Observables}

Notre but est de tester si la configuration générique issue du scénario du 
trempage est de type DCC. En d'autres termes, nous voulons analyser certains
détails de l'ensemble statistique décrivant l'état final, en particulier les
polarisations des modes et leurs correlations. Pour ce faire, nous allons 
construire certains outils permettant de formaliser et de quantifier les 
notions dont nous avons besoin, en particulier la polarisation des ondes 
$\bfvarphi$. Nous prenons le parti de construire le modèle le plus simple 
possible, de façon à ne pas obscurcir notre argument. 
\par
Dans ce modèle simple, bien que la taille de la boite soit fixée, nous avons 
en tête un système en expansion rapide (l'hypothèse du trempage est une 
idéalisation de l'effet de l'expansion). Dans cette image, après
un certain temps, le système est si dilué que les modes de Fourier découplent
les uns des autres et évoluent librement. La modélisation la plus 
simple\footnote{Nous négligeons la durée finie de la période de découplage,
   pour laquelle une modélisation plus précise du phénomène est nécessaire.
   Nous supposons que tous les modes découplent au même instant $t_f$, qui
   peut être vu comme la fin d'une période commune de découplage, laquelle
   est implicitement supposée être très courte devant l'échelle de temps
   caractéristique $\sim 1$~fm.}
de cet effet consiste à arrêter ``à la main'' l'évolution couplée (régie par
les équations du mouvement (\ref{motion})) à un temps de découplage $t_f$
(freeze-out). La configuration du champ à l'instant $t_f$ joue le rôle 
d'une source classique rayonnant des pions, on peut la voir comme la 
condition initiale de l'évolution ultérieure libre, les fonctions 
$\varphi_j (\vec k,t>t_f)$ représentant les ondes associées à la 
propagation des pions rayonnés par cette source. Les propriétés de 
ceux-ci sont donc entièrement déterminées par la configuration du champ 
à l'instant de découplage $t_f$. 
\par
Donnons un aspect plus formel à cette image. Les quanta émis par une source 
classique sont dans un état cohérent~\cite{Itzuber} (dans la suite, nous nous
limitons au secteur des pions)
\beq
\label{coherent}
 | \alpha , t_f\rangle = \exp \left(\int d^3k \,\,
 \bfalpha(\vec k,t_f) \cdot \mbox{\bf a}^\dagger(\vec k) \right)|0\rangle \, ,
\end{equation}\noindent
avec 
\eq
 \bfalpha \cdot \mbox{\bf a}^\dagger =
 \sum_{j=1}^3 \alpha_j a_j^\dagger \, ,
\eq
où $a_j^\dagger(\vec k)$ est l'opérateur de création d'un pion libre, de composante
d'isospin $j$ et d'impulsion $\vec k$, et $\alpha_j(\vec k,t_f)$ est la valeur
propre de l'opérateur d'anihilation correspondant ($t_f$ est un paramètre), 
correspondant à l'état propre (\ref{coherent}). C'est aussi la composante
de Fourier de la source classique qui rayonne les pions~\cite{ABL}, elle 
est déterminée par la configuration du champ classique à l'instant $t_f$ à 
travers la relation\footnote{On peut 
    voir la configuration classique du champ à l'instant $t_f$ comme la valeur
    moyenne du champ quantique correspondant dans l'état cohérent 
    (\ref{coherent}).} 
\beq
\label{alpha}
 \bfalpha(\vec k,t_f) = 
 \frac{i\dot \bfvarphi (\vec k,t_f) + \omega_k \bfvarphi (\vec k,t_f)}
 {\sqrt{2 \omega_k }} \, \, ,
\end{equation}\noindent
où $\omega_k=\sqrt{k^2 + m_\pi^2}$. Nous avons déjà mentionné le fait que 
l'état (\ref{coherent})-(\ref{alpha}) doit être vu comme la condition initiale
pour l'évolution ultérieure ($t>t_f$), supposée libre et, de ce fait, contient 
toute l'information utile. En fait, l'idée de l'évolution ultérieure libre
n'est pas nécessaire, l'état du système à l'instant $t_f$ étant tout ce qui nous
intéresse. Il est cependant instructif d'introduire cette idée qui nous permettra 
de nous faire une image physique claire et simple des quantités pertinentes à 
mesurer. Pour alléger les notations,
nous omettrons dorénavant d'indiquer explicitement la dépendance en $t_f$.
Nous écrirons par exemple, $\bfalpha (\vec k)$ pour $\bfalpha (\vec k,t_f)$, et 
$\bfalpha (\vec k,t)$ pour $\bfalpha (\vec k,t>t_f)$. De plus, tant que nous
focaliserons notre attention sur un mode $\vec k$ donné, nous omettrons 
l'indice $\vec k$, lequel sera réintroduit quand ce sera nécessaire.
L'évolution ultérieure s'écrit donc
\beq
\label{wave}
 \bfalpha(t) = \bfalpha \, e^{- i \omega \, (t-t_f)} \, .
\end{equation}\noindent
Le nombre moyen de quanta d'isospin $j$ associé à cette onde est 
indépendant du temps et vaut 
\beq
\label{number}
 \bar n_j = \langle \alpha | a_j^\dagger a_j  | \alpha \rangle = 
 |\alpha_j (t)|^2 = |\alpha_j|^2 \, .
\end{equation}\noindent
Enfin, la fraction $f$ de pions neutres dans un mode $\vec k$ donné est
($0 \le f \le 1$)
\beq
\label{ratio}
 f = \frac{\bar n_3}{\bar n_1 + \bar n_2 + \bar n_3} \, .
\end{equation}\noindent
\par
Pour caractériser la structure d'isospin de la configuration du champ
à l'instant $t_f$, nous introduisons le concept de polarisation des
ondes sortantes dans l'espace d'isospin : c'est la trajectoire
décrite par l'extrémité du vecteur $\bfvarphi^{out}(t)\equiv\bfvarphi(t>t_f)$
dans l'espace d'isospin. On a
\bearn
\label{rewave}
 \bfvarphi^{out} (t) & = & 
 \sqrt{\frac{2}{\omega}} \, \mbox{Re} \, \bfalpha (t) \, \\
\label{imwave}
 \dot\bfvarphi^{out} (t) & = & 
 \sqrt{2\omega} \, \mbox{Im} \, \bfalpha (t) \, .
\eearn
Tout d'abord, il est facile de voir que ce 
mouvement est planaire : en effet, le vecteur ${\bf I} = \bfvarphi^{out} \times
\dot\bfvarphi^{out}$, qui est l'analogue du moment angulaire en mécanique 
du point\footnote{${\bf I}_{\vec k}$ est la composante $\vec k$ du générateur
   des rotations dans l'espace d'isospin 
   $\int d^3x \, \bfphi(\vec x,t) \times
   \dot\bfphi(\vec x,t) \propto \int d^3k \, \bfvarphi(\vec k,t) \times 
   \dot\bfvarphi(\vec k,t)$, lequel est conservé. Pour $t>t_f$ les modes 
   sont découplés et chaque composante ${\bf I}_{\vec k}$ est conservée.}, 
est indépendant du temps. La trajectoire de $\bfvarphi^{out}$ est donc une 
ellipse\footnote{C'est la trajectoire périodique caractérisée par une échelle 
   de temps unique (ici $1/\omega$) la plus générale} dans le plan 
   perpendiculaire à ${\bf I}$.
Appellons ${\bf u}$ et $L$ (${\bf v}$ et $l$) la direction et la longueur du 
grand (petit) demi-axe de cette ellipse, comme indiqué sur le 
schéma de la Fig~\ref{fig_ellipse} (${{\bf u}}^2 = {{\bf v}}^2 = 1$,
${\bf u}.{\bf v} = 0$, ${\bf I} = \omega l L  \, {\bf u} \times {\bf v}$).
On a alors
\beq
\label{ellipse}
 \bfalpha (t) = \sqrt{\frac{\omega}{2}} 
 \left( L \, {\bf u} + i l \, {\bf v} \right) \, 
 e^{-i ( \omega \, (t-t_f) + \eta)} \, ,
\end{equation}\noindent
où $\eta$ est un facteur de phase à déterminer à partir de l'Eq.~(\ref{wave}),
ainsi d'ailleurs que les longueurs $L$ et $l$, et les directions ${\bf u}$ et 
${\bf v}$.
\begin{figure}[htbp]
\epsfxsize=4.in \centerline{ \epsfbox{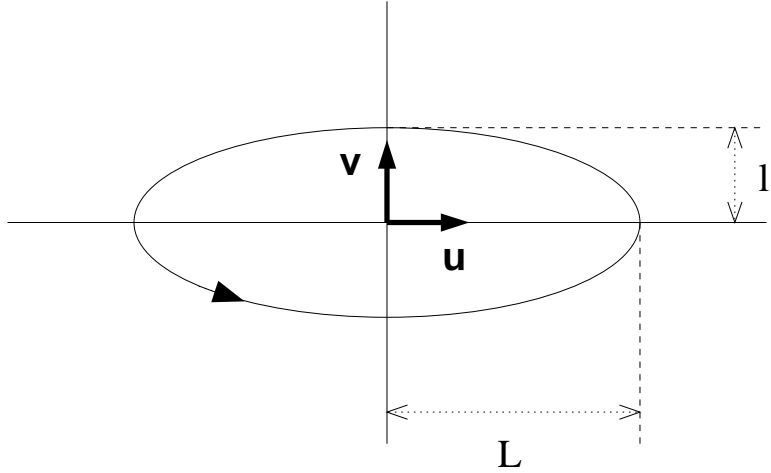}}
\caption{\small Trajectoire elliptique du vecteur $\bfvarphi^{out}$ dans le 
   plan perpendiculaire à ${\bf I} = \bfvarphi^{out} \times \dot\bfvarphi^{out}$.} 
\label{fig_ellipse}
\end{figure}
Avec cette paramétrisation, il vient, pour le nombre d'occupation moyen 
(\ref{number})
\eq
 \bar n_j = \frac{\omega}{2} \, 
 \left( L^2 \, {u_j}^2 + l^2 \, {v_j}^2 \right) \, ,
\eq
et la fraction de pions neutres (\ref{ratio})
\beq
\label{ratio2}
 f = \frac{L^2 \, {u_3}^2 + l^2 \, {v_3}^2}{L^2 + l^2} \, .
\end{equation}\noindent
Une mesure directe de la polarisation de l'onde sortante $\bfvarphi^{out}$
est donnée par l'excentricité\footnote{Pour se faire une image du sens
   physique de l'excentricité, il suffit de penser à une expérience de
   polarisation d'ondes lumineuses (dispositif de type polariseur-analyseur). 
   Dans le cas des ``ondes de pions'', on décompose la polarisation sur 
   la base des états de polarisation linéaire $|\pi_- \rangle$, $|\pi_0 \rangle$ 
   et $|\pi_+ \rangle$, les rôles des analyseurs étant joués par des détecteurs
   sensibles à ces états de charges respectifs.}
de l'ellipse (\ref{ellipse}), définie comme le rapport des longueurs des petit 
et grand demi-axes : $e=l/L$ ($0 \le e \le 1$). Le calcul de cette quantité à 
partir de la configuration à l'instant $t_f$ est simple. Le nombre total de 
pions dans le mode $\vec k$, $\bar n = \sum_{j=1}^3 \bar n_j$, est égal à 
l'énergie dans le mode $\vec k$ divisée par l'énergie d'un pion :
\eq
 \bar n = \frac{\dot\varphi^2 (t_f) + \omega^2 \, \varphi^2 (t_f)}{2 \omega} \, ,
\eq
où l'on a noté $\phi^2 = \bfphi \cdot \bfphi$ et de même pour $\dot\varphi^2$.
En utilisant la paramétrisation (\ref{ellipse}) à $t=t_f$, on obtient les
relations
\eq
 l^2 + L^2 = \frac{2}{\omega} \,  \bar n \, \, \, , \, \, \,
 l \, L = \frac{I}{\omega} \, \, ,
\eq
où $I$ est la longueur du vecteur d'isospin ${\bf I}=\bfvarphi (t_f) \times
\dot\bfvarphi (t_f)$ défini plus haut. On en déduit les longueurs
\eq
 l^2 = \frac{\bar n - \sqrt{{\bar n}^2 - I^2}}{\omega} \, \, \, , \, \, \,
 L^2 = \frac{\bar n + \sqrt{{\bar n}^2 - I^2}}{\omega} \, \, ,
\eq
et l'excentricité
\beq
\label{excent}
 e^2 = \frac{\bar n - \sqrt{{\bar n}^2 - I^2}}{\bar n + \sqrt{{\bar n}^2 - I^2}} \, .
\end{equation}\noindent
\par 
Avec ces définitions en main, considérons le cas de la polarisation
rectiligne (ou linéaire). L'onde $\vec k$ est  polarisée linéairement 
si le vecteur $\bfvarphi_{\vec k}^{out}$ oscille le long d'un axe donné : 
$\bfvarphi_{\vec k}^{out}$ et $\dot\bfvarphi_{\vec k}^{out}$ 
sont collinéaires. La polarisation rectiligne est donc 
caractérisée par le fait que ${\bf I}_{\vec k} = {\bf 0}$, ou encore : 
$l_{\vec k} = 0$, $e_{\vec k} = 0$.
On a donc 
\bearn
 \bfalpha_{linéaire} (\vec k,t) & = & \alpha (\vec k) \, 
 e^{- i \omega_k \, (t-t_f)} \, {\bf u}_{\vec k} \, , \nonumber \\
 f_{linéaire}(\vec k) & = & \cos^2 \theta_{\vec k} \, ,
 \label{linear}
\eearn
où $\theta_{\vec k}$ est l'angle entre la direction d'oscillation 
${\bf u}_{\vec k}$ et l'axe $\pi_3$ de l'espace d'isospin. Aussi bien
la dynamique (Eq.~\ref{motion}) que l'ensemble statistique décrivant
l'état initial sont invariants sous le groupe $O(3)$ des rotations
dans l'espace d'isospin : il n'existe aucune direction privilégiée. 
Donc, si une telle onde, polarisée linéairement, est produite de façon 
générique dans l'état final, la distribution des valeurs de la fraction
neutre $f(\vec k)$ sera donnée\footnote{Dans un espace à trois dimensions,
   la direction aléatoire ${\bf u}$ est reperée par les deux angles 
   $(\theta,\varphi)$. Si toutes les directions sont équiprobables, la 
   probabilité pour que la direction ${\bf u}$ soit dans l'angle solide
   $d^2 \Omega = \sin \theta \, d\theta \, d\varphi$ centré autour de la
   direction $(\theta,\varphi)$ est $d^2 \Omega /4 \pi$. La probabilité 
   pour que $\cos^2 \theta$ soit compris entre $f$ et $f+df$ 
   s'écrit $P(f) df$, avec
   \eq
    P(f) = \int_0^\pi \sin \theta \, d\theta \, \delta(f-\cos^2 \theta) = 
    \int_0^1 dx \, \delta(f-x^2) = \frac{1}{2\sqrt{f}} \, .
   \eq} 
par la loi en $1/\sqrt{f}$ 
(cf. Eq.~(\ref{dccdist})). Ceci reste approximativement vrai dans le cas,
plus réaliste, où les ondes sortantes ont une polarisation quasi-rectiligne
($l_{\vec k} \ll L_{\vec k}$). Il est important de remarquer
que, s'il est toujours possible d'écrire la fraction neutre $f(\vec k)$
(Eq.~(\ref{ratio})) comme le carré du cosinus d'un certain angle $\chi_{\vec k}$,
celui-ci n'a pas, en général, la signification d'une orientation
aléatoire uniforme dans l'espace d'isospin. La loi en $1/\sqrt{f}$ 
n'est valable que dans le cas de la {\em polarisation linéaire} (voir
par exemple l'Annexe~\ref{neutral}). 
\par
Considérons maintenant la configuration DCC ``idéale'', telle qu'elle
a été proposée au début des années 1990 : tous les $\bfvarphi_{\vec k}^{out}$
sont linéairement polarisés, et oscillent
{\em dans la même direction}\footnote{Le DCC est un état d'isospin total nul : 
   $\int d^3k \, {\bf I}_{\vec k} = {\bf 0}$. Ceci est clairement spécifié 
   dans la réf.~\cite{ABL}, et est une hypothèse, plus ou moins explicite, 
   dans tous les articles originaux, où l'idée du DCC a été proposée.} $\bf u$ :
\eq
 \bfalpha_{DCC} (\vec k,t) = \alpha (\vec k) \, 
 e^{- i \omega_k \, (t-t_f)} \, {\bf u} \, .
\eq
Définissant la fraction du nombre total de pions neutres
\beq
\label{dcc}
 f^{tot} =  \frac{N_3}{N_1 + N_2 + N_3} \, ,
\end{equation}\noindent
avec $N_j = \int d^3k \, \bar n_j (\vec k)$, on a
\eq
 f_{DCC}^{tot} = \cos^2 \theta \, ,
\eq
où $\theta$ est l'angle entre la direction $\bf u$ et l'axe $\pi_3$.
Pour un DCC idéal, la fraction neutre {\em totale} est distribuée selon
la loi en $1/\sqrt f$. Dans une situation plus réaliste, on s'attend à ce 
que seuls les modes amplifiés contribuent à la structure du DCC. Il faudrait
donc ne considéder que les modes de grande longueur d'onde dans (\ref{dcc}).
De plus, il est peu probable que les ondes $\bfvarphi_{\vec k}^{out}$ aient
une polarisation strictement rectiligne, ni même qu'elles soient strictement
alignées les unes avec les autres, et on s'attend à des déviations à la loi
idéale. Le point clef de la configuration DCC est le fort degré de corrélation 
entre les modes. Introduisons la fonction de corrélation
\beq
\label{recorrel}
 C_{\mathcal O} (\vec k,\vec k') =
 \frac{\langle \, \mathcal O (\vec k) \, \mathcal O (\vec k') \, \rangle_c}
 {\sqrt{\langle \, \mathcal O^2 (\vec k) \, \rangle_c \, 
 \langle \, \mathcal O^2 (\vec k') \, \rangle_c}} \, ,
\end{equation}\noindent
où $\langle A \, B \rangle_c = \langle A \, B \rangle - 
\langle A \rangle \, \langle B \rangle$, $\langle ... \rangle$ désigne
la moyenne statistique, et $\mathcal O$ représente une observable quelconque. 
Les observables pertinentes pour l'étude des corrélations entre polarisations, 
sont la fraction neutre $f$ (Eq.~(\ref{ratio})) et l'excentricité $e$ 
(Eq.~(\ref{excent})).

\section{Les résultats}

Nous travaillons avec un réseau cubique de taille $N=64$. Les équations
du mouvement (\ref{motion}) sont résolues numériquement à l'aide de
l'algorithme dit ``straggered leap-frog''~\cite{NumRec} avec un incrément 
de temps $\Delta t = 0.04 a$. Les composantes de Fourier 
(\ref{ampl})-(\ref{ampldot}) sont calculées numériquement par une méthode
de type ``Fast Fourier Transform'', adaptée aux conditions aux bords utilisées
ici~\cite{NumRec} (voir Annexe~\ref{neumann}). Avec une configuration initiale
donnée, on peut suivre l'évolution temporelle des $\bfvarphi (\vec k,t)$.
Nous reproduisons complètement le résultat de la Réf.~\cite{RW}, que  nous 
rappellons brièvement ici (voir aussi le Chap.~\ref{INTRO}). Nous 
calculons à chaque instant la moyenne de $\varphi_j^2 (\vec k,t)$ sur le 
volume de l'espace des phases délimité par les sphères de rayons
$k \pm \delta k/2$, où $k = ||\vec k||$, avec $\delta k=0.057a^{-1}$. 
L'évolution temporelle de cette quantité est tracée, pour différentes valeurs 
de $k$, sur la Fig.~\ref{fig_RW1} dans le Chap.~\ref{INTRO}\footnote{Les
   Figs.~\ref{fig_RW1} et \ref{fig_RW2} ont été obtenues avec des conditions 
   aux bords périodiques, pour des raisons techniques : le processus de 
   moyennage de la Réf.~\cite{RW}, décrit dans le texte, est alors plus 
   facile à implémenter. La quantité à moyenner est alors $|\varphi_j|^2$. 
   De plus, la normalisation des coefficients de Fourier a été choisie 
   identique à celle de~\cite{RW} pour faciliter la comparaison.}.
Dans les directions d'isospin ($j=1,2,3$), les modes de basse fréquence
sont fortement amplifiés et oscillent en phase avec une période 
$\pi/\omega_k \simeq \pi/m_\pi$. Ni amplification, ni oscillations collectives
ne sont observées dans la direction $\sigma$ ($j=4$). Ce comportement moyen
s'explique, de manière qualitative, à l'aide des arguments de champ 
moyen~\cite{RW} exposés dans le Chap.~\ref{INTRO}. En résumé : pour
les temps courts ($t \lesssim 10 a$), la courbure du potentiel effectif, 
c'est à dire la masse effective au carré, est négative (cf. Fig.~\ref{fig_RW2}), 
et les modes de grande longueur d'onde, dont la fréquence est imaginaire pure, 
sont amplifiés. C'est l'instabilité spinodale, que nous avons eu maintes fois
l'occasion de discuter. L'amplification ultérieure ($10a \lesssim  t \lesssim 50a$)
est due aux oscillations régulières de la valeur moyenne du champ autour de sa 
valeur asymptotique $f_\pi$. C'est la résonnance paramétrique
\cite{param1,param2,param3}. Le phénomène d'amplification est évidemment 
transitoire et, pour des temps suffisament longs ($t \gtrsim 100a$), l'énergie
initiale est équi-répartie entre tous les modes, le système est dans un 
état d'équilibre stationnaire. 
\par
Ici, notre but est l'analyse statistique détaillée de la structure d'isospin 
de l'état final, où le champ a été fortement amplifié, la question étant de 
savoir si des corrélations entre les polarisations des différents modes
ont été générées avec l'amplification. Au vu de la discussion précédente, 
il est intéressant de discerner entre les différents mécanismes responsables
de l'amplification. Dans la suite, nous présentons donc des résultats pour
deux valeurs du temps de découplage : $t_f=10a$ (avec une statistique de
$21 \times 10^3$ évènements), qui correspond à la fin de la période 
d'instabilité spinodale, et $t_f=56a$ ($10.9 \times 10^3$ évènements), 
temps auquel l'amplification moyenne (cf. Eq.~(\ref{amplifac}) plus bas) 
est maximale. Par souci de clarté, nous ne considérerons que les 
modes\footnote{En pratique
   nous calculons les modes $(n,0,0)$, $(0,n,0)$ et $(0,0,n)$ et prenons
   la moyenne des trois contributions, exploitant l'invariance du problème
   sous les rotations spatiales et augmentant ainsi la statistique. Le 
   raisonnement ne s'appliquant pas pour $n=0$, la statistique pour ce mode
   est trois fois moindre que pour les autres.}
$\vec k = (k = n \Delta k,0,0)$, où $\Delta k = \pi/Na \approx 10$~MeV.
De plus, le modèle $\sigma$-linéaire étant une théorie effective 
aux échelles $\lesssim 100$~MeV, nous ne considérons que la fenêtre
$0 \le n \le 15$.
\par
Définissons le facteur d'amplification du mode $k$ à l'instant $t_f$ :
\beq
\label{amplifac}
 \mathcal A (k,t_f) = \frac{P (k,t_f)}{P (k,0)} \, ,
\end{equation}\noindent
où
\eq
 P (k,t) = \omega_k \, \sum_{j=1}^3 \bar n_j (k,t)
\eq
est la densité moyenne d'énergie dans le mode $k$ à l'instant $t$. 
Les nombres d'occupation moyens $\bar n_j (k,t)$ correspondant
sont calculés à partir de la configuration du champ à l'instant $t$,
au moyen des Eqs.~(\ref{alpha}) et (\ref{number}). La dépendance
avec $k$ de la moyenne statistique du facteur d'amplification
$\langle \mathcal A(k,t_f) \rangle$ est présentée Fig.~\ref{fig_Mamplif}.
Le comportement observé est bien celui attendu : les amplitudes des modes
de grande longueur d'onde sont amplifiés par rapport aux autres. 
Pour $t_f=10a$ on note un accord semi-quantitatif avec les arguments 
de champ moyen (voir le Chap.~\ref{INTRO}, Eqs.~(\ref{MFeom})-(\ref{effmass}))
qui prévoient une amplification monotone des modes
tels que $k \lesssim f_\pi$, amplification d'autant plus importante que $k$
est petit. La fenêtre des modes amplifiés est considérablement
diminuée et l'amplification moyenne augmentée dans le cas $t_f=56a$.

\begin{figure}[htbp]
\epsfxsize=5.5in \centerline{ \epsfbox{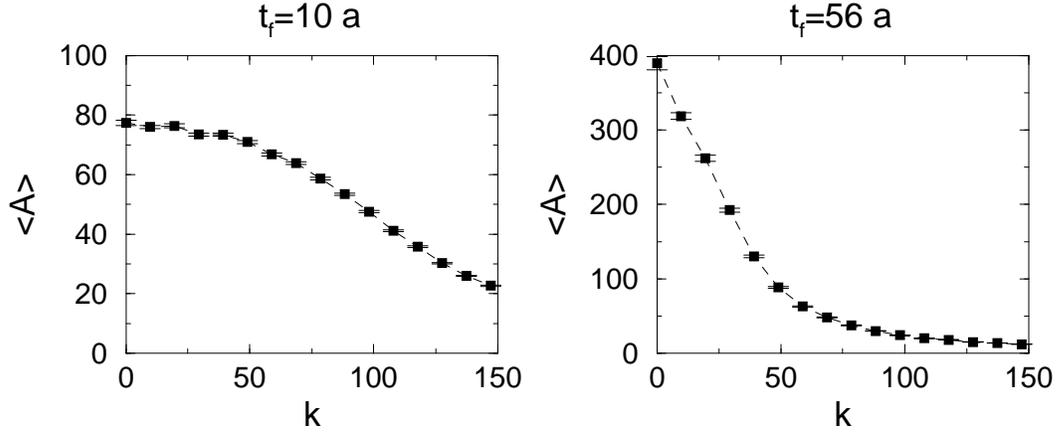}}
\caption{\small Le facteur d'amplification moyen 
    $\langle \mathcal A (k,t_f) \rangle$ en fonction de l'impulsion $k$ (en MeV)
    pour $t_f=10a$ (à gauche) et $t_f=56a$ (à droite). Les barres d'erreur 
    représentent l'incertitude statistique et les lignes en traits tirés servent
    de guides pour l'\oe il. L'amplification moyenne est une fonction
    monotone décroissante de l'impulsion, comme le prévoit l'approximation
    de champ moyen.} 
\label{fig_Mamplif}
\end{figure}

Le comportement qualitatif de l'amplification moyenne en tant que fonction
de l'impulsion $k$, en particulier la décroissance monotone, est donc assez 
bien décrit par l'approximation de champ moyen. Il est intéressant d'aller
plus loin et de caractériser plus précisément la distribution des valeurs 
du facteur d'amplification. Les histogrammes de cette distribution pour 
le mode zéro, le plus amplifié en moyenne, sont représentés Fig.~\ref{fig_Hamplif},
aux instants $t_f=10a$ et $t_f=56a$. Dans les deux cas on observe de très 
fortes fluctuations du facteur d'amplification autour de sa valeur moyenne.
Supposons que, pour un évènement donné, l'amplification $\mathcal A(k)$ dans le
mode $k$ soit supérieure à sa valeur moyenne $\langle \mathcal A(k) \rangle$.
L'amplification $\mathcal A(k+\delta k)$ dans un mode voisin est-elle aussi
supérieure à sa valeur moyenne ? Autrement dit, le facteur d'amplification
est-il, évènement par évènement, une fonction décroissante de l'impulsion ?
La réponse est non, comme le montre la Fig.~\ref{fig_Camplif} qui représente 
la fonction de corrélation $\mathcal C_{\mathcal A}$ définie par
l'Eq.~(\ref{recorrel}) avec $\mathcal O \equiv \mathcal A$. La décroissance 
monotone du facteur d'amplification avec l'impulsion n'est qu'une propriété 
moyenne. Cette absence de corrélations entre les amplifications dans différents 
modes\footnote{La longueur de corrélation est inférieure à la coupure 
   infrarouge $\Delta k$.} 
n'a pas de conséquences très importantes du point de vue phénoménologique.
En effet, en pratique on mesure le nombre total (chargés et neutres) de pions 
moyenné sur des bin dans l'espace des phases. Les amplitudes totale des modes 
étant indépendantes, ce moyennage sur des bins est équivalent à la moyenne
sur l'ensemble statistique pour des bins suffisament grands. La quantité
pertinente, du point de vue de l'amplitude totale, est donc l'amplification
moyenne. On retrouve le fait (Fig.~\ref{fig_Mamplif}) que le scénario du 
trempage rempli très efficacement le premier point du cahier des charges,
décrit dans l'introduction de ce chapitre, concernant la formation d'un 
champ de pion fort (classique).

\begin{figure}[htbp]
\epsfxsize=5.5in \centerline{ \epsfbox{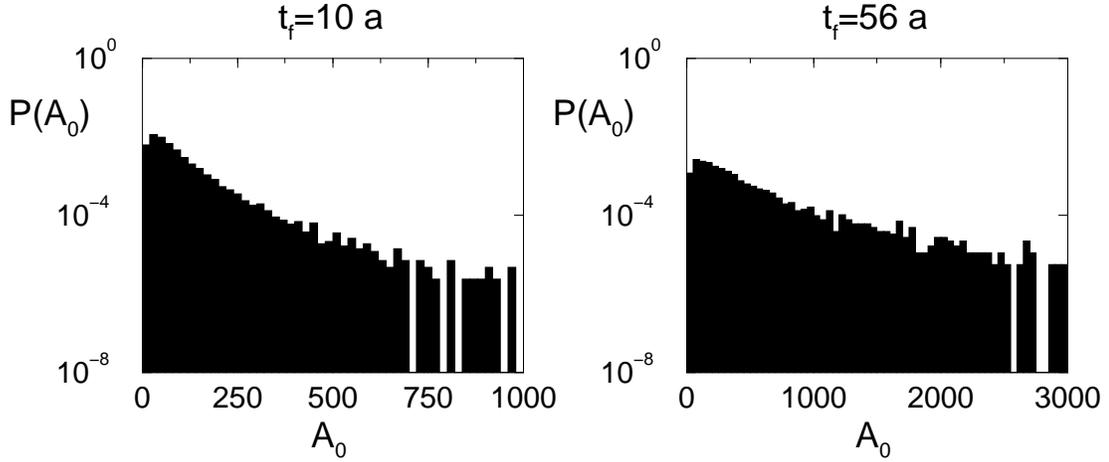}}
\caption{\small Histogrammes des valeurs de l'amplification dans le mode $k=0$,
    dont l'amplification moyenne est la plus grande, pour $t_f=10a$ (à gauche) 
    et $t_f=56a$ (à droite). Les fluctuations de l'amplification autour de
    sa valeur moyenne sont très grandes pour les modes les plus amplifiés
    en moyenne. Ces histogrammes sont les analogues de ceux de la
    Fig.~\ref{fig_histo} du Chap.~\ref{PROBA}. Dans le cas présent 
    cependant, on sélectionne les évènements intéressant (amplifiés) 
    en trempant artificiellement le système.} 
\label{fig_Hamplif}
\end{figure}
\begin{figure}[htbp]
\epsfxsize=5.5in \centerline{ \epsfbox{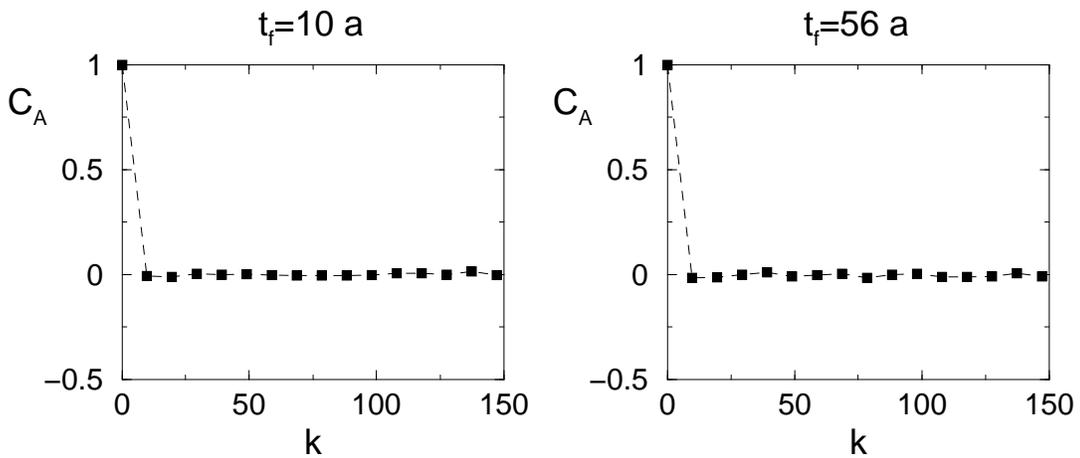}}
\caption{\small La fonction de corrélation réduite $C_A (0,k)$ 
    (Eq.~(\ref{recorrel})) en fonction de $k$ (en MeV) pour $t_f=10a$ 
    (à gauche) et $t_f=56a$ (à droite). Les fonctions de corrélation
    du type $C_A (k_0,k)$ et $C_A(k_0,p)$, où $k_0$ et $k$ sont des impulsions
    dans la direction $\hat k_x$ tandis que $p$ est une impulsion dans
    une direction différente, montrent des profils similaires : les 
    amplifications des différents modes sont statistiquement indépendantes.} 
\label{fig_Camplif}
\end{figure}

Venons-en maintenant à notre étude proprement dite, c'est à dire à
l'analyse de la structure d'isospin des configurations dans l'état final
et commençons par le second point de notre cahier des charges :
l'état de polarisation des modes individuels. 
Les distributions des valeurs de la fraction neutre $f(k)$ sont 
représentées pour différentes valeurs de $k$ sur les Figs.~\ref{fig_Hfrac1} 
et \ref{fig_Hfrac2}, correspondant respectivement à $t_f=10a$ et $t_f=56a$. 
Bien que toutes présentent de fortes fluctuations autour de la valeur 
moyenne $\langle f \rangle=1/3$, on remarque un changement de profil 
lorsqu'on passe des modes amplifiés à ceux qui ne le sont pas. En fait
pour tous les modes en dehors de la fenêtre d'amplification ($n \le 15$
pour $t_f=10a$, $n \le 4$ pour $t_f=56a$), les distribution des valeurs
de $f$ présentent le même profil linéaire. Il est intéressant de remarquer
que cette loi linéaire est précisémment celle que l'on obtient dans le
cas d'un ensemble thermique de pions (voir l'Eq.~(\ref{thermalf})) de
l'Annexe~\ref{neutral}). Cette observation conforte l'idée selon laquelle
ces modes sont déjà thermalisés~\cite{RW,Rajrev}. En ce qui concerne les modes 
amplifiés, qui sont ceux qui nous intéressent, la distribution en $f$ est
sensiblement la même que dans l'état initial (cf. Eq.~(\ref{indist}) de 
l'Annexe~\ref{neutral}). 

\begin{figure}[htbp]
\epsfxsize=5.5in \centerline{ \epsfbox{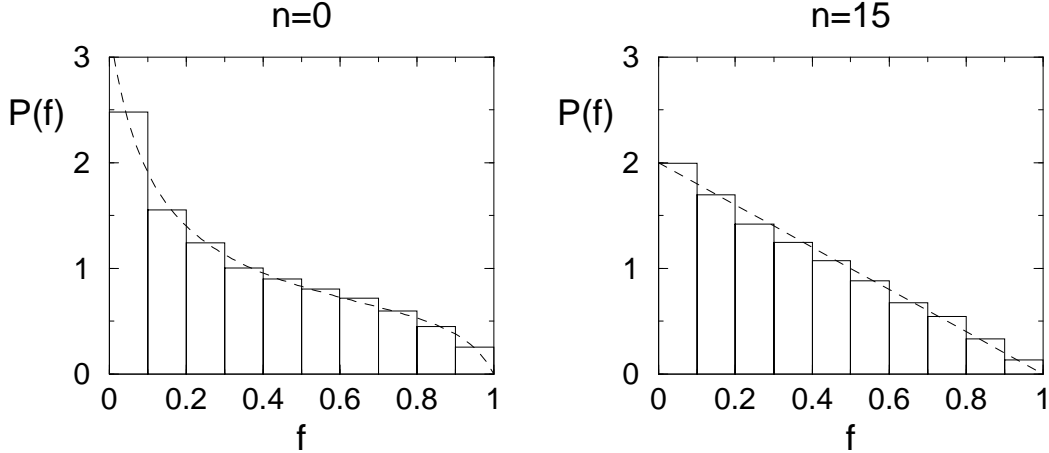}}
\caption{\small Histogrammes des valeurs des fractions neutres dans les modes
   $k=n \Delta k$ ($\Delta k \approx 10$~MeV) pour $n=0$ (à droite) et $n=15$ 
   (à gauche), à l'instant $t_f=10a$. Les histogrammes correspondant aux modes
   $n \le 14$ sont tous compatibles avec la distribution correspondante dans
   l'état initial (indiquée en traits tirés pour $n=0$), tandis que pour $n\ge15$
   on observe une distribution linéaire, à comparer avec la loi $2(1-f)$ 
   (indiquée en traits tirés) obtenue dans le cas d'un ensemble thermique 
   (cf. Eq.~(\ref{thermalf}) de l'Annexe~\ref{neutral}).} 
\label{fig_Hfrac1}
\end{figure}
\begin{figure}[htbp]
\epsfxsize=5.5in \centerline{ \epsfbox{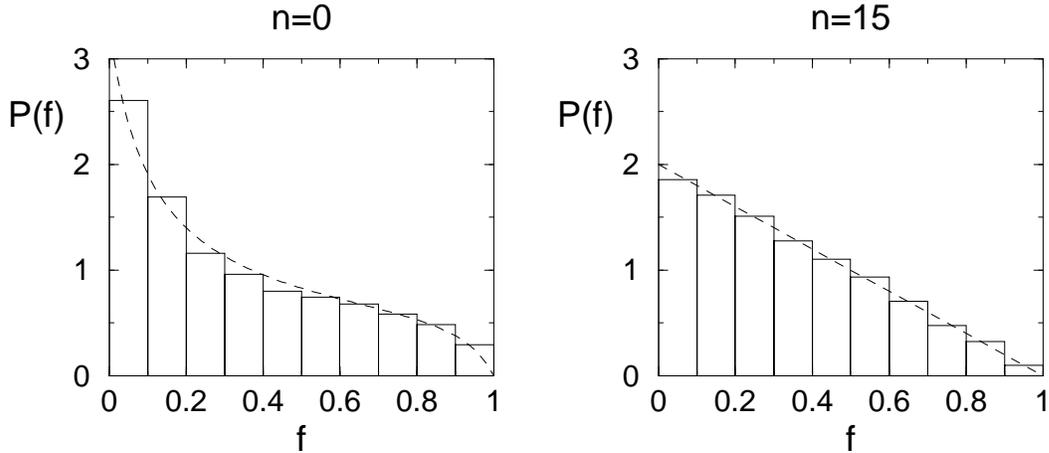}}
\caption{\small Même chose que sur la Fig.~\ref{fig_Hfrac1} pour
   $t_f=56a$. Les histogrammes correspondant aux modes $n \le 3$ sont 
   compatibles avec la distribution initiale. Tous les modes $n \ge 4$ 
   présentent une distribution linéaire.} 
\label{fig_Hfrac2}
\end{figure}

L'analyse des excentricités des polarisations elliptiques permet d'étudier
plus précisément ce dernier point. La Fig.~\ref{fig_Mex} représente l'excentricité 
moyenne $\langle e(k,t_f) \rangle$ en fonction de $k$ à différents instants. 
Dans les deux cas $t_f=10a$ et $t_f=56a$, les modes pour lesquels l'amplification
moyenne est importante ont une excentricité moyenne très proche, bien que 
légèrement inférieure, de la valeur correspondante dans l'état initial. 
Dans le cas $t_f=56a$, on voit que l'excentricité moyenne des modes qui 
ne font pas, ou plus, partie de la fenêtre des modes amplifiés ($k \gtrsim 50$~MeV)
est indépendante de $k$ et très supérieure à la valeur initiale correspondante. 
Cette observation s'avère vraie, non seulement en moyenne, mais évènement 
par évènement, au niveau des distributions des excentricités des différents 
modes, représentées aux instants initial et final, $t_f=10a$ et $t_f=56a$, 
sur les Figs.~\ref{fig_Hex0}, \ref{fig_Hex1} et \ref{fig_Hex2} respectivement. 
Pour les modes amplifiés, les distributions finales sont légèrement
décallées vers les petites valeurs de $e$, à l'inverse, les modes thermalisés
ont une polarisation plus circulaire (grandes valeurs de l'excentricité) que
dans l'état initial. Pour ces derniers, la distribution est indépendante de $k$.
Bien que les modes de grande longueur d'onde aient une polarisation
de plus en plus linéaire à mesure qu'ils sont amplifiés, il s'agit d'un
effet très faible. On voit par exemple sur les Figs.~\ref{fig_Hex0}, 
\ref{fig_Hex1} et \ref{fig_Hex2}, que la proportion d'évènements pour
lesquels l'excentricité du mode zéro est inférieure à $0.1$ est de :
$13\%$ dans l'état initial, $16\%$ à $t_f=10a$, et $18\%$ à $t_f=56a$.

\begin{figure}[htbp]
\epsfxsize=3.5in \centerline{ \epsfbox{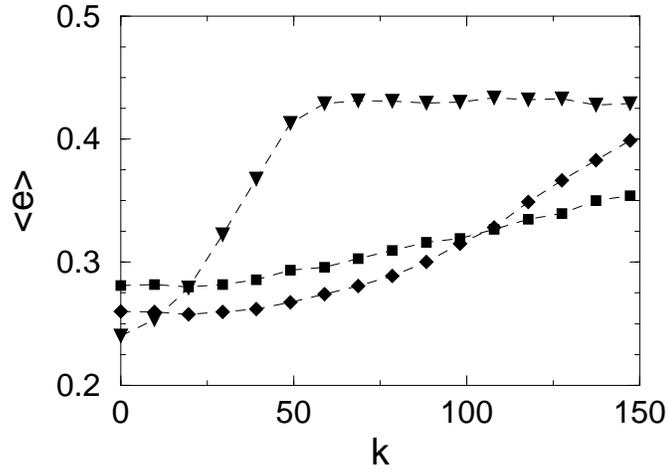}}
\caption{\small L'excentricité moyenne $\langle e(k,t_f) \rangle$ en fonction
   de $k$ (en MeV) dans l'état initial (carrés), ainsi que pour $t_f=10a$
   (losanges) et $t_f=56a$ (triangles). Les lignes sont des guides pour l'\oe il.} 
\label{fig_Mex}
\end{figure}

Les modes qui, à un instant donné, sont dans la fenêtre 
d'amplification ne voient pas leur polarisation notablement modifiée,
tandis que ceux qui ne ressentent plus l'amplification et thermalisent,
se retrouvent avec une polarisation ``thermique'', d'autant plus 
différente de leur polarisation initiale que leur fréquence est grande.
Le mécanisme responsable de l'amplification ne modifie donc pas l'état de
polarisation des modes amplifiés, la distribution de la 
fraction de pions neutres dans ces modes est essentiellement donnée 
par la distribution correspondante dans l'état initial (cf. 
Figs.~\ref{fig_Hfrac1} et~\ref{fig_Hfrac2}).

\begin{figure}[htbp]
\epsfxsize=5.5in \centerline{ \epsfbox{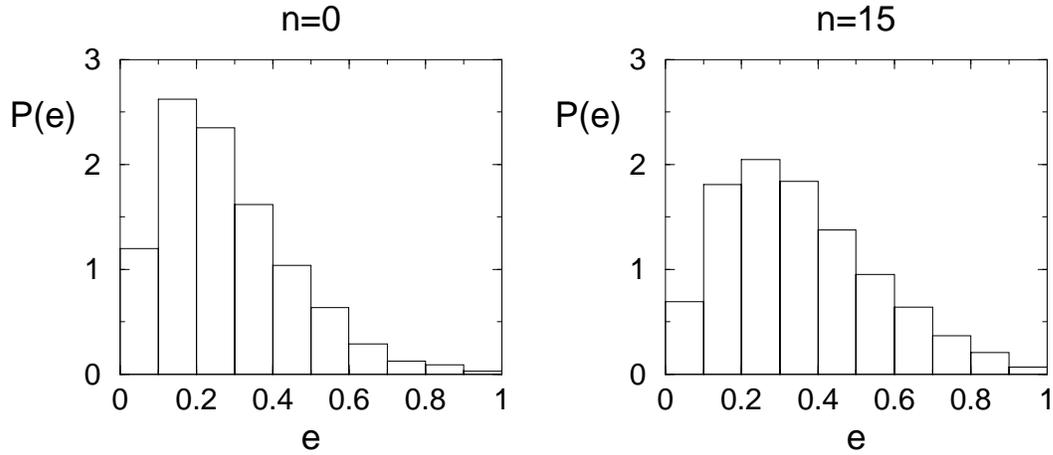}}
\caption{\small Histogrammes des valeurs de l'excentricité (Eq.~(\ref{excent}))
   dans l'état initial, pour les modes $n=0$ (à droite) et $n=15$ (à gauche).} 
\label{fig_Hex0}
\end{figure}
\begin{figure}[htbp]
\epsfxsize=5.5in \centerline{ \epsfbox{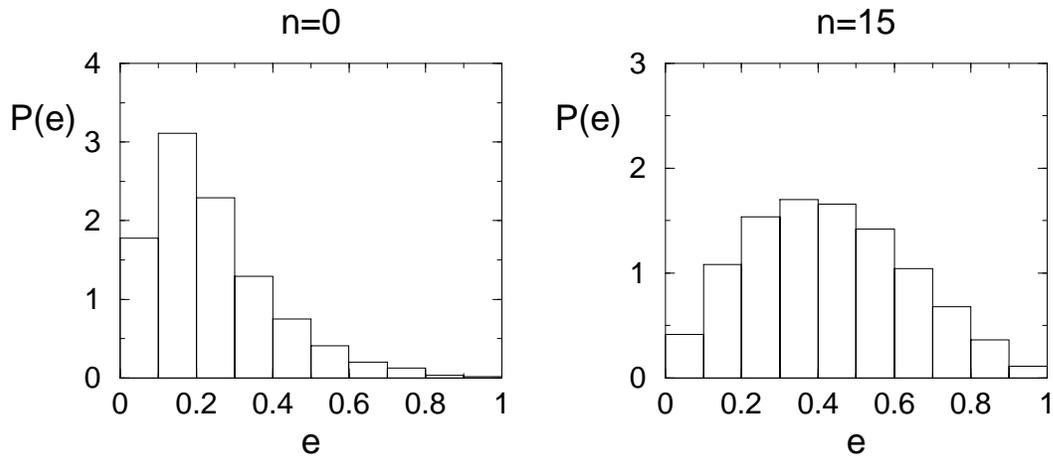}}
\caption{\small Même chose que sur la Fig.~\ref{fig_Hex0}, cette fois
    dans l'état final à $t_f=10a$. Les distributions correspondant aux modes 
    tels que $n < 11$ sont déplacées vers les petites valeurs de $e$ par rapport
    à l'état initial. Ce décalage est le plus accentué pour $n=0$, et
    s'estompe à mesure que $n$ augmente pour s'inverser à partir de $n=11$ :
    au delà les distributions sont décalées vers les grandes valeurs de $e$.} 
\label{fig_Hex1}
\end{figure}
\begin{figure}[htbp]
\epsfxsize=5.5in \centerline{ \epsfbox{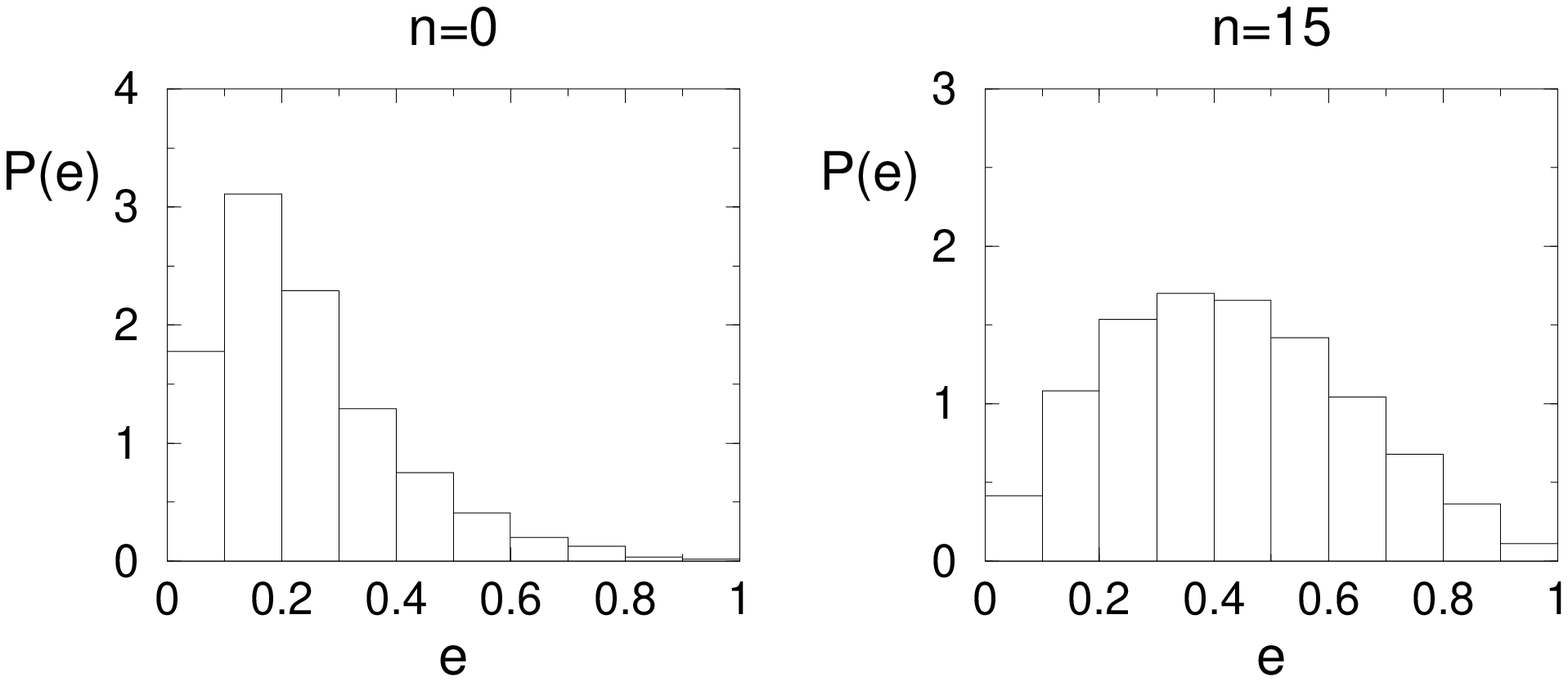}}
\caption{\small Même chose que sur la Fig.~\ref{fig_Hex1}, pour $t_f=56a$.
   Le même phénomène de décalage de la distribution est observé : décalage
   vers la gauche pour $n \le 2$, et vers la droite au delà. Ce décalage
   vers les grande valeurs de $e$ pour les modes de grande fréquence se 
   stabilise cependant : les distributions correspondant aux modes $n \ge 7$
   sont toutes identiques (voir la distribution de $n=15$ ci-dessus).} 
\label{fig_Hex2}
\end{figure}

Il est cependant intéressant de noter que, bien que les modes soient
loin d'être polarisés strictement linéairement, cela n'entraine que
de petits écarts à la loi en $1/\sqrt f$. En effet ici, de même que dans 
le cas idéal, la fraction de pions neutres dans un mode $k$ donné fluctue 
très fortement d'un évènement à l'autre, si bien que du point de vue
d'un expérimentateur la différence n'est pas très importante.

\begin{figure}[htbp]
\epsfxsize=5.5in \centerline{ \epsfbox{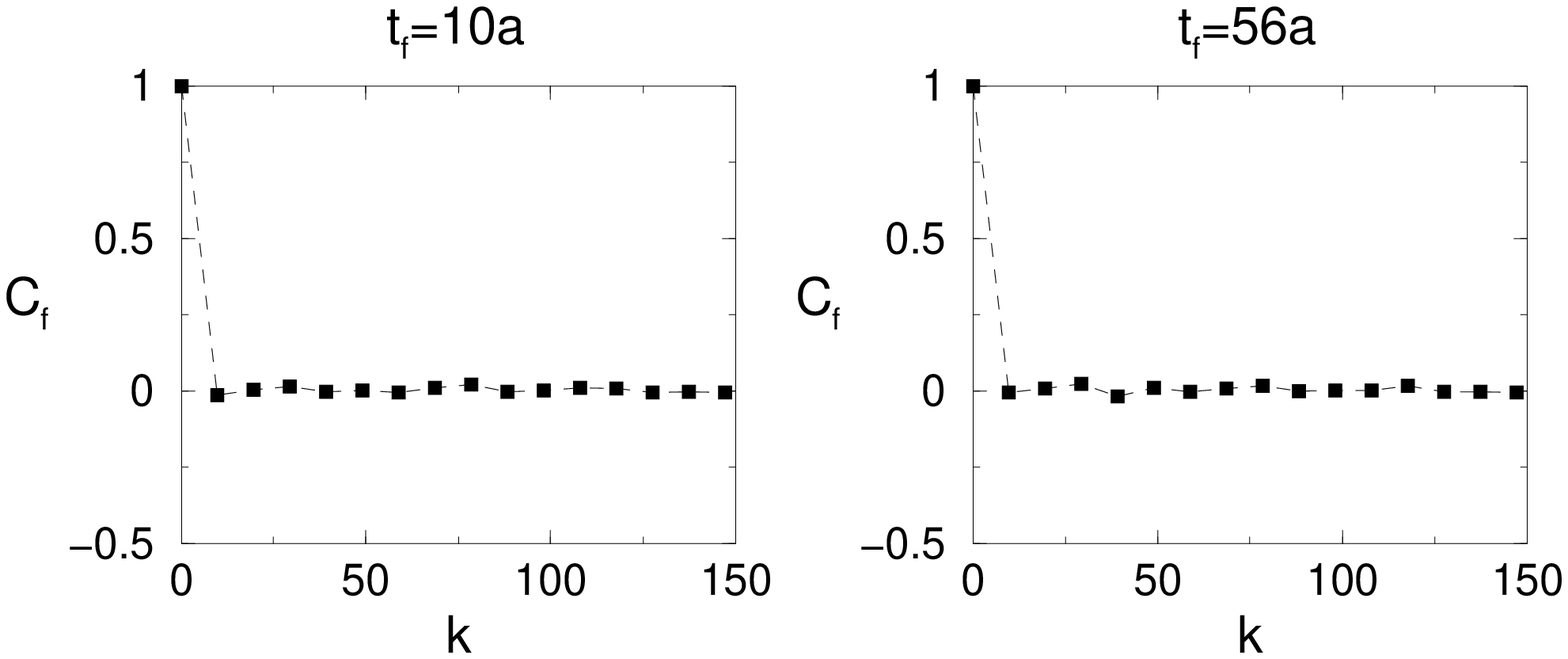}}
\caption{\small La fonction de corrélation réduite $C_f (0,k)$ 
    (Eq.~(\ref{recorrel})) en fonction de $k$ (en MeV) pour $t_f=10a$ 
    (à gauche) et $t_f=56a$ (à droite). De même que pour l'amplification
    (Fig.~\ref{fig_Camplif}), on a calculé les fonctions de corrélation entre
    différents modes, pour la fraction neutre ainsi que pour l'excentricité.
    Il en ressort que les polarisations des différents modes sont
    statistiquement indépendantes.} 
\label{fig_Cfrac}
\end{figure}

Le dernier point de notre étude concerne la question de l'alignement entre
les directions d'oscillations des différents modes dans l'espace d'isospin.
La Fig.~\ref{fig_Cfrac} représente la fonction de corrélation
$C_f (0,k)$ (Eq.~(\ref{recorrel})) en fonction
de $k$. Pour les deux valeurs de $t_f$ étudiées, les fractions neutres
des différents modes sont complètement indépendantes les unes des autres
dans l'état final. Il en va de même pour les excentricités. Bien qu'en moyenne 
les amplitudes $\varphi_j (\vec k,t)$ des modes amplifiés oscillent
en phase dans chacune des directions d'isospin, les vecteurs 
$\bfvarphi (\vec k,t)$ oscillent dans des directions complètement 
indépendantes : dans un certain sens, les différents modes sont comme 
autant de DCC indépendants. La conséquence phénoménologique immédiate
est l'atténuation, voire la disparition des fluctuations de la 
fraction neutre du nombre total de pions dans un bin d'espace des phases. 
Ceci est illustré sur la Fig.~\ref{fig_Htot} qui représente la 
distribution de la fraction neutre lorsque les contribution de 
seulement quelques modes équivalents (c'est à dire ayant la même 
distribution individuelle) sont pris en compte\footnote{Il s'agit du rapport 
   $\displaystyle \frac{N_3}{N_1 + N_2 + N_3}$, où 
   $\displaystyle N_j =  \sum_{k \in bin} \bar n_j (k)$.}.

\begin{figure}[htbp]
\epsfxsize=5.5in \centerline{ \epsfbox{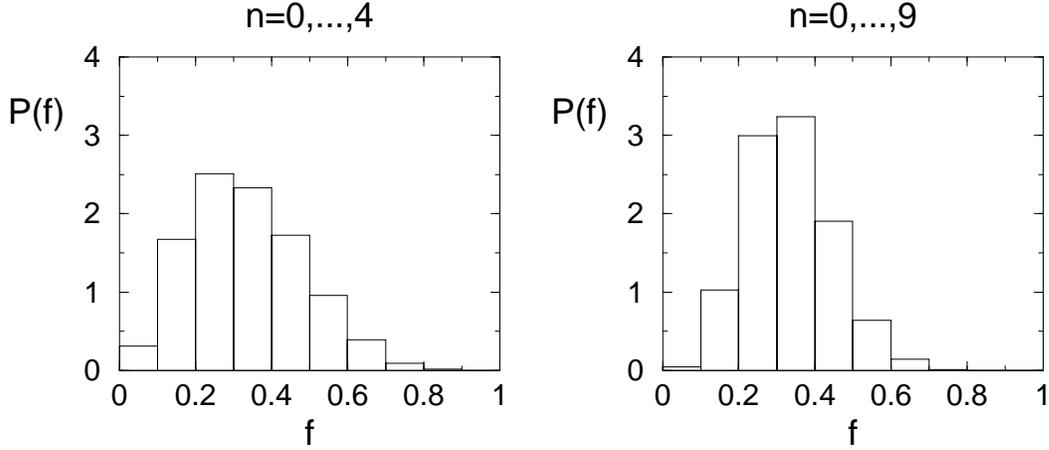}}
\caption{\small Histogrammes des valeurs des fractions neutres provenant de
    la somme des contributions des modes $n=0,...,4$ (à droite) et 
    $n=0,...,9$ (à gauche), à l'instant $t_f=10a$. Dans ce cas tous les modes
    considérés ont des distributions individuelles de $f$ essentiellement
    identiques. On voit que le signal est rapidement détruit par l'interférence
    entre les contributions statistiquement indépendantes.} 
\label{fig_Htot}
\end{figure}

\section{Discussion et conclusion}

\subsection{Résumé}

Partant d'un ensemble symétrique pour le champ chiral 
$\bfphi = (\bfpi,\sigma)$, nous résolvons numériquement les équations 
du mouvement du modèle $\sigma$-linéaire, qui peut être vu comme une théorie 
effective décrivant la dynamique des excitations bosoniques de basse énergie
de QCD. Le trempage des fluctuations de l'état initial est un mécanisme très 
efficace pour générer des oscillations cohérentes de grande longueur d'onde 
et de grande amplitude~\cite{RW}.
\par
L'analyse statistique détaillée de l'état final montre que la structure 
d'isospin du champ de pion y est sensiblement la même que dans l'état 
initial. Bien que différentes de la loi en $1/\sqrt f$, propre à une 
polarisation rectiligne, les distributions de la fraction neutre dans
les modes individuels sont très larges, ce qui est le point important 
pour la phénoménologie. Cependant, les polarisations des différents 
modes sont complètement indépendantes les unes des autres. La conséquence
phénoménologique importante est la suppression des grandes
fluctuations de la fraction de pions neutres lorsque les contributions
de plusieurs modes sont prises en compte, par exemple dans un bin de 
l'espace des impulsions.
\par
Le point clef est de réaliser que l'hypothèse d'un état initial spatialement
désordonné implique que les modes de Fourier du champ sont initialement 
indépendants les uns des autres. Notre résultat peut donc être énoncé comme 
suit : les non-linéarités de la dynamique {\em ne génèrent pas} de 
corrélations entre les modes. Le chaos que l'on a postulé à l'instant 
initial se retrouve dans l'état final. Cela contredit l'idée largement
répandue, selon laquelle la configuration générée dans le modèle le plus 
simple utilisé jusqu'à présent, est un DCC, tel qu'il a été proposé à 
l'origine. Ce modèle permet d'expliquer l'aspect classique du DCC,
pas sa polarisation hypothétique.
 
\subsection{Comparaison avec des travaux antécédents}

De nombreuses études concernant les implications phénoménologiques
du scénario du trempage ont été réalisées depuis~\cite{RW}.
Parmi les auteurs de ces travaux, certains~\cite{GGP,Raj,RAN2}, 
s'intéressant à la distribution de la fraction de pions neutres 
dans l'état final, ont obtenus des résultats analogues à celui de 
la Fig.~\ref{fig_Htot}. 
\par
Etudiant l'état final du modèle de Rajagopal et Wilczek en termes de
domaines d'orientation donnée (par analogie avec les domaines 
d'aimantation formés lors du trempage d'un feromagnétique), 
S.~Gavin, A.~Gocksch et R.~D.~Pisarski~\cite{GGP} arrivent à la
conclusion suivante : le système est formé d'un grand nombre de petits
domaines orientés aléatoirement et dont la taille est de l'ordre
de la longueur de corrélation, ce qui se traduit par une distribution
binomiale de la fraction neutre du nombre {\em total} de pions\footnote{Les
   auteurs en question étudient différents régimes de couplage. La
   conclusion décrite ici est celle obtenue dans le régime de couplage fort,
   pertinent à la physique des collisions d'ions lourds qui nous intéresse ici.}.
Par la suite, K.~Rajagopal a souligné le fait que le système en question, 
étant hors d'équilibre, ne peut être caractérisé par une échelle de 
longueur unique~\cite{Raj}, de sorte que l'image de la Réf.~\cite{GGP}
n'est pas appropriée. Dans ce chapitre, nous avons montré que les différents 
{\em modes} du champ se comportent comme des DCC indépendants : ils sont 
amplifiés (ce qui justifie l'approximation classique), ont une orientation 
différente de celle du vide physique, mais leurs polarisations dans l'espace
d'isospin sont statistiquement indépendantes. Bien sûr, en prenant la transformée 
de Fourier inverse, on arrive à la même conclusion dans l'espace des 
configurations : les orientations du champ $\bfphi$ sont indépendantes 
d'un site à l'autre. Il n'est cependant pas correct de dire
que l'on a produit une pléthore de petits DCC (domaines), chacun de la taille 
du pas du réseau. En effet, les modes de grande longueur d'onde, qui ont été
fortement amplifiés et qui sont, par conséquent les degrés de liberté 
intéressants, occupent {\em tout} le volume de la collision où le champ a été 
trempé. De plus, la représentation dans l'espace des impulsions donne une 
image plus proche de ce que les expérimentateurs voient. Quoi qu'il en 
soit, les ``domaines'' sont éphémères, les détecteurs enregistrent finalement 
les impulsions des particules produites.
\par
Dans la Réf.~\cite{Raj}, Rajagopal étudie les implications phénoménologiques
de son modèle, notamment quant à la distribution de la fraction de pions neutres
de basse énergie, en découpant l'espace des phases comme dans une expérience
réelle. Il calcule, pour un évènement donné, la distribution sur l'ensemble 
des bins, de la fraction neutre des pions dont l'impulsion 
$||\vec k|| \lesssim 300$~MeV. Il obtient une distribution relativement 
piquée autour de $f=1/3$, ce qu'il interprète comme : ``an admixture of a 
$1/\sqrt f$ distribution.'' Il a en tête l'image d'un condensat chiral 
désorienté, formé par la superposition cohérente des modes de grande 
longueur d'onde, plongé dans le bain thermique incohérent formé par les 
autres modes. Cette image dans la représentation en impulsion, 
est plus satisfaisante. De plus, l'observation du fait que des modes 
de différente nature (amplifiés ou thermalisés) contribuent de façon 
incohérente, supprimant ainsi le signal, est parfaitement en accord 
avec notre résultat. Cependant, nous avons mis en évidence le fait que 
les modes de basse fréquence n'agissent pas de concert pour former un 
condensat cohérent, c'est le fait d'avoir moyenné les contributions 
d'un grand nombre de modes dans chaque bin qui est responsable de la 
suppression des fluctuations de la fraction neutre.
\par
Finalement, J.~Randrup~\cite{RAN2} calcule la distribution
de la fraction neutre de pions dont les impulsions sont inférieures à 
différentes coupures. Il n'observe une loi large (en $1/\sqrt f$) que dans 
le cas où le seul mode zéro contribue. L'accord entre les résultat du calcul
classique exact et de l'approximation de champ moyen le conduit à dire : 
``each $\vec k$ contributes pions having an independant orientation in
isospin space.'' C'est précisémment ce que nous avons obtenu ici. 
Cependant Randrup travaille dans un modèle avec expansion, c'est à dire 
qu'il modélise l'ensemble de la collision. Par conséquent, la plupart 
des évènements qu'il considère n'ont pas été significativement amplifiés 
(cf. Chap.~\ref{PROBA}). Dans ce sens, son analyse est incomplète et ne 
permet pas de conclure quant à la question que nous nous sommes posé ici.
\par
Certains indices concernant la structure d'isospin générée dans le
scénario du trempage ont été relevés dans la littérature, en particulier
l'absence de fluctuations significatives de la fraction de pions neutres.
Cependant, l'analyse détaillée de la configuration du champ issu du 
trempage n'avait encore jamais été réalisée, en fait la question 
de la correspondance entre cette configuration et le DCC n'avait jamais 
été soulevée. Le travail présenté dans ce chapitre permet de clarifier
la situation.

\subsection{Spéculations}

Le scénario du trempage sous sa forme la plus simple ne permet
pas de générer une configuration de type DCC. L'inclusion
d'effets quantiques dans l'approximation de champ moyen, ou encore
de l'expansion, n'altèrent probablement pas cet état de faits, 
les premiers par construction, et la dernière n'étant qu'une façon 
sophistiquée de modéliser le trempage lui-même ainsi que le découplage 
entre les modes. Notre étude indique que c'est l'absence de corrélations
dans l'état initial qui est à l'origine du problème.
\par
En fait, si le phénomène du trempage des fluctuations initiales
semble assez naturel dans le contexte des collisions d'ions
lourds à haute énergie, l'hypothèse selon laquelle le système
est complètement thermalisé à l'instant initial est loin d'être
justifiée à l'heure actuelle. En particulier, dans ces systèmes de 
petite taille, les modes de grande longueur d'onde peuvent ne pas
avoir eu suffisamment de temps pour thermaliser, avant que le
phénomène d'instabilité spinodale ne se produise. On peut imaginer
que des corrélations, présentes dans l'état initial, soient
amplifiées lors de l'évolution hors d'équilibre ultérieure. En effet,
on s'attend, d'après les arguments de champ moyen, à ce que le phénomène
d'amplification par instabilité spinodale se produise pour une grande 
classe de conditions initiales\footnote{La principale restriction étant
   que les fluctuations initiales ne soient pas trop importantes, de façon
   à ce que le système puisse entrer dans la région d'instabilité.}.
Dans ce cas, le problème se ramène à celui, hautement non-trivial, de la 
construction d'un état initial réaliste, ce qui nécessite la description
des premiers instants de la collison\footnote{Le modèle de la Réf.~\cite{Anomaly}
   est une possibilité intéressante.}. 
\par
A l'inverse, si des corrélations étaient présentes dans l'état initial,
elles pourraient ne pas survivre à la dynamique non-linéaire. A nouveau,
l'état final serait constitué d'une superposition incohérente de modes 
désorientés. Dans cette situation, pour mesurer de grandes 
fluctuations\footnote{Ces fluctuations sont le signal d'un état semi-classique.}
de la fraction de pions neutres dans un bin donné, il est nécessaire
d'isoler aussi proprement que possible les modes individuels. Le volume 
du système doit alors ètre suffisamment grand pour que l'on ait une 
statistique satisfaisante, et suffisamment petit pour que deux modes 
voisins soient bien séparés, la coupure infrarouge étant inversement 
proportionnelle à la taille du système.

\subsection{Conclusion}

Nous avons montré que l'idée selon laquelle le mécanisme du trempage 
permet d'expliquer la formation d'un condensat chiral désorienté lors 
d'une collision d'ions lourds, idée très largement répandue, est incorrecte.
L'hypothétique structure d'isospin du DCC tel qu'il a été proposée à l'origine, 
n'est pas générée par la dynamique. Il n'existe à ce jour aucun modèle
microscopique capable de décrire la formation d'un DCC. C'est là le résultat 
essentiel de ce chapitre. 
\par
D'un autre côté, il est possible que l'image originale du DCC soit par trop 
idéalisée. Son extrême opposé est celle d'une superposition incohérente 
des modes du champ de pion classique. Nous avons vu que la manifestation la 
plus frappante de ces ``ondes classiques d'isospin'' reste la largeur 
inhabituelle de la distribution de leurs fractions neutres individuelles.
\par
Pour aller plus loin, des développements sont nécessaires,
notamment en ce qui concerne la description de l'état initial, 
à moins que des données expérimentales ne viennent trancher
la situation.

{\appendix

\chapter{Création de particules dans une \\
         géométrie en expansion}
\label{EXPANSION}

Dans un espace de géométrie (riemannienne) quelconque, certaines difficultés
apparaissent, comme par exemple le problème de la définition d'une énergie 
conservée, qui empêchent la généralisation de concepts usuels en théorie 
quantique des champs. Dans le problème qui nous concerne au Chap.~\ref{PROBA}, 
nous travaillons dans l'espace plat de Minkowski et le probléme de l'énergie 
est simplement une conséquence triviale du fait que nous avons choisi un 
système de coordonnées curviligne : l'énergie dans un co-volume décroît du 
fait de l'expansion (provoquant le trempage des fluctuations). Bien entendu,
l'énergie totale du système est conservée. Cependant tout n'est pas aussi 
facile. En particulier, nous allons voir qu'il n'est point besoin de courbure 
de l'espace pour que le concept de particule devienne ambigü (ce qui est 
embêtant car nous cherchons précisément à calculer le nombre de particules). 
Ce problème a une interpretation physique très simple : nous sommes dans un
référentiel accéléré. D'après le principe d'équivalence d'Einstein, cela 
revient à dire que nous sommes dans un certain champ de gravitation, lequel 
est couplé au champ de matière (par exemple par l'intermédiaire du terme de 
masse $\sqrt{-g} \, \phi^2$) et agit à la manière d'une source classique en 
créant des quanta de ce dernier. Définir le concept de particule revient à 
définir le concept de vide. Choisissons une définition du vide
à un instant $\tau_0$, ce ``vide'' est rempli de ``particules'' à un instant
ultérieur $\tau$. Nous allons discuter ce problème sur l'exemple d'un champ 
scalaire massif libre. L'action classique s'écrit
\eq
 \mathcal S = \frac{1}{2} \, \int d^4x \, \sqrt{-g}
 \left( g_{\mu \nu} \, \p^\mu \phi \, \p^\nu \phi + m^2 \phi^2 \right) \, .
\eq
Le tenseur d'énergie impulsion est~\cite{LANDAU,Birrell}
\bear
 T_{\mu\nu} & = & - \frac{2}{\sqrt{-g}} \, 
 \frac{\delta \mathcal S}{\delta g_{\mu\nu}} \\
 & = & \nabla_\mu \phi \, \nabla_\nu \phi - 
 g_{\mu\nu} \left( \frac{1}{2} \nabla_\alpha \phi \, \nabla^\alpha \phi -
 \frac{1}{2} m^2 \phi^2 \right) \, ,
\eear
et satisfait à l'équation
\eq
 \nabla^\mu T_{\mu\nu} =0 \, . 
\eq
En utilisant la propriété $\nabla^\alpha g_{\mu\nu} = 0$, on en déduit 
l'équation du mouvement pour le champ
\eq
 \left( \nabla_\mu \nabla^\mu + m^2 \right) \phi = 0 \, .
\eq
On remarque que la trace $T_\mu^\mu \neq 0$ dans la limite $m\rightarrow0$.
Cette forme du tenseur énergie-impulsion n'est pas invariante sous les 
dilatations d'échelle dans la limite où la théorie l'est. Pour
remédier à ce problème, on utilise le fait que le tenseur énergie-impulsion 
est défini à un terme de divergence nulle près. On remplace donc 
$T_{\mu\nu} \rightarrow T_{\mu\nu} + \delta T_{\mu\nu}$, 
avec\footnote{Ceci revient à remplacer $m^2 \rightarrow m^2 + \xi R$ dans 
   l'action, où $R$ est le scalaire de Ricci~(voir \cite{Birrell}). On obtient 
   alors directement le tenseur $T_{\mu\nu} + \delta T_{\mu\nu}$ en utilisant 
   les formules suivantes : 
   \bear
    \delta g^{\mu\nu} & =&  - g^{\mu\rho} g^{\nu\sigma} 
         \delta g_{\rho\sigma} \, \\
    \delta \sqrt{-g} & = & \frac{1}{2} \sqrt{-g} \, 
       g^{\mu\nu} \delta g_{\mu\nu} \, ,\\
    \delta R & = & R^{\mu\nu} \delta g_{\mu\nu} + 
       g^{\rho\sigma} g^{\mu\nu} (\nabla_\mu \nabla_\nu \delta g_{\rho\sigma} +
       \nabla_\sigma \nabla_\nu \delta g_{\rho\mu} ) \, .
    \eear
   Dans notre cas $R^{\mu\nu}=0$ et $R=0$, les équations du mouvement ne sont
   donc pas modifiées. On peut voir que, dans le cas d'une géométrie de courbure
   non-nulle, le terme $\xi R \phi^2$ assure l'invariance d'une théorie de masse
   nulle sous les transformations conformes de l'espace, la valeur du coefficient
   $\xi$ dépend de la dimension de l'espace, ici $d=4$ et $\xi=1/6$~\cite{Birrell}.}
\eq
 \delta T_{\mu\nu} = \frac{1}{6} \left( g_{\mu\nu} \nabla_\alpha \nabla^\alpha -
 \nabla_\mu \nabla_\nu \right) \phi^2 \, .
\eq
En utilisant la propriété $\nabla^\alpha g_{\mu\nu} =0$, on obtient
\eq
 \delta T_{\mu\nu} = \frac{1}{6} \, \nabla^\alpha 
 \left( g_{\mu\nu} \nabla_\alpha - g_{\alpha\mu} \nabla_\nu \right) \phi^2 \, . 
\eq
Le terme entre parenthèses est antisymétrique en $\alpha$, $\nu$, tandis
que, dans le cas d'un espace plat où les dérivées covariantes 
commutent\footnote{Le tenseur de courbure $R^{\mu\nu}$ est proportionnel 
   au commutateur $[\nabla^\mu \, ; \, \nabla^\nu]$.}, 
$\nabla^\mu \nabla^\alpha$ est symétrique. On a donc
bien $\nabla^\mu \delta T_{\mu\nu} = 0$.
En utilisant l'équation du mouvement, on obtient $T_\mu^\mu = m^2 \phi^2$, qui
s'annule bien dans la limite conforme.
\par
Le Hamiltonien du système s'écrit $H (\tau) = \tau^3 \, \int d^3x \, 
\sqrt h \, T_{\tau \tau}$, où $\tau^3 \, \sqrt h$ est le déterminant 
de la métrique décrivant la pseudo-sphère de rayon $\tau$
\beq
 H (\tau) = \tau^3 \, \int d^3x \, \sqrt{h} \, 
 \left( \frac{1}{2} (\p_\tau \hat\phi)^2 - 
 \frac{1}{2\tau^2} \phi \, \Delta^{(3)} \phi +
 \frac{m^2}{2} \phi^2 + 
 \frac{1}{2\tau} (\phi \p_\tau \phi + \p_\tau \phi \phi) \right) \, .
 \label{App1_ham}
\end{equation}\noindent

\subsubsection*{Quantification}

Le moment conjugué du champ $\phi$ est 
\eq
 \Pi = \frac{\delta \mathcal L}{\delta \p_\tau \phi} = 
 \sqrt{-g} \, \p_\tau \phi \, ,
\eq
et les relations de commutation canoniques s'écrivent
\beq
 \left[ \phi (\tau , \vec x) \, ; \, 
 \Pi (\tau , \vec x \, ') \right] = i \, \delta^{(3)} (\vec x - \vec x \, ') 
 \, ,
 \label{App1_commut1}
\end{equation}\noindent
les autres commutateurs à temps propres égaux étant nuls.
Dans le système de coordonnées de Minkowski, la théorie libre s'exprime 
comme un ensemble d'oscillateurs harmoniques indépendants. Dans notre
système de coordonnées curvilignes (et, plus généralement dans 
une géométrie courbe), le champ ``libre'', c'est à dire sans
auto-interaction, est couplé à un champ classique externe : 
la métrique. Nous allons voir cependant que le problème peut
être ramené à un ensemble d'oscillateurs indépendants dont les 
fréquences dépendent du temps. Introduisons tout d'abord la variable de temps
adimensionnée $u= \ln (m \tau)$ ($u$ est aussi appellé le ``temps conforme''), 
dans la suite toutes les fonctions exprimées comme fonction de $u$ portent un 
tilde : $f(\tau) \equiv \tilde f (u)$. Définissons les quantités adimensionnées
\eq
 \tvarphi (u,\vec x) = \tau \phi (\tau,\vec x) \, \, , \, \,
 \tpi (u,\vec x) = \tau \p_\tau \varphi (\tau,\vec x) = 
 \p_u \tvarphi (u,\vec x) \, \, , \, \,
 \tH (u) = \tau H (\tau) \, ,
\eq
ainsi que les projections sur les fonctions propres du laplacien 
tri-dimensionnel $\Delta^{(3)}$ (cf. Eq.~(\ref{eigenfunc}) du 
Chap.~\ref{PROBA})
\bearn
 \tvarphi_{\vec s} (u) & = & \int d^3 x \, \sqrt{h} \, 
 \mathcal Y_{\vec s}^* (\vec x) \, \tvarphi (u,\vec x) \, , \\
 \label{App1_modephi}
 \tpi_{\vec s} (u) & = & \int d^3 x \, \sqrt{h} \, 
 \mathcal Y_{\vec s}^* (\vec x) \, \tpi (u,\vec x) \, ,
 \label{App1_modepi}
\eearn
qui sont telles que $\tvarphi_{\vec s}^{\dagger} = (-1)^m \, \tvarphi_{-\vec s}$
et $\tpi_{\vec s}^{\dagger} = (-1)^m \, \tpi_{-\vec s}$. En termes de ces 
quantités, les modes $\vec s$ et $-\vec s$ apparaissent couplés. Pour rendre
manifeste le découplage des modes, introduisons les variables 
normales~\cite{DIRAC}, définies ci-dessous pour $m>0$
\bear
 \tq_{\vec s} = \frac{1}{\sqrt{2}} \, \left( \tvarphi_{\vec s} + 
 \tvarphi_{\vec s} ^{\dagger} \right) \, \, & , & \, \,
 \tp_{\vec s} = \frac{1}{\sqrt{2}} \, \left( \tpi_{\vec s} + 
 \tpi_{\vec s} ^{\dagger} \right) \, \, , \\
 \tq_{-\vec s} = \frac{i}{\sqrt{2}} \, \left( \tvarphi_{\vec s} -
 \tvarphi_{\vec s} ^{\dagger} \right) \, \, & , & \, \,
 \tp_{-\vec s} = \frac{-i}{\sqrt{2}} \, \left( \tpi_{\vec s} -
 \tpi_{\vec s} ^{\dagger} \right) \, \, .
\eear
En utilisant les Eqs.~(\ref{determinant}),~(\ref{ortho1}),~(\ref{ortho2}), et
(\ref{App1_commut1})), on obtient
\eq
 \left[ \tq_{\vec s} (u) \, ; \, \tp_{\vec s \, '}^{\dagger} (u) \right] =
 i \, \delta^{(3)} (\vec s - \vec s \, ') \, ,
\eq
les autres commutateurs étant nuls. Le hamiltonien s'écrit 
\eq
 \tH (u) = \int d^3s \left( \frac{1}{2} \, \tp_{\vec s}^2 + 
 \frac{\tomega_s^2 (u)}{2} \, \tq_{\vec s}^2 \right) \, ,
\eq
où $\tomega_s^2 (u) = s^2 + \tau^2 \, m^2$.
On a donc, comme annoncé, une superposition d'oscillateurs indépendants dont
les fréquences dépendent du temps\footnote{Dans le Chap.~\ref{PROBA}, la 
   dépendance temporelle des fréquences a deux origines : la géométrie, qui
   donne lieuy au terme $s/\tau$ dans l'expression~(\ref{App1_dispersion})
   des fréquences physiques, et l'auto-interaction du champ, qui donne lieu
   à la dépendance temporelle de la masse effective des quasi-particules 
   dans l'approximation de champ moyen.}. Les équations de Heisenberg s'écrivent
\bear
 \tq' (u) & = & -i \, \left[ \tH (u) \, ; \, \tq (u)  \right] = \tp (u) \, , \\
 \tp' (u) & = & -i \, \left[ \tH (u) \, ; \, \tp (u) \right] = 
 - \tomega (u) \, \tq (u) \, ,
\eear
où l'on a noté $\tilde f' (u) = \dd \tilde f / \dd u$.
La fréquence physique de nos oscillateurs est 
\beq
 \omega_s (\tau) = \frac{\tomega_s (u)}{\tau} = 
 \sqrt{\frac{s^2}{\tau^2} + m^2}\, .
 \label{App1_dispersion}
\end{equation}\noindent
L'effet du (pseudo) champ gravitationnel, c'est à dire
de l'expansion, est de décaler les fréquences vers le rouge au fur et à 
mesure que le temps avance. On voit apparaitre le problème de la définition
de la notion de particule. Nous allons maintenant préciser ce point.

\subsubsection*{Création de particules, approximation adiabatique}

Introduisons les opérateurs de création et d'anihilation $\ta_{\vec s}^{\dagger}$
et $\ta_{\vec s}$
\bearn
 \tvarphi_{\vec s} (u) & = & \tpsi_s (u) \, \ta_{\vec s} + 
 \tpsi_s^* (u) \, (-1)^m \, \ta_{-\vec s}^{\dagger} \, , \\
 \label{App1_Heisphi1}
 \tpi_{\vec s} (u) & = & \tpsi_s' (u) \, \ta_{\vec s} + 
 \tpsi_s^{*'} (u) \, (-1)^m \, \ta_{-\vec s}^{\dagger} \, ,
 \label{App1_Heispi1}
\eearn
qui satisfont les relations de commutation
\beq
 \left[ \ta_{\vec s} \, ; \, \ta_{\vec s'}^{\dagger} \right] = 
 \delta^{(3)} (\vec s - \vec s^{'}) \, ,
 \label{App1_commut2}
\end{equation}\noindent
les autres commutateurs étant nuls.
La fonctions complexe $\tpsi_s$ est solution de l'équation
\beq
 f'' + \tomega_s (u) \, f = 0 \, .
 \label{App1_modeeq}
\end{equation}\noindent
Les relations (\ref{App1_commut1}) et(\ref{App1_commut2}) sont équivalentes 
pourvu que\footnote{Le Wronskien $\tW_s$ est conservé : $\tW' = 0$.}
\beq
 \tW_s = \tpsi_s (u) \, \tpsi_s^{*'} (u) - \tpsi_s^* (u) \, \tpsi_s' (u) =
 i \, .
 \label{App1_wronsk}
\end{equation}\noindent
Soit $u_0$ (ou $\tau_0$), l'instant de référence, auquel les représentations
de Heisenberg et de Schr\"odinger coïncident, on a
\bear
 \tvarphi_{\vec s} & = & \tpsi_s (u_0) \, \ta_{\vec s} + 
 \tpsi_s^* (u_0) \, (-1)^m \, \ta_{-\vec s}^{\dagger} \, , \\
 \tpi_{\vec s} & = & \tpsi_s' (u_0) \, \ta_{\vec s} + 
 \tpsi_s^{*'} (u_0) \, (-1)^m \, \ta_{-\vec s}^{\dagger} \, ,
\eear
et dénotons par $\tU(u,u_0)$ l'opérateur unitaire qui connecte ces deux
représentations\footnote{Cet opérateur peut être construit 
   explicitement~\cite{Combescure}.}
\bearn
\label{App1_Heisphi2}
 \tvarphi_{\vec s} (u) & = & 
 \tU (u,u_0) \, \tvarphi_{\vec s} \, \tU^{-1} (u,u_0) \nonumber \\
 & = & \tpsi_s (u_0) \, \ta_{\vec s} (u) + 
 \tpsi_s^* (u_0) \, (-1)^m \, \ta_{-\vec s}^{\dagger} (u) \, , \\
\label{App1_Heispi2}
 \tpi_{\vec s} (u) & = & 
 \tU (u,u_0) \, \tpi_{\vec s} \, \tU^{-1} (u,u_0) \nonumber \\
 & = & \tpsi_s' (u_0) \, \ta_{\vec s} (u) + 
 \tpsi_s^{*'} (u_0) \, (-1)^m \, \ta_{-\vec s}^{\dagger} (u) \, ,
\eearn
où $\ta_{\vec s} (u) = \tU (u,u_0) \, \ta_{\vec s} \, \tU^{-1} (u,u_0)$. 
D'après (\ref{App1_Heisphi1})-(\ref{App1_Heispi1}) et 
(\ref{App1_Heisphi2})-(\ref{App1_Heispi2}), et en utilisant (\ref{App1_wronsk})
\beq
 \ta_{\vec s} (u) = \tmu_s (u) \ta_{\vec s} + 
 \tnu_s (u) \, (-1)^m \, \ta_{-\vec s}^{\dagger} \, ,
 \label{App1_bogo}
\end{equation}\noindent
avec
\bearn
\label{App1_alphabogo}
 i \, \tmu_s^* (u) & = & \tpsi_s (u_0) \, \tpsi_s^{*'} (u) - 
 \tpsi_s' (u_0) \, \tpsi_s^* (u) \, , \\
\label{App1_betabogo}
 i \, \tnu_s^* (u) & = & \tpsi_s (u_0) \, \tpsi_s' (u) - 
 \tpsi_s' (u_0) \, \tpsi_s (u) \, .
\eearn
La transformation de Bogoliubov (\ref{App1_bogo}) est unitaire :
\eq
 | \tmu_s (u) |^2 - | \tnu_s (u) |^2 = 1 \, .
\eq
Supposons que le système soit initialement (à l'instant $u_0$) dans
l'état $| 0,u_0 \rangle$ défini par la relation 
$a_{\vec s} \, | 0,u_0 \rangle = 0 \, , \forall \vec s$. Le nombre de particules
inital dans le mode $\vec s$ est donc $\langle 0,u_0 | \, a_{\vec s}^{\dagger} \,
a_{\vec s} \, | 0,u_0 \rangle = 0$. Le nombre de particules dans le mode $\vec s$
à l'instant $u$ est facilement obtenu à l'aide de (\ref{App1_bogo}), 
(\ref{App1_alphabogo}) et (\ref{App1_betabogo}) :
\eq
 \tilde n_s (u) = 
 \langle 0,u_0 | \, \ta_{\vec s}^{\dagger} (u) \, 
 \ta_{\vec s} (u) \, | 0,u_0 \rangle = | \tnu_s (u) |^2 \, .
\eq
On voit explicitement le problème mentionné au début de cette partie :
différents choix de la fonction $\tpsi_s$ correspondent à différents
opérateurs $a_{\vec s}$ et donc différents vides $| 0,u_0 \rangle$. De façon
générale, aucun choix de $a_{\vec s}$ ne correspond à l'opérateur 
d'anihilation de particules physiques, le nombre de particules n'étant
pas constant dans une métrique dépendant du temps. Le problème de la
dépendance en temps du vide est assez sérieux : par exemple, il est habituel, 
en théorie quantique des champs, de soustraire au Hamiltonien sa valeur moyenne 
dans le vide, infinie, les seules quantités physiquement observables 
étant les différences d'énergie entre états. Si la définition
du vide dépend du temps, cette procédure n'a plus de sens car on doit
alors soustraire des quantités différentes à des instants différents,
on ne peux pas comparer deux mesures faites à des instants différents.
\par
Notons de plus que la dépendance temporelle du nombre moyen de particules
rend la mesure de cette quantité tout à fait incertaine, autrement dit,
le nombre moyen de particules n'a pas une signification physique
bien définie~\cite{Parker}. En effet, supposons que notre observateur en
co-mouvement cherche à faire une mesure de cette quantité, mesure 
nécessitant une durée $\Delta \tau$. Le résultat de la mesure est entâché 
de l'incertitude
\eq
 \Delta N \gtrsim (m \Delta \tau )^{-1} + |A| \Delta \tau \, ,
\eq
où le premier terme vient du principe d'incertitude de Heisenberg et
où $A$ est le taux moyen de production de particule par unité de temps 
dans le co-volume considéré. L'incertitude minimale possible est
$\Delta N_{min} = 2(|A|/m)^{1/2}$, avec $\Delta \tau = (m|A|){-1/2}$. 
Il n'existe pas de valeur de $\Delta \tau$ pour laquelle $\Delta N_{min}$
s'annule.
\par
L. Parker a étudié ce problème vers la fin des années 1960~\cite{Parker} et 
a proposé de définir les particules physiques, ou, de manière équivalente, 
les fonctions modes $\tpsi_s (u)$, comme étant celles qui minimisent 
$\Delta N_{min}$, c'est à dire qui minimisent le taux de production $A$.
Cela revient à faire un développement dans le taux d'expansion, c'est à 
dire dans un paramètre décrivant la ``vitesse'' à laquelle la métrique 
change avec le temps. Intuitivement, il est clair que plus l'expansion 
est lente, plus on est proche du cas statique (Minkowski), et plus le 
taux de production de particules est faible : c'est le développement 
adiabatique. Considérons, avec L.~Parker et S.~A.~Fulling~\cite{Fulling}, 
les solutions ``de fréquence positive'' généralisées des Eqs.~(\ref{App1_modeeq}) 
et (\ref{App1_wronsk}) :
\beq
 \tg_s (u) = \frac{1}{\sqrt{2 \tOmega_s (u)}} \, 
 \exp \left( -i \, \int_{u_0}^u dv \, \tOmega_s (v) \right) \, ,
 \label{App1_adiab}
\end{equation}\noindent
où $\tOmega_s (u)$ est une fonction réelle satisfaisant l'équation
(cf.~(\ref{App1_modeeq}))
\eq
 \tOmega_s^2 (u) = \tomega_s^2 (u) - 
 \frac{1}{2} \frac{\tOmega_s''}{\tOmega_s} + 
 \frac{3}{4} \left( \frac{\tOmega_s'}{\tOmega_s} \right)^2 \, .
\eq
L'approximation adiabatique d'ordre $0$ est obtenue en identifiant, à 
un instant de référence $u_0$,
\eq
 \tpsi_s (u_0) = \tg_s^{(0)} (u_0) \, \, \, , \, \, \, 
 \tpsi_s' (u_0) = \tg_s^{(0)'} (u_0) \, ,
\eq
où $\tg_s^{(0)}$ est donnée par l'Eq.~(\ref{App1_adiab}) en remplaçant 
$\tOmega_s \rightarrow \tOmega_s^{(0)} = \tomega_s$ :
\bear
 \tg_s^{(0)} (u_0) & = & \frac{1}{\sqrt{2 \tomega_s (u_0)}} \, , \\
 \tg_s^{(0)'} (u_0) & = & - \left[ \frac{\tomega_s' (u_0)}{2 \tomega_s (u_0)} +
 i \, \tomega_s (u_0) \right] \, \tg_s^{(0)} (u_0) \, .
\eear
Les approximations successives, $\tg_s^{(p)}$, sont obtenues par
récurence
\eq
  \left( \tOmega_s^{(p+1)} (u) \right)^2 = \tomega_s^2 (u) - 
 \frac{1}{2} \frac{\tOmega_s^{(p)''}}{\tOmega_s^{(p)}} + 
 \frac{3}{4} \left( \frac{\tOmega_s^{(p)'}}{\tOmega_s^{(p)}} \right)^2 \, .
\eq
Mentionnons enfin le fait que cette procédure de quantification permet 
de renormaliser la théorie de façon non-ambig\"ue, c'est à dire de manière indépendente du temps~\cite{Lamperthesis,Fulling,Birrell}. On voit en 
effet que l'approximation adiabatique est d'autant meilleure que $s$ est 
grand : dans la limite $s \rightarrow +\infty$ on a $\tOmega_s \simeq s$, 
qui ne dépend pas du temps. On comprend donc qu'il soit possible de traiter 
les divergences ultraviolettes les plus sévères une fois pour toutes à 
l'instant de référence $u_0$. En fait on peut montrer~\cite{Fulling,Birrell} 
qu'en identifiant les fonctions modes $\tpsi_s$ avec $\tg_s^{(p)}$ à
l'instant $u_0$, le taux de création de particules dans le mode $\vec s$ à 
un instant ultérieur $u$ décroît comme $1/s^{p+1}$.

\chapter{Equilibre thermique et équilibre thermique local}
\label{THERMAL}

Dans cette Annexe, nous présentons le modèle 
$\sigma$-linéaire dans l'approximation grand $N$, dans le 
cas où le système est à l'équilibre thermodynamique à la 
température $T$. Nous calculons ensuite les variances 
des distributions gaussiennes utilisées pour échantillonner les 
conditions initiales au Chap.~\ref{PROBA}. Nous utilisons
des notations conventionnelles : $\vec x=(x,y,z)$ désigne
le vecteur position dans le système de coordonnées carthésiennes
et $d^3x=dxdydz$ l'élément de volume infinitésimal.

\section{Equilibre thermodynamique}

Considérons un volume infini où le champ chiral est à l'équilibre 
thermodynamique, la matrice densité du système s'écrit
\beq
\label{TH_density}
 \rho = \frac{1}{\mathcal Z} \, \mbox{e}^{-H/T} \, ,
\end{equation}\noindent
où $H$ est le Hamiltonien et $\mathcal Z = Tr[\rho]$ est la fonction
de partition. Dans cet état, le paramètre d'ordre est uniforme et constant
\bearn
\label{TH_phi}
 \langle \, \Phi_a (t,\vec x) \, \rangleT & = & \phiT^a \, , \\
\label{TH_phidot}
 \langle \, \dot\Phi_a (t,\vec x) \, \rangleT & = & 0 \, , 
\eearn
avec la notation $\langle \mathcal O \rangleT = Tr[ \rho \, \mathcal O ]$.
On définit le champ de fluctuation autour de cette valeur 
moyenne\footnote{La fluctuation $\bfdphi$ définie
   ici, est différente de celle définie au Chap.~\ref{PROBA}}
\bearn
\label{TH_dphi}
 \dphi_a (t,\vec x) & = & \Phi_a (t,\vec x) - \phiT^a \, , \\
\label{TH_dphidot}
 \dot{\dphi}_a (t,\vec x) & = & \dot\Phi_a (t,\vec x) \, .
\eearn
Dans l'approximation de champ moyen grand $N$, le système de particules
en interaction est remplacé par un ensemble de ``quasi-particules'' indépendantes 
de masse effective $\sqrt{\chiT}$. En d'autres termes, le hamiltonien est 
diagonnal dans la base ``nombre'' associée à ces quasi-particules. En dénotant respectivement par $a_{a,\vec k}^\dagger$ et 
$a_{a,\vec k}$ les opérateurs de création et d'anihilation d'une 
quasi-particule ``thermique'' d'impulsion $\vec k$ et de composante 
chirale $a$, on a, dans la représentation de Heisenberg,
\bearn
\label{TH_fourier}
 \dphi_a (t,\vec x) & = & \int \frac{d^3k}{(2 \pi)^3} \, 
 \frac{1}{\sqrt{2 E_k (\chiT)}} \, 
 \left[ a_{a,\vec k} \, \mbox{e}^{-ikx} +
 a_{a,\vec k}^\dagger \, \mbox{e}^{ikx} \right] \, , \\
\label{TH_fourierdot}
 \dot{\dphi}_a (t,\vec x) & = & -i \, \int \frac{d^3k}{(2 \pi)^3} \, 
 \sqrt{\frac{E_k (\chiT)}{2}} \, 
 \left[ a_{a,\vec k} \, \mbox{e}^{-ikx} -
 a_{a,\vec k}^\dagger \, \mbox{e}^{ikx} \right] \, ,
\eearn
où $E_k (\chiT)=\sqrt{k^2 + \chiT}$ et $kx = E_k (\chiT)t - \vec k \cdot \vec x$.
Avec notre choix de normalisation, les relations de commutation sont
\beq
 \left[ a_{a,\vec k} \, ; \, a_{b,\vec k \, '}^\dagger \right] = 
 (2 \pi)^3 \, \delta_{ab} \, \delta^{(3)} (\vec k - \vec k \, ') \, .
\end{equation}\noindent
Le hamiltonien décrivant les fluctuations autour du paramètre d'ordre s'écrit
\beq
 H = \int \frac{d^3k}{(2 \pi)^3} \, E_k (\chiT) 
 \sum_{a=1}^N \left( a_{a,\vec k}^\dagger \, a_{a,\vec k} + \frac{1}{2} \right) \, ,
\end{equation}\noindent
On obtient aisément
\beq
\label{TH_state}
 \langle \, a_{a,\vec k}^\dagger \, a_{b,\vec k \, '} \, \rangleT = 
 (2 \pi)^3 \, \delta_{ab} \, \delta^{(3)} (\vec k - \vec k \, ') \, N_k (\chiT)
 \, \, \, , \, \, \, 
 \langle \, a_{a,\vec k} \, a_{b,\vec k \, '} \, \rangleT = 0 \, \, ,
\end{equation}\noindent
où $N_k (\chiT)$ est la distribution de Bose-Einstein pour les quasi-particules : 
\beq
 N_k (\chiT) = \frac{1}{\mbox{e}^{E_k (\chiT)/T} - 1} \, .
\end{equation}\noindent
La masse effective $\sqrt{\chiT}$ des quasi-particules satisfait la relation
d'auto-cohérence (équation de gap, cf. (\ref{chi}))
\beq
\label{TH_chi}
 \frac{\chiT}{\lambda} = \phiT^2 - v^2 + N \int_0^\Lambda \frac{k^2 dk}{2 \pi^2} \,
 \frac{2 N_k(\chiT) + 1}{2 E_k (\chiT)} \, ,
\end{equation}\noindent
où $\phiT^2 = \bfphiT \cdot \bfphiT$. La valeur du
paramètre d'ordre $\phiT$ est telle que le potentiel effectif est minimum 
(cf. Eq.~(\ref{mean})) et dépend elle-même de $\chiT$
\beq
\label{TH_mean}
 \chiT \bfphiT = H {\bf n}_\sigma \, .
\end{equation}\noindent
Les divergences apparaissant dans l'Eq.~(\ref{TH_chi}) sont absorbées dans
les paramètres nus $v$ et $\lambda$ de la façon exposée au Chap.~\ref{PROBA}.
Dans la limite chirale $H=0$, on voit, d'après l'eq.~(\ref{TH_mean}), que
dans la phase brisée\footnote{Dans l'aproximation grand $N$, 
   la transition de phase chirale est du second ordre et la température
   critique est donnée par~\cite{Lamperthesis,DJ} $T_c=\sqrt{(12/N)}f_\pi 
   \simeq 160$ MeV pour $N=4$.} ($T<T_c$), où $\phiT \neq 0$, la masse effective des
pions est nulle, tandis que dans la phase symétrique où $\phiT = 0$, celle-ci
est {\it a priori} non-nulle. Dans la cas où $H \neq 0$, cette transition
de phase disparait, mais la discussion précédente reste approximativement 
valable : on parle d'une ``phase'' de haute température, où $\phiT \ll f_\pi$,
et d'une ``phase'' de basse température, où $\phiT \sim f_\pi$. La variation 
de $\chiT$ avec la température est montrée sur la Fig.~\ref{TH_fig_chiT} pour
différentes valeurs de la coupure $\Lambda$. Comme attendu, à mesure que la 
valeur de la température se rapproche de celle de la coupure, la présence de 
celle-ci se fait de plus en plus ressentir\footnote{Notre modèle n'a de sens 
physique que pour $T \ll \Lambda$.}. L'étude présentée au Chap.~\ref{PROBA} 
concerne la valeur $T=200$ MeV, où la coupure n'a que peu d'influence, le 
point de comparaison $T=400$ MeV est à la limite de validité du modèle 
(voir la discussion au Chap.~\ref{PROBA}).

\begin{figure}[h]
\label{TH_fig_chiT}
\epsfxsize=4.in \centerline{ \epsfbox{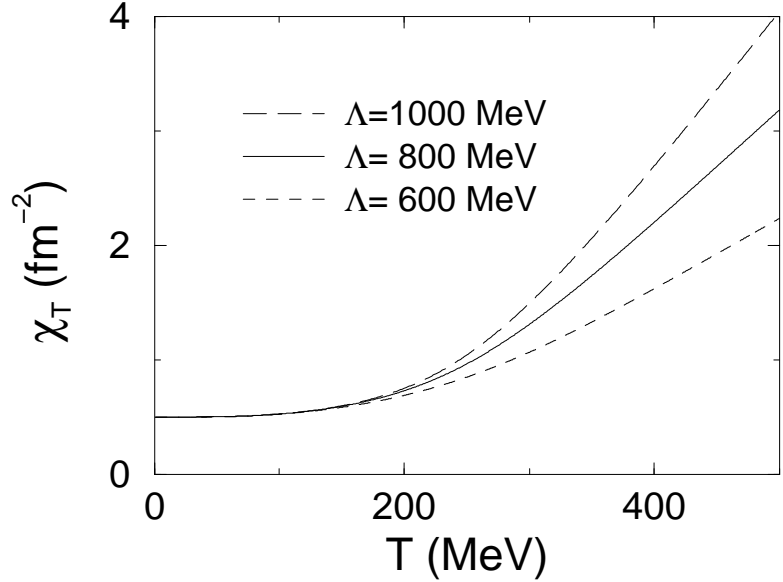}}
\caption[fig1]{\small La masse effective au carré $\chiT$ en fonction de $T$,
pour différentes valeurs de la coupure ultra-violette $\Lambda$.} 
\end{figure}

%
%
%
%
%
%

\section{Equilibre thermique local}

Considérons un tout autre système : un petit volume $V_0$, sphérique,
où le champ est à l'équilibre thermique local à la température $T$. 
Cette notion signifie que les fluctuations statistiques dans le volume 
$V_0$ sont les mêmes que si celui-ci était une sous-partie du système 
considéré dans la partie précédente (à la même température). 
Il est bien évident que ceci n'a de sens que si ce qui se passe à 
l'intérieur du volume $V_0$ ne dépend pas, ou peu, de ce qui se 
passe à l'extérieur, ce qui veut dire que la taille de notre système 
doit être de l'ordre de grandeur de la longueur de corrélation au moins. 
Dans ce qui suit, nous nous intéressons à la distribution de probabilité 
des valeurs possibles des moyennes spatiales
du champ et de sa dérivée sur le volume $V_0$. Ces quantités s'interprètent,
pour un observateur vivant à l'intérieur de $V_0$, comme les valeurs
du paramètre d'ordre et de sa dérivée temporelle. Utilisant la définition
de l'équilibre local, nous délimitons une petite boule $V_0$ dans le 
système en équilibre thermodynamique décrit plus haut, et calculons 
la distribution des fluctuations statistiques dans ce sous-volume.
Pour éviter d'écrir inutilement la variable de temps, nous travaillons 
dans la représentation de Scr\"odinger.
\par
On cherche donc la distribution de probabilité des résultats
possibles $\bar\bfphi$ et $\bar{\dot\bfphi}$ de la mesure des 
observables associées aux opérateurs
\bearn
\label{TH_SA1}
 \bar\Phi  & = & \frac{1}{V_0} \int_{V_0} d^3x \, \Phi (\vec x) \, ,\\
\label{TH_SA2}
 \bar{\dot\Phi} & = & \frac{1}{V_0} \int_{V_0} d^3x \, \dot\Phi (\vec x) \, .
\eearn
Celle-ci est déterminée par la fonction caractéristique
\beq
 Z ({\bf p},{\bf q}) = 
 \langle \, \exp i \, \left\{ {\bf p} \cdot \bar\Phi + 
 {\bf q} \cdot \bar{\dot\Phi} \right\} \, \rangleT \, .
\end{equation}\noindent
Le Hamiltonien étant quadratique, la distribution cherchée, qui est en fait 
l'exact analogue de la distribution de Wigner, bien connue en mécanique 
quantique, est gaussienne. Les moments du logarithme de $Z ({\bf p},{\bf q})$, 
ou cumulants, d'ordre supérieur à $2$ sont nuls. Les valeurs moyennes, ou 
cumulants d'ordre $1$, sont aisément obtenues à l'aide des Eq.~(\ref{TH_phi}),
(\ref{TH_phidot}) et (\ref{TH_mean}) :
\bearn
\label{TH_moyphi}
 \mbox{E} \left[ \bar\phi_a \right] & = &  
 \langle \, \bar\Phi_a \, \rangleT = \frac{H}{\chiT} \, \delta_{a0} \, , \\
\label{TH_moyphidot}
 \mbox{E} \left[ \bar{\dot\phi}_a \right] & = &  
 \langle \, \bar{\dot\Phi}_a \, \rangleT = 0 \, .
\eearn
Les variances et covariances (cumulants d'ordre $2$) s'écrivent 
(cf. (\ref{TH_dphi}) et (\ref{TH_dphidot}))
\bearn
\label{TH_varphi}
 \mbox{Var} \left[ \bar\phi_a \right] & = & \frac{1}{V_0^2} \, 
 \int_{V_0} d^3x \, d^3y \, 
 \langle \, \dphi_a (\vec x) \dphi_a (\vec y) \, \rangleT
 \, , \\
\label{TH_varphidot}
 \mbox{Var} \left[ \bar{\dot\phi}_a \right] & = &\frac{1}{V_0^2} \, 
 \int_{V_0} d^3x \, d^3y \, \langle \, \dot{\dphi}_a (\vec x)  
 \dot{\dphi}_a (\vec y) \, \rangleT \, , \\
\label{TH_cov}
 \mbox{Cov} \left[ \bar\phi_a , \bar{\dot\phi}_a \right] & = & \frac{1}{2 V_0^2} \,
 \int_{V_0} d^3x \, d^3y \, \langle \, \dphi_a (\vec x) \dot{\dphi}_a (\vec y) +
 \dot{\dphi}_a (\vec y) \dphi_a (\vec x)  \, \rangleT \, ,
\eearn
les termes non-diagonnaux dans les indices chiraux sont nuls. 
Le hamiltonien étant quadratique dans le champ
$\dphi$ et sa dérivée $\dot{\dphi}$ et ne contenant pas de termes croisés
du type $\dphi \, \dot{\dphi}$, il est clair que la fonction de corrélation
apparaissant dans l'Eq.~(\ref{TH_cov}) est nulle. Calculons la variance de la 
moyenne spatiale du champ, Eq.~(\ref{TH_varphi}). La fonction de corrélation 
apparaissant dans cette formule est
\beq
 \langle \, \dphi_a (\vec x) \dphi_a (\vec y) \, \rangleT = 
 \int \frac{d^3 k}{(2\pi)^3} \, 
 \frac{2 N_k (\chiT) + 1}{2 E_k (\chiT)} \, \,
 \mbox{e}^{i \, \vec k \cdot (\vec x - \vec y)} \, .
\end{equation}\noindent
Introduisons les coordonnées sphériques $\vec x = (x,\theta_x,\varphi_x)$
et $\vec y = (y,\theta_y,\varphi_y)$, où les angles azimutaux $\theta_x$
et $\theta_y$ sont repérés par rapport à la direction du vecteur $\vec k$.
Les intégrations sur les orientations possibles des vecteurs $\vec k$, 
$\vec x$ et $\vec y$ sont triviales. Après les changements d'échelle
$x \rightarrow x/R$ et $y \rightarrow y/R$, il vient ($k=|\vec k|$)
\beq
\label{TH_sigma1}
 V_0^2 \, \sigma_1^2 = 4 \, R_0^4 \, \int_0^{+\infty} 
 \frac{dk}{E_k} \, \coth \left( \frac{E_k}{2T} \right) \, \mathcal F (k R_0) \, ,
\end{equation}\noindent
avec
\beq
 \mathcal F (q) = 
 \int_0^{1} dx \, \int_{0}^{1} dy \, x  y \, \sin (q x) \, \sin (q y) 
 = \frac{( \sin q - q \, \cos q )^2}{q^4} \, .
\end{equation}\noindent
Ci-dessus on a dénoté $\sigma_1^2$ la variance cherchée 
(Eq.~(\ref{TH_varphi})). On obtient de la même façon la variance
(\ref{TH_varphidot}), que l'on dénote par $\sigma_2^2$ : 
\beq
\label{TH_sigma2}
 V_0^2 \, \sigma_2^2 = 4 \, R_0^4 \, \int_0^{+\infty} 
 dk \, E_k \, \coth \left( \frac{E_k}{2T} \right) \, \mathcal F (k R_0) \, .
\end{equation}\noindent
Réécrivons les formules (\ref{TH_sigma1}) et (\ref{TH_sigma2}) sous la forme
générique ($\mathcal F (q) = \mathcal F (-q)$)
\beq
 V_0^2 \, \sigma^2 = 2 \, R_0^3 \, \int_{-\infty}^{+\infty} dy \,
 g \left( \frac{y}{R_0} \right) \, \mathcal F(y) \, ,
\end{equation}\noindent
où $g(k)$ est une fonction de $E_k = \sqrt{k^2 + \chiT}$.
Dans la limite où $R_0 \sqrt{\chiT} \gg 1$, on obtient le premier terme
du développement asymptotique de l'expression ci-dessus :
\beq
 V_0^2 \, \sigma^2 \sim 2 \, R_0^3 \, g(0) \, 
 \int_{-\infty}^{+\infty} dy \, \mathcal F(y) \, .
\end{equation}\noindent
En utilisant
\beq
 \int_{-\infty}^{+\infty} dy \, \mathcal F(y) = \frac{\pi}{3} \, ,
\end{equation}\noindent
on obtient (cf. (\ref{varphiapprox}) et (\ref{varphidotapprox})) : 
\beq
\label{TH_asymptotique}
 \sigma^2 \sim \frac{1}{2V_0} \, g(0) \, ,
\end{equation}\noindent
Cette formule est valable quand le rayon $R_0$ de la bulle est suffisament 
grand devant la longueur de corrélation $\lambda_T = 1/\sqrt{\chiT}$, la 
dépendance en $1/V_0$ traduisant le fait que les fluctuations à l'intérieur du 
volume $V_0$ ne dépendent pas de son environnement : diverses cellules 
de taille $R_0 \gg \lambda_T$ sont statistiquement indépendantes.
\par
Dans le Chap.~\ref{PROBA}, nous sommes intéressés à des valeurs de $R_0$
aussi petites qu'il est permis physiquement, c'est à dire 
$R_0 \gtrsim \lambda_T$. A strictement parler, notre modèle est défini
en présence d'une coupure ultraviolette $\Lambda$, et les intégrales
(\ref{TH_sigma1}) et (\ref{TH_sigma2}) doivent être calculées avec cette
coupure (notons que la coupure intervient aussi dans le calcul de $\chiT$).
Cela n'influe que très peu sur la valeur de la dispersion $\sigma_1$ 
(Eq.~(\ref{TH_sigma1})), tant que $\Lambda R_0 \gg 1$. En effet, la moyenne 
spatiale (\ref{TH_SA1}) supprime les contributions des modes de petite 
longueur d'onde, seuls les modes tels que $k R_0 \lesssim 1$ survivent.
La Fig.~\ref{TH_fig_disp1} montre la variation de $\sigma_1$ en fonction
du rapport $R_0/\lambda_T = R_0 \sqrt{\chiT}$ pour $T=200$ MeV et pour
diverses valeurs de la coupure. 
La présence de celle-ci se fait plus ressentir dans le calcul de la dispersion
$\sigma_2$ qui, à cause du facteur $E_k$ au numérateur dans l'intégrand
de (\ref{TH_sigma2}), diverge logarithmiquement dans l'ultra-violet. 
Cependant, pour des valeurs physiques de la coupure ($\Lambda \sim 1$ GeV), 
cette dépendance logarithmique n'a que peu d'influence sur la valeur de 
$\sigma_2$. La Fig.~\ref{TH_fig_disp2} est l'analogue de la Fig.~\ref{TH_fig_disp1}, 
pour la dispersion $\sigma_2$ des fluctuations de la dérivée temporelle 
du paramètre d'ordre. Dans les deux cas, les courbes en pointillés sont 
obtenues à partir de la formule asymptotique (\ref{TH_asymptotique}).
Les valeurs des dispersions $\sigma_1$ et $\sigma_2$ utilisées dans le
calcul du Chap.~\ref{PROBA} sont résumées dans le tableau~\ref{TH_tab}.

\begin{table}[htb]
\begin{center}
\begin{tabular}{|c|c|c|c|c|}
\hline
$T$ & $\chiT$ & $R_0/\lambda_T$ & $\sigma_1$ & $\sigma_2$ \\
(MeV) & (fm$^{-2}$) & & (fm$^{-1}$) & (fm$^{-2}$) \\
\hline
200 & 0.73 & 1 & 0.21 & 0.40 \\
\hline
200 & 0.73 & 2 & 0.11 & 0.14 \\
\hline
400 & 2.2 & 1 & 0.37 & 1.07 \\
\hline
400 & 2.2 & 2 & 0.19 & 0.42 \\
\hline
\end{tabular}
\end{center}
\caption{Les différentes valeurs des dispersions utilisées dans le
   calcul présenté au Chap.~\ref{PROBA} ($\Lambda = 800$ MeV).}
\label{TH_tab}
\end{table}

\newpage

\begin{figure}[t]
\epsfxsize=4.in \centerline{ \epsfbox{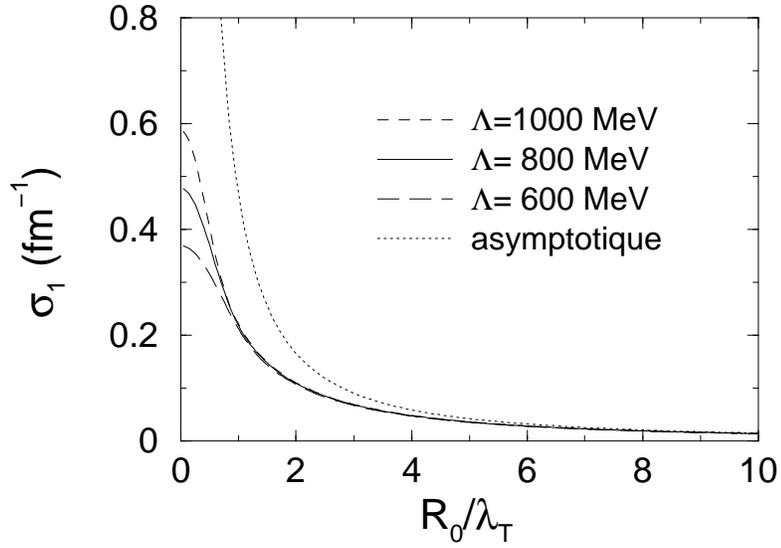}}
\caption[fig1]{\small La dispersion $\sigma_1$ (en fm$^{-1}$) en fonction 
   du rapport $R_0/\lambda_T$ pour $T=200$  MeV, et pour différentes valeurs
   de la coupure $\Lambda$. La courbe en pointillé représente la formule 
   asymptotique (\ref{TH_asymptotique}).} 
\label{TH_fig_disp1}
\end{figure}

\begin{figure}[b]
\epsfxsize=4.in \centerline{ \epsfbox{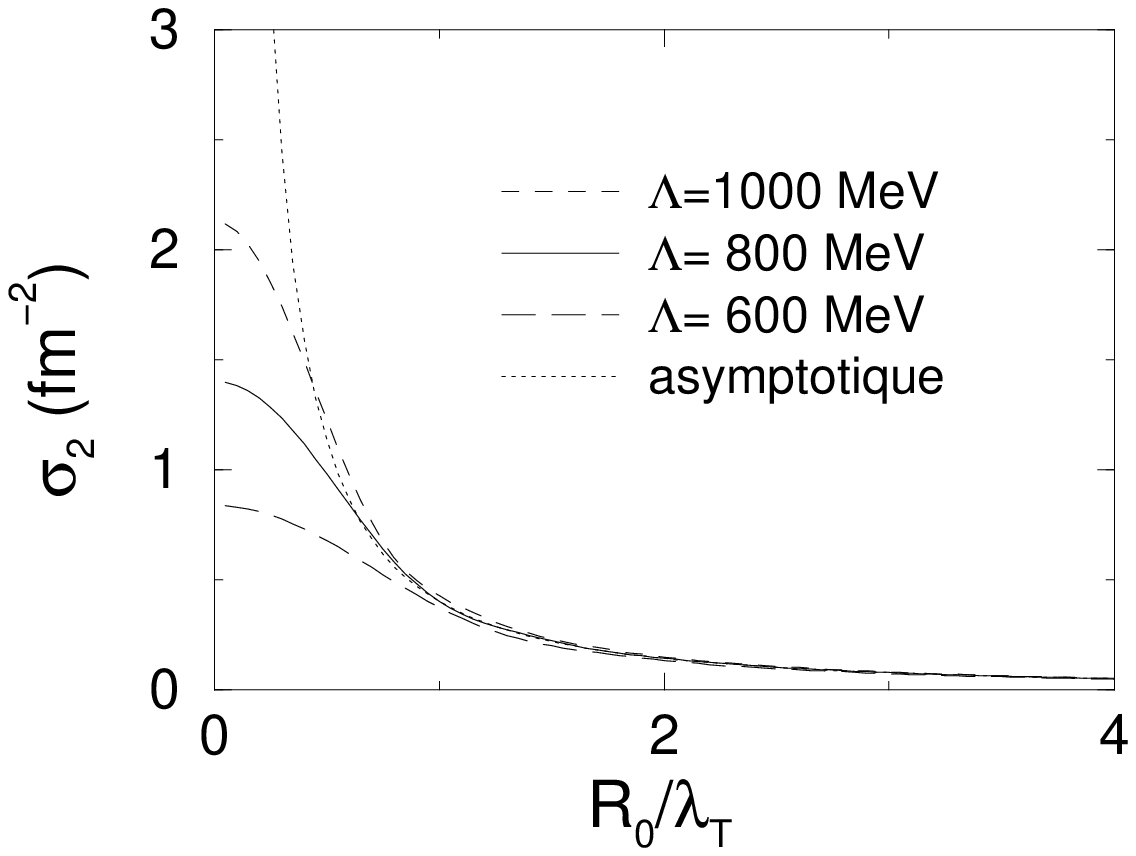}}
\caption[fig1]{\small L'analogue de la Fig.~\ref{TH_fig_disp1}, pour la
   dispersion $\sigma_2$.} 
\label{TH_fig_disp2}
\end{figure}

\chapter{Calcul d'intégrales}
\label{INTEGRALE}

Dans cet annexe, nous calculons les intégrales utilisées
pour renormaliser l'équation de gap au Chap.~\ref{PROBA}
 (cf.~Réf.~\cite{Lamperthesis}).
\par
Définissons les fonctions
\beq
 I_n(\Lambda,\mu) = \int_0^\Lambda \frac{k^2 dk}{( k^2 + \mu^2)^{n/2}} \, .
 \label{fonction}
\end{equation}\noindent
Il est facile de voir que
\beq
 \frac{1}{\mu} \, \frac{\p I_n(\Lambda,\mu)}{\p \mu} = 
 - n I_{n+2} (\Lambda,\mu) \, .
 \label{relation}
\end{equation}\noindent
Commençons par le calcul de $I_1$, nécessaire dans la renormalisation de la masse.
En écrivant
\eq
 \frac{k^2}{\sqrt{k^2 + \mu^2}} = k \, \frac{\dd}{\dd k} \sqrt{k^2 + \mu^2} \, ,
\eq
et en intégrant par parties, on obtient
\eq
 I_1 (\Lambda,\mu) = \Lambda \sqrt{\Lambda^2 + \mu^2} -
 \int_0^\Lambda dk \, \sqrt{k^2 + \mu^2} \, .
\eq
En écrivant 
\eq
 \sqrt{k^2 + \mu^2} = \frac{k^2 + \mu^2}{\sqrt{k^2 + \mu^2}} \, ,
\eq
on obtient aisément
\eq
 2 I_1 (\Lambda,\mu) = \Lambda^2 \sqrt{ 1 + x^2 } - 
 \mu^2 \sinh^{-1} \left(\frac{1}{x}\right) \, ,
\eq
où $x=\mu/\Lambda$. En écrivant
\eq
 \ln (\sinh a + \cosh a) = \ln \mbox{e}^a = a = \sinh^{-1} (\sinh a) \, ,
\eq
d'où l'on tire
\eq
 \sinh^{-1} a = \ln ( a + \sqrt{1 + a^2} ) \, ,
\eq
on réécrit l'expression précédente sous la forme
\beq
 I_1 (\Lambda,\mu) = \frac{\Lambda^2}{2} \sqrt{ 1 + x^2 } -
 \frac{\mu^2}{2} \, \ln \left( \frac{1 + \sqrt{ 1 + x^2 }}{x} \right) \, .
 \label{I1}
\end{equation}\noindent
Dans la limite où $x \ll 1$, on a
\beq
 I_1 (\Lambda,\mu) \simeq \frac{1}{2} 
 \left( \Lambda^2 - \mu^2 \ln \frac{\Lambda}{\mu} \right) + ... \, \, ,
 \label{I1_asympt}
\end{equation}\noindent
où $...$ désigne la partie non-divergente dans la limite 
$\Lambda \rightarrow +\infty$.
Les résultats précédents ont été obtenus en supposant $\mu^2 > 0$
Dans le Chap.~\ref{PROBA} on a ausi besoin, pour isoler les divergences 
à chaque instant, du comportement asymptotique de $I_1$ dans la limite 
$x \ll 1$ pour $\mu^2 = \chi < 0$. Dans ce cas, on écrit $\mu = i |\mu|$ et, 
en séparant ($\Lambda > \mu$) 
\eq
 \int_0^\Lambda \equiv \int_0^{|\mu|} + \int_{|\mu|}^\Lambda \, ,
\eq
il est facile de vérifier que la formule (\ref{I1}) est toujours valable
avec la prescritpion $\ln z = \ln |z| + i \pi/2$. Le comportement 
(\ref{I1_asympt}) est toujours valable en remplaçant $\mu \rightarrow |\mu|$.
\par
Pour la renormalisation de la constante de couplage, nous avons besoin de 
l'intégrale $I_3$. En utilisant (\ref{relation}) on obtient immédiatement
\beq
 I_3 (\Lambda,\mu) = \frac{-1}{\sqrt{1 + x^2}} + 
 \ln \left( \frac{1 + \sqrt{ 1 + x^2}}{x} \right)\, .
\end{equation}\noindent
Dans la limite où $x \ll 1$
\beq
 I_3 (\Lambda,\mu) \simeq  \ln \frac{\Lambda}{\mu}  + ... \, \, .
\end{equation}\noindent

\chapter{Scénario du trempage : l'état initial}
\label{neumann}

Dans cette annexe, nous détaillons la discrétisation de la transformée
de Fourier avec les conditions au bords de Neumann, puis nous calculons
la distribution de probabilité des valeurs des composantes de Fourier
discrètes dans l'état initial du modèle de Rajagopal et Wilczek~\cite{RW}
 (cf. Chap.~\ref{QUENCH}).

\section*{Conditions de Neumann}

Considérons, pour simplifier les notations, le cas à une dimension.
Nous donnerons plus loin la généralisation (triviale) à trois dimensions.
Soit $H(x)$ une fonction de la variable $x$ sur le segment $[0,L]$, 
avec des conditions aux bords de Neumann : $H'(x=0)=H'(x=L)=0$,
où le prime désigne la dérivation par rapport à $x$.
Il est clair que, de la décomposition de cette fonction en onde
plane $\exp(ikx)$, ne susbsistent que les $\cos(kx)$ qui, contrairement
aux $\sin(kx)$, sont compatibles avec ces conditions aux bords.
\par
Pour implémenter ces conditions aux bords et la décomposition de Fourier
correspondante, définissons la fonction $G(x)$, vivant sur l'axe
réel $-\infty < x < +\infty$, périodique de période $2L$, paire, et
telle que $G(x)=H(x)$ sur le segment $[0,L]$. Les composantes
de Fourier de $G(x)$ sont
\eq
 g_n = 
 \frac{1}{\sqrt{2L}} \, \int_{-L}^{+L} dx \, G(x) \, 
 \exp \left( 2 i \pi n \frac{x}{2L} \right) =
 \sqrt{\frac{2}{L}} \, \int_0^{+L} dx \, H(x) \, \cos (kx) \equiv h(k) \, ,
\eq
où $k=n \, \pi/L$. Les fonctions {\em réelles} $h(k)$ sont les composantes
de Fourier de $H(x)$, elles sont paires ($h(-k)=h(k)$).

\subsection*{Discrétisation}

La discrétisation du problème ci-dessus peut être faite de plusieur
façons, la plus simple consistant à échantillonner les valeurs de nos
fonctions $G$ et $H$ à intervalles réguliers $x = ja$ où $a$ est le pas
du réseau de taille $L=Na$ ($N$ est le nombre de sites). Cependant, nous 
aimerions répéter la construction précédente pour les composantes 
de Fourier de $H$. Pour cela, il est judicieux d'échantillonner 
les valeurs des fonctions aux points $x = a/2 + ja$ : $H_j=H(x = (j+1/2)a)$.
L'intervalle $-L \le x \le +L$ correspond alors à $-N \le j \le N-1$, la
périodicité de la fonction $H$ s'écrit $G_{N+j} = G_{j-N}$, tandis que la
propriété de parité prend la forme $G_{j} = G_{-(j+1)}$. On a, pour
les composantes de Fourier,  
\eq
 g_n = \sqrt{\frac{a}{2N}} \, \sum_{j=-N}^{N-1} \, G_j \, 
 \exp \left[ \frac{2i\pi}{2N} n (j+\frac{1}{2}) \right] = 
 \sqrt{\frac{2a}{N}} \, \sum_{j=0}^{N-1} \, H_j \, 
 \cos \left( \frac{\pi n (j+1/2)}{N}  \right) = h_n \, .
\eq
Les composantes $h_n$ satisfont à la relation ``d'anti-périodicité''
$h_{n-N}=-h_{n+N}$, ce qui limite l'espace de Fourier (ou espace 
réciproque) utile à la première zone de Brillouin\footnote{La première
   zone de Brillouin s'étend de $-N$ à $N$, autrement dit : 
   $-\pi/a \le k \le \pi/a$. Dans le cas présent, on a $h_N=0$.} : 
$-(N-1) \le n \le N-1$. La relation de parité $h_{-n}=h_n$ ramène
le nombre de degrés de liberté indépendants à $N$ : $0 \le n \le N-1$,
comme il se doit (dans l'espace direct, on a $N$ nombres réels $\phi_j$
indépendants). La dernière égalité ci-dessus peut être réécrite sous la 
forme matricielle (dans la suite nous  prendrons $a=1$)
\beq
\label{fourier1D}
 h_n = \sum_{j=0}^{N-1} \mathcal W_{nj} \, H_j \, \Leftrightarrow 
 h = \mathcal W \cdot H \, ,
\end{equation}\noindent
avec 
\beq
\label{W}
 \mathcal W_{nj} = 
 \sqrt{\frac{2}{N}} \, \cos \left( \frac{\pi n (j+1/2)}{N}  \right)\, .
\end{equation}\noindent
\par
Certaines propriétés de la transformation (\ref{fourier1D}) nous
serons utiles pour le calcul de la distribution des valeurs $f_n$
dans un modèle gaussien. 

\subsection*{Transformation inverse}

On définit la matrice $\mathcal W^\dagger$, transposée de la matrice 
$\mathcal W$ par la relation habituelle
\eq
 \mathcal W^\dagger_{jn} = \mathcal W_{nj} \, .
\eq
Calculons les éléments de la matrice $\mathcal W \, \mathcal W^\dagger$.
Les éléments non-diagonnaux sont nuls : 
\eq
 (\mathcal W \, \mathcal W^\dagger)_{nm} =
 \frac{2}{N} \, \sum_{j=0}^{N-1} \, \mathcal W_{nj} \, \mathcal W_{mj} =
 \frac{1}{2N} \left( \mathcal A_{n+m} + \mathcal A_{-(n+m)} +
 \mathcal A_{n-m} + \mathcal A_{-(n+m)} \right) = 0 \, ,
\eq
où
\eq
 \mathcal A_p =
 \sum_{j=0}^{N-1} \exp \left( i \frac{\pi p (j+1/2)}{N}  \right) = 
 \frac{1}{2i} \, \frac{(-1)^p - 1}{\sin \left(\frac{\pi p}{2N} \right)} = 
 \mathcal A_{-p}
\eq
 ($0 < p = n \pm m < 2N \Rightarrow \sin(\pi p/2N) \neq 0$).
Dans le calcul des éléments diagonnaux, il faut distinguer les cas
$n=m=0$ et $n=m\neq 0$. Ceci est relié au fait que la relation de
symétrie $f_n=f_{-n}$ est triviale pour $n=0$. Au final on obtient
\beq
\label{WWd}
 (\mathcal W \, \mathcal W^\dagger)_{nm} = 
 \delta_{nm} \, (1 + \delta_{n0}) \, .
\end{equation}\noindent
En utilisant les identités $2 \cos (a) \, \cos (b) = \cos (a+b) + \cos(a-b)$,
et
\eq
 \sum_{n=0}^{N-1} \cos \left( \frac{\pi n p}{N} \right) = 
 \frac{1}{2} \, \left( 1 - (-1)^p \right) \, ,
\eq
on obtient
\beq
\label{WdW}
 (\mathcal W^\dagger \, \mathcal W)_{jk} = 
 \delta_{jk} + \frac{1}{N} \, .
\end{equation}\noindent
A partir de cette relation on déduit
\eq
 (h \cdot \mathcal W )_j = \sum_{n=0}^{N-1} h_n \, \mathcal W_{nj} =
 (\mathcal W^\dagger \cdot h)_j = 
 \sum_{k=0}^{N-1} (\mathcal W^\dagger \, \mathcal W)_{jk} \, H_k =
 H_j + \frac{1}{N} \, \sum_{j=0}^{N-1} H_j \, .
\eq
En écrivant la somme dans le membre de gauche sous la forme 
(cf. Eqs.~(\ref{fourier1D})-(\ref{W}))
\eq
 \frac{1}{N} \, \sum_{j=0}^{N-1} H_j = \frac{1}{2} \, h_0 \, \mathcal W_{0j} \, ,
\eq
on obtient la formule d'inversion~\cite{NumRec}
\beq
\label{inverse}
 H_j = (\mathcal W^{-1} \cdot h)_j =
 \left( \, \sum_{n=0}^{N-1} h_n \, \mathcal W_{nj} \right)^\prime \, ,
\end{equation}\noindent
où le prime signifie que le terme $n=0$ ne contribue à la somme
que par la moitié de sa valeur. En utilisant l'une des relations
précédentes, il est facile de voir que
\beq
\label{parseval}
 H \cdot H = \sum_{j=0}^{N-1} H_j^2 = 
 \left( \, \sum_{n=0}^{{N-1}} h_n^2 \right)^\prime \, .
\end{equation}\noindent
Enfin, on a ($\mbox{Det} [ \mathcal W^\dagger ] = \mbox{Det} [ \mathcal W ]$)
\beq
\label{detW}
 \mbox{Det} [ \mathcal W \, \mathcal W^\dagger ] = 
 (\mbox{Det} [ \mathcal W ])^2 = 2 \, .
\end{equation}\noindent

\section*{Distribution gaussienne}

Dans le modèle du Chap.~\ref{QUENCH}, l'état initial est
caractérisé par un ensemble statistique gaussien : les valeurs de chacune des 
composantes chirales du champ en chaque n\oe ud du réseau direct sont des 
nombres aléatoires gaussiens indépendants. Il en est de même pour les valeurs 
des composantes de la dérivée temporelle du champ. Imaginons que la fonction 
$H(x)$ représente une des composantes chirales du champ ou de sa dérivée.
Les $H_j$, $j=0,...,N-1$ sont donc des nombres gaussiens indépendants
de variance $\langle H^2 \rangle$.
Qu'en est-il des composantes de Fourier $h_n$ ?
\par 
La relation (\ref{WWd}) implique que les composantes $h_n$
sont des variables aléatoires gaussiennes indépendantes. En effet,
on a ($\langle H_i \, H_j \rangle = \delta_{ij} \, \langle H^2 \rangle$)
\beq
\label{largeur}
 \langle h_n \, h_m \rangle = 
 \langle H^2 \rangle \, (\mathcal W \, \mathcal W^\dagger)_{nm} 
 \propto \delta_{nm} \, ,
\end{equation}
ainsi que
\bear
 \langle h_n \, h_m \, h_p \, h_q \rangle & = &
 \sum_{ijkl} \mathcal W_{ni} \, \mathcal W_{mj} \, 
 \mathcal W_{pk} \, \mathcal W_{ql} \, 
 \langle H_i \, H_j \, H_k \, H_l \rangle \\
 & = & \langle H^2 \rangle^2 \, \left[
 (\mathcal W \, \mathcal W^\dagger)_{nm} \, 
 (\mathcal W \, \mathcal W^\dagger)_{pq} +
 (\mathcal W \, \mathcal W^\dagger)_{np} \, 
 (\mathcal W \, \mathcal W^\dagger)_{mq} +
 (\mathcal W \, \mathcal W^\dagger)_{nq} \, 
 (\mathcal W \, \mathcal W^\dagger)_{mp} \right] \\
 & = & \langle h_n \, h_m \rangle \, 
 \langle h_p \, h_q \rangle +
 \langle h_n \, h_p \rangle \, 
 \langle h_m \, h_q \rangle + 
 \langle h_n \, h_q \rangle \, 
 \langle h_m \, h_p \rangle \, ,
\eear
et ainsi de suite.
\par
De façon plus précise, et plus concise, on écrit la  probabilité
\eq
 P_H (H) \, \prod_{j=0}^{N-1} dH_j = 
 P_h (h) \, \prod_{n=0}^{N-1} dh_n \, ,
\eq
d'où l'on tire la densité de probabilité (cf. Eq.~(\ref{detW}))
\beq
\label{direct}
 P_h (h) = 
 \frac{1}{\sqrt 2} \, P_H (\mathcal W^{-1} \cdot h) \, .
\end{equation}\noindent
Le facteur $1/\sqrt 2$ est dû au fait que le mode zéro a une variance
égale au double de celle des autres modes. Comme on l'a remarqué plus haut,
cela vient de ce que la relation de parité $h_n = h_{-n}$ est
trivialement satisfaite pour le mode zéro : ce degré de liberté est en
quelque sorte moins contraint et fluctue plus. Dans le cas qui nous 
intéresse, on a 
\eq
 P_H (H) = \frac{1}{\left( 2 \pi \langle H^2 \rangle \right)^{N/2}} \,
 \exp \left( - \frac{H \cdot H}{2 \langle H^2  \rangle} \right) \, ,
\eq
et donc
\eq 
 P_h (h) = 
 \prod_{n=0}^{N-1} 
 \frac{1}{\sqrt{2 \pi \sigma_n^2}} \,
 \exp \left(-\frac{h_n^2}{2\sigma_n^2} \right) \, ,
\eq
avec $\sigma_n^2=\langle H^2 \rangle \, ( 1 + \delta_{n0} )$.

\section*{Le cas tri-dimensionnel}

La généralisation des considérations précédentes à trois dimensions est
immédiate. La fonction $H(\vec x)$, sur le cube $0 \le x,y,z \le L=Na$ de
volume $V=L^3$, est échantillonnée aux points $\vec x = a(1/2+i,1/2+j,1/2+k)$,
où sa valeur est notée $H_{\vec j} \equiv H_{i,j,k}$. On implémente
les conditions aux bords de Neumann en introduisant une fonction 
$G(\vec x)$ ayant les même propriétés de périodicité et parité que dans
le cas uni-dimensionnel, par rapport à chacune des trois variables $x,y,z$.
Considérant le cas d'une des composantes chirales du champ, les composantes
de Fourier s'écrivent ($a=1$)
\beq
\label{fourier3D}
 h_{\vec n} \equiv h_{l,m,n} =
 \sum_{i=0}^{N-1} \sum_{j=0}^{N-1} \sum_{k=0}^{N-1}
 \mathcal W_{li} \, \mathcal W_{mj} \, \mathcal W_{nk} \, 
 H_{i,j,k} \, ,
\end{equation}\noindent
où $\mathcal W$ est donnée par (\ref{W}).
Comme précédemment, les degrés de libertés indépendants, au nombre 
de $N^3$ sont : $0 \le l,m,n \le N-1$. Les relations de parité
sont
\eq
 h_{l,m,n}=h_{l,m,-n}=h_{l,-m,n}=h_{-l,m,n}=
 h_{-l,-m,n}=h_{-l,m,-n}=h_{l,-m,-n}=h_{-l,-m,-n} \, .
\eq
On voit que la quantité d'information non-triviale contenue dans ces 
relations varie selon qu'aucun, un, deux ou trois des indices $l,m,n$ 
sont nuls. Comme dans le cas uni-dimensionnel, on s'attend à ce que cela
se répercute sur les largeurs des gaussiennes correspondantes.
En effet, on voit que (cf. Eq.~(\ref{largeur}))
\bear
 \langle h_{\vec n} \, h_{\vec n \, '} \rangle & = &
 \langle H^2 \rangle \,
 (\mathcal W \, \mathcal W^\dagger)_{ll'} \, 
 (\mathcal W \, \mathcal W^\dagger)_{mm'} \,
 (\mathcal W \, \mathcal W^\dagger)_{nn'} \\
 & = & \langle H^2 \rangle \, \delta_{\vec n \vec n\, '} \,
 (1+\delta_{l0}) \, (1+\delta_{m0}) \, (1+\delta_{n0}) \, .
\eear
On vérifie aisément\footnote{Faisons le calcul
   dans le cas bi-dimensionnel : on écrit
   \bear
    \sum_{m,n=0}^{N-1} h_{m,n}^2 & = &
    \sum_{i,j=0}^{N-1} \, \sum_{k,l=0}^{N-1} 
    (\mathcal W^\dagger \, \mathcal W)_{ik} \, 
    (\mathcal W^\dagger \, \mathcal W)_{jl} \,
    H_{i,j} \, H_{k,l} \\
    & = & \sum_{i,j} H_{i,j}^2 + 
    \frac{1}{N} \, \sum_{i,j} \sum_k H_{i,j} \, H_{k,j} +
    \frac{1}{N} \, \sum_{i,j} \sum_l H_{i,j} \, H_{i,l} +
    \frac{1}{N^2} \, \sum_{i,j} \sum_{k,l} H_{i,j} \, H_{k,l} \, .
   \eear
   On a, de même,
   $\displaystyle
    \sum_{m=0}^{N-1} h_{m,0}^2 = 
    \frac{2}{N} \, \sum_{i,j} \sum_l H_{i,j} \, H_{i,l} +
    \frac{2}{N^2} \, \sum_{i,j} \sum_{k,l} H_{i,j} \, H_{k,l}
   $,
   et
   $\displaystyle
    \frac{1}{N^2} \, \sum_{i,j} \sum_{k,l} H_{i,j} \, H_{k,l} = 
    \frac{1}{4} \, h_{0,0}^2
   $,\\
   d'où l'on tire
   \eq
    \frac{1}{N} \, \sum_{i,j} \sum_l H_{i,j} \, H_{i,l} =
    \frac{1}{2} \, \sum_{m=0}^{N-1} h_{m,0}^2 - \frac{1}{4} \, h_{0,0}^2 = 
    \frac{1}{2} \, \sum_{m=1}^{N-1} h_{m,0}^2 + \frac{1}{4} \, h_{0,0}^2 \, .
   \eq
   De la même façon on obtient
   $\displaystyle
    \frac{1}{N} \, \sum_{i,j} \sum_k H_{i,j} \, H_{k,j} =
    \frac{1}{2} \, \sum_{n=1}^{N-1} h_{0,n}^2 + \frac{1}{4} \, h_{0,0}^2
   $.
   En combinant le tout
   \eq
    \sum_{i,j} H_{i,j}^2 = 
    \sum_{m,n=1}^{N-1} h_{m,n}^2 +
    \frac{1}{2} \, \sum_{m=1}^{N-1} h_{m,0}^2 +
    \frac{1}{2} \, \sum_{n=1}^{N-1} h_{0,n}^2 +
    \frac{1}{4} \, h_{0,0}^2 \, 
   \eq
   CQFD.}
que la généralisation de l'Eq.~(\ref{parseval}) est
\beq
 H \cdot H = \sum_{i,j,k=0}^{N-1} H_{i,j,k}^2 = 
 \left( \, \sum_{l,m,n=0}^{N-1} h_{l,m,n}^2 \right)^\prime \, .
\end{equation}\noindent
où le prime signifie que, dans la somme, les termes dont un des trois
indices $l,m,n$ est nul sont comptés avec un facteur $1/2$ supplémentaire,
ceux dont deux des trois indices sont nuls avec un facteur $1/4$ et celui
dont les trois indices sont nuls, avec un facteur $1/8$. Il est alors
facile de voir que la distribution des $h_{\vec n}$ est, dans le cas gaussien,
\eq 
 P_h (h) = 
 \prod_{l,m,n=0}^{N-1} 
 \frac{1}{\sqrt{2 \pi \sigma_{\vec n}^2}} \,
 \exp \left(-\frac{h_{\vec n}^2}{2\sigma_{\vec n}^2} \right) \, ,
\eq
avec $\sigma_{\vec n}^2=\sigma_{l,m,n}^2=
\langle H^2 \rangle \, (1+\delta_{l0})\, (1+\delta_{m0}) \, (1+\delta_{n0})$.

\section*{Calcul numérique des coefficient $h_{\vec n}$}

Dans le Chap.~\ref{QUENCH}, nous devons calculer les coefficients de Fourier
$\bfvarphi_{\vec n}$ et $\dot\bfvarphi_{\vec n}$ (Eq.~(\ref{fourier3D})) des 
composantes d'isospin du champ $\bfphi$ et de sa dérivée temporelle $\dot\bfphi$,
à différent instants. En pratique, nous nous limitons aux composantes $(n,0,0)$,
$(0,n,0)$ et $(0,0,n)$, qui se calculent comme des transformées de Fourier
uni-dimensionnelles (cf. Eq.~(\ref{fourier1D})). Par exemple
\eq
 \bfvarphi_n \equiv \bfvarphi_{n,0,0} = 
 \frac{2}{N} \, \sum_{i=0}^{N-1} \mathcal W_{ni} \, \tilde{\bfphi}_i \, ,
\eq
où 
\eq
 \tilde{\bfphi}_i = \sum_{j,k=0}^{N-1} \bfphi_{i,j,k} \, .
\eq
Le calcul de la transformée de Fourier uni-dimensionnelle ci-dessus
se fait aisément à l'aide d'un programme de type ``Fast Fourier
Transform''~\cite{NumRec}.

\chapter{Modèle gaussien}
\label{neutral}

Il est instructif de calculer la distribution de la fraction $f$ de pions 
neutres dans le cas où les $\bfvarphi_{\vec k} \equiv \bfvarphi (\vec k,t_f)$ 
et $\dot\bfvarphi_{\vec k} \equiv \dot\bfvarphi (\vec k,t_f)$ sont des
variables aléatoires gaussiennes, indépendantes, de moyennes nulles et 
de dispersions respectives $\sigma_1$ et $\sigma_2$ 
(celles-ci sont les mêmes pour toutes les direction d'isospin $j=1,2,3$ et 
pour tous les modes $\vec k$). On écrit alors la probabilité 
($c \equiv {j,\vec k}$)
\beq
\label{ensemble}
 \mbox{Proba} \left( \{ \bfvarphi_{\vec k} \},\{\dot \bfvarphi_{\vec k}\} \right) = 
 \prod_c \, P_{\varphi} (\varphi_c) \, P_{\dot\varphi} (\dot\varphi_c) \,
 d\varphi_c \, d\dot\varphi_c \, ,
\end{equation}\noindent
\bearn
\label{gauss1}
 P_{\varphi} (x) & = & \frac{1}{\sqrt{2 \pi \sigma_1^2}} \, 
 \exp \left(  - \frac{x^2}{2 \sigma_1^2} \right) \, , \nonumber\\
\label{gauss2}
 P_{\dot\varphi} (x) & = & \frac{1}{\sqrt{2 \pi \sigma_2^2}} \, 
 \exp \left(  - \frac{x^2}{2 \sigma_2^2} \right) \, ,
\eearn
Les modes étant indépendants les uns des autres, nous focalisons notre 
attention sur l'un d'entre eux et omettons l'indice $\vec k$ correspondant. 
D'après l'Eq.~(\ref{alpha}) du Chap.~\ref{PROBA}, on a, pour chaque 
composante d'isospin,
\bearn
\label{re}
 \varphi_j & = & \sqrt{\frac{2}{\omega}} \, \mbox{Re} \, \alpha_j =
 A_j \, \cos \gamma_j  \, ,\nonumber \\
\label{im}
 \dot\varphi_j & = & \sqrt{2\omega} \, \mbox{Im} \, \alpha_j =
 \omega A_j \, \sin \gamma_j \, ,
\eearn
où $\gamma_j$ est défini par la relation 
$\alpha_j = \sqrt{\bar n_j} \, e^{i \gamma_j}$ et $A_j=\sqrt{2\bar n_j/\omega}$.
Les différentes directions d'isospin étant indépendantes, la distribution de
probabilité pour les amplitudes $A_j$ et les phases $\gamma_j$ est le produit 
des distributions individuelles $P_{A,\gamma} (A_j,\gamma_j)$, où
\beq
\label{polardist}
 P_{A,\gamma} (A,\gamma) = 
 \omega A \, P_{\varphi} (A \cos \gamma) \,
 P_{\dot\varphi} (\omega A \sin \gamma) \, .
\end{equation}\noindent
On en déduit la distribution de probabilité pour les valeurs de l'amplitude
dans une direction $j$ donnée :
\bearn
 P_A (A) & = & \int_{0}^{2\pi} d\gamma \, P_{A,\gamma} (A,\gamma) \nonumber\\
 & = & \frac{\omega}{2\pi \sigma_1 \sigma_2} \, A \, e^{- A^2 \Delta_{+}} \, 
 \int_0^{2\pi} d\gamma \, e^{-A^2 \Delta_{-} \cos (2\gamma)} \nonumber\\
 & = & \frac{\omega}{\sigma_1 \sigma_2} \, A \, e^{- A^2 \Delta_{+}} \, I_0(A^2 \Delta_{-})
\label{radial}
\eearn
où $I_\nu(x)$ est la fonction de Bessel modifiée de première espèce, et
\beq
\label{delta}
 \Delta_{\pm} = 
 \frac{1}{4} \left| \frac{1}{\sigma_1^2} \pm \frac{\omega^2}{\sigma_2^2} \right| \, .
\end{equation}\noindent

\section*{Calcul de la distribution de la fraction de pions neutres}

La distribution de probabilité de la fraction neutre 
$f = {A_3}^2/({A_1}^2 + {A_2}^2 + {A_3}^2)$ est donnée par
\beq
 P_f (f) = \int_0^{+ \infty} dx dy dz \, 
 P_A(x) \, P_A(y) \, P_A(z) \, 
 \delta \left( f - \frac{{z}^2}{x^2 + y^2 + z^2} \right)
\end{equation}\noindent
En introduisant les coordonnées sphériques $(r,\theta,\phi)$, où 
$0 \le r < +\infty$, $0 \le \theta \le \pi/2$, et $0 \le \phi \le \pi/2$,
et en posant $u=\cos \theta$, on peut integrer la fonction delta  
($\delta (f-u^2) \equiv \delta (u-\sqrt{f})/2\sqrt{f}$ pour $0 \le u \le 1$). 
On obtient
\beq
\label{etape1}
 P_f (f) = \frac{1}{2} \left( \frac{\omega}{\sigma_1 \sigma_2} \right)^3 (1-f) 
 \int_0^{+ \infty} dr \, r^5 \, I_0(r^2 f \Delta_{-}) \, e^{-r^2 \Delta_{+}} \,
 \Phi \left( r^2 (1-f) \Delta_{-} \right) \, ,
\end{equation}\noindent
avec
\bear
 \Phi (z) & =  & \int_0^{\pi/2} d\phi \, \sin \phi \, \cos \phi \, 
 I_0(z \cos^2 \phi) \, I_0(z \sin^2 \phi) \\
 & = & \frac{1}{2z} \int_0^z dx \, I_0(z-x) \, I_0(x) \, .
\eear
Cette dernière intégrale est un produit de convolution au sens de la transformée
de Laplace. En utilisant le théorème de convolution\footnote{La transformée 
   de Laplace du produit de convolution de deux fonctions $F$ et $G$ est égal 
   au produit des transformées de Laplace de ces deux fonctions : 
   \eq
    \mathcal L \left\{ \int_0^t dz \, F(t-z) \, G(z) \right\} (s) =
    \mathcal L \{ F(t) \} (s) \, \, \mathcal L \{ G(t) \} (s) \, ,
   \eq},
et la transformée de Laplace de la fonction de Bessel $I_0$
\beq
\label{Laplace}
 \mathcal L \{ I_0(t) \} (s) = \int_0^{+ \infty} dt \, e^{-st} \, I_0(t) =
 \frac{1}{\sqrt{s^2 - 1}} \, \, \, , \, \, \, s > 1 \, \, ,
\end{equation}\noindent 
on obtient
\eq
 \Phi (z) = 
 \frac{1}{2 z} \, \mathcal L^{-1} \left\{ \frac{1}{s^2 - 1} \right\} (z) = 
 \frac{\sinh z}{2z} \, .
\eq
On écrit alors la distribution de probabilité cherchée sous la forme
\beq
\label{etape2}
 P_f (f) = F_1(\Delta_{+},f \Delta_{-},(1-f) \Delta_{-}) \, ,
\end{equation}\noindent
où l'on a défini les fonctions ($x=r^2$)
\eq 
 F_p(a,b,c) = \int_0^{+ \infty} dx \,x^p \, e^{- a x} \, I_0(bx) \, \sinh (cx) \, .
\eq
qui satisfont de manière évidente à la relation de récurrence
\eq
 F_p(a,b,c) = - \frac{\partial}{\partial a} \, F_{p-1}(a,b,c) \, .
\eq
\par 
Le calcul de $F_0(a,b,c)$ est aisé : en explicitant la fonction $\sinh (x)$
sous l'intégrale, on voit que
\eq
 F_0(a,b,c) = \frac{1}{2} \mathcal L \{ I_0(bx) \} (a-c) -
 \frac{1}{2} \mathcal L \{ I_0(bx) \} (a+c) \, ,
\eq
d'où on déduit, en utilisant (\ref{Laplace})
\beq
 F_0(a,b,c) = \frac{1}{2} \left( \frac{1}{\sqrt{{(a-c)}^2- b^2}} -
 \frac{1}{\sqrt{{(a+c)}^2- b^2}} \right) \, \, \, , \, \, \,
 a \pm c > b \, .
\end{equation}\noindent
Il vient 
\beq
 F_1(a,b,c) = \frac{1}{2} \left( 
 \frac{a-c}{\left[ \, {(a-c)}^2- b^2 \, \right]^{3/2}} -
 \frac{a+c}{\left[ \, {(a+c)}^2- b^2 \, \right]^{3/2}} \right) \, \, \, , \, \, \,
 a \pm c > b \, ,
\end{equation}\noindent

\noindent
Finalement (cf. Eq.~(\ref{etape2}), notez que pour $a=\Delta_{+}$, 
$b=f \Delta_{-}$ et $c=(1-f) \Delta_{-}$ on a bien $a \pm c>b$),

\beq
\label{distrib0}
 P_f (f) = \frac{1}{2} \left( F_\Omega (f) + F_{-\Omega} (f) \right)
\end{equation}\noindent
avec
\beq
\label{distrib1}
 F_\Omega (f) = \left( \Omega - (1-f) \right) 
 \left( \frac{\Omega + 1}{\Omega - (1-2f)} \right)^{3/2}
\end{equation}\noindent
et
\beq
\label{distrib2}
 \Omega = \frac{\Delta_{+}}{\Delta_{-}} = \left|
 \frac{\sigma_2^2 + \omega^2 \sigma_1^2}
 {\sigma_2^2 - \omega^2 \sigma_1^2} \right| \, .
\end{equation}\noindent
 
\section*{Calcul de la distribution du nombre total de pions}

De la même façon, on calcule la distribution du nombre total de pions 
dans le mode $\vec k$ : $\bar n = \omega A^2 /2$, où $A^2 = \sum_{j=1}^3 A_j^2$
\beq
\label{nbtotal}
 P_n (\bar n) = \frac{2}{\omega} \, 
 P_{A^2} \left( \frac{2}{\omega} \bar n \right) \, 
\end{equation}\noindent
avec
\beq
 P_{A^2} (A^2) = \int_0^{+ \infty} dx dy dz \, 
 P_A(x) \, P_A(y) \, P_A(z) \, 
 \delta \left( A^2 - x^2 + y^2 + z^2 \right)
\end{equation}\noindent
Comme précédemment, on intègre la fonction delta en passant en coordonnées 
shériques. En utilisant (\ref{radial}) il vient ($u=\cos \theta=x^2$)
\beq
 P_{A^2} (A^2) = \frac{1}{8\Delta_-} \, 
 \left( \frac{\omega}{\sigma_1 \sigma_2} \right)^3 \,
 A^2 \, \mbox{e}^{-A^2 \Delta_+} \, \Psi (A^2 \Delta_-) \, ,
\end{equation}\noindent
où
\bear
 \Psi (z) & = & \int_0^1 dx \, I_0(zx) \, \sinh (z(1-x)) \\
 & = & \frac{1}{z} \, \int_0^z dy \, I_0(y) \, \sinh (z-y) \, .
\eear
En utilisant le théorème de convolution de la transformée de Laplace,
ainsi que les transformées de Laplace de $I_0$ (Eq.~(\ref{Laplace})) et
\eq
 \mathcal L \left\{ \sinh t \right\} (s) = \frac{1}{s^2-1} \, ,
\eq
on obtient 
\eq
 \Psi (z) = \frac{1}{z} \, 
 \mathcal L^{-1} \left\{ \frac{1}{(s^2-1)^{3/2}} \right\} (z) = I_1(z) \, .
\eq
Finalement, la distribution du nombre total de pions dans le mode $\vec k$
est donnée par l'Eq.~(\ref{nbtotal}), avec
\beq
 P_{A^2} (A^2) = \frac{1}{8\Delta_-} \, 
 \left( \frac{\omega}{\sigma_1 \sigma_2} \right)^3 \,
 A^2 \, \mbox{e}^{-A^2 \Delta_+} \, I_1(A^2 \Delta_-) \, .
\end{equation}\noindent

\section*{Cas particuliers}

Il est instructif d'étudier la formule (\ref{distrib0}) pour 
quelques exemples particuliers de la classe d'ensemble statistiques 
gaussiens (\ref{ensemble})-(\ref{gauss2}), et en particulier les
deux cas limites $\Omega \rightarrow 1$ et $\Omega \rightarrow +\infty$.
\par
Le cas $\Omega = 1$, ou encore $\Delta_+ = \Delta_-$, correspond aux deux
situations $\sigma_1=0$ ou bien $\sigma_2=0$. Cela signifie qu'un des deux 
vecteurs, $\bfvarphi$ ou $\dot\bfvarphi$, est fixé, et donc nul. Le vecteur
$\bfalpha$ est alors proportionnel à un vecteur orienté aléatoirement dans 
l'espace d'isospin et on doit retrouver la loi en $1/\sqrt{f}$ pour la 
distribution de la fraction neutre $f$. En effet, considérons par exemple
la limite $\sigma_2 \rightarrow 0$, c'est à dire $\dot\bfvarphi = {\bf 0}$. 
Alors $\bfalpha \propto \bfvarphi$ et la distribution de la fraction neutre
tends vers\footnote{Dans cette limite, $\Delta_+ \sim \Delta_- \sim
   \omega^2/4\sigma_2^2 \rightarrow +\infty$, mais 
   $\Delta_+ - \Delta_- = 1/2\sigma_1^2$. La distribution (\ref{radial}) tend, 
   comme il se doit, vers la distribution (\ref{gauss1}) : 
   \eq
    P_A (A) \rightarrow \frac{2}{\sqrt{2\pi\sigma_1^2}} \, 
     \exp \left(  - \frac{A^2}{2 \sigma_1^2} \right) = 2 \, P_\varphi (A) \, ,
   \eq
   où l'on a utilisé le développement asymptotique 
   $\displaystyle
    I_\nu (z) \sim \frac{\mbox{e}^z}{\sqrt{2\pi z}} \, .
   $
   Ci-dessus, le facteur $2$ supplémentaire vient du fait que $A=|\varphi|$.}
 ($\Omega \rightarrow 1$)
\beq
 P_f (f) = \frac{1}{2 \sqrt f} \, . 
\end{equation}\noindent
\par
Un deuxième cas intéressant est la sous-classe d'ensembles tels que
$\sigma_2 = \omega \, \sigma_1$ ($\Omega \rightarrow + \infty$). En termes de
la distribution de probabilité $P_\alpha (\alpha_j)$ pour les nombres
complexes $\alpha_j$ (cf.~Eq.~(\ref{alpha}) du Chap.~\ref{QUENCH})
\eq
 P_\alpha (\alpha) \, d^2 \alpha = P_{A,\gamma} (A,\gamma) \, dA \, d\gamma \, ,
\eq
on a, d'après les Eqs.~(\ref{re})-(\ref{polardist}) (voir aussi (\ref{radial}))
\beq
\label{thermal}
 P_\alpha (\alpha) = \frac{1}{\pi \sigma^2} \, \mbox{e}^{-|\alpha|^2/\sigma^2} \, ,
\end{equation}\noindent
où 
\eq
 \sigma^2 = \sigma_1 \, \sigma_2 = \langle \bar n \rangle = 
 \int d^2\alpha \, P_{\alpha} (\alpha) \, |\alpha|^2 
\eq
 ($|\alpha|^2 = \bar n$ est le nombre moyen de quanta dans l'état 
cohérent~$| \alpha \rangle$ et~$\langle \bar n \rangle$ est la moyenne
statistique de ce nombre dans l'ensemble (\ref{thermal})).
La fonction $P_\alpha (\alpha)$ est appellée la 
représentation de Glauber~\cite{Glauber} de la matrice 
densité\footnote{On peut toujours représenter la matrice densité sur
   la base des états cohérents. Une très grande classe d'opérateur densité
   peuvent s'écrire $\rho = \int d^2 \alpha \, 
   P_\alpha (\alpha) | \alpha \rangle \langle \alpha |$.}. 
Pour la classe d'ensembles (\ref{thermal}) la distribution de la fraction
neutre $f$ est
\beq
\label{thermalf}
 P_f (f) = 2 \, (1-f) \, .
\end{equation}\noindent
Cette distribution ne dépend pas de la largeur $\sigma$. Un exemple
particulièrement intéressant de la distribution (\ref{thermal}) est
celui d'un ensemble de quanta de fréquences $\omega$ en équilibre
thermique~\cite{Glauber}\footnote{On a alors
    $\sigma^2=(\exp (\omega/T) -1)^{-1} \simeq T/\omega$, la dernière
    égalité étant valable pour les modes de basse fréquence ($\omega \ll T$)
    pour lesquels l'approximation de champ classique est justifiée.}.
\par
Enfin, un troisième cas particulier intéressant est celui de l'ensemble
initial du le modèle de Rajagopal et Wilzcek~\cite{RW},
utilisé au Chap.~\ref{QUENCH}. Les variables $\phi_j (\vec x)$ et 
$\dot\phi_j (\vec x)$ sont des nombres gaussiens de variances respectives
$\langle \phi_j^2 \rangle$ et $\langle \dot\phi_j^2 \rangle$, distribués
indépendamment sur les n\oe uds du réseau. Cela implique (cf. Annexe~\ref{neumann})
que les composantes de Fourier $\varphi_j (\vec k)$ et $\dot\varphi_j (\vec k)$ 
sont des nombres gaussiens indépendants de variances respectives (en unités 
du pas du réseau $a$)
\eq
 \sigma_1 = \mathcal N \, \langle \phi_j^2 \rangle = 
 \mathcal N \, \frac{v^2}{16} \, \, \, , \, \, \,
 \sigma_2 = \mathcal N \, \langle \dot\phi_j^2 \rangle = 
 \mathcal N \, \frac{v^2}{4} \, ,
\eq
où $\mathcal N$ est une constante de normalisation dépendant de $\vec k$.
La distribution de la fraction neutre dans chacun des modes $\vec k$ est
alors donnée par les Eqs.~(\ref{distrib0})-(\ref{distrib2}) avec
\beq
\label{indist}
 \Omega_k = \left| \frac{4 + \omega_k}{4 - \omega_k} \right| \, .
\end{equation}\noindent
La Fig.~\ref{fig_distrib0} représente la distribution de la fraction
neutre $f_k$ dans les modes $\vec k = (k=n \Delta k,0,0)$, dans l'ensemble 
initial du scénario du trempage, pour les deux valeurs extrêmes de la 
fenêtre d'impulsion étudiée au Chap.~\ref{QUENCH} : $0 \le k \lesssim 150$~MeV
($0 \le n \le 15$).

\begin{figure}[htbp]
\epsfxsize=5.5in \centerline{ \epsfbox{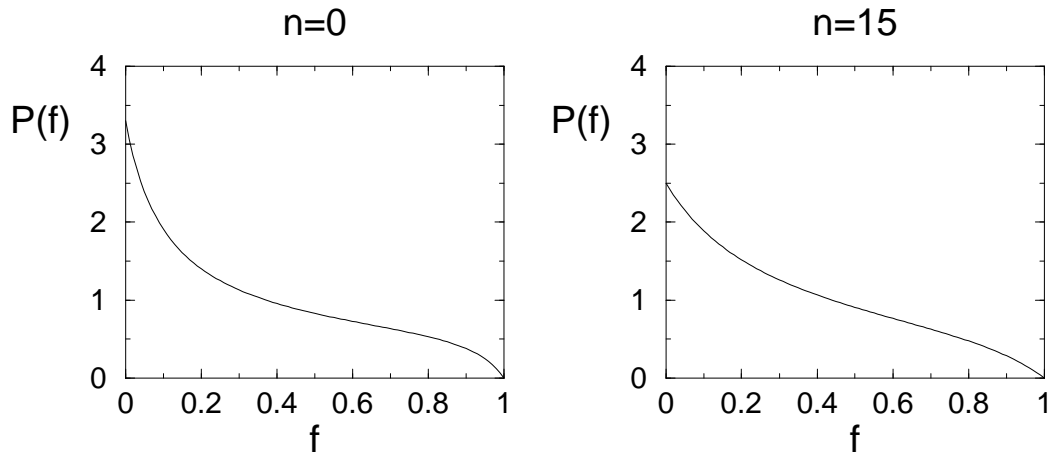}}
\caption[fig1]{\small Distribution de la fraction $f_k$ de pions neutres
    dans le mode $\vec k = (k=n \Delta k,0,0)$ ($\Delta k = \pi/Na \approx 10$~MeV),
    pour $n=0$ (à droite) et $n=15$ (à gauche), dans l'état initial
    (Eqs.~(\ref{ensemble})-(\ref{gauss2})).} 
\label{fig_distrib0}
\end{figure}

\section*{Remarque à propos des conditions aux bords périodiques}

Nous reprenons ici la remarque du bas de la page \pageref{ftnote_bc}, 
concernant l'inconvénient des conditions aux bords périodiques, et en 
donnons une illustration. 
\par
Avec des conditions périodiques, les $\bfvarphi$ sont des vecteurs d'isospin 
complexes : $\bfvarphi = {\bf a} + i {\bf b}$. Il est facile de montrer que
dans l'état initial, les composantes $a_j$ et $b_j$ des parties réelle et 
imaginaire sont des variables aléatoires gaussiennes indépendantes et de même
variance\footnote{Excepté bien sûr, pour le mode zéro qui, lui, est réel :
   $\bfvarphi_{\vec k}^* = \bfvarphi_{-\vec k}$.}. 
En d'autres termes, la distribution de probabilité du vecteur
$\bfvarphi$ est une loi du type (\ref{thermal}). Ceci implique que la 
probabilité pour que le carré du cosinus de l'angle $\theta$ entre le
vecteur réel $|\bfvarphi|$ et l'axe $\pi_3$ soit compris entre 
$\cos^2 \theta = f$ et $f+df$ est égale\footnote{La loi en $1/\sqrt f$, obtenue
   avec des conditions périodiques par l'auteur de la Réf.~\cite{RAN2}, n'est 
   valable que pour le mode zéro, réel. Attention, le $f$ dont il est question
   ici ne représente pas la fraction de pions neutres, mais l'orientation 
   du vecteur $|\bfvarphi|$.} 
à $2(1-f)df$ (cf. Eq.~(\ref{thermalf})). Ceci est un artefact des conditions
aux bords périodiques : les parties réelles et imaginaires sont deux degrés
de liberté indépendants décrivant le même mode $\vec k$. Dans le cas où 
les orientations dans l'espace d'isospin sont toutes equiprobables, on 
attend la distribution caractéristique en $1/\sqrt f$. Les conditions 
de Neumann sont donc mieux adaptées à l'étude de l'orientation 
dans l'espace d'isospin.

}

\part{VERS LA FORMATION D'UN PLASMA DE MATIÈRE DÉCONFINÉE}

\setcounter{chapter}{3}
\chapter{Equilibration thermique des gluons}
\label{QGP}

Un des enjeux principaux de l'étude des collisions nucléaires à haute
énergie est la possibilité de former un plasma de matière 
hadronique déconfinée, c'est à dire un système de quarks et de gluons en 
équilibre thermodynamique (local). Les données accumulées durant les dix 
dernières années à l'AGS (BNL) et au SPS (CERN) fournissent diverses
indications~\cite{QM99} en faveur de cette hypothèse, mais il est encore 
difficile de parler de preuves. L'avènement des collisions à très haute 
énergie, actuellement réalisées à RHIC (BNL), et, à partir de 2005, au 
LHC (CERN), ouvre une nouvelle ère dans ce domaine
de recherche. En particulier les collaborations PHENIX à RHIC et ALICE à LHC,
prévoient d'étudier les photons et les paires de leptons de grande
énergie transverse (jusqu'à $p_t \sim 10$~GeV), qui sont des signatures 
directes de la matière déconfinée~\cite{PHENIX,ALICE}. 
\par
Dans la région centrale d'une collision d'ions lourds à suffisament haute 
énergie, on s'attend à ce que la majeure partie de l'énergie transverse 
soit produite sous forme d'un grand nombre de partons de relativement 
grande impulsion transverse ($p_t \sim 1-2$~GeV). Cependant qu'il est 
rapidement dilué par la forte expansion longitudinale, le système ainsi
produit tend à s'équilibrer localement du fait des interactions mutuelles 
entre ses constituants. Lorsque l'intervalle de temps moyen entre deux 
collisions successives devient supérieur au temps typique de formation 
d'un hadron, le système de quarks et de gluons cesse d'exister en tant 
que tel et laisse place à un gaz de hadrons en interaction. C'est une 
question d'importance considérable pour l'interprètation prochaine des 
données que de savoir si le système de partons initialement produits 
a le temps d'atteindre un état d'équilibre local avant la phase 
hadronique. En effet, jusqu'à présent les calculs théoriques concernant 
les signatures de la matière déconfinée reposent sur 
l'hypothèse selon laquelle celle-ci est en équilibre thermique local. 

\section*{Vers l'équilibre local}

Une première étude du processus d'équilibration thermique dans les
premiers instants de la collision a été faite par G.~Baym en 1984~\cite{Baym}.
Dans cet article, l'auteur suppose que l'évolution du système de partons
est gouvernée par une équation de Boltzmann qu'il modélise par une 
approximation de temps de relaxation où le temps de relaxation $\theta$
est constant. Baym montre alors de façon analytique que malgré l'effet
de l'expansion, le système atteint effectivement le régime hydrodynamique 
de Bjorken~\cite{Bjor0}, sur une échelle de temps donnée par $\theta$. 
Plus tard, différents travaux consacrés à l'étude des propriétés de 
transport d'un plasma de quarks et de gluons (voir par 
exemple~\cite{BMPR,HeisPet}, voir aussi~\cite{LeBellac}), ont permis 
d'estimer l'échelle de temps caractérisant la relaxation cinétique 
d'un tel système. Le résultat obtenu est typiquement de l'ordre de $1$~fm, 
ce qui donne à penser que l'équilibre thermique est rapidement atteint.
\par
Avec cette idée en tête, différents auteurs (voir en particulier~\cite{Biro}, 
et plus récemment~\cite{Elliott}) étudient alors le processus d'équilibration
chimique dans les collisions d'ions lourds. Le système, en expansion 
longitudinale rapide est supposé localement isotrope à chaque instant
(équilibre thermique local), et des équations de transport
sont dérivées, qui décrivent l'évolution temporelle des variables macroscopiques
comme les densités moyennes d'énergie et de particules (gluons, quarks 
et antiquarks) par unité de volume. Ces équations sont résolues numériquement
avec des conditions initiales directement tirées de codes Monte Carlo
simulant la production de partons (minijets, voir plus loin) dans les 
tout premiers instants de la collision. Les conclusions de ces études 
sont que le système n'atteint pas l'équilibre chimique, les densités de 
quarks et de gluons sont inférieures à leurs valeurs à l'équilibre. 
Ceci a des implications importantes pour la phénoménologie, ainsi
que pour la théorie : d'une part il faut tenir compte de l'aspect
``hors d'équilibre chimique'' dans les calculs d'observables~\cite{BDRS,SMM}, 
d'autre part il est nécessaire de bien décrire l'évolution du système 
pour faire des prédictions viables, ce qui implique une bonne connaissance
des processus d'équilibration (ici chimique), ainsi qu'une bonne description
de l'état initial.
\par
Au milieu des années 1990, certains auteurs réexaminent le problème de 
la thermalisation dans le même cadre que l'étude de Baym, mais en 
faisant l'hypothèse plus réaliste d'un temps de relaxation dépendant du 
temps~\cite{HeisWang1,HeisWang2,Wong}. En effet, du fait de la forte 
expansion, le système est rapidement dilué, et les collisions sont de plus 
en plus rares, il est donc naturel que l'échelle caractérisant la 
relaxation vers l'équilibre croisse avec le temps\footnote{Par exemple, 
   au voisinnage de l'équilibre local, la seule quantité dimensionnée est 
   la température locale $T$, et le temps de relaxation est donc inversement
   proportionnel à $T$. Cependant, pour un système en expansion longitudinale 
   dans le régime hydrodynamique, $T \propto t^{-1/3}$~\cite{Bjor0}, et donc 
   $\theta \propto t^{1/3}$.}.
Dans ces conditions, il n'est pas clair que l'équilibre puisse être atteint,
le point clef réside dans la compétition entre les effets respectifs de 
l'expansion et des collisions. Par exemple, dans un modèle où $\theta \propto t^p$,
le système n'atteint jamais le régime hydrodynamique si $p>1$~\cite{HeisWang2}.
Dans un modèle plus réaliste, le temps de relaxation à l'instant $t$ dépend de 
l'état du système à l'instant $t$ qui dépend lui-même de la valeur du temps
de relaxation à l'instant antécédent, et ainsi de suite. Contrairement au 
cas étudié par Baym, où le temps d'équilibration est essentiellement donné 
par le temps de relaxation (constant), lui-même déterminé par la nature 
intrinsèque du système étudié, dans le cas où le temps de relaxation 
dépend de l'histoire du système, le temps {\em effectif} d'équilibration
dépend en plus d'un aspect extrinsèque : la condition initiale. 
Pour estimer la durée du régime transitoire il est donc nécessaire de
suivre l'évolution du système à partir d'une condition initiale donnée.
Dans le cas des collisions d'ions lourds, qui nous intéresse ici, cela
signifie que l'on doit avoir une description réaliste de l'état initial
du système de partons ainsi que des processus qui gouvernent son évolution
ultérieure. 

\section*{Production de gluons dans les premiers instants de 
la collision : le problème de la condition initiale.}

Le problème de la caractérisation de l'état initial du système de gluons
produits lors d'une collision d'ions lourds ultra-relativistes est très
délicat et constitue un domaine de recherche en soi. Deux scénarios ont 
été proposés dans la littérature : le scénario des minijets et le scénario 
de saturation. Dans ce qui suit nous les décrivons de manière très 
schématique, notre but étant de donner une image intuitive de ces scénarios.

Plaçons-nous dans le référentiel du centre de masse de la collision. 
Avant l'impact, chacun des noyaux
incidents peut être vu comme un disque très aplati dans la direction 
de son mouvement, à l'intérieur duquel sont confinés les quarks de 
valence des nucléons, et avec lequel se déplace un ``nuage'' de partons 
virtuels : les quarks de la mer et les gluons, les plus nombreux étant 
ceux dont l'impulsion longitudinale est très faible comparée à celle 
des noyaux incidents (la fraction $x$ d'impulsion longitudinale emportée 
par le parton est~$\ll 1$). Au moment de la collision, les deux noyaux se 
traversent l'un l'autre, et les partons de leurs nuages partoniques 
respectifs interagissent les uns avec les autres. Du fait de ces 
interactions, certains de ces partons virtuels (mais quasi-réels) sont 
``projetés'' sur leur couche de masse, on dit qu'ils 
sont ``libérés'' lors de la collision : ils ne font plus partie intégrante 
des noyaux qui les ``transportaient'' initialement. Un grand nombre de 
partons réels d'impulsion transverse $p_t$ sont ainsi produits pendant 
un intervalle de temps $\sim 1/p_t$, qui est aussi l'échelle de 
temps typique pour que les partons d'impulsions longitudinales différentes 
se séparent physiquement les uns des autres~\cite{MB,AHM1}. 

\subsection*{Le scénario des minijets}

Si l'impulsion typique du parton produit est suffisamment grande devant
l'échelle typique des interactions fortes ($p_t \gg \Lambda_{QCD} \sim 200$~MeV),
le processus de production relève du régime perturbatif ($\alpha_S (p_t) \ll 1$), 
où le taux de production peut être calculé de façon contrôlée
(voir Fig.~\ref{fig_minijet}). La section efficace de production
décroissant comme $1/p_t^4$ (voir par exemple~\cite{Eskola}), la 
contribution dominante à l'énergie transverse initialement produite 
vient des gluons d'énergie intermédiaire (suffisamment faible pour 
que la probabilité de production soit considérable, et suffisamment 
grande pour que le calcul perturbatif ait un sens), $p_t \sim 1-2$~GeV : 
les minijets~\cite{KLL,MB,Eskola}.

\begin{figure}[htbp]
\epsfxsize=4.in \centerline{ \epsfbox{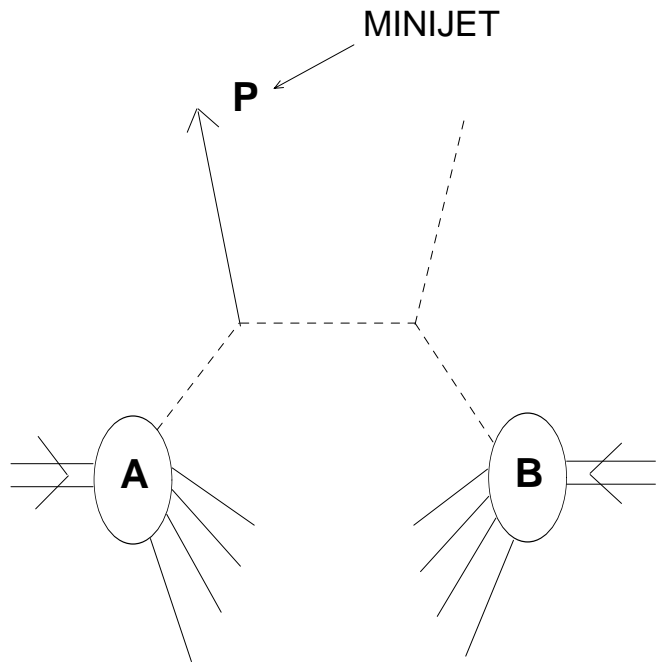}}
\caption{\small Diffusion élastique entre deux partons des noyaux incidents
lors d'une collision nucléaire à haute énergie. Le taux de production 
perturbatif est dominé par les partons d'impulsion 
$p_t \sim 1-2$~GeV, les minijets.} 
\label{fig_minijet}
\end{figure}

\subsection*{Le scénario de saturation}

Le scénario de saturation propose une origine différente pour la production
initiale d'énergie transverse. Dans les noyaux incidents, la densité de gluons 
de petit $x$ par unité d'espace des phases transverse $\frac{dN}{d^2b d^2 p_t}$, 
où $\vec b$ mesure la position dans le plan transverse, augmente quand
$p_t$ diminue, et atteint un maximum pour des impulsions transverses 
$p_t \lesssim Q_s$, où $Q_s$ est l'impulsion de
saturation~\cite{Musat1,Musat2,McLsat}. 
Les gluons de faible impulsion transverse étant donc en très grand nombre, 
on s'attend à ce que leur contribution à la production d'énergie transverse 
soit non-négligeable. Cependant, le calcul perturbatif n'a pas de sens dans 
ce cas, et il n'existe à ce jour aucun calcul solide du taux de production 
de ces gluons. On peut cependant se faire un image du processus de production : 
les gluons considérés ici étant très nombreux, on peut les voir comme formant 
un champ classique\footnote{Exactement comme nous l'avons
   fait pour le champ de pions dans le chapitre précédent.} généré par les 
quarks de valence des nucléons, en mouvement rapide dans le référentiel
considéré : c'est l'équivalent non-abélien du champ de
Weizsäcker-Williams~\cite{McLVen}. Dans cette image, la collision nucléaire est 
vue comme une ``collision'' entre les champs associés à chacun des noyaux 
incidents. Sur la Fig.~\ref{fig_saturation}, on a représenté très schématiquement 
cette collision : un gluon virtuel d'un des noyaux est libéré (projeté 
sur couche de masse) par interaction avec le champ classique de l'autre 
noyau. En pratique, on peut décrire cette collision en résolvant 
(numériquement) les équation de champ classique~\cite{KraVent,KraVenc}. 
On peut se convaincre avec A.~H.~Mueller~\cite{AHM1}, que la majorité des 
gluons libérés sont ceux dont l'impulsion $p_t \sim Q_s$, et ce durant un 
intervalle de temps $\sim 1/Q_s$. La valeur de l'impulsion de saturation 
$Q_s$ dépend de la fraction $x$ d'impulsion longitudinale des gluons 
considérés, et donc de l'énergie de la collision. Aux énergies de RHIC 
($\sqrt s = 200$~AGeV), $Q_s \approx 1$~GeV, et à LHC ($\sqrt s = 5.5$~ATeV), 
$Q_s \approx 2-3$~GeV.

\begin{figure}[htbp]
\epsfxsize=4.in \centerline{ \epsfbox{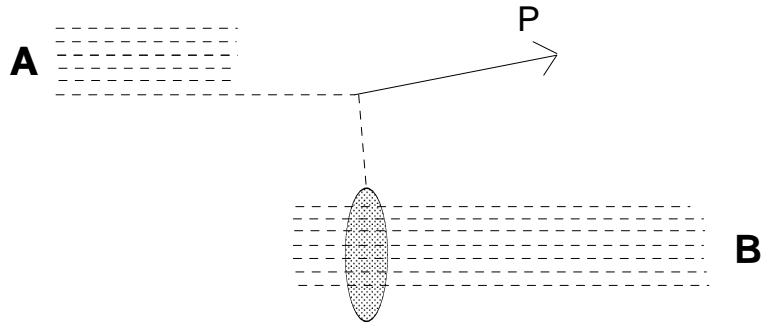}}
\caption{\small Dans le scénario de saturation, les gluons virtuels de 
petite impulsion transverse des noyaux incidents peuvent être vus comme 
un champ classique de couleur associé aux quarks de valence en mouvement 
rapide. Dans la représentation schématique de la collision, ci-dessus, on 
suit un gluon du champ du noyau provenant de la gauche. Celui-ci est 
libéré lors de la collision, en interagissant avec l'ensemble des gluons 
de l'autre noyau, c'est à dire avec le champ classique associé.} 
\label{fig_saturation}
\end{figure}

Dans ces deux scénarios la contribution dominante à la production initiale
d'énergie transverse provient des partons dont l'impulsion transverse est
de l'ordre de $1-2$~GeV. Cependant, l'origine de ces partons
est très différente dans les deux cas, ce qui conduit à des prédictions 
très différentes pour l'état initial du système. La Fig.~\ref{fig_gluon}, 
où on a représenté schématiquement la densité de gluons produits par unité 
d'espace des phases transverse, résume ces deux scénarios. 

\begin{figure}[htbp]
\epsfxsize=4.in \centerline{ \epsfbox{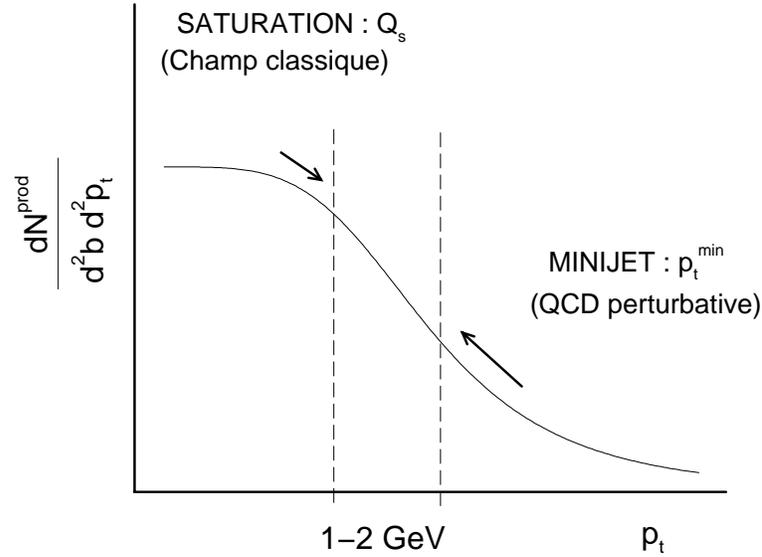}}
\caption{\small Les densités de gluons produits par unité d'espace des phases 
transverse dans les scénarios de saturation et des minijets. Dans les deux
cas la région qui domine la production d'énergie transverse est la même : 
ce sint les gluons d'impulsion transverse $p_t \sim 1-2$~GeV. L'origine de
ces gluons est cependant très différente dans les deux cas, ce qui conduit
à des prédictions différentes pour l'état initial du système.} 
\label{fig_gluon}
\end{figure}

Récemment, les auteurs des Réfs.~\cite{AHM1,AHM2,Raju} d'une part 
et~\cite{Dumitru} d'autre part ont étudié, indépendamment et dans 
des approches différentes, la question de la thermalisation du gaz de 
gluons respectivement dans les scénarios de saturation et des minijets.
Dans~\cite{AHM1,AHM2}, A.~H.~Mueller considère l'effet des collisions 
élastiques dans l'approximation des petites déviations. En supposant 
une forme idéalisée pour la distribution initiale des gluons, il est 
capable de suivre analytiquement l'évolution du système durant les 
premiers instants de la collision. Bien que ses approximations ne lui 
permettent pas de suivre le gaz de gluons jusqu'à l'équilibre, il
semble que le processus d'équilibration commence très tôt. Dans la 
Réf.~\cite{Raju}, J.~Bjoraker et R.~Venugopalan reprennent ce modèle
et résolvent numériquement l'équation de Boltzmann proposée par Mueller.
Il apparait clairement que le système tend vers l'équilibre malgré l'expansion.
En comparant la dépendance en temps de différentes quantités macroscopiques
(température et densité d'entropie par unité de volume) avec celle attendue
dans le régime hydrodynamique, ces auteurs mesurent des temps d'équilibration 
de l'ordre de $3-4$~fm à RHIC et $2-3$~fm à LHC. Parallèlement, G.~C.~Nayak, A.~Dumitru, L.~Mc~Lerran et W.~Greiner~\cite{Dumitru} étudient la thermalisation 
du gaz de minijets dans une approximation de temps de relaxation et concluent 
à l'équilibration avec des temps caractéristiques de l'ordre de $4-5$~fm aussi 
bien aux énergies de RHIC que de LHC. 
\par
Les approches utilisées dans~\cite{Raju} d'une part et dans~\cite{Dumitru}
d'autre part sont cependant très différentes, et il est difficile de comparer
ces résultats les uns avec les autres. En particulier les auteurs 
de~\cite{Dumitru} identifient le temps de relaxation avec le libre parcours
moyen, et il n'est pas clair que la dynamique qui résulte de cette hypothèse
arbitraire soit directement reliée à l'équation de Boltzmann résolue 
dans~\cite{Raju}.Dans ce chapitre, nous reprenons l'étude de la thermalisation 
du gaz de gluons et comparons les scénarios de saturation et
des minijets dans une seule et même approche. Suivant Mueller~\cite{AHM2},
nous considérons l'effet des collisions élastiques dans la limite des petites
déviations (voir l'Annexe~\ref{landaucoll}). Nous traitons le terme de collision
de l'équation de Boltzmann correspondante dans l'approximation du temps de 
relaxation et calculons le temps de relaxation de façon auto-cohérente 
en faisant directement intervenir la dynamique de l'équation de Boltzmann 
sous-jacente (voir aussi~\cite{HeisWang1,HeisWang2,Wong}). Dans le cas de 
l'intégrale de collision de Landau-Mueller, nous sommes capables de calculer
analytiquement toutes les intégrales sur l'espace des impulsions, ce qui
diminue considérablement l'effort numérique à fournir. En comparant nos 
résultats avec la solution exacte de~\cite{Raju}, on peut tester notre 
approche, qui s'avère être remarquablement bonne compte tenu de son
degré de simplicité. 
\par
Nous arguons que le critère utilisé dans~\cite{Raju} pour caractériser 
l'équilibre n'est pas satisfaisant et proposons plutôt de tester le degré
d'isotropie de différentes observables, critère plus intuitif d'une part
et plus directement relié à la phénoménologie d'autre part. En conséquence,
nos conclusions sont qualitativement différentes de celles de~\cite{Raju},
en particulier nous montrons qu'aux énergies de RHIC les collisions 
élastiques ne sont pas suffisantes pour permettre au système d'atteindre 
l'équilibre thermique avant la phase hadronique. En ce qui concerne le
scénario des minijets, nous arguons que les résultats de~\cite{Dumitru} 
ne sont pas crédibles, les auteurs n'ayant pas correctement implémenté les 
lois de conservation. Ici encore, nous montrons que l'équilibre n'est pas
atteint à RHIC.
\par
De même que dans les chapitres précédents, nous tenterons de mettre en avant
les idées physiques, les détails techniques étant relégués en Annexe. Le 
travail présenté dans ce chapitre a été réalisé en collaboration avec 
Dominique Schiff~\cite{DSJS}.

\newpage
\section{Le modèle}

Considérons une collision entre deux noyaux lourds à très haute énergie.
Pour fixer les idées, choisissons deux noyaux
identiques, sphériques de rayon $R$. 
Dans le référentiel du centre de masse, du fait 
de la contraction de Lorentz, chacun de ces noyaux a l'aspect d'une galette
de rayon $R$ et d'épaisseur $l = R \sqrt{1-\beta^2}$ dans la direction
$z$ de l'axe de la collision, où $\beta \leq 1$ est la vitesse du noyau en 
unité de la vitesse de la lumière. Dans une collision à très haute énergie,
on peut modéliser les deux noyaux incidents par deux disques infiniment plats
($\beta \simeq 1$, $l \simeq 0$) et de rayon infini\footnote{A l'instant  
   de l'impact ($t=0$), toute l'énergie de la collision est localisée dans 
   une région de taille longitudinale nulle. Cette énergie étant finie, la 
   région en question doit avoir une extension transverse infinie~\cite{BK1}. 
   Cette idéalisation est raisonnable dans la région centrale de la collision
   ($z \simeq 0$, $\|\vec x_t \| \ll R$, où $\vec x_t$ repère la position
   dans le plan transverse à l'axe de la collision) et pour des 
   temps au moins inférieurs à la taille transverse du système : $t \lesssim R$.}
Après la collision, la matière baryonique étant  très peu ralentie~\cite{CYWONG},
les deux disques, qui s'éloignent maintenant l'un de l'autre avec des vitesses
proches de celle de la lumière, laissent derrière eux une région de matière
non-baryonique hautement excitée. Nous adopterons l'idéalisation de
Bjorken~\cite{Heis,Bjor0}, où les deux disques s'éloignent à la vitesse de 
la lumière. Le problème est alors invariant sous les transformations de 
Lorentz suivant l'axe de la collision.
\par
Nous considérerons ici des collisions à très haute énergie (RHIC,LHC),
et nous supposerons que la constante de couplage des interactions fortes
$\alpha_S \ll 1$.
Durant les tous premiers instants après l'impact, un grand nombre de 
partons, essentiellement des gluons, sont produits.
L'expansion dilue très rapidement le système, et les nombres d'occupation
(les densités de partons par unité d'espace des phases : 
$f \propto d^6N/d^3p \, d^3x$)
deviennent rapidement inférieurs à l'unité\footnote{On peut se convaincre 
   que la description en termes d'une équation de Boltzmann est valable 
   dès que $f \lesssim 1/\alpha_S$. Quand $f \lesssim 1$, on peut
   négliger les effets de statistique quantique.}. 
On peut alors traiter le système
comme un gaz de particules 
classiques\footnote{Par opposition à l'approximation de {\em champ classique},
   valide quand les nombres d'occupation sont grands devant l'unité.},
et décrire le taux de variation par unité de temps de la distribution $f$ 
par une équation de Boltzmann : 
\eq
 \frac{\dd f}{\dd t} = \mathcal C [f] \, ,
\eq
où le membre de gauche désigne la variation due au mouvement libre des
particules du gaz : en l'absence de champ externe, $\frac{\dd f}{\dd t} = 
\p_t f(\vec p,\vec x,t) + \vec v \cdot \vec \nabla_x f(\vec p,\vec x,t)$, avec
$\vec v = \vec p/p$. Le membre de droite décrit l'effet des collisions entre
les particules du gaz. 
\par
La distribution $f(\vec p,\vec x,t)$ décrit la 
densité de gluons {\em réels} (sur couche de masse : $P^2=0$) par unité d'espace 
des phases à l'instant $t$, et dépend en général de l'état de couleur des 
gluons considérés. Ici, nous cherchons à décrire la relaxation
vers l'équilibre cinétique (local), c'est à dire de relaxation vers une
distribution (localement) isotrope. L'échelle de temps caractéristique 
de ce processus étant grande devant celle du processus de relaxation de 
couleur\footnote{Au voisinage de l'équilibre, les échelles
   de temps caractéristiques des processus de relaxation de couleur et de
   relaxation cinétique sont bien séparées : $\tau_{couleur}^{-1} \sim
   \tau_{cin\acute{e}tique}^{-1}/\alpha_S$, la première, étant beaucoup plus
   sensible à la physique de grande longueur d'onde (voir par 
   exemple~\cite{BIancu}).}, 
on peut négliger ce dernier sur les échelles de temps qui nous intéressent.
Bien que la dérivation du terme de collision que nous allons utiliser repose
sur l'hypothèse d'une distribution individuelle {\em indépendante} de la 
couleur, l'équation de Boltzmann obtenue doit être vue comme une équation 
effective pour les échelles de temps caractéristiques de la relaxation
cinétique. Dans cette logique la distribution $f$ s'interprète comme
une distribution {\em moyennée} sur les états de couleur (et de polarisation).
\par
Mentionnons de plus que la description en terme d'une équation de Boltzmann
``standard'' n'a de sens que si le temps moyen entre deux collisions 
(le libre parcours moyen $\bar l$) successives est suffisamment grand 
devant la longueur d'onde des particules considérées $\ell \sim 1/p$.
L'échelle typique d'impulsion des gluons décrits ici\footnote{Les excitations
   de grande longueur d'onde ($\ell \gg 1/\bar\epsilon$, où $\bar\epsilon$
   est l'énergie moyenne par particule) sont des modes collectifs (ils
   ``contiennent'' un grand nombre d'excitations dures 
   ($\ell \sim 1/\bar\epsilon$)) et agissent comme un champ externe classique
   (qui n'est autre que le champ moyen créé par le mouvement des particules
   dures) sur la dynamique des modes durs~(voir par exemple~\cite{BIancu}).
   Dans ce chapitre, nous négligeons ce champ externe.} 
est donc $p \gtrsim {\bar l}^{-1}$.
\par
Du fait de la géométrie du problème, la physique est invariante sous les
transformations de Lorentz le long de l'axe $z$ de la 
collision~\cite{Baym}\footnote{L'échelle caractérisant les inhomogénéités
   spatiales dans la direction transverse est $\sim R$. Dans la région 
   centrale on peut donc négliger la dépendance en $\vec x_t$ de la 
   distribution $f$. De plus, par symétrie, $f$ ne dépend que de 
   $p_t=\| \vec p_t \|$ et est paire en $p_z$.}
\eq
 f(\vec p_t,p_z,z,t) \equiv f(\vec p_t,\tilde p_z,\tau) \, ,
\eq
où $\tau=\sqrt{t^2-z^2}$ et $\tilde p_z = \gamma (p_z - u \, p)$, avec 
$p^2 = p_t^2 + p_z^2$, $u=z/t$ et $\gamma = (1-u^2)^{-1/2}=t/\tau$. 
Il suffit donc de se limiter à la tranche $z=0$. Il est facile de voir que
\eq
 \left. \frac{\dd f}{\dd t} \right|_{z=0} = 
 \p_t f(\vec p,t) - \frac{p_z}{t} \, \p_{p_z} f(\vec p,t) = 
 \left. \p_t f(\vec p,t) \right|_{p_zt=cte} \, .
\eq
En ce qui concerne le terme de collision de l'équation de Boltzmann, nous 
nous limiterons, comme nous l'avons mentionné dans l'introduction, aux 
collisions élastiques entre gluons. La section efficace différentielle de 
diffusion $gg \rightarrow gg$ est fortement piquée vers l'avant, et
diverge pour les diffusion à très petit angle de déviation, exactement
comme dans le cas de la diffusion entre particules chargées : c'est la 
divergence de Rutherford due à la portée infinie de l'interaction 
(l'échange d'un quantum de masse nulle). Les collisions à petite déviation 
étant les plus fréquentes, elles dominent la dynamique et nous ne retiendrons 
que leur contribution à l'intégrale de collision. En développant l'intégrand 
autour des petites valeurs de l'angle de déviation~\cite{AHM2,LanLif} (voir 
aussi l'Annexe~\ref{landaucoll}), la première contribution non nulle est égale 
à la divergence du flux de particules dans l'espace des impulsions : l'effet des 
collisions élastiques à petite déviation est équivalent à un processus 
diffusif dans l'espace des impulsions. En explicitant ce terme, on obtient, 
dans la région centrale ($z=0$) et dans l'approximation de particules 
classiques ($f \ll 1$) (cf. Eq.~(\ref{Boltzcl})),
\beq
\label{BE}
 \left. \p_t f \right|_{p_zt} = 
 \mathcal B \, N_0 \nabla_p^2 f + 
 2 \mathcal B N_{-1} \vec\nabla_p (\vec v \, f) \, ,
\end{equation}\noindent
où l'on a défini les moments\footnote{Nous introduisons ici une notation
   différente de celle habituellement utilisée dans la littérature :
   $\langle m \rangle = \int d^3p \, m (\vec p) \, f(\vec p,t)$
   désigne la moyenne par unité de volume et 
   $\overline{\langle m \rangle} = \langle m \rangle / n$
   la moyenne par particule ($n=\langle 1 \rangle$ est la densité moyenne 
   de particules par unité de volume).} 
($\int_{\vec p} \, \equiv \, 2(N_c^2-1) \int d^3p/(2\pi)^3$)
\beq
\label{Ns}
 N_s(t)= \langle p^s \rangle = \int_{\vec p} \, p^s \, f(\vec p,t)
\end{equation}\noindent
 ($N_0 = n$ est la densité moyenne de particules par unité de volume, 
$N_1 = \epsilon$ est la densité moyenne d'énergie par unité de volume).
La divergence de Rutherford de la section efficace de diffusion gluon-gluon
se manifeste\footnote{Les tout premiers termes du développement dans 
   l'angle de déviation $\theta$ donnent, en principe, des contributions 
   plus singulières. Cependant, ces termes sont nuls dans le cas d'une
   distribution indépendante des degrés de liberté de couleur. C'est là
   l'origine de la hiérarchie des échelles de temps caractérisant les 
   processus de relaxation de couleur et cinétique. Cette dernière est 
   en fait sensible à la physique de l'écrantage (chromo-)~électrique, 
   c'est à dire aux échelles de distance de l'ordre de la longueur de 
   Debye (voir Eq.~(\ref{logdiv1})).} 
par la présence du ``grand logarithme'' $L=2 \int d\theta/\theta$ : 
\eq
 \mathcal B = \pi \alpha_S^2 \frac{N_c^2}{N_c^2-1} \, L \, ,
\eq
Cette divergence logarithmique est régularisée par le phénomène d'écrantage
de Debye :
\beq
\label{logdiv1}
 L = \ln \left( \frac{\underline p^2}{m_D^2}\right) \, ,
\end{equation}
où $m_D$ est la masse d'écran de Debye, et où $\underline p$ désigne 
l'ordre de grandeur de l'impulsion typique des particules du milieu 
($m_D^2 \ll \underline{p}$) (voir les Eqs.(\ref{minangle}) et (\ref{logdiv0}) 
de l'Annexe~\ref{landaucoll}).

\subsection{Lois de conservation}

Dans la région centrale, le tenseur énergie-impulsion s'écrit
\eq
 T^{\mu\nu} = \langle \, \frac{p^\mu p^\nu}{p} \, \rangle =
 \int_{\vec p} \, \frac{p^\mu p^\nu}{p} \, f(\vec p,t) \equiv
 \mbox{diag} (\epsilon,P_T,P_T,P_L) \, ,
\eq
et est diagonal, du fait de la symétrie du problème ($f(\vec p,z=0,t) = 
f(-\vec p,z=0,t)$). Ses composantes représentent respectivement les
densités moyennes d'énergie et de pression (transverse et longitudinale) 
dans la région centrale à l'instant $t$ : 
\eq
 \epsilon (t) = \langle \, p \, \rangle  = N_1 (t) \, \, \, , \, \, \,
 P_{T,L} (t) = \langle \, p_{\perp,z}^2/p \, \rangle  \, ,
\eq
où $p_\perp^2 = p_t^2/2 = (p_x^2 + p_y^2)/2$. Les particules du milieu
étant de masse nulle (${T^\mu}_\mu = 0$), on a $\epsilon = P_L + 2P_T$.
\par
Les lois de conservation de l'impulsion et du moment angulaire sont trivialement 
satisfaites par symétrie. L'énergie étant conservée dans les collisions 
individuelles, le taux de variation par unité de temps de la densité d'énergie
induit par les collisions est nul :
\eq
 \int_{\vec p} \, p \, \, \mathcal C [f] (\vec p,t) = 0 \, .
\eq
On en déduit l'équation d'évolution de la densité d'énergie 
($\dot\epsilon =\p_t \epsilon$) : 
\beq
\label{consen}
 \dot\epsilon + \frac{\epsilon + P_L}{t} = 0 \, .
\end{equation}
De plus, il est clair que les collisions élastiques conservent le nombre total
de particules, ce qui implique que le taux de variation collisionnel de la
densité moyenne de particules est nul. On a donc l'équation 
($n (t) = \langle 1 \rangle = N_0 (t)$)
\eq
 \dot n + \frac{n}{t} = \int_{\vec p} \mathcal C [f] (\vec p,t) = 0 \, ,
\eq
dont la solution est
\beq
\label{consnb}
 n(t) = n(t_0) \frac{t_0}{t} \, .
\end{equation}
\par 
Remarquons que ces deux équations, ne faisant pas intervenir le terme
de collision, gardent la même forme quelles que soient les approximation 
faites sur ce dernier.

\subsection{Expansion .{\it vs}. Collisions}
\label{page_disc}

Partant d'une condition initiale donnée, c'est à dire d'une distribution
initiale $f_0(\vec p)$, l'évolution du système est gouvernée par la 
compétition entre deux phénomènes : d'un coté (à gauche dans l'Eq.~(\ref{BE}))
l'expansion, de l'autre les collisions.
\par
L'expansion longitudinale rapide du système tend à rendre la distribution 
fortement anisotrope : à impulsion $(\vec p_t,p_z)$ donnée, le nombre moyen 
de particules qui quittent la région centrale librement (c'est à dire par 
leur mouvement propre, sans l'aide des collisions) est supérieur au nombre 
de particules qui y pénetrent de la même manière, et ceci d'autant plus que 
l'impulsion longitudinale considérée est grande. L'expansion a donc tendance 
à ``vider'' la région centrale de son impulsion longitudinale.
\par
L'effet des collisions est, schématiquement, de créer des particules 
d'impulsions diverses, d'orientations aléatoires et tend donc, à l'inverse
de l'expansion, à rendre la distribution isotrope (dans la région centrale). 
Dans l'approximation des petites déviations, il s'agit d'un processus diffusif 
dans l'espace des impulsions~\cite{LanLif} (voir la discussion plus haut, 
ainsi que l'Annexe~\ref{landaucoll}) : de la même manière qu'un gaz tend, 
par diffusion, à remplir l'espace (des configurations) de manière isotrope 
par diffusion, on a ici un gaz qui se diffuse dans l'espace des impulsions, 
et tend à occuper celui-ci de manière isotrope. 
\par
Il est clair que les influences respectives de chacun de ces effets sur 
l'évolution du système dépend de l'état initial de celui-ci. L'objet
de ce chapitre est précisément l'étude de cette compétition pour 
différentes conditions initiales. A cette fin, il est instructif de 
caractériser les deux régimes limites correspondant chacun à la 
victoire d'un effet sur l'autre.

\subsubsection{Le cas libre}

Dans le cas où l'influence des collisions est complètement insignifiante
devant celle de l'expansion, on peut négliger le membre de droite
de l'équation de Boltzmann, ce qui est formellement équivalent au 
cas où les particules du gaz sont libres ($\alpha_S=0$) : 
\eq
 \left. \p_t f_{libre}(\vec p,t) \right|_{p_zt} = 0 \, .
\eq
ou encore
\beq
\label{streaming}
 f_{libre} (\vec p,t) = f_0 (\vec p_t,p_z\frac{t}{t_0}) \, .
\end{equation}\noindent
La distribution devient rapidement $\sim \delta(p_z)$, on voit par exemple 
que $\langle p_z^2 \rangle / \langle p_t^2 \rangle \sim 1/t^2$. 

\subsubsection{Le régime hydrodynamique}

La situation à l'extrême opposé est celle où l'effet des collisions
est suffisament fort pour maintenir l'isotropie à chaque instant. 
Dans ce cas, c'est le terme de gauche de l'Eq.~(\ref{BE}) qui est
négligeable et la forme de la distribution est déterminée par l'équation
\beq
\label{hydroeq}
 \mathcal C [f_{eq}] = 0 \, .
\end{equation}\noindent
La solution isotrope la plus générale est de la forme
\beq
\label{disteq}
 f_{eq} (\vec p,t) = \lambda (t) \, \mbox{e}^{-p/T(t)} \, ,
\end{equation}\noindent
où les fonctions du temps $\lambda (t)$ et $T (t)$ (fugacité et température 
locales) sont déterminées par les équations de conservation qui, ne faisant
pas intervenir le terme de collision, gardent leurs formes respectives 
(\ref{consen}) et (\ref{consnb}) dans la limite (\ref{hydroeq}). 
On a ($P_L = P_T = \epsilon/3$)
\eq
 \epsilon \sim \lambda \, T^4 \sim t^{-4/3} \, \, \, , \, \, \,
 n \sim \lambda \, T^3 \sim t^{-1} \, ,
\eq
d'où on déduit, pour la fugacité et la température dans le régime 
hydrodynamique,
\beq
\label{hydro}
 \lambda_{hydro} \sim \mbox{cte} \, \, \, , \, \, \, T_{hydro} \sim t^{-1/3} \, .
\end{equation} 
 
Sur une échelle de temps $\Delta t \ll \theta_{relax}$, où $\theta_{relax}$ 
est l'échelle de temps caractérisant le processus de relaxation vers l'équilibre 
local, la dynamique est celle du régime libre. A l'inverse, sur des échelles 
de temps $\Delta t \gg \theta_{relax}$, la dynamique est décrite par le régime
hydrodynamique. Enfin, pour des échelles de temps intermédiaires 
$\Delta t \sim \theta_{relax}$, le système est dans un régime transitoire,
la dynamique est décrite par l'Eq.~(\ref{BE}). Dans la suite, l'échelle
de temps à laquelle on s'intéresse est donnée par la durée de vie du système
de partons, lequel cesse d'exister en tant que tel quand la densité d'énergie
moyenne par unité de volume devient inférieure à la densité critique
$\epsilon_c \sim 1$GeV/fm$^3$. La question est de savoir si le système 
atteint le régime hydrodynamique avant la fin du temps qui lui est imparti.
Autrement dit, nous voulons savoir comment se compare la durée du régime
transitoire avec la durée de vie. 
\par
Dans une situation statique, l'ordre de grandeur de la durée du régime 
transitoire est simplement donné par $\theta_{relax}$. Dans le cas présent
cependant, la question posée est moins triviale qu'elle n'en a l'air. 
En effet, du fait de l'expansion, le nombre de particules par unité de volume
diminue et les collisions sont donc de moins en moins fréquentes. Autrement
dit, l'échelle de temps $\theta_{relax}$ augmente avec le temps : le système
approche de l'équilibre local de plus en plus lentement à mesure que le temps
passe. La durée du régime transitoire ``effectif'' dépend donc de l'histoire 
du système (et donc aussi de son état initial) et ne peut être obtenue 
qu'en suivant son évolution au cours du temps, c'est à dire en résolvant 
l'équation de Boltzmann~(\ref{BE}).

\section{L'approximation du temps de relaxation}

La résolution numérique de l'Eq.~(\ref{BE})~\cite{Raju} s'avère relativement 
lourde en regard de la question posée, essentiellement qualitative (le modèle 
très simple, décrit précédemment, ne peut être crédible au niveau quantitatif). 
Pour cette raison, nous allons encore simplifier la dynamique en remplaçant 
le terme de collision par un terme linéaire dans la différence $f-f_{eq}$. 
Autrement dit, nous modélisons l'effet des collisions sur le taux de variation 
de la distribution $f$ par unité de temps par un terme de relaxation 
exponentielle sur une échelle de temps $\theta$ : le temps 
de relaxation (voir par exemple~\cite{Baym}) : 
\beq
\label{RTA}
 \left. \p_t f(\vec p,t) \right|_{p_zt} = 
 - \frac{f(\vec p,t)-f_{eq}(\vec p,t)}{\theta(t)} \, ,
\end{equation}\noindent
où (cf. Eq.~(\ref{disteq}))
\eq
 f_{eq} (\vec p,t) = \lambda (t) \, \mbox{e}^{-p/T(t)} \, ,
\eq
les fonctions du temps $\lambda$ et $T$ devant être déterminée
à l'aide des lois de conservation\footnote{Comme nous l'avons vu plus haut,
   dans le cas où seules les collisions élastiques sont prises en compte,
   le nombre de particules est conservé. Il est alors nécessaire d'introduire
   le paramètre $\lambda$ pour prendre en compte cette loi de conservation, 
   ce qui a été omis par les auteurs de la Réf.~\cite{Dumitru}.}
 (\ref{consen})-(\ref{consnb}).
\par
L'approximation de temps de relaxation consiste à paramétriser 
la distribution $f$, solution de l'équation de Boltzmann, à l'aide
des trois quantité $\lambda (t)$, $T(t)$ et $\theta (t)$, et ce à chaque 
instant $t$. On ne doit pas s'attendre à ce que la distribution soit
bien décrite par cette paramétrisation. Par exemple, dans cet ansatz
on voit que toutes les impulsions relaxent vers l'équilibre avec le 
même taux, ce qui n'a {\it a priori} aucune raison d'être le cas. 
En ce sens, l'approximation de relaxation est une sorte
d'approximation de champ moyen : les différentes impulsions sont effectivement
découplées les unes des autres et relaxent toutes avec un même taux moyen
que nous déterminerons dans la suite de façon auto-cohérente.

\subsection{La méthode des moments}

On peut calculer $\theta$ en identifiant le taux de variation collisionnel 
de certains moments de la distribution $f$ avec son analogue dans l'ansatz
(\ref{RTA}) (voir par exemple~\cite{HeisWang2,Wong}). Le temps de 
relaxation ainsi obtenu contient l'information concernant l'effet des 
collisions sur la dynamique. Nous appliquons ici cette méthode de manière
systématique pour différents moments caractérisant différentes échelles 
d'impulsion. Le choix du moment le plus adéquat dépend de la physique à 
laquelle on s'intéresse. Dans le cas du terme de collision de l'Eq.~(\ref{BE}), 
nous pourrons calculer analytiquement toutes les intégrales sur l'espace des
impulsions, ce qui simplifie considérablement la résolution numérique.
\par
Prise au pied de la lettre, l'approximation du temps de relaxation consiste
à identifier le terme de collision de l'équation de Boltzmann avec le membre
de gauche de l'Eq~(\ref{RTA}) : 
\eq
 - \frac{f-f_{eq}}{\theta} \equiv \mathcal C[f] \, .
\eq
Cette équation, prise au sens fort, c'est à dire telle qu'elle est écrite 
ci-dessus, pour tout $\vec p$, est trop contraignante et n'a pas de solution. 
On doit en fait la prendre dans un sens plus faible, en en prenant la valeur 
moyenne sur les impulsions avec un certain poids :
\beq
\label{momeq}
 - \frac{\langle m \rangle - \langle m \rangle_{eq}}{\theta_m} =
 \int_{\vec p} m(\vec p) \, \mathcal C[f] (\vec p,t) \, ,
\end{equation}\noindent
où 
\beq
\label{moment}
 \langle m \rangle_{(eq)} (t) = 
 \int_{\vec p} m(\vec p) \, f_{(eq)} (\vec p,t) 
\end{equation}\noindent
la fonction $m(\vec p)$ étant {\it a priori} quelconque, et où $\theta_m$ est 
le temps de relaxation correspondant. Nous voyons  encore une fois que
l'approximation du temps de relaxation (\ref{RTA}) n'est pas, à strictement 
parler, une approximation pour la dynamique microscopique, mais plutôt 
une modélisation de la relaxation du système dans sa globalité.
\par
Les équations de conservation conduisent aux relations\footnote{
   $\displaystyle
    N_s^{eq} (t) = \int_{\vec p} \, p^s \, f_{eq}(\vec p,t) =
    (s+2)! \, \frac{N_c^2-1}{\pi^2} \, \lambda(t) \, T^{s+3}(t) \, .
   $}
\bearn
\label{numb}
 n(t) & = & n_{eq} (t) = 
 2 \frac{N_c^2-1}{\pi^2} \, \lambda(t) \, T^3(t) \, , \\
\label{engy}
 \epsilon (t) & = & \epsilon_{eq} (t) = 
 6 \frac{N_c^2-1}{\pi^2} \, \lambda(t) \, T^4(t) \, .
\eearn
A ce point, il est important de faire la remarque suivante : les paramètres
$\lambda$ et $T$ n'ont pas, en général, les significations physiques de
la fugacité\footnote{On peut écrire $\lambda = \exp (- \mu/T)$. A l'équilibre 
   (local), $\mu = \mu_{hydro}$ est le potentiel chimique (local).}
et de la température du système, au sens thermodynamique (ou plutôt
hydrodynamique dans notre cas) de ces termes. En effet, les notions de 
fugacité et de température (locales) ne sont bien définies que pour un 
système à l'équilibre (local), auquel cas elles s'identifient avec les 
paramètres $\lambda$ et $T$ : $\lambda=\lambda_{hydro}$ et $T=T_{hydro}$.

\subsection{Procédure de résolution}

Les paramètres inconnus $\theta$, $\lambda$ et $T$ sont obtenus, à chaque instant,
à partir des Eqs. (\ref{momeq}), (\ref{numb}) et (\ref{engy}). Illustrons la
procédure de résolution sur un exemple simple : 
\eq
 m (\vec p) = p_z^2 - p_\perp^2 \, ,
\eq
où $p_\perp^2 = p_t^2/2$.
On a donc $\langle m \rangle_{eq} = 0$, et, après quelques intégrations
par parties, on obtient, pour l'Eq.~(\ref{momeq}),
\beq
\label{Mexple}
 \frac{\langle p_z^2 - p_\perp^2 \rangle}{\theta} = 
 4 \mathcal B \, N_{-1} \, (P_L - P_T) \, .
\end{equation}\noindent
Il nous faut donc calculer les quantités (nous laissons de côté le
calcul de $\mathcal B$ pour le moment) $\langle p_z^2 - p_t^2/2 \rangle$,
$N_{-1}$, $P_L - P_T$, et $\epsilon$ à chaque instant $t$. Pour ce faire, 
nous écrivons la solution formelle de l'Eq.~(\ref{RTA})~\cite{Baym,HeisWang1}
\beq
\label{RTABaym}
 f(\vec p,t)=f_0(\vec p_t,p_z\frac{t}{t_0}) \, \mbox{e}^{-x(t)} +
 \int_{t_0}^t dt' \, \frac{\mbox{e}^{x(t')-x(t)}}{\theta(t')} \,
 f_{eq} (\vec p_t,p_z\frac{t}{t'},t') \, ,
\end{equation}\noindent
avec
\eq
 x(t) = \int_{t_0}^t \frac{dt'}{\theta (t')} \, .
\eq
Pour une distribution initiale donnée, on en déduit, pour chacune des quantités 
pré-citées, une équation du type
\beq
\label{MBaym}
 M (t) =  M(t_0) \, \mathcal F_M^{(0)} (t_0/t) \, \mbox{e}^{-x} + 
 \int_{t_0}^t dt' \, \frac{\mbox{e}^{x'-x}}{\theta'} \,
 \mathcal F_M^{(eq)} (t'/t) \, M_{eq} (t') \, ,
\end{equation}\noindent
où $x \equiv x(t)$, $x' \equiv x(t')$, $\theta' \equiv \theta (t')$ et
$M \equiv \langle p_z^2\rangle$, $\langle p_\perp^2 \rangle$, $N_{-1}$, $P_L$,
$P_T$, ou $\epsilon$, selon le cas. Les fonctions $\mathcal F_M^{(0,eq)}$ 
ne dépendent que de la géometrie des distributions $f_0$ et $f_{eq}$ dans
l'espace des impulsions, elles sont données dans l'Annexe~\ref{intm} pour 
les différents $M$ dont nous aurons besoin dans ce chapitre. Les moments 
$M_{eq}$, calculés avec la distribution $f_{eq}$, ne dépendent que de 
$\lambda$ et $T$.
\par
On résoud numériquement le système d'équations (\ref{numb}), (\ref{engy}), 
(\ref{Mexple}), (\ref{MBaym}) par itérations successives, selon la
méthode proposée dans la Réf.~\cite{Dumitru}. Dans le cas présent,
toutes les intégrales sur les impulsions peuvent être calculées
analytiquement. Les intégrales temporelles (cf. Eq.~(\ref{MBaym}))
sont calculées par la méthode de Simpson (corrigée par un trapèze dans les 
cas où le nombre de points sous l'intégrale discrète est pair)~\cite{NumRec}.
A chaque pas de temps, la procédure d'itération (décrite dans~\cite{Dumitru}) 
est initialisée par une première estimation des quantités $M$, obtenue en 
calculant les intégrales temporelles par la méthode des rectangles, pour 
laquelle il n'est pas nécessaire de connaitre la valeur de l'intégrand 
pour la borne supérieure de l'intégrale. Cette procédure d'initialisation 
n'est pas strictement nécessaire, mais améliore la convergence de la 
procédure d'itération. On vérifie la bonne précision de la méthode en 
résolvant, en parallèle avec notre système d'équations, l'Eq.~(\ref{MBaym}) 
dans des cas solubles analytiquement, par exemple en calculant $n(t)$, non 
pas à l'aide de l'Eq.~(\ref{consnb}), mais à partir de 
l'Eq.~(\ref{MBaym}) avec $\mathcal F_n (a) = a$.
   
\subsection{Le choix du moment}

Comme nous l'avons remarqué plus haut, dans l'approximation du temps de 
relaxation la distribution $f$ relaxe vers la distribution $f_{eq}$
d'une manière indépendante de $\vec p$ : toutes les échelles d'impulsions
relaxent avec le même taux moyen $\theta$ (cf Eq.~(\ref{RTA})). D'un autre 
coté, nous avons vu que le calcul de ce temps de relaxation nécessite de 
choisir un moment particulier de la distribution, c'est à dire une 
fonction $m(\vec p)$ (cf. Eq.~(\ref{momeq})). Il est clair que les 
différents moments $\langle m \rangle$ sont sensibles à différentes parties
de la distribution $f$, c'est à dire à différentes échelles d'impulsion,
et il en est par conséquent de même pour les temps de relaxation $\theta_m$
correspondant. Ceci est illustré sur la Fig.~\ref{fig_scale}, où on a 
représenté l'évolution temporelle du rapport $P_L/P_T$ des pressions 
longitudinale et transverse pour différents choix de $m$, pour une 
condition initiale donnée.

\begin{figure}[htbp]
\epsfxsize=5.5in \centerline{ \epsfbox{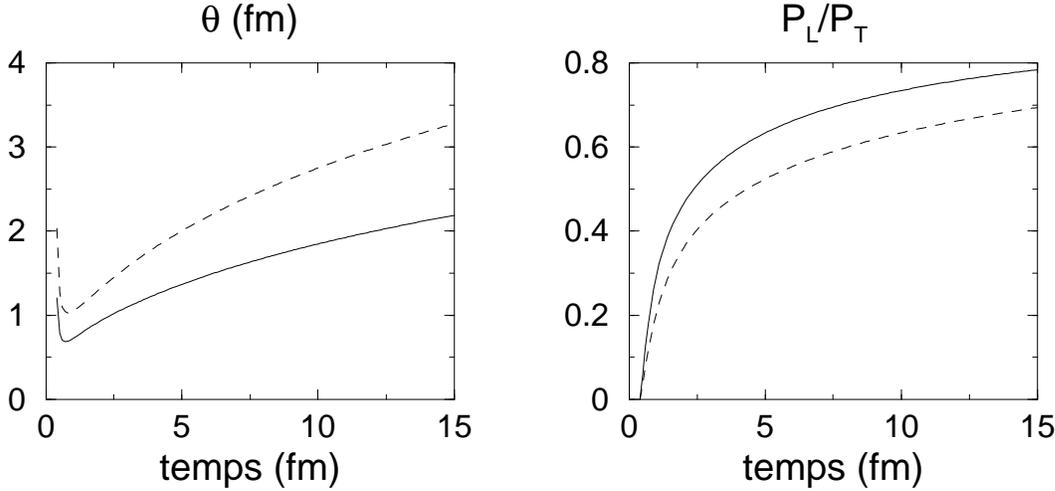}}
\caption{\small Le rapport des pressions longitudinale et transverse (à droite),
ainsi que le temps de relaxation en fm (à gauche) en fonction du temps 
(en fm), pour différents choix du moment $\langle m \rangle$ : 
$\langle m_1 \rangle = P_L - P_T$ (courbes en trait-plein) et 
$\langle m_2 \rangle = \langle p_z^2 - p_\perp^2 \rangle$ (courbes en trait-tiret).
Le premier, étant sensible à une échelle d'impulsion plus faible que le 
second, correspond à une relaxation plus rapide vers le régime hydrodynamique. 
Ceci est dû au fait que les impulsions les plus grandes sont les plus sensibles 
à l'expansion (voir la discussion page~\pageref{page_disc}). Ces courbes ont
été obtenues dans le scénario de saturation à RHIC (voir la section Résultats).} 
\label{fig_scale}
\end{figure}

On peut donc préciser la signification de l'approximation
du temps de relaxation : on remplace la dynamique microscopique par une
relaxation globale vers l'équilibre, avec un taux {\em caractéristique d'une
certaine échelle d'impulsion}. Pour un problème donné, il est donc judicieux
de choisir le moment $m$ selon l'échelle typique d'impulsion qui caractérise 
la physique à laquelle on s'intéresse. En ce qui concerne la question de la 
relaxation vers le régime hydrodynamique, l'échelle d'impulsion typique qui 
nous intéresse est celle qui caractérise les variables macroscopiques telles 
que la pression ou la densité d'énergie. Le choix $\langle m \rangle = \epsilon$
(c'est à dire $m(\vec p)=p$) correspondant déjà à l'équation de conservation de 
l'énergie, nous opterons pour\footnote{On obtient des résultats identiques 
   à ceux présentés dans la suite en choisissant 
   $\langle m \rangle = P_L$ ou $\langle m \rangle = P_T$. La différence
   des pressions n'est autre que la viscosité de cisaillement 
   $\zeta$~\cite{Landhydro} : $P_L - P_T = 2 \zeta/3t$.} 
$\langle m \rangle = P_L - P_T$, c'est à dire
\eq
 m(\vec p) = \frac{p_z^2 - p_\perp^2}{p} \, .
\eq
Après quelques intégrations par parties, on obtient, à partir de 
l'Eq.~(\ref{momeq}),
\beq
\label{momeq1}
  \frac{P_L - P_T}{\theta} = 
  4 \, \mathcal B \, N_0 \, \langle \frac{p_z^2 - p_\perp^2}{p^3} \rangle +
  2 \, \mathcal B \, N_{-1} \, \langle \frac{p_z^2 - p_\perp^2}{p^2} \rangle \, .
\end{equation}\noindent

\section{Conditions initiales et résultats}

Comme nous l'avons expliqué dans l'introduction, notre but est ici
d'étudier de façon systématique le processus de relaxation vers
l'équilibre local (le régime hydrodynamique) pour différentes conditions
initiales proposées dans la littérature pour décrire l'état du système
immédiatement après l'impact dans les collisions d'ions lourds à RHIC 
et LHC : le scénario de saturation et le scénario des minijets.
Bien que chacun de ces deux cas ait été étudié 
récemment~\cite{AHM1,AHM2,Raju,Dumitru}, ils n'ont jamais été comparés 
l'un à l'autre dans une seule et même approche. 
En comparant nos résultats avec ceux de la Réf.~\cite{Raju}, où l'Eq.~(\ref{BE}) 
est résolue numériquement dans le scénario de saturation, nous pouvons tester 
notre approche, qui s'avère être remarquablement bonne, compte tenu de son 
caractère très qualitatif. Nous verrons que le critère utilisé dans~\cite{Raju}
pour caractériser l'écart à l'équilibre n'est pas satisfaisant. Dans cette
étude, nous proposons de mesurer le degré d'isotropie de différentes 
observables, en conséquence de quoi nos conclusions sont qualitativement
différentes de celles de~\cite{Raju}. Nous étudions ensuite le scénario
des minijets, récemment étudié dans la Réf.~\cite{Dumitru}, mais où les 
auteurs n'ont pas implémenté correctement les lois de conservation. Enfin, 
nous examinons la robustesse des résultats par rapport aux détails du 
calcul, en particulier en ce qui concerne les approximations de couplage 
faible et logarithmique.

\subsection{``Temps de vie''} 

Dans ce qui suit, nous examinons la vitesse avec laquelle différentes 
configurations initiales relaxent vers le régime hydrodynamique. 
Du fait de l'expansion, le système est rapidement
dilué et, après un certain temps, la description en termes des 
degré de liberté partoniques n'a plus de sens. Le point important
est donc de comparer le temps caractéristique d'approche de 
l'équilibre local avec le temps de vie du système. Autrement
dit, nous voulons estimer le degré d'isotropie de la distribution
$f$ à l'instant $t_{max}$ où le système de partons cesse d'exister 
en tant que tel. Nous donnons ci-dessous des arguments très
qualitatifs\footnote{En particulier,
   les estimations ci-dessous reposent sur l'hypothèse que le plasma
   de gluons peut être vu en première approximation comme une collection 
   de particules de masse nulle, ce qui n'est pas le cas (voir par
   exemple~\cite{Entropy}).}
qui n'ont de sens qu'{\it a fortiori}, les ordres de grandeur obtenus
étant en accord avec les résultats des calculs sur réseaux (voir par
exemple~\cite{Lattice}). 
\par
Une description en termes des degrés de liberté
partoniques n'a de sens que si la distance typique entre les 
excitations du système est inférieure à la taille typique d'un hadron
$\sim 1$~fm, autrement dit, si la densité moyenne $n$ d'excitations par 
unité de volume est supérieure à la densité critique $n_c \sim 1$/fm$^3$. 
Dans le cas où le système de gluons est à l'équilibre 
thermodynamique\footnote{Pour un ensemble de gluons à l'équilibre 
   thermodynamique à la température $T_{eq}$, la distribution est donnée
   par la distribution de Bose-Einstein, pour laquelle on a :
   $n_{eq} = 2 (N_c^2-1) \, \zeta (3) \,T_{eq}^3 / \pi^2$, 
   $\epsilon_{eq} = 6 (N_c^2-1) \, \zeta (4) \,T_{eq}^4 / \pi^2$
   où $\zeta(3) \simeq 1.202$ et $\zeta (4) = \pi^4/90 \simeq 1.1$.
   Pour $N_c=3$, on a $(N_c^2-1)/\pi^2 \simeq 1$. Attention, les 
   relations précédentes sont obtenues avec une distribution de 
   Bose-Einstein et ne doivent pas être confondues avec les 
   Eqs.~(\ref{numb}) et~(\ref{engy}). Cependant, on note que la 
   confusion n'aurait rien changé à nos estimations.}, 
la densité de particules est reliée à la température $T_{eq}$ par la 
relation $n_{eq} \simeq 2 \, T_{eq}^3$. On en déduit la température 
critique $T_c \sim 1$~fm$^{-1} \sim 200$~MeV, et la densité d'énergie 
correspondante $\epsilon_c \simeq 6 \, T_c^4 \sim 1$~GeV/fm$^3$.
\par
Plus la densité de particules est élevée initialement, 
plus il faudra de temps pour la diluer et donc plus la durée de vie
sera longue. Dans le cas présent, où le nombre total de particules est
conservé, on peut déduire l'ordre de grandeur du temps de vie du système
à partir des seules conditions initiales : en utilisant l'Eq.~(\ref{consnb}),
il vient
\beq
\label{tfinal}
 \frac{t_{max}}{t_0} = \frac{n (t_0)}{n_c} \, ,
\end{equation}\noindent 
où on a défini $t_{max}$ par la relation $n(t_{max})=n_c$. 
\par
Dans chacun des deux scénarios considérés dans la suite, l'estimation de 
cette durée de vie à partir du critère $\epsilon \gtrsim \epsilon_c$ donne
grossièrement le même ordre de grandeur que celui obtenu à partir du critère
ci-dessus. En revanche, le critère qui consiste à comparer le paramètre $T$ 
(Eq.~(\ref{disteq})) avec la température critique $T_c$ n'a de sens que 
si le système est à l'équilibre. En effet, en général, le paramètre $T$ 
mesure l'énergie moyenne par particule ($T \propto \epsilon/n$) et n'a 
la signification physique d'une température, au sens thermodynamique 
(hydrodynamique) du terme, que si le système est à l'équilibre (local).
\par 
Il est clair que ces considérations très qualitatives ne servent qu'à donner
un ordre d'idée. Cependant les critères $n \gtrsim n_c$ et 
$\epsilon \gtrsim \epsilon_c$ délimitent le domaine de cohérence physique 
de notre description. 

\subsection{Le scénario de saturation}

Nous suivons ici les auteurs de la Réf.~\cite{Raju} pour caractériser l'état
initial des gluons produits dans le scénario de saturation. 
La distribution initiale idéalisée (\ref{insat}) ci-dessous a été proposée 
par A.~H.~Mueller~\cite{AHM1,AHM2} : les gluons de la fonction d'onde du
noyau incident dont l'impulsion transverse est $\sim Q_s$ sont 
simplement libérés dans la collision. En résolvant numériquement les
équations de champ classique pour l'interaction entre les champs
de Weizsäcker-Williams non-abélien des noyau incidents, A.~Krasnitz et
R.~Venugopalan~\cite{KraVent} calculent le temps de production des 
gluons réels. Ils obtiennent $t_i \approx 0.3$~fm à RHIC ($Q_s = 1$~GeV) et
$t_i \approx 0.13$~fm à LHC ($Q_s=2-3$~GeV).
\par
On voit que le temps de formation est typiquement $\sim 1/Q_s$. Cet
ordre de grandeur correspond aussi au temps nécessaire pour 
que les gluons de différentes impulsions longitudinales (plus précisément 
dans différentes unités de rapidité) se séparent physiquement les uns
des autres~\cite{MB,AHM1}. De même, au bout d'un temps $\sim 1/Q_s$, les
gluons de grande impulsion longitudinale ont quitté la région centrale, les
gluons restant ont une impulsion longitudinale $p_z \ll Q_s$. 
De plus, du fait du phénomène de saturation, la densité de gluons libérés 
par unité d'espace des phases est $\sim 1/\alpha_S N_c$. Le nombre d'occupation
dans la région centrale à l'instant $t_i$ peut s'écrire
\eq 
 f_{sat} (\vec p,t_i) = \frac{c}{\alpha_S N_c} \, \delta (p_z t_i) \,
 \Theta (Q_s^2 - p_t^2) \, ,
\eq
où $c$ est une constante estimée $\sim 1$ par Mueller~\cite{AHM1}, et calculée
numériquement pour un groupe de jauge $SU(2)$ dans la Réf.~\cite{KraVenc} :
$c=1.3$. La fonction delta modélise une distribution très étroite dans la
direction longitudinale : $p_z/Q_s \sim p_z t_i \ll 1$. On choisi l'instant
initial $t_0$ comme dans la Réf.~\cite{Raju} : on laisse le système évoluer
librement jusqu'à ce que le nombre d'occupation, initialement 
$\sim 1/\alpha_S N_c$, devienne $\sim 1$ :
\eq
 t_0 = \frac{c}{\alpha_S N_c} \, t_i \, .
\eq
L'instant $t_0$ est le temps à partir duquel l'approximation de particules 
classiques commence à être raisonnable, c'est l'instant initial pour notre étude.
On a alors~\cite{Raju} 
\beq
\label{insat}
 f_0 (\vec p) = f_{sat} (\vec p,t_0) = 
 \delta (p_z t_i) \, \Theta (Q_s^2 - p_t^2) \, ,
\end{equation}\noindent
Ici nous prendrons $\alpha_S=0.3$, $N_c=3$. Les valeurs des différents
paramètres correspondant aux énergies de RHIC et LHC sont résumées dans
le tableau~\ref{tab_sat}. 
\par 
On peut estimer le temps au bout duquel la densité de particules par
unité de volume devient $\sim n_c \sim 1$~fm$^{-3}$ directement à partir
des conditions initiales à l'aide de l'Eq.~(\ref{tfinal}). On
obtient $t_{max} \simeq 7$~fm à RHIC et $t_{max} \simeq 30$~fm à LHC.
Cette dernière valeur dépasse la limite au delà de laquelle l'hypothèse 
d'expansion longitudinale n'est plus raisonnable\footnote{Cette hypothèse
   est certainement valable pour $t \lesssim R$, où $R$ est le rayon
   des noyaux incidents (pour un noyau de Plomb ($A=208$), on estime 
   $R \sim 1.2 \, A^{1/3}$~fm~$\sim 6$~fm). Pour notre propos, à caractère
   essentiellement qualitatif, la limite supérieure $t \lesssim 15 - 20$~fm
   est optimiste mais pas déraisonnable.}.
En suivant l'évolution du système au cours du temps (voir ci-dessous),
on voit que le critère $\epsilon \gtrsim \epsilon_c$ donne des estimations
du même ordre de grandeur ($\simeq 4$~fm à RHIC, et $\simeq 15$~fm à LHC).
Dans la suite, nous prendrons ces valeurs comme limites supérieures 
(voir le tableau~\ref{tab_sat}).

\begin{table}[htbp]
\begin{center}
\begin{tabular}{|c||c|c|c|c|c|}
\hline
\small{\bf SATURATION} & $Q_s$ (GeV) & $t_0$ (fm) & $n(t_0)$ (fm$^{-3}$) & 
 $\epsilon (t_0)$  (GeV/fm$^{3}$) & $t_{max}$ (fm) \\
\hline
\hline
RHIC &  1. & 0.4  &  18.1 & 12.0 & $\sim 10$ \\
\hline
LHC  &  2. & 0.18 & 163.4 & 217.9 & $\sim 30$ \\
\hline
\end{tabular}
\end{center}
\caption{Valeurs des paramètres correspondant aux énergie de RHIC et LHC dans
le scénario de saturation (voir la Réf.~\cite{Raju}). Le temps $t_{max}$, 
est la limite au delà de laquelle notre description n'a certainement plus
aucun sens.}
\label{tab_sat}
\end{table}

Pour aller plus loin, nous devons préciser la forme du logarithme $L$
(Eq.~(\ref{logdiv1})). Dans les premiers instant après $t_0$, les particules 
de la région centrale ont des impulsions essentiellement transverses et 
interagissent entre elles en échangeant des gluons d'impulsion transverse. 
Dans ce cas, la masse de Debye (cf. Eq.~(\ref{logdiv1})) se réduit à la masse 
d'écran transverse~\cite{Raju} (voir l'Annexe~\ref{landaucoll}) :
\beq
\label{debye}
 m_T^2 (t) = \frac{\alpha_S N_c}{\pi^2} \, \int \frac{d^3p}{p} \, f(\vec p,t) = 
 \frac{4 \pi \alpha_S N_c}{N_c^2-1} \, N_{-1} (t) \, ,
\end{equation}
et la quantité $\underline p$ dans l'Eq.~(\ref{logdiv1}) est l'échelle
d'impulsion transverse typique. Dans la suite, nous prendrons
\beq
\label{logdiv2}
 L = \ln \left( \frac{\overline{\langle p_t^2 \rangle}}{m_T^2} \right) \, ,
\end{equation}\noindent
où $\overline{\langle ... \rangle} = \langle ... \rangle/n$ désigne la
moyenne par particule.
Bien  que ce choix soit motivé par la forme de la distribution initiale
(\ref{insat}), hautement anisotrope, on peut voir qu'il reste raisonnable 
dans le cas d'une distribution  isotrope, où il est plus approprié de 
choisir $L_{iso} = \ln (\overline{\langle p^2 \rangle}/m_D^2)$.
En effet, pour une distribution isotrope $m_D^2 = 2 \, m_T^2$ 
(voir Annexe~\ref{landaucoll}), $\langle p_t^2 \rangle = 
2 \, \langle p^2 \rangle/3$, et on a $L - L_{iso} = \ln 4/3 \simeq 0.3$.
\par
Il est important de remarquer que la validité de l'approximation logarithmique
est rendue très marginale par le fait que la valeur $\alpha_S=0.3$ est
à la limite du régime de couplage faible. En effet, de façon générale,
si $\underline p$ est l'échelle typique d'énergie du problème (ici 
$\underline p \sim Q_s$), on aura $m_T^2 \sim \alpha_S \underline p^2$, 
et $L \sim \ln (c/\alpha_S)$, où $c \sim 1$. Ici, $\alpha_S$ n'est pas 
réellement petite devant l'unité (en particulier $\alpha_S N_c \approx 1$)
et $L\ll 1$\footnote{Certains choix 
   pour $L$ peuvent même être tels que $L<0$ à l'instant initial. C'est
   le cas par exemple de l'Eq.~(27) de la Réf.~\cite{Raju}, pour les valeurs 
   des paramètres correspondant aux énergies de LHC.}. 
C'est la valeur de la constante $c$ qui contrôle la valeur de $L$. Autrement
dit, la valeur de $L$ dépend des détails du problème, ce qui traduit le
fait qu'à strictement parler, l'approximation logarithmique ne s'applique 
pas. Avec notre choix (\ref{logdiv2}), il est facile de voir qu'à l'instant 
initial, pour la distribution (\ref{insat})
\eq
 L (t_0) = \ln \left( \frac{\pi}{4 \alpha_S N_c} \, Q_s t_i \right) \, ,
\eq
où $t_i$ est le temps de formation des partons de l'état initial 
($Q_s t_i \sim 1$), le facteur $\pi/4\alpha_S N_c \approx 0.8 < 1$
est donné par la {\em géometrie} de la distribution initiale dans l'espace
des impulsions. On voit que la valeur précise de $Q_s t_i$ est cruciale pour 
que $L > 0$. 
\par
Pour la distribution (\ref{insat}), $L \ll 1$ à l'instant initial,
quelque soit le choix que l'on fait pour l'échelle $\underline p$
et la masse d'écran. Bien que la situation s'améliore 
par la suite, comme le montre la Fig.~\ref{fig_logsat}, on reste constamment à la limite du domaine de validité de l'approximation logarithmique.

\begin{figure}[htbp]
\epsfxsize=3.5in \centerline{ \epsfbox{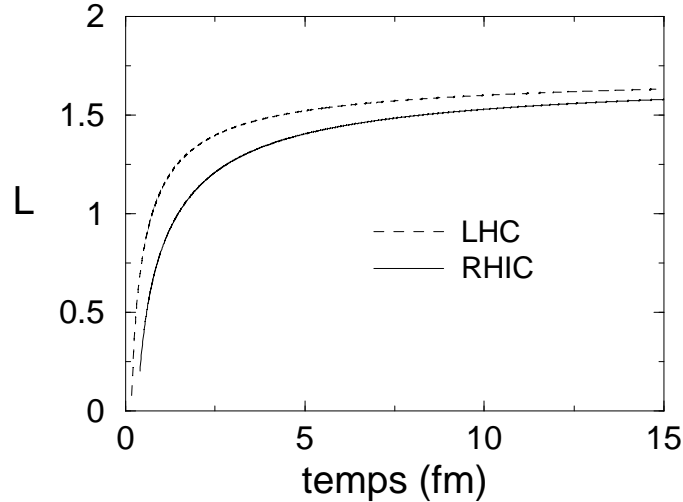}}
\caption{\small Le ``grand'' logarithme (\ref{logdiv2}) en fonction du 
temps (en fm), pour les valeurs des paramètres correspondant aux énergies 
de RHIC et LHC dans le scénario de saturation.} 
\label{fig_logsat}
\end{figure}

Pour tester notre approche, nous comparons nos résultats avec la solution
exacte de la Réf.~\cite{Raju}. Le premier graphe de la Fig.~\ref{fig_raju}
représente l'évolution des quantités $\epsilon$, $3P_L$ et $3P_T$ pour
les valeurs des paramètres correspondant aux énergies de RHIC 
(voir tableau~\ref{tab_sat}), les deux suivant représentent les  
moyennes par particule des carrés des impulsions longitudinale et
transverse, $\overline{\langle p_z^2 \rangle} (t)$ et 
$\overline{\langle p_\perp^2 \rangle} (t)$, pour RHIC et LHC respectivement.
Ces courbes sont les analogues de celles des Figs. 7 et 9 de la Réf.\cite{Raju},
on a choisi les échelles et les unités de manière à faciliter la comparaison :
non seulement l'approche à l'équilibre est correctement reproduite, mais on
voit que dans l'ensemble, nos résultats sont en accord semi-quantitatif avec 
la solution exacte, ce qui est remarquable compte tenu du degré de simplicité 
de notre approche. 
   
\begin{figure}[htbp]
\epsfxsize=4.in \centerline{ \epsfbox{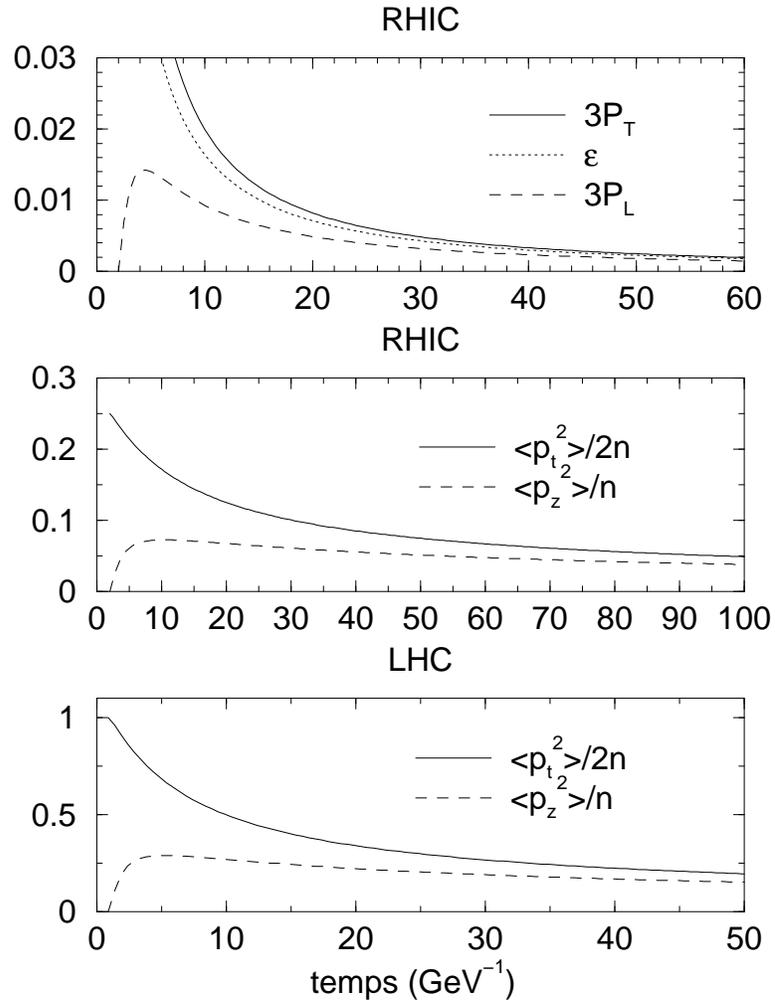}}
\caption{\small Reproduction de différentes courbes de la Réf.~\cite{Raju}.
Les différentes quantités présentées ici sont obtenues dans l'approximation 
du temps de relaxation. Les échelles et les unités ont été choisies de manière
à faciliter la comparaison. En se reportant à~\cite{Raju} on trouve un accord
semi-quantitatif.}
\label{fig_raju}
\end{figure}

Venons-en maintenant aux résultats de notre étude du scénario de saturation. 
Pour caractériser l'écart à l'équilibre cinétique local, c'est à dire le degré
d'anisotropie de la distribution, nous mesurons les rapports
\beq
\label{Rk}
 R_k = \frac{\langle p_z^2 / p^k \rangle}{\langle p_\perp^2 / p^k \rangle} \, ,
\end{equation}\noindent
en fonction du temps. Pour une distribution complètement
isotrope, $R_k^{iso}=1 \, , \, \forall k$. Les moments du type
$\langle p_{z,\perp}^2 / p^k \rangle$ sont sensibles à des échelles
d'impulsion de plus en plus faible à mesure que $k$ augmente, les 
rapports $R_k$ (plus précisément l'écart $1-R_k$) mesurent donc le degré 
d'anisotropie des différentes parties de la distribution, c'est à dire 
des différentes échelles d'impulsions. La Fig.~\ref{fig_relaxsat} 
représente l'évolution temporelle des rapports $R_k (t)$ pour $k=0,...,3$, 
à RHIC et LHC. On observe le comportement qualitatif attendu : les moments 
sensibles aux échelles d'impulsions les plus grandes sont ceux qui relaxent 
le moins vite vers l'équilibre. Cette hiérarchie de vitesses de relaxation, 
déjà observée dans~\cite{Raju}, est un simple effet de la géométrie du système\footnote{Les 
   particules de grande impulsion longitudinale désertent rapidement la 
   région centrale du fait de la seule expansion. Il s'ensuit que les 
   particules ``dures'' (de grande énergie) de cette région ont une 
   impulsion essentiellement transverse. Il faut donc plusieurs collisions 
   à petite déviation pour créer une particule de grande impulsion longitudinale,
   cela prend beaucoup de temps, plus qu'il n'en faut à cette dernière pour 
   quitter la région centrale.}, 
laquelle est prise en compte exactement dans l'Eq.~(\ref{RTA}).

\begin{figure}[htbp]
\epsfxsize=5.5in \centerline{ \epsfbox{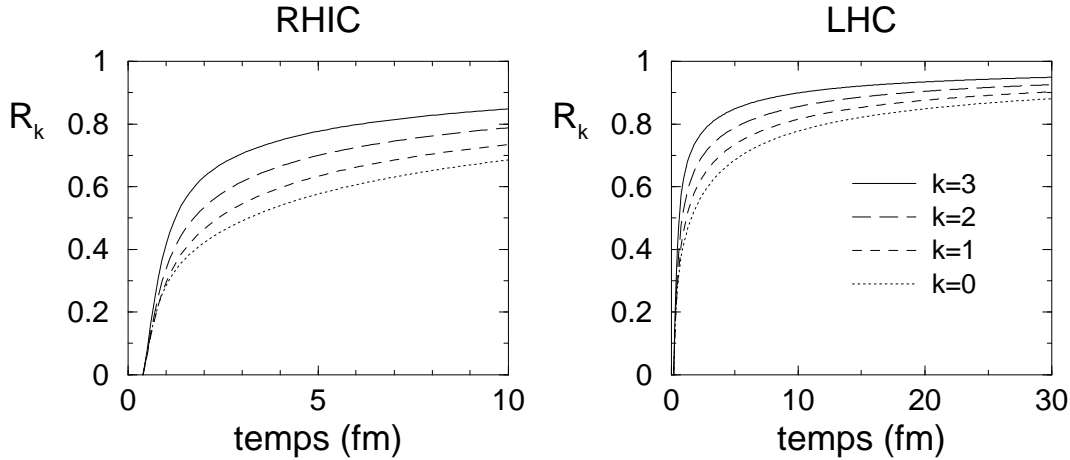}}
\caption{\small Evolution temporelle des rapports $R_k$ définis par 
l'Eq.~(\ref{Rk}) pour le scénario de saturation à RHIC (à gauche) et 
LHC (à droite). Les écarts $1-R_k$ mesurent le degré d'anisotropie de la
distribution microscopique $f(\vec p,t)$, pour différentes échelles 
d'impulsions.}
\label{fig_relaxsat}
\end{figure}

Le processus de relaxation est bien plus rapide à LHC et, la durée de vie 
y étant plus grande, le degré d'isotropie atteint est bien meilleur. 
On voit clairement que pour $t \sim t_{max}$, le système est encore 
très anisotrope à RHIC, tandis qu'à LHC le degré d'isotropie est très bon. 
Dans ce dernier cas, le degré d'isotropie est déjà satisfaisant 
pour $t \gtrsim 15$~fm où $R_{k=0,...,3} \gtrsim 0.8$. 
\par
Ces résultats semblent en complète contradiction avec ceux de la 
Réf.~\cite{Raju}, où  les auteurs mesurent les ``temps d'équilibration''
$t_{eq} = 3.24$~fm et $t_{eq} = 2.36$~fm pour les valeurs des paramètres
correspondant à RHIC ($\alpha_S = 0.3$, $Q_s = 1$~GeV) et LHC 
($\alpha_S = 0.3$, $Q_s = 2$~GeV) respectivement. Cette contradiction
apparente est due à ce que les critères utilisés pour caractériser
l'écart à l'équilibre ne sont pas équivalents : dans~\cite{Raju}, 
$t_{eq}$ est défini comme le temps auquel les quantités $T \, t^{1/3}$, 
et $s/n$, l'entropie moyenne par particule, atteignent $90 \%$ de leurs 
valeurs respectives à l'équilibre (ces deux quantités sont constantes dans 
le régime hydrodynamique). Par ailleurs, on peut voir sur les Figs.~7 et~9 
de~\cite{Raju} (voir aussi notre Fig.~\ref{fig_raju}), que les valeurs des 
rapports $R_1 = P_L/P_T$ et $R_0 = 
\langle p_z^2 \rangle/\langle p_\perp^2 \rangle = 
\overline{\langle p_z^2 \rangle} \, / \, \overline{\langle p_\perp^2 \rangle}$ 
au cours du temps sont en accord avec celles de notre Fig.~\ref{fig_relaxsat}.
En particulier, pour $t=t_{eq} = 3.24$~fm, on a $P_L/P_T \simeq 1/2$. 
Le critère que nous avons utilisé pour définir l'équilibre cinétique,
c'est à dire l'isotropie\footnote{Il est facile de montrer que l'Eq.~(\ref{BE})
   ne peut avoir de solution isotrope que dans le cas où le membre de gauche 
   est négligeable, ce qui correspond, par définition, au régime
   hydrodynamique (cf. Eq.~(\ref{hydroeq}))}, est d'une part plus intuitif et 
d'autre part plus précis. En effet les valeurs asymptotiques des rapports $R_k$
sont bien connues : $R_k^{hydro} = 1$, tandis que celles des quantités 
utilisées dans~\cite{Raju} dépendent de l'histoire du système. 
De plus le degré d'isotropie du système est l'observable importante du point 
de vue phénoménologique.

\subsection{Le scénario des minijets}
\label{sec_jet}

Nous suivons ici le choix des auteurs de la Réf.~\cite{Dumitru} pour 
caractériser la distribution initiale dans le scénario des minijets. 
Dans la région centrale, celle-ci est prise de la forme~(\ref{disteq}) : 
\beq
\label{injet}
 f_0 (\vec p) = f_{jet} (\vec p,t_0=1/p_0) = 
 \lambda_{jet} \, \mbox{e}^{-p/T_{jet}} \, ,
\end{equation}\noindent
où les paramètres $\lambda_{jet} = \lambda (t_0)$ et $T_{jet} = T(t_0)$ 
sont déterminés à partir les densités de particules $n_{jet}=n(t_0)=n_{eq} (t_0)$ 
et d'énergie $\epsilon_{jet} = \epsilon (t_0) = \epsilon_{eq} (t_0)$ du système 
de partons initialement produit, à l'aide des Eqs.~(\ref{numb}) et (\ref{engy}).
La contribution dominante vient des partons d'énergie $p_0$, dont
le temps de formation est $t_0 = 1/p_0$. Les valeurs des paramètres de 
ce scénario pour des collisions frontales Au-Au à RHIC et Pb-Pb à LHC, 
sont données dans le tableau~\ref{tab_jet}. Dans les deux cas, les limites
supérieures $t_{max}$ pour les temps auxquels $n \sim n_c$ et 
$\epsilon \sim \epsilon_c$ sont du même ordre de grandeur que celles 
obtenues dans le scénario de saturation.

\begin{table}[htbp]
\begin{center}
\begin{tabular}{|c||c|c|c|c|c|c|}
\hline
\small{\bf MINIJET} & $t_0$ (fm) & $n_{jet}$ (fm$^{-3}$) & $\epsilon_{jet}$
(GeV/fm$^{3}$) &  $\lambda_{jet}$ & $T_{jet}$ (GeV) & $t_{max}$ (fm) \\
\hline
\hline
RHIC & 0.18 & 34.3 & 56.0 & 1.0 & 0.535 & $\sim 10$ \\
\hline
LHC  & 0.09 & 321.6 & 1110.0 & 1.0 & 1.13 & $\sim 30$ \\
\hline
\end{tabular}
\end{center}
\caption{Valeurs des paramètres correspondant aux énergie de RHIC et LHC dans
le scénario des minijets (voir la Réf.~\cite{Dumitru}).}
\label{tab_jet}
\end{table}

De même que dans l'étude du scénario de saturation, nous prendrons 
$\alpha_S=0.3$ et ferons le choix~(\ref{logdiv2}) pour le logarithme 
$L$. En ce qui concerne la validité de l'approximation 
logarithmique, bien que toujours à la limite, la situation est plus 
confortable ici que dans le cas du scénario de saturation : dans 
l'intervalle de temps étudié on a $1.5 \lesssim L \lesssim 5$.
En particulier, il est facile de voir qu'à l'instant initial
\eq
 L(t_0) = 
 \ln \left( \frac{7 \pi}{4 \alpha_S N_c} \frac{1}{\lambda_{jet}} \right) \, .
\eq
Encore une fois le profil de la distribution joue un rôle important (c'est
lui qui détermine le facteur $7 \pi/4 \alpha_S N_c$).

\begin{figure}[htbp]
\epsfxsize=5.5in \centerline{ \epsfbox{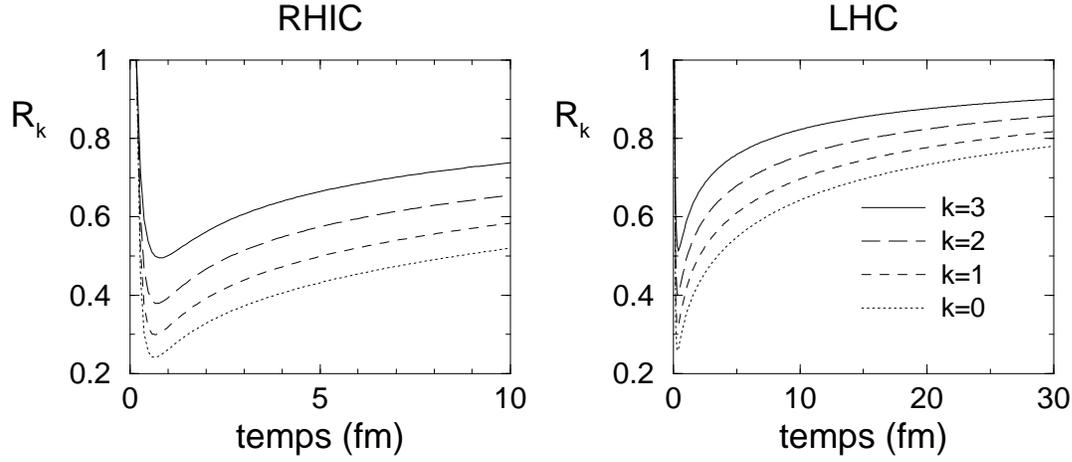}}
\caption{\small Evolution temporelle des rapports $R_k$ (Eq.~(\ref{Rk}) 
pour le scénario des minijets à RHIC et LHC.}
\label{fig_relaxjet}
\end{figure}

On voit sur la Fig.~\ref{fig_relaxjet} que l'isotropie initiale est
rapidement détruite dans les premiers instants de l'évolution. C'est
la phase initiale de régime libre dont nous avons parlé plus haut : 
à l'instant initial, le temps de relaxation est non-nul\footnote{Mentionnons 
   à ce propos un détail technique concernant le calcul du temps de
   relaxation à l'instant initial dans le scénario des minijets : pour la 
   distribution (\ref{injet}), l'Eq.~(\ref{momeq}) est triviale ($0=0$) 
   et ne permet donc pas de calculer le temps de relaxation $\theta(t_0)$. 
   On contourne la difficulté en exploitant le fait que le système est 
   initalement dans un régime libre (voir l'Annexe~\ref{thetajet}).}
et il faut attendre un temps $\Delta t = t-t_0 \lesssim \theta (t_0) \sim 1$~fm) 
avant que l'effet des collisions ne se fasse sentir, l'isotropie est alors 
lentement reconstruite. 
\par
Après que l'isotropie initiale ait été détruite par l'expansion, la distribution
est très piquée autour de $p_z=0$, et l'évolution ultérieure est très semblable 
à celle du scénario de saturation. On voit cependant que le processus de 
relaxation est plus lent dans le cas présent et les degrés d'isotropie 
atteints sont par conséquent nettement inférieurs, aussi bien à RHIC qu'à LHC 
(les temps de vie sont du même ordre de grandeur dans les deux scénarios 
(voir Tabs.~(\ref{tab_sat}) et~(\ref{tab_jet})). Dans ce dernier cas, bien que 
le degré d'isotropie finisse par être relativement bon, il faut attendre 
$t \sim 30$~fm pour avoir $R_{k=0,...,3} \gtrsim 0.8$, soit deux fois plus 
longtemps que dans le scénario de saturation. Sur la base de ce critère, les 
temps d'équilibration obtenus dans le scénario des minijets sont $\gg t_{max}$ 
à RHIC, $\sim t_{max}$ à LHC.
\par
Ces conclusions sont en désaccord qualitatif avec celles des auteurs de 
la Réf.~\cite{Dumitru}, qui mesurent des temps d'équilibration similaires
$\sim 4-5$~fm à RHIC et LHC. Cependant l'approche utilisée dans cet article
n'est pas satisfaisante, en particulier à cause du fait que bien que seuls
les processus élastiques y soient considérés, le nombre total de particules 
n'est pas conservé. En fait il augmente avec le temps (la densité moyenne 
par unité de volume $n$ décroit moins vite que $1/t$), ce qui diminue les 
temps d'équilibration (les collisions étant plus fréquentes). Le fait
que ceux-ci soient du même ordre de grandeur à RHIC et à LHC vient de ce 
que dans~\cite{Dumitru}, la constante de couplage n'est pas constante, mais
augmente avec le temps, sa valeur étant calculée à chaque instant en fonction 
de l'échelle typique d'impulsion des particules du milieu. Celle-ci étant
plus petite à RHIC qu'à LHC, l'intensité de l'interaction, et donc la 
fréquence des collisions y est augmentée. Dans la partie suivante, où 
nous considérons le cas d'une constante de couplage variable, nous verrons
cependant que cet effet ne modifie pas nos conclusions présentes.

\subsection{Robustesse des résultats}

Pour obtenir une information la plus fiable 
possible dans l'approche utilisée ici, il est important de tester la 
sensibilité des résultats précédent par rapport aux détails de la 
description, comme par exemple la valeur de $\alpha_S$, ou encore le 
choix de l'échelle d'impulsion $\underline{p}$ sous le logarithme $L$ 
(cf. Eqs.~(\ref{logdiv1})). En effet, étant à la limite du domaine de 
validité de l'approximation logarithmique, on s'attend à ce que de petits 
changement sous le logarithme aient des effets non-négligeables. 

\subsubsection*{Constante de couplage variable}

Il est intéressant de considérer le cas où la valeur de la constante de 
couplage $\alpha_S$ augmente avec le temps, à mesure que décroît l'énergie 
typique des particules du milieu. Comme nous l'avons mentionné plus haut, 
on s'attend alors à ce que les particules interagissent plus fortement à 
RHIC qu'à LHC, l'énergie moyenne par particule y étant plus faible.
\par
On calcule la constante de couplage à chaque instant en fonction 
de l'énergie moyenne par particule : $\alpha_S (t) \equiv 
\alpha_S (\mu = \bar\epsilon = 3T)$, avec la parametrisation
\beq
\label{running}
 \alpha_S (\mu) = \alpha_S (M_Z) \, 
 \frac{\ln(M_Z/\Lambda_{QCD})}{\ln(\mu/\Lambda_{QCD})} \, ,
\end{equation}\noindent
où $M_Z \simeq 90$~GeV est la masse du boson $Z$, $\alpha_S (M_Z) = 0.1$, 
et où on a pris $\Lambda_{QCD} \simeq 200$~MeV. 
\par
Dans le scénario des minijets, à LHC, on a $0.2 < \alpha_S (t) < 0.4$ sur
l'intervalle de temps considéré et on obtient un résultat très similaire à 
celui de la partie~\ref{sec_jet}, obtenu avec $\alpha_S = 0.3$. Bien que, par 
rapport au cas où $\alpha_S$ est constante, la situation s'améliore à 
RHIC, où $0.3 < \alpha_S (t) < 0.5$, l'équilibre n'est pas atteint.
Les courbes représentant l'évolution de $R_1$ sont présentées dans la 
partie suivante. 
\par
Dans le scénario de saturation, l'énergie moyenne par particule est 
plus faible et la constante de couplage devient rapidement $\gtrsim 0.5$. 
Les résultats obtenus n'ont pas de sens physique. Ceci doit être pris comme
une manifestation de la fragilité du calcul perturbatif. 

\subsubsection*{Estimation des incertitudes}

Pour nous faire une idée plus générale de l'ordre de grandeur des 
incertitudes dues aux détails de notre approche, nous remplaçons simplement 
$L \rightarrow 2 \, L$, puis $L \rightarrow L/2$,
dans le terme de collision\footnote{Le changement $L \rightarrow a \, L$ 
   revient à remplacer le temps de relaxation $\theta$ par $\theta/a$ 
   dans l'Eq.~(\ref{RTABaym}). Quand $a=2$ ($a=1/2$), le système relaxe 
   plus rapidement (lentement).}
 (on travaille ici avec $\alpha_S = \mbox{cte} = 0.3$). 
Les courbes des parties précédentes deviennent des ``bandes'' dont la 
largeur représente le degré de confiance avec lequel interprèter nos 
résultats. Les bandes correspondant à l'observable 
$R_1 = P_L/P_T$ (cf. Eq.~(\ref{Rk})) pour les scénarios de saturation et 
des minijets sont représentées sur les Fig.~\ref{fig_robsat} et~\ref{fig_robjet}
respectivement. Sur la Fig.~\ref{fig_robjet}, on a aussi représenté les courbes
correspondant au cas où la constante de couplage dépend du temps (voir la 
partie précédente), on voit que celles-ci sont à l'intérieur du `domaine
d'incertitude'' décrit ci-dessus. Les observations qualitatives des parties
précédentes restent valables. Cependant, à cause des incertitudes liées à 
notre description, on ne peut pas dire si l'équilibre est atteint à LHC
dans le scénario des minijets. Dans le scénario de saturation, où l'équilibre
est certainement atteint, il est est impossible d'estimer précisémment
la durée du régime transitoire.

\begin{figure}[htbp]
\epsfxsize=5.5in \centerline{ \epsfbox{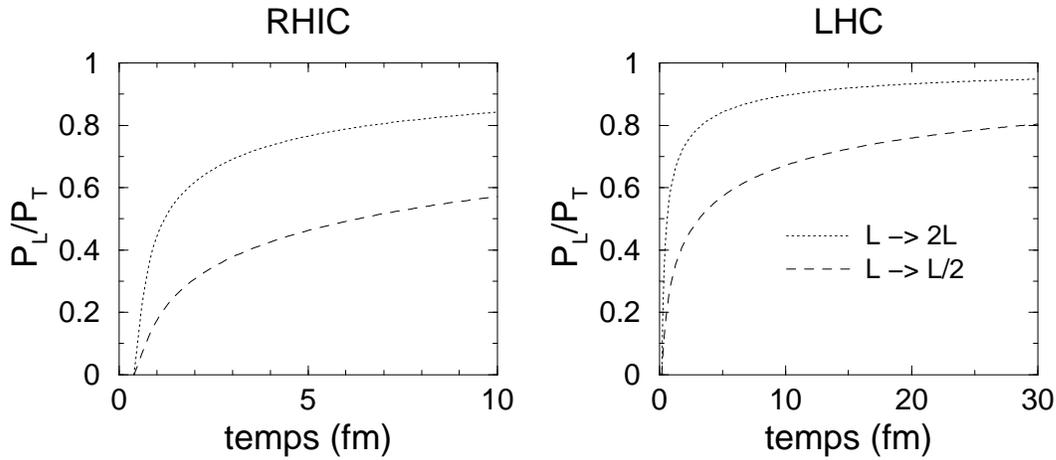}}
\caption{\small Dépendance des résultats par rapport aux détails de la
description pour le scénario de saturation. Les courbes délimitent 
la marge de variation de l'évolution temporelle du rapport $R_1 = P_L/P_T$ 
quand on change $L\rightarrow 2L$ (pointillés) et $L\rightarrow L/2$ 
(tirets).} 
\label{fig_robsat}
\end{figure}
\begin{figure}[htbp]
\epsfxsize=5.5in \centerline{ \epsfbox{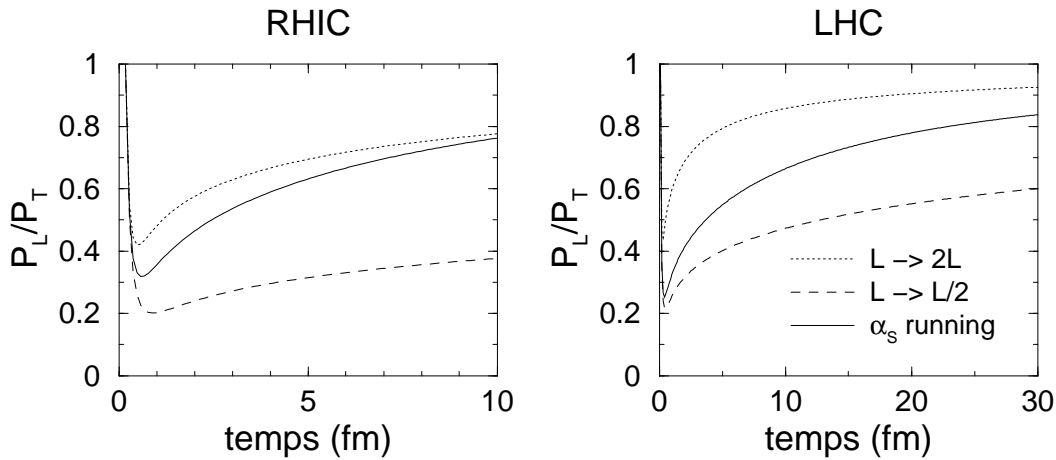}}
\caption{\small Dépendance des résultats par rapport aux détails de la
description pour le scénario des minijets. Les courbes représentent la marge 
de variation de l'évolution temporelle du rapport $R_1 = P_L/P_T$ quand 
on change $L\rightarrow 2L$ (pointillés) et $L\rightarrow L/2$ 
(tirets). Les courbes en trait continu sont obtenues avec une 
constante de couplage variable, dépendant de l'énergie moyenne 
$\bar\epsilon = 3T$ des particules du milieu (voir Eq.~(\ref{running})).} 
\label{fig_robjet}
\end{figure}

\subsection{Conclusions et perspectives}

En résumé, nous avons étudié l'évolution du système de gluons, produit 
dans les tout premiers instant d'une collision nucléaire à très haute énergie,
et en particulier la contribution des collisions élastiques de petite déviation
au processus d'équilibration cinétique. Nous avons étudié et comparé deux
scénarios différents pour l'état initial du système, en résolvant les équations
cinétiques dans une approximation de temps de relaxation  ``auto-cohérente'',
laquelle donne des résultats en bon accord avec la solution exacte obtenue
dans~\cite{Raju} pour le scénario de saturation. Nos conclusions sont résumées 
dans le tableau~\ref{tab_hydro}. Dans les deux scénarios étudiés, les processus
élastiques ne sont pas suffisamment efficaces pour rendre la distribution 
isotrope avant la phase hadronique aux énergies de RHIC ($t_{eq} \gg t_{max}$).
Dans le scénario de saturation, pour LHC, le régime hydrodynamique est atteint. 
La détermination exacte du temps d'équilibration n'est cependant pas possible 
étant données les incertitudes de la description utilisée. Dans le scénario
minijet, il apparait même difficile de dire si l'équilibre local est atteint.

\begin{table}[htbp]
\begin{center}
\begin{tabular}{|lcr|lcr|lcr||lcr|lcr|}
\cline{4-15}
\multicolumn{3}{l}{} & \multicolumn{6}{|c||}{\small{\bf SATURATION}} &
\multicolumn{6}{|c|}{\small{\bf MINIJET}} \\

\multicolumn{3}{l|}{}  & \multicolumn{3}{c}{RHIC} & \multicolumn{3}{c||}{LHC} &
\multicolumn{3}{c}{RHIC} & \multicolumn{3}{c|}{LHC}\\
\hline
& $t_{max}$ $(fm)$ & & & $\sim 10$ & & & $\sim 30$ & & & $\sim 10$ & & &
$\sim 30$ & \\
\hline
& $R_1$ & & & $\lesssim 0.8$ & & & $\gtrsim 0.8$ & & & $\lesssim 0.8$ & & & 
$\gtrsim 0.6$ & \\
\hline
\multicolumn{3}{|c|}{Régime hydrodynamique ?} & & non & & & oui & &
& non & & & ? & \\
\hline
\end{tabular}
\end{center}
\caption{Le système a-t-il atteint le régime hydrodynamique ?}
\label{tab_hydro}
\end{table}

Les auteurs de la Réf.~\cite{Branching} ont récemment argué que les processus 
inélastiques de branchement ($g  + \mbox{milieu} \leftrightarrow gg +
\mbox{milieu}$), contribuant au même même ordre que les processus élastiques 
dans l'approximation logarithmique, jouent un rôle important dans le processus
d'équilibration cinétique, et accélèrent considérablement celui-ci.
Il est fort probable que ces processus inélastiques permettent de 
rendre le temps d'équilibration négligeable à LHC (quel que soit
le scénario le plus adapté). La situation est moins claire à RHIC.
Il est très important, pour l'interprètation des données 
expérimentales actuellement accumulées à RHIC, de savoir si le
régime transitoire peut être négligé, ou si l'aspect hors équilibre
du système doit être pris en compte dans les calculs concernant
les signatures directes du plasma.

{\appendix

\setcounter{chapter}{5}
\chapter{L'intégrale de collision de Landau}
\label{landaucoll}

Dans le Chap.~\ref{QGP} nous nous intéressons à la matière
produite immédiatement après l'impact, dans la région centrale
d'une collision d'ions lourds ultrarelativistes. Nous supposons 
que la densité d'énergie est suffisament élevée pour que cette
matière puisse être décrite par un ensemble de partons 
(essentiellement des gluons) interagissant faiblement les uns 
avec les autres. Autrement dit, nous supposons que la constante 
de couplage forte est très petite : $\alpha_S \ll 1$. 
Nous supposons de plus qu'après quelques instants, le système
est suffisament dilué pour pouvoir être décrit par une équation 
de Boltzmann pour la densité de gluons par unité d'espace des phases,
que nous supposons {\em indépendante} de l'état de polarisation et de 
couleur des gluons\footnote{A strictement parler, la construction du
   terme de collision qui suit n'est valable que dans le cas d'une
   distribution indépendante de la couleur et de la polarisation. 
   Cependant, on peut voir l'équation de Boltzmann obtenue comme
   décrivant l'évolution du système sur des échelles de temps 
   grandes devant l'échelle caractérisant la relaxation de couleur.
   En effet, cette dernière est beaucoup plus rapide que l'échelle
   de relaxation vers l'équilibre cinétique à laquelle on s'intéresse 
   ici. La distribution $f$ doit être vue comme une moyenne sur les 
   états de polarisation et de couleur.}
\eq
 f (\vec p,\vec x,t) = \frac{(2\pi)^3}{2(N_c^2-1)}
 \frac{d^6 N}{d^3p \, d^3x} \, ,
\eq
Construisant le terme de collision de l'équation de Boltzmann 
correspondante, nous nous limiterons 
à la contribution des collisions élastiques $gg \rightarrow gg$, 
à l'ordre dominant en $\alpha_S$. Nous devons considérer tous les 
processus du type ($\epsilon=+,-$ désigne la polarisation et 
$a=1,...,N_c^2-1$ la couleur) : 
\eq
 (\vec p,\epsilon,a) + (\vec p \, ', \epsilon',a') \rightarrow
 (\vec p_1,\epsilon_1,a_1) + (\vec p_1 \, ', \epsilon_1',a_1') \, ,
\eq
ainsi que le processus inverse. Dans les deux cas il nous faut sommer 
sur tous les états d'impulsion, de polarisation et de couleur des gluons 
autres que celui auquel on s'intéresse (le premier ci-dessus), et moyenner 
sur les états de polarisation et de couleur de ce dernier, schématiquement :
\eq
 \frac{1}{2(N_c^2-1)} \sum_{\epsilon,a} 
 \left[ \sum_{\epsilon',a'} \sum_{\epsilon_1,a_1} \sum_{\epsilon_1',a_1'}
 \right] \, .
\eq
Il est commode de rétablir une certaine symétrie entre les deux gluons 
$(\vec p,\epsilon,a)$ et $(\vec p \, ', \epsilon',a')$ en exprimant 
la probabilité de transition apparaissant sous l'intégrale de collision 
en termes du carré de l'amplitude  de diffusion 
$|\mathcal M_{gg \rightarrow gg}|^2$, 
sommé sur les états de polarisation et de couleur de l'état final et 
moyenné sur ceux des {\em deux gluons} de l'état initial. Ceci 
revient à multiplier et diviser par $2(N_c^2-1)$ :  
\eq
 2(N_c^2-1) \left\{\frac{1}{2(N_c^2-1)} \sum_{\epsilon,a} \, 
 \frac{1}{2(N_c^2-1)} \sum_{\epsilon',a'}
 \left[ \sum_{\epsilon_1,a_1} \sum_{\epsilon_1',a_1'} \right] \right\} \, .
\eq
L'intégrale de collision s'écrit alors 
\bearn
 \mathcal C [f] (\vec p) = \frac{2(N_c^2-1)}{2} \, 
 \int && \!\!\!\!\!\!\!\!\!\!\!
 \frac{d^3 p'}{(2\pi)^3} \, \frac{d^3 p_1}{(2\pi)^3} \,
 \frac{ d^3 p_1'}{(2\pi)^3} \,
 w(\vec p \, \vec p \, ' \rightarrow \vec p_1 \, \vec p_1 \, ')  \nonumber \\
 && \left[ f_1 \, f_1' \, (1+f) \, (1+f') - 
 f \, f' \, (1+f_1) \, (1+f_1') \right] \, ,
\label{intcoll}
\eearn
où $f\equiv f(\vec p)$, $f'\equiv f(\vec p \, ')$, 
$f_1\equiv f(\vec p_1)$ et $f_1'\equiv f(\vec p_1 \, ')$ (toutes les 
fonctions aparaissant dans le terme de collision sont prises au même 
point d'espace-temps, nous avons donc omis d'indiquer la dépendance
en $\vec x$ et $t$) et où
\eq
 w(\vec p \, \vec p \, ' \rightarrow \vec p_1 \, \vec p_1 \, ') = 
 w(\vec p_1 \, \vec p_1\, ' \rightarrow \vec p \, \vec p \, ') = 
 (2\pi)^4 \delta^{(4)} (P + P' - P_1 - P_1') \,
 \frac{|\mathcal M_{gg \rightarrow gg}|^2}{16 \, p \, p' \, p_1 \, p_1'} \, .
\eq
Le fait que les gluons sont des bosons indiscernables est pris en 
compte dans le calcul de l'amplitude de transition 
$|\mathcal M_{gg \rightarrow gg}|^2$ (terme d'échange), 
et résulte dans le fait que la probabilité de transition $w$ est symétrique
dans l'échange $\vec p_1 \leftrightarrow \vec p_1 \, '$ (voir la symétrie sous 
l'échange $u \leftrightarrow t$ dans l'expression~(\ref{amplitrans})). 
Après avoir intégré sur toutes les valeurs possibles de $\vec p_1$ et
$\vec p_1 \, '$, on a donc compté deux fois l'ensemble des configurations
possibles, d'où le facteur $1/2$ devant l'intégrale.
\par
Dans le cas qui nous intéresse ici, il est possible de calculer la 
contribution dominante à cette intégrale en exploitant le fait
que l'amplitude de diffusion diverge fortement dans la limite 
des diffusions à petit angle de déviation~\cite{LanLif}.

\section*{La section efficace de diffusion $gg \rightarrow gg$}

La section efficace différentielle de diffusion s'écrit
\beq
\label{crsec1}
 d \sigma = (2\pi)^4 \, \delta^{(4)} (P + P' - P_1 - P_1') \,
 \frac{|\mathcal M_{gg \rightarrow gg}|^2}{4 \, P \cdot P'} \, 
 \frac{d^3 p_1}{(2\pi)^3 2 p_1} \, \frac{d^3 p_1'}{(2\pi)^3 2 p_1'} \, ,
\end{equation}\noindent
où $|\mathcal M_{gg \rightarrow gg}|^2$ est le carré de l'amplitude 
de diffusion $gg \rightarrow gg$ sommé sur les états de polarisation
et de couleur des gluons de l'état final, et moyenné sur ceux de 
l'état initial. En termes des variables de Mandelstam
\eq
 s = (P + P')^2 \, \, \, \, , \, \, \, \,
 t = (P - P_1)^2  \, \, \, \, , \, \, \, \,
 u = (P - P_1')^2 =  - s - t \, ,
\eq
on a (voir par exemple~\cite{AHM1})
\beq
\label{amplitrans}
 |\mathcal M_{gg \rightarrow gg}|^2 = 4 g^4 \frac{N_c^2}{N_c^2-1} 
 \left( 3 - \frac{ut}{s^2} - \frac{us}{t^2} - \frac{st}{u^2} \right) \, ,
\end{equation}\noindent
avec $g^2=4\pi \, \alpha_S$. 
Dans le référentiel du centre de masse de la collision $gg \rightarrow gg$:
\bear
 P^\mu & = & p \, (1,0,0,1) \, \, \, \, , \, \, \, \, 
 P_1^\mu =  p \, (1,0,\sin \theta , \cos \theta)  \, \, ,\\
 (P')^\mu & = & p \, (1,0,0,-1) \, \, \, \, , \, \, \, \,
 (P_1')^\mu = p \, (1,0,-\sin \theta , -\cos \theta) \, \, ,
\eear
on a 
\eq
 s = 4 \, p^2 \, \, \, \, , \, \, \, \,
 t = -2 \, p^2 \, (1 -  \cos \theta) \, \, \, \, , \, \, \, \,
 u = -2 \, p^2 \, (1 +  \cos \theta)  \, \, .
\eq
La section efficace (\ref{crsec1}) diverge fortement pour les petites valeurs 
de $t$ ($\theta \approx 0$) et, par symétrie, pour les petites valeurs de $u$
($\theta \approx \pi$). Ces deux régions de l'espace des phases dominent donc
l'intégrale de collision, et leurs contributions respectives sont identiques
par symétrie. Il suffit donc d'en calculer une et de multiplier le résultat
par un facteur $2$, qui vient se simplifier avec le facteur de symétrie $1/2$ 
devant l'intégrale (Eq.~(\ref{intcoll})). Tout se passe alors comme si les
gluons étaient des particules discernables. Dans la suite nous omettrons dès 
le départ les facteurs de symétrie et calculerons la contribution de la région
$\theta \approx 0$ (diffusion à petit angle de déviation), où l'on a
\beq
\label{crsec2}
 d \sigma \simeq 4\pi \alpha_S^2 \frac{N_c^2}{N_c^2-1} \, 
 \frac{d \theta^2}{p^2 \theta^4} \, \frac{d \varphi}{2 \pi} \, .
\end{equation}

\section*{L'intégrale de collision dans la limite des petits angles}

Du fait de la conservation de l'impulsion, l'intégrale sur $\vec p_1 \, '$ est 
triviale, et la probabilité de transition $w$ ne dépend que de trois vecteurs.
Il est utile de l'exprimer comme une fonction des demi-sommes des impulsions
initiales et finales des deux particules : $(\vec p + \vec p_1)/2$ et
$(\vec p \, ' + \vec p_1 \, ')/2$. Remplaçant l'intégrale sur $\vec p_1$ par
une intégrale sur l'impulsion échangée $\vec q = \vec p_1 - \vec p$, on écrira
\eq
 w(\vec p \, \vec p \, ' \rightarrow \vec p_1 \, \vec p_1 \, ') \equiv 
 w(\vec p + \frac{\vec q}{2}, \vec p \, ' - \frac{\vec q}{2} ; \vec q)
 \simeq w + \frac{q^i}{2} \left( \p_i w - \p_i' w \right) \, ,
\eq
où $w=w(\vec p , \vec p \, ' ; \vec q \approx \vec 0)$, et
$\p_i \equiv \p/\p p^i$, le prime désignant
la dérivation par rapport aux composantes de $\vec p \, '$.
On développe de même les poids statistiques : 
\bear
 f_1 & \simeq & f + q^i \p_i f + 
 \frac{q^i q^j}{2} \, \p_{ij} f \\
 f_1' & \simeq & f' - q^i \p_i' f' + 
 \frac{q^i q^j}{2} \, \p_{ij}' f' \, ,
\eear
où $\p_{ij} \equiv \p^2/\p p^i \p p^j$.
Dans le développement de la combinaison des poids statistiques 
apparaissant sous l'intégrale de collision, le premier terme non-nul 
est $q^i \, (F' \p_i f - F \p_i' f') \, w$, où $F=f(1+f)$. 
La probabilité de transisiton 
$w (\vec p \, \vec p \, ' \rightarrow \vec p_1 \, \vec p_1 \, ')$ étant symétrique 
dans l'échange des états initial et final (autrement dit, c'est une fonction
paire de $\vec q$), ce terme ne survit pas à 
l'intégration sur $d^3q$. La première contribution non-nulle à 
l'intégrale de collision vient du terme du second ordre dans le 
développement de l'intégrand
\bear
 \mathcal C & \simeq & \int_{\vec p \, '} \,
 \left[ (\p_i \mathcal B^{ij}) - (\p_i' \mathcal B^{ij}) \right] 
 \left[ F' \, (\p_j f) - F \, (\p_j' f') \right] \\
 & + & \int_{\vec p \, '} \,
 \left\{ F' \, (\p_{ij} f) + F \, (\p_{ij}' f') - \left[ 1 + f + f' \right]
 \left[ (\p_i f) \, (\p_j' f') + (\p_j f) \, (\p_i' f') \right] \right\} \, 
 \mathcal B^{ij} \, ,
\eear
où 
\beq
\label{Bij1}
 \mathcal B^{ij} = \frac{1}{2} \int \frac{d^3 q}{(2\pi)^3} \, q^i q^j \, 
 w (\vec p,\vec p \, ';\vec q)\, ,
\end{equation}\noindent
et où l'on a noté 
\eq
 \int_{\vec p \, '} \, \, \equiv \, \, 2(N_c^2-1) \int \frac{d^3p'}{(2\pi)^3} \, .
\eq
En utilisant la relation $B^{ij}=B^{ji}$, on réécrit
le dernier terme sous la seconde intégrale : 
\eq
 \left[ 1 + f + f' \right]
 \left[ (\p_i f) \, (\p_j' f') + (\p_j f) \, (\p_i' f') \right] \, \mathcal B^{ij}
 = \mathcal B^{ij} \, 
 \left[ (\p_i F) \, (\p_j' f') + (\p_j f) \, (\p_i' F') \right] \, ,
\eq
et, en réorganisant les différents termes, on obtient (les dérivées agissent
sur tout les termes entre accolades)
\eq
 \mathcal C \simeq - \int_{\vec p \, '} \, (\p_i - \p_i') \, 
 \left\{ \mathcal B^{ij} \, 
 \left[ F \, (\p_j' f')-  F' \, (\p_j f) \right] \right\} \, .
\eq
Le second terme est l'intégrale d'une divergence totale et est donc
nul. On obtient alors (voir~\cite{LanLif})
\beq
\label{div1}
 \mathcal C \simeq - \p_i \, s^i \, ,
\end{equation}\noindent
où
\beq
\label{div2}
 s^i = \int_{\vec p \, '} \, \mathcal B^{ij} \, 
 \left[ F \, (\p_j' f') - F' \, (\p_j f) \right] \, .
\end{equation}\noindent
\par
Il n'est pas surprenant que le terme de collision s'exprime 
comme une divergence dans la limite des petits
angle : dans l'espace des impulsions, la particule 
est repérée par le vecteur $\vec p$.
Après une collision typique, elle se retrouve à la ``position'' 
voisine repérée par le vecteur $\vec p + \vec{\delta p}$. 
On voit donc que les particules se déplacent le long de ``trajectoires'' 
continues dans l'espace des impulsions. Il s'agit d'un processus 
diffusif dans cet espace et, par conséquent, le terme de collision
correspondant doit s'exprimer comme la divergence du flux de particule
dans l'espace des impulsions. On peut construire explicitement ce 
flux~\cite{LanLif}, le résultat obtenu co\i ncide\footnote{Ceci est fait
   explicitement dans la Réf.~\cite{LanLif} pour le cas d'un gaz de 
   particules classiques. Il est facile de répéter la discussion en prenant 
   en compte la statistique quantique.} avec l'Eq.~(\ref{div2}).
\par
Il nous reste à calculer les composantes du tenseur $\mathcal B^{ij}$. Par 
définition de la section efficace (Eq.~(\ref{crsec1})), on a
\eq
 w(\vec p \, \vec p \, ' \rightarrow \vec p_1 \, \vec p_1 \, ') \, 
 \frac{d^3p_1}{(2\pi)^3} \, \frac{d^3 p_1'}{(2\pi)^3}  =
 v_{rel} \, d\sigma \, ,
\eq
avec 
\eq
 v_{rel} = \frac{P \cdot P'}{p p'} =
 \sqrt{ (\vec v - \vec v \, ')^2 - (\vec v \times \vec v \, ')^2} \, .
\eq
On a donc ($q_i=-q^i$)
\eq
 \mathcal B^{ij} = \mathcal B_{ij} = 
 \frac{1}{2} \int q_i q_j \, v_{rel} \, d \sigma \, .
\eq
En suivant la discussion de la Réf.~\cite{LanLif}, il est facile de 
déduire la structure générale de ce tenseur. Pour des particules de masse 
nulle, on obtient, en utilisant l'Eq.~(\ref{crsec2}),
\beq
\label{Bij2}
 \mathcal B_{ij} = \pi \alpha_S^2 \, \frac{N_c^2}{N_c^2-1} \, L \, 
 \left[ (1-\vec v \cdot \vec v \, ') \delta_{ij} + v_i v_j' + v_j v_i' \right] \, ,
\end{equation}\noindent
où 
\eq
 L = \int \frac{d\theta^2}{\theta^2} \, .
\eq
Cette intégrale sur les angles diverge logarithmiquement. La divergence 
correspondant aux petites valeurs de $\theta$ a pour origine physique 
la portée infinie de l'interaction entre les deux partons (gluons), 
dans la limite de couplage faible (échange d'un gluon de masse nulle). 
En réalité la portée de cette interaction est écrantée par la présence 
des autres partons du milieu, c'est le phénomène d'écrantage de 
Debye : le gluon échangé, de type électrique, a une masse 
effective (il est habillé par les interactions avec le milieu) 
$m_D$, la masse de Debye.
L'intégrale doit donc être limitée à la valeur minimale $\theta_{min}$,
correspondant à la valeur minimale de la quadri-impulsion échangée, égale
à la masse de Debye : 
\beq
\label{minangle}
 -t_{min} = m_D^2 \sim \underline{p}^2 \, \theta_{min}^2 \, ,
\end{equation}\noindent 
où $\underline{p}$ désigne l'ordre de grandeur de l'impulsion typique
des partons du milieu. La divergence de l'intégrale $L$ pour les 
grandes valeurs de $\theta$ est non-physique et reflète le fait que les 
formules précédentes ont été obtenues dans l'approximation des petits 
angles, qui cesse d'être valable quand $\theta \sim 1$. On a donc, à 
l'approximation logarithmique (les termes négligés dans le développement
doivent être petit devant le logarithme $L$),
\beq 
\label{logdiv0}
 L = \ln \left( \frac{1}{\theta_{min}^2} \right) \, .
\end{equation}\noindent
\par

\section*{Calcul du terme de collision}

En développant des formules (\ref{div1})-(\ref{div2}) il vient, après 
quelques intégrations par parties,
\bear
 \mathcal C & = & (\p_{ij} f) \, \int_{\vec p \, '} \, \mathcal B^{ij} F' +
 \p_i F \, \int_{\vec p \, '} \, f' \, (\p_j' \mathcal B^{ij}) \\
 && + F \, \int_{\vec p \, '} f' (\p_j' \p_i \mathcal B^{ij}) +
 (\p_j f) \, \int_{\vec p \, '} \, F' (\p_i \mathcal B^{ij}) \, ,
\eear
avec $\p_i \mathcal B^{ij} = 2 \mathcal B \, {v'}^j/p$ et 
$\p_j' \p_i \mathcal B^{ij} = 4 \mathcal B / p p'$ (on a noté 
$\mathcal B = \pi \alpha_S^2 \, \frac{N_c^2}{N_c^2-1} \, L$). 
Dans la région centrale de la collision (voir le Chap.~\ref{QGP}) la symétrie 
du problème est telle que la distribution $f(\vec p)$ est une fonction paire 
de $\vec p$. Les intégrales du type $\int d^3p \, f \, v^i$ sont donc nulles 
et l'expression ci-dessus se simplifie considérablement. En remarquant que
$\p_i v^i = \vec \nabla \cdot \vec v = 2/p$, on obtient finalement
\beq
\label{BoltzQ}
 \left. \mathcal C [f] \right|_{z=0} = \mathcal B \, M_0 \, \nabla^2 f +
 2 \mathcal B \, N_{-1} \, \vec \nabla \left[ \vec v \, f(1+f) \right] \, ,
\end{equation}\noindent
où l'on a défini les moments
\bearn
\label{momMs}
 M_s & = & 2 (N_c^2 -1) \, \int \frac{d^3p}{(2\pi)^3} \, p^s \, f(1+f) \, , \\
\label{momNs}
 N_s & = & 2 (N_c^2 -1) \, \int \frac{d^3p}{(2\pi)^3} \, p^s \, f \, .
\eearn
\par
L'Eq.~(\ref{BoltzQ}) ci-dessus se simplifie encore dans le cas 
d'un gaz non-dégénéré, c'est à dire d'un gaz classique ($f \ll 1$), 
où on a 
\beq
\label{Boltzcl}
 \left. \mathcal C [f] \right|_{z=0} = \mathcal B \, N_0 \, \nabla^2 f +
 2 \mathcal B \, N_{-1} \, \vec \nabla ( \vec v \, f ) \, .
\end{equation}\noindent
Nous utilisons cette dernière forme du terme de collision dans l'étude
présentée au Chap.~\ref{QGP}. Cette approximation est motivée par le fait 
que le système est rapidement dilué à cause de l'expansion. Les partons 
peuvent alors être considérés comme des {\em particules} classiques
(par opposition à l'approximation de {\em champ} classique du 
Chap.~\ref{QUENCH}, valable quand $f \gg 1$).

\section*{Masse de Debye et masse d'écran transverse}

On peut comprendre simplement l'origine physique du phénomène d'écrantage 
de Debye de la façon suivante : supposons que deux particules, plongées 
dans un milieu, tentent d'interagir l'une avec l'autre en échangeant 
un quanta virtuel. Celui-ci est émis par la particule $1$ et se propage
``en direction'' de la particule $2$. Il est clair que plus la distance 
à parcourir est grande, plus la probabilité pour que celui-ci soit absorbé 
par une particule du milieu avant d'arriver à destination augmente : 
la portée de l'interaction, infinie dans le vide, est limitée du fait
de la présence du milieu. Ceci se traduit par le fait
que le quanta échangé correspond à un champ effectivement massif,  
``habillé'' par ses interactions avec les particules du milieu. La masse
correspondant à l'écrantage de la partie électrique du champ correspondant
(la seule qui joue un rôle dans notre cas) est appellée masse de Debye.
Dans le cas d'un système de gluon, les auteurs de~\cite{Debyemass} obtiennent
l'expression\footnote{La dérivation de cette formule n'est pas présentée
   dans la référence citée. On peut obtenir une formule identique, au facteur
   de couleur près, dans le cas d'un plasma QED, dans la théorie de la 
   réponse linéaire. Nous pensons que la formule~(\ref{mdebye}) est obtenue 
   avec le même type d'approximations pour un système de gluons} : 
\beq
\label{mdebye}
 m_D^2 = - \frac{\alpha_S N_c}{\pi^2} \, \lim_{q \rightarrow 0} \,
 \int d^3p \, 
 \frac{\vec q \cdot \vec\nabla_p \, f(\vec p,t)}{\vec q \cdot \vec v} \, ,
\end{equation}\noindent
où $\vec v = \vec p/p$ et $q=||\vec q||$, $\vec q$ étant l'impulsion du 
gluon échangé. On voit que, de façon générale, la masse d'écran dépend
de la direction de cette dernière. Dans le cas d'un gluon d'impulsion 
transverse $\vec q \equiv (\vec q_t,q_z=0)$ et pour une distribution 
indépendante de la direction dans la direction transverse 
$f(\vec p,t) \equiv f(p_t,p_z,t)$, on obtient, après intégration par 
parties,
\beq
\label{mdebyetrans}
 \left. m_D^2 \right|_{trans} = \frac{\alpha_S N_c}{\pi^2} \, 
 \int \frac{d^3p}{p} \, f(\vec p,t)  + 
 \frac{2\alpha_S N_c}{\pi} \int_{-\infty}^{+\infty} dp_z \, |p_z| \, f(p_t=0,p_z,t) 
 \, .
\end{equation}\noindent
Dans le Chap.~\ref{QGP}, nous aurons à considérer le cas où la largeur de la
distribution dans la direction longitudinale est négligeable : 
$f(\vec p,t) \propto \delta(p_z)$. On obtient alors l'expression de 
la masse d'écran transverse
\beq
\label{mtrans}
 m_T^2 = \frac{\alpha_S N_c}{\pi^2} \, \int \frac{d^3p}{p} \, f(\vec p,t) \, .
\end{equation}\noindent
Nous aurons aussi besoin du cas où la distribution $f$ est isotrope. On a alors $\vec\nabla_p f(p,t) = \vec v \, f'(p,t)$, où $f' \equiv \p f/\p p$, et la masse 
d'écran (\ref{mdebye}) est indépendante de la direction du gluon échangé, 
comme il se doit (il est facile de voir que l'expression~(\ref{mdebyetrans}),
obtenue pour une distribution générale, se réduit à la formule ci-dessous dans
le cas d'un distribution isotrope). On obtient, après intégration par partie,
\beq
\label{miso}
 m_D^2 = \frac{2\alpha_S N_c}{\pi^2} \, \int \frac{d^3p}{p} \, f(p,t) \, .
\end{equation}\noindent
\par
Bien que l'expression (\ref{mtrans}) de la masse d'écran transverse ait été
obtenue pour le cas d'une distribution $\propto \delta(p_z)$, il nous sera
pratique de l'extrapoler au cas où la distribution est isotrope.
On a alors $m_T^2 = m_D^2/2$.

\chapter{Le calcul des moments}
\label{intm}

Dans cette annexe, nous dérivons les formules utilisées au Chap.~\ref{QGP}
pour calculer le temps de relaxation dans la méthode des moments.
Dans le cas du terme de collision de Landau (voir Annexe~\ref{landaucoll}),
il est possible de calculer explicitement toutes les intégrales sur l'espace
des impulsions pour tous les moments dont nous avons besoin au Chap.~\ref{QGP}.
\par
La solution formelle de l'équation du temps de relaxation s'écrit 
(cf. Eq.~(\ref{RTABaym}))
\beq
\label{RTABaym1}
 f(\vec p,t)=f_0(\vec p_t,p_z\frac{t}{t_0}) \, \mbox{e}^{-x(t)} +
 \int_{t_0}^t dt' \, \frac{\mbox{e}^{x(t')-x(t)}}{\theta(t')} \,
 f_{eq} (\vec p_t,p_z\frac{t}{t'},t') \, ,
\end{equation}\noindent
où
\eq
 x(t) = \int_{t_0}^t \frac{dt'}{\theta (t')} \, ,
\eq
et où $f_0(\vec p)$ est la distribution initiale et 
$f_{eq} (\vec p,t) = \lambda (t) \, \mbox{e}^{-p/T(t)}$.
\par
De façon générale, nous voulons calculer les intégrales du type 
\eq
 M (t) = \langle m \rangle = \int_{\vec p} \, m(\vec p) \, f(\vec p,t) \, ,
\eq
avec $\int_{\vec p} \equiv 2(N_c^2-1) \int d^3p/(2\pi)^3$.
Considérons par exemple le cas $M = N_s = \langle p^s \rangle$ 
(cf. Eq.~(\ref{Ns})), et concentrons nous d'abord sur le second terme du membre
de droite de l'Eq.~(\ref{RTABaym1}). On doit calculer l'intégrale
\eq
 I_s^{eq} (t,t') = \int_{\vec p} \, p^s \, 
 f_{eq} (\vec p_t,p_z\frac{t}{t'},t')) \, .
\eq
En faisant le changement de variable $p_zt/t' \rightarrow p_z$ et en utilisant 
un système de coordonnées sphériques pour $\vec p$, on obtient aisément
\eq
 I_{eq} (t,t') = \frac{t'}{t} \, N_s^{eq} (t') \,
 h_s \left( \frac{t'}{t} \right) \, ,
\eq
avec 
\eq
 N_s^{eq} (t) = \int_{\vec p} \, p^s \, f_{eq} (\vec p,t) =
 (s+2)! \, \frac{N_c^2-1}{\pi^2} \, \lambda (t) \, T^{s+3} (t) \, ,
\eq
et où la fonction
\beq
\label{function0}
 h_s (a) = \int_0^1 dx \, \left[ 1 - (1-a^2) x^2 \right]^{s/2}  \, .
\end{equation}\noindent
Pour calculer la contribution du premier terme du membre de droite de 
l'Eq.~(\ref{RTABaym1}), on doit spécifier la condition initiale. 
Dans le Chap.~\ref{QGP}, on considère deux type de conditions initiales :
le scénario des minijets, où $f_0 (\vec p) = f_{eq} (\vec p,t)$, et le scénario
de saturation, où $f_0 (\vec p) = \delta (p_z) \, g(p_\perp)$.
Dans la suite, nous définissons la notation
\eq
 \left\{ A || B \right\} =
 \left\{
 \begin{array}{rl}
 A & \mbox{si } f_0 (\vec p) = f_{eq} (\vec p,t_0) \\
 B & \mbox{si } f_0 (\vec p) = \delta (p_z) g(p_\perp) 
 \end{array}
 \right.
\eq
On obtient finalement
\beq
\label{moment0}
 t \, \mbox{e}^{x} \, N_s (t) = t_0 \, N_s (t_0) \, 
 \left\{ h_s (t_0/t) || 1 \right\} + 
 \int_{t_0}^t \frac{dt'}{\theta'} \,
 t' \, \mbox{e}^{x'} \, N_s^{eq} (t') \, h_s (t'/t) \, ,
\end{equation}\noindent
où $x=x(t)$, $x'=x(t')$, $\theta' = \theta (t')$.
\par
On calcule de manière analogue les moments ``longitudinaux'' 
$N_s^z =\langle p_z^2 \, p^{s-2} \rangle$ (par exemple, la pression
longitudinale $P_L = \langle p_z^2/p \rangle = N_1^z$) :
\beq
\label{momentz}
 t^3 \, \mbox{e}^{x} \, N_s^z (t) = t_0^3 \, N_s (t_0) \, 
 \left\{ h_s^z (t_0/t) || 0 \right\} + 
 \int_{t_0}^t \frac{dt'}{\theta'} \,
 (t')^3 \, \mbox{e}^{x'} \, N_s^{eq} (t') \, h_s^z (t'/t) \, ,
\end{equation}\noindent
avec 
\beq
\label{functionz}
 h_s^z (a) = \int_0^1 dx \, x^2 \, \left[ 1 - (1-a^2) x^2 \right]^{s/2-1}  \, .
\end{equation}\noindent
Enfin, les moments ``transverse'' $N_s^\perp =\langle p_\perp^2 \, p^{s-2} \rangle$
sont aisément obtenus à partir des précédents : 
\eq
 N_s = N_s^z + 2 \, N_s^\perp \, .
\eq
\par
Nous donnons ci-dessous les expressions des fonctions $h_s$ et $h_s^z$ pour
$s=-2,...,2$. Les intégrales (\ref{function0}) et (\ref{functionz}) 
se calculent aisément. On peut aussi utiliser la relation suivante :
\beq
\label{func0z}
 h_s^z (a) = \frac{2}{s} \, \frac{\dd}{\dd a^2} h_s (a) \, .
\end{equation}\noindent

\vspace{0.5cm}
En notant $A=\sqrt{1-a^2}$,
\begin{center}
\begin{tabular}{|llr|llr|}
\hline
 &&&&& \\
 & $\displaystyle h_s (a) = 
        \int_0^1 dx \, \left[ 1 - (1-a^2) x^2 \right]^{s/2}$ & &
 & $\displaystyle h_s^z (a) = 
        \int_0^1 dx \, x^2 \, \left[ 1 - (1-a^2) x^2 \right]^{s/2-1}$ & \\
 &&&&& \\
 \hline
 &&&&& \\
 & $\displaystyle h_2 (a) = \frac{2+a^2}{3}$ & &
 & $\displaystyle h_2^z (a) = \frac{1}{3}$ & \\
 &&&&& \\
 & $\displaystyle h_1 (a) = 
       \frac{1}{2} \left( a + \frac{\arcsin A}{A} \right)$ & &
 & $\displaystyle h_1^z (a) = 
       \frac{1}{2A^2} \, \left( \frac{\arcsin A}{A} -a \right) $ & \\
 &&&&& \\
 & $\displaystyle  h_0 (a) = 1$ & &
 & $\displaystyle  h_0^z (a) = 
       \frac{1}{A^2} \, \left( \frac{\arctan A}{A} - 1 \right)$ & \\
 &&&&& \\
 & $\displaystyle  h_{-1} (a) = \frac{\arcsin A}{A}$ & &
 & $\displaystyle  h_{-1}^z (a) = 
       \frac{1}{A^2} \, \left( \frac{1}{a} - \frac{\arcsin A}{A} \right)$ & \\
 &&&&& \\
 & $\displaystyle  h_{-2} (a) = \frac{\arctan A}{A}$ & &
 & $\displaystyle  h_{-2}^z (a) = 
       \frac{1}{2A^2} \, \left( \frac{1}{a^2} - \frac{\arctan A}{A} \right)$ & \\
 &&&&& \\
 \hline
\end{tabular} 
\end{center}

\par
Dans le Chap.~\ref{QGP}, nous avons résumé les Eqs.~(\ref{moment0}) et
(\ref{momentz}) ci-dessus sous la forme (cf. Eq.~(\ref{MBaym})
\eq
 M (t) =  M(t_0) \, \mathcal F_M^{(0)} (t_0/t) \, \mbox{e}^{-x} + 
 \int_{t_0}^t dt' \, \frac{\mbox{e}^{x'-x}}{\theta'} \,
 \mathcal F_M^{(eq)} (t'/t) \, M_{eq} (t') \, ,
\eq
où $M$ désigne un moment quelconque. On a
\bear
 \mathcal F_{N_s}^{(0)} (a) & = & a \, \left\{ h_s (a) || 1 \right\}
 \, \, \, , \, \, \, 
 \mathcal F_{N_s}^{(eq)} (a) = a \, h_s (a) \, \\
 \mathcal F_{N_s^z}^{(0)} (a) & = & a^3 \, \left\{ h_s^z (a) || 1 \right\}
 \, \, \, , \, \, \, 
 \mathcal F_{N_s^z}^{(eq)} (a) = a^3 \, h_s^z (a) \, .
\eear

\chapter{Calcul de $\theta (t_0)$ dans le scénario des minijets}
\label{thetajet}

La méthode des moments, exposée au Chap.~\ref{QGP}, ne permet
pas de calculer le temps de relaxation dans le cas où $f=f_{eq}$,
comme c'est le cas dans le scénario des minijets, à l'instant inital. 
Pour calculer $\theta (t_0)$, il suffit cependant de s'écarter
de $f_{eq}$ de façon infinitésimale en expoitant le fait que dans 
les tout premiers instants, le système ne ressent pas encore l'effet 
des collisions et est dans un régime quasi-libre. Nous détaillons ce 
calcul dans cette annexe.
\par
La distribution initiale est donnée par 
($\lambda_0=\lambda_{jet}$ et $T_0=T_{jet}$)
\eq
 f (\vec p,t_0) = \lambda_0 \, \mbox{e}^{-p/T_0} \, .
\eq
Le temps de relaxation étant non-nul à l'instant initial, il
est clair que pour les temps $t$ tels que $t-t_0 \ll \theta(t_0)$, 
le système est dans un régime essentiellement libre. L'évolution 
est alors donnée par l'Eq.~(\ref{streaming}), et on a 
\eq
 f \left(\vec p, t_0 (1 + \varepsilon) \right) \simeq 
 f(\vec p,t_0) \, \left(1 - \frac{\varepsilon}{T_{jet}} \, 
 \frac{p_z^2}{p} \right) \, .
\eq
On en déduit, pour les moments $N_s$, $N_s^z$ et $N_s^\perp$, définis
dans l'annexe~\ref{intm},
\eq
 N_s(t_0+\delta t) = N_s (t_0) - \frac{\varepsilon}{3T_0} \, N_{s+1} (t_0) \, ,
\eq
et
\eq
 N_s^z(t_0+\delta t) - N_s^\perp(t_0+\delta t) = 
 - \frac{\varepsilon}{15T_0} \, N_{s+1} (t_0) \, ,\\
\eq
où l'on a utilisé le fait que la distribution est isotrope à $t_0$.
Avec ces notations, l'Eq.~(\ref{momeq1}) s'écrit :
\eq
  \frac{N_1^z - N_1^\perp}{\theta} = 
  4 \, \mathcal B \, N_0 \, \left( N_{-1}^z - N_{-1}^\perp \right) +
  2 \, \mathcal B \, N_{-1} \, \left( N_0^z - N_0^\perp \right) \, ,
\eq
où 
\eq
 \mathcal B = \pi \alpha_S^2 \frac{N_c^2}{N_c^2-1} \, L \,
\eq
et $L$ est donné par les Eqs.~(\ref{debye})-(\ref{logdiv2}).
En se plaçant à l'instant $t=t_0(1+\varepsilon)$, et en utilisant les
expressions ci-dessus, il vient, au premier ordre non-trivial en $\epsilon$
(on écrit $\theta \left(t_0(1+\varepsilon)\right) = \theta (t_0) + 
\varepsilon \dot\theta (t_0)$),
\eq
 \frac{N_2}{\theta_0} = 
  4 \, \mathcal B \, ( N_0 )^2 +
  2 \, \mathcal B \, N_{-1} \, N_1 \, ,
\eq
où toutes les quantités sont calculées à $t_0$ ($\theta_0 = \theta (t_0)$).
En utilisant 
\eq
 N_s (t_0) = (s+2)! \, \frac{N_c^2-1}{\pi^2} \, \lambda_0 \, T_0^{s+3} \, .
\eq
On obtient finalement
\eq
 \frac{1}{\theta_0} = \frac{7(\alpha_S N_c)^2}{6 \pi} \, 
 \ln \left( \frac{7 \pi}{4 \alpha_S N_c \lambda_0} \right) \, \lambda_0 \, T_0 \, .
\eq
Avec les valeurs des paramètres $\alpha_S =0.3$, $N_c=3$ et $\lambda_0 = 1$, 
correspondant au scénario des minijets (voir le Chap.~\ref{QGP}), on obtient
\eq
 \theta_0 \simeq \frac{2}{T_0} \, ,
\eq
c'est à dire $\theta_0 \simeq 0.7$~fm à RHIC, et $\theta_0 \simeq 0.3$~fm à LHC.

}

\chapter*{Conclusion}
\addcontentsline{toc}{chapter}{Conclusion}
\pagestyle{plain}

\vspace{0.5cm}

Dans la première partie de cette thèse, nous avons étudié la
possibilité qu'un condensat chiral désorienté soit formé lors
du passage rapide de la transition de phase chirale dans une
collision d'ions lourds. Nous avons proposé une méthode
originale d'échantillonnage des conditions initiales pour
le champ de pion en équilibre thermique local dans une petite
bulle formée lors de la collision. Dans les travaux antécédents,
ces conditions initiales sont habituellement choisies arbitrairement. 
Nous avons ainsi pu calculer une toute première estimation de la 
probabilité de formation d'une configuration classique du champ. 
En fait nous en obtenons une limite supérieure, qui s'avère relativement
faible, typiquement de l'ordre de $10^{-3}$. Ce résultat a des 
conséquences importantes aussi bien expérimentales que théoriques.
En particulier une stratégie de détection évènement-par-évènement 
est préférable.
\par
Nous avons ensuite
étudié la structure d'isospin de la configuration 
du champ produite par trempage des fluctuations initiales, 
et avons montré que, contrairement à ce qui a été cru jusqu'à 
maintenant, celle-ci n'est pas une configuration DCC : les directions 
d'oscillation des différents modes dans l'espace d'isospin sont 
statistiquement indépendantes les unes des autres. Le modèle 
le plus simple utilisé jusqu'à présent permet d'expliquer 
la nature classique du DCC, pas sa polarisation collective 
hypothétique. Nous avons montré que la dynamique ne génère
pas les corrélations cherchées entre les modes, une description 
plus réaliste de l'état initial pour le champ de pion est 
nécessaire pour trancher la question de la possibilité de 
formation d'un DCC dans le scénario du trempage.

\vspace{0.5cm}

Dans la deuxième partie de la thèse, nous avons étudié la
question de la thermalisation des gluons produits dans les 
tout premiers instants de la collision. Nous avons considéré
les collisions élastiques de petite déviation et modélisé
l'équation de Boltzmann correspondante par une approximation
de temps de relaxation où le temps de relaxation est calculé
de façon auto-cohérente. Nos résultats sont en accord 
semi-quantitatif avec la solution exacte récemment obtenue pour une
condition initiale donnée.
Nous avons comparé différents scénarios proposés dans la 
littérature pour décrire l'état initial. Nous arguons que les 
critères utilisés dans les travaux précédents pour caractériser
l'écart à l'équilibre local n'est pas satisfaisant. En mesurant
plutôt le degré d'anisotropie de différentes observables, nous 
arrivons à des conclusions qui contredisent celles précédemment 
obtenues. En particulier, nous montrons que les collisions
élastiques ne sont pas suffisantes pour thermaliser le système
aux énergies de RHIC. A LHC, l'équilibre est atteint dans le
scénario de saturation, les incertitudes de l'approche, en 
particulier celles liées à la fragilité du calcul perturbatif,
ne permettent pas de conclure dans le scénario des minijets.
\par
Il est nécessaire d'aller plus loin et d'inclure les 
contributions des processus inélastiques de branchement.
Cependant, notre étude indique qu'il faut d'ores et déjà 
s'attendre à ce que le régime transitoire ne soit pas complètement 
négligeable à RHIC. Nous prévoyons d'inclure les quarks dans 
la description afin d'estimer l'influence de ces effets 
hors d'équilibre sur les spectres en énergie transverse 
des dileptons produits, qui seront étudiés par les 
collaborations PHENIX à RHIC et ALICE à LHC.

\renewcommand{\bibname}{Références}
\bibliographystyle{unsrt}
\bibliography{BIBLI_DCC_0,BIBLI_DCC_1,BIBLI_QGP}
\addcontentsline{toc}{chapter}{Références}
\pagestyle{fancy}

\end{document}